\documentclass[a4paper,11pt]{article}

\usepackage{jheppub}
\usepackage{amsmath}
\numberwithin{equation}{section}
\usepackage{graphicx}
\usepackage{xcolor}
\usepackage{soul}
\usepackage{cancel}
\usepackage{braket}
\usepackage{multicol}
\usepackage{subfig}
\usepackage{bm}
\usepackage{empheq}
\usepackage{booktabs}

\newcommand{\redeqn}{equation}
\newcommand{\dbar}{d\hspace*{-0.08em}\bar{}\hspace*{0.1em}}
\newcommand{\al}{\alpha}
\newcommand{\nn}{\nonumber}

\newcommand{\mm}{\mathcal{M}}
\newcommand{\eps}{\epsilon}
\newcommand{\tts}{\mathbf{T}_t^2}
\newcommand{\tsu}{\mathbf{T}_{s-u}^2}

\newcommand{\T}{{\bf T}}
\newcommand{\beq}{\begin{\redeqn}}
\newcommand{\eeq}{\end{\redeqn}}
\newcommand{\beqa}{\begin{eqnarray}}
\newcommand{\eeqa}{\end{eqnarray}}

\newcommand{\Ts}{{\bf T}_s^2}
\newcommand{\Tu}{{\bf T}_u^2}

\newcommand{\CA}{C_A}
\newcommand{\nf}{n_f}
\newcommand{\LO}{\text{LO}}

\newcommand{\ratio}[1]{ \left(\frac{p^2}{p_{#1}^2}\right)^{\!\eps}}

\newcommand{\mTree}{\hat{\mathcal{M}}_{\text{tree}}}
\newcommand{\mtree}{\mathcal{M}_{\text{tree}}}
\newcommand{\ag}[1]{\hat{\alpha}_g^{(#1)}}
\newcommand{\kusp}[1]{K^{(#1)}}

\newcommand{\mExpP}[2]{\hat{\mathcal{M}}^{(+,#1,#2)}}
\newcommand{\mExpM}[2]{\hat{\mathcal{M}}^{(-,#1,#2)}}
\newcommand{\dd}{\mathbf{\Delta}}

\newcommand{\ord}{\mathcal{O}}

\newcommand{\as}{\alpha_s}

\newcommand{\MM}{\mathcal{M}}
\newcommand{\M}{\mathcal{M}}

\newcommand{\MMtree}{\mathcal M^{\text{tree}}}
\newcommand{\Mreduced}{\hat{\mathcal{M}}}

\allowdisplaybreaks
\title{Scattering amplitudes in the Regge limit and the soft anomalous dimension through four loops}

\author[a]{Giulio Falcioni,}
\author[a]{Einan Gardi,}
\author[a]{Niamh Maher,}
\author[b]{Calum Milloy,}
\author[b,c]{Leonardo Vernazza}

\affiliation[a]{Higgs Centre for Theoretical Physics, School of Physics and Astronomy,\\ The University of Edinburgh, Edinburgh EH9 3FD, Scotland, UK}
\affiliation[b]{Dipartimento di Fisica and Arnold-Regge Center, Universit\`{a} di Torino,\\
  and INFN, Sezione di Torino, Via P. Giuria 1, I-10125 Torino, Italy}
\affiliation[c]{Theoretical Physics Department, CERN, Geneva 1211, Switzerland}

\emailAdd{Giulio.Falcioni@ed.ac.uk}
\emailAdd{Einan.Gardi@ed.ac.uk}
\emailAdd{N.Maher@sms.ed.ac.uk}
\emailAdd{Calum.Milloy@to.infn.it}
\emailAdd{Leonardo.Vernazza@to.infn.it}

\abstract{
    Using rapidity evolution equations we study two-to-two gauge-theory scattering amplitudes in the Regge limit. We carry out explicit computations at next-to-next-to-leading logarithmic accuracy through four loops and present new results for both infrared-singular and finite contributions to the amplitude.  New techniques are devised in order to derive the colour structure stemming from three-Reggeon exchange diagrams in terms of commutators of channel operators, obtaining results that are valid for any gauge group, and apply to scattered particles in any colour representation.
    We also elucidate the separation between contributions to the Regge cut and Regge pole in the real part of the amplitude to all loop orders. We show that planar contributions due to multiple-Reggeon exchange diagrams can be factorised as a Regge pole along with the single-Reggeon exchange, and when this is done, the singular part of the gluon Regge trajectory is directly determined by the cusp anomalous dimension. We explicitly compute the Regge cut component of the amplitude through four loops and show that it is non-planar.
    From a different perspective, the new results provide important information on soft singularities in general kinematics beyond the planar limit: by comparing the computed corrections to the general form of the four-loop soft anomalous dimension we derive powerful constraints on its kinematic dependence, opening the way for a bootstrap-based determination.    
}

\date{June 2021}

\begin{document}

\begin{flushright}
CERN-TH-2021-200 
\vspace*{-25pt}
\end{flushright}

\maketitle

\section{Introduction}
The high-energy limit of scattering amplitudes has been a fascinating avenue in exploring the strong interactions already before the discovery of QCD, see e.g.~\cite{Eden:1966dnq,Veneziano:1968yb,Collins:1977jy,Mandelstam:1963cw,Gribov:1961fr,Gribov:1967vfb}. 
The behaviour of amplitudes at high energy was described in terms of fundamental objects in Regge theory associated with specific exchanges in the scattering process, which are characterised by their singularities in the complex angular momentum plane~\cite{Gribov:1961fr,Gribov:1967vfb,Collins:1977jy}, namely Regge cuts and Regge poles.
With the development of perturbative QCD it became clear that this limit is also an extraordinary laboratory for investigating the behaviour of gauge interactions to high perturbative orders, and resumming them.

The discovery of gluon Reggeization in non-abelian gauge theories~\cite{Grisaru:1973wbb,Grisaru:1973vw,Grisaru:1973ku,Lipatov:1976zz}, made it possible to predict the high-energy behaviour of amplitudes to all orders in the coupling in the BFKL approach~\cite{Fadin:1975cb,Kuraev:1976ge,Kuraev:1977fs,Balitsky:1978ic} and provided a direct link between concepts of Regge theory and QCD. In particular, focusing on $2\to 2$ scattering, perturbative corrections are dominated at high energies by large logarithms of the ratio between the centre-of-mass energy $s$ and the momentum transfer $-t$. Each perturbative loop order comes with an additional power of this logarithm, hence the expansion must be resummed to all orders in the strong coupling.   

At the leading logarithmic (LL) approximation this resummation amounts to a simple exponentiation. This is best understood as a consequence of factorisation of the amplitude in rapidity. 
Factorisation translates into evolution equations, such as the above mentioned BFKL equation \cite{Fadin:1975cb,Kuraev:1976ge,Kuraev:1977fs,Balitsky:1978ic} and its non-linear generalisation, the Balitsky-JIMWLK 
equation~\cite{Balitsky:1995ub,Kovchegov:1999yj,JalilianMarian:1996xn,JalilianMarian:1997gr,Iancu:2001ad}, whose solutions resum the dependence on $s/(-t)$. 
BFKL theory can also be seen as describing the dynamics of effective degrees of freedom, the Reggeised gluons, or Reggeons, which propagate in the two-dimensional transverse space. The aforementioned simplicity of the LL approximation can thus be seen as due to the propagation of a single Reggeon in the $t$-channel. 

Beyond the LLs, the solutions of rapidity evolution equations develop a rich structure, which features Regge cuts, in addition to the Regge pole. These Regge cuts can be understood as due to the exchange of multiple Reggeons.
In the Next-to-Leading Logarithmic (NLL) approximation of the amplitude, the pole and the cut contributions are well understood. The former determines the real part of the amplitude in terms of a single Reggeon exchange, similarly to the LLs~\cite{Fadin:2006bj,Ioffe:2010zz,Fadin:2015zea}. In turn, the Regge cut is associated with the exchange of a pair of Reggeons, which enters the imaginary part of the amplitude. This tower of corrections has been recently computed by solving the BFKL equation iteratively through 13 loops \cite{Caron-Huot:2013fea,Caron-Huot:2017zfo,Gardi:2019pmk,Caron-Huot:2020grv}. This example makes it clear that the special features of the Regge limit allow to uncover structures of the perturbative expansion of gauge theories, far beyond the reach of fixed-order calculations in general kinematics. Furthermore, it holds the promise of resumming the expansion.

In this work we focus on the tower of Next-to-Next-to-Leading Logarithms (NNLL) in the real part of the amplitude. Here, contributions of both a single-Reggeon exchange 
and a triple-Reggeon exchange become important and their interplay generates both a Regge pole and a Regge cut. These effects have been studied using different approaches \cite{Bartels:1980pe,Kwiecinski:1980wb,DelDuca:2001gu,DelDuca:2013ara,DelDuca:2013dsa,DelDuca:2014cya,Fadin:2016wso,Fadin:2017nka,Caron-Huot:2017fxr,Falcioni:2020lvv,Fadin:2021csi}, but the all-order expression for the NNLL amplitude is not yet known. 
While the Balitsky-JIMWLK 
equation~\cite{Balitsky:1995ub,Kovchegov:1999yj,JalilianMarian:1996xn,JalilianMarian:1997gr,Iancu:2001ad} has been shown to apply 
\cite{Caron-Huot:2013fea,Caron-Huot:2017fxr,Falcioni:2020lvv}
also in this case, its exact solution is beyond the reach of present methods.
Significant simplification may be gained by considering the planar limit, where the exchange of any number of Reggeons can be solved exactly~\cite{Lipatov:1993yb,Faddeev:1994zg,Derkachov:2001yn,Derkachov:2002wz}. Indeed, it is long known that Regge cuts arise from non-planar diagrams~\cite{Mandelstam:1963cw,Eden:1966dnq,Collins:1977jy} and therefore their contribution is expected to disappear in the planar limit, drastically simplifying the amplitude, which can then be expressed as a Regge pole. We are interested in the structure of the amplitude in general colour, and we follow the approach of refs.~\cite{Caron-Huot:2013fea,Caron-Huot:2017fxr,Caron-Huot:2017zfo,Gardi:2019pmk,Caron-Huot:2020grv,Falcioni:2020lvv} developing an iterative, order-by-order solution of the Balitsky-JIMWLK equations. Our focus here is on the NNLL tower through four loops.

Our motivation is twofold. On the one hand we wish to explore the structure of the perturbative high-energy amplitude itself, notably the interplay of Regge pole and Regge cut contributions at NNLL accuracy. 
On the other hand, we use the high-energy limit as a tool to investigate the long-distance singularity structure of non-planar $2\to 2$ amplitudes in general kinematics in QCD and in ${\cal N}=4$ super Yang-Mills (SYM), beyond the accuracy of state-of-the-art fixed-order calculations, see e.g.~\cite{Caola:2021rqz}.

In studying the high-energy limit we elucidate the separation between contributions to the Regge cut and Regge pole in the real part of the amplitude to all loop orders. We explain some of the surprising findings of ref.~\cite{Caron-Huot:2017fxr} where the ``cut'' contribution was identified directly as the component arising for multiple-Reggeon exchange diagrams. 
Here we show that planar contributions due to multiple-Reggeon exchange diagrams can be factorised as a Regge pole along with the single-Reggeon exchange. Moreover, we find that when this is done, the singular part of the gluon Regge trajectory is directly determined by the cusp anomalous dimension, in line with the prediction by Korchemskaya and Korchemsky~\cite{Korchemskaya:1994qp,Korchemskaya:1996je}. We explicitly compute the Regge cut component of the amplitude through four loops and show that it is non-planar.

A central part of this paper constitutes the study of infrared singularities of gauge-theory amplitudes, using information  from the Regge limit, continuing studies along similar lines at lower orders in  refs.~\cite{DelDuca:2011ae,DelDuca:2011wkl,Caron-Huot:2013fea,Caron-Huot:2017fxr,Caron-Huot:2017zfo,Falcioni:2020lvv}.
As is well known, infrared singularities are generated by a universal soft anomalous dimension \cite{Catani:1998bh,Sterman:2002qn,Aybat:2006mz,Aybat:2006wq,Gardi:2009qi,Gardi:2009zv,Becher:2009cu,Becher:2019avh,Almelid:2015jia}, currently known to three 
loops~\cite{Almelid:2015jia}. We contrast the results of our explicit computations in the Regge limit through four loops, with predictions based on the factorisation and  exponentiation of infrared singularities. In this way, we extract information regarding the soft anomalous dimension at four loops as well as on the infrared-finite parts of the four-loop amplitude in the high-energy limit.
Furthermore, we use these results to constrain a general parametrisation of the four-loop soft anomalous dimension for multi-leg massless scattering~\cite{Becher:2019avh}, extracting explicitly the high-energy limit of the relevant kinematic functions. This information opens the way for bootstrapping these functions in general kinematics, similarly to what was done at three loops in ref.~\cite{Almelid:2017qju}.

The outline of the paper is as follows: in section~\ref{sec:formalism} we 
introduce the theoretical background regarding the high-energy limit of $2\to2$ scattering including the treatment of colour, infrared divergences and the separation between the Regge pole and Regge cut components of the amplitude. Finally, we also  
define the so-called \emph{reduced amplitude}, which is most directly accessible within our approach.
Next, in section~\ref{sec:JIMWALK_to_NNLL} we first review the formalism we use to compute the Regge limit of $2\to2$ amplitudes based on an iterative solution of the Balitsky-JIMWLK equation~\cite{Caron-Huot:2013fea,Caron-Huot:2017fxr} and then we characterise the entire tower of NNLL of the reduced amplitude to all orders. Section~\ref{sec:computation} describes the techniques we use to carry out the calculation, in particular to evaluate the colour structure for general representation of the scattering partons. The complete results are given in section~\ref{sec:results}, where we also investigate the separation between the Regge pole and Regge cut contributions at the NNLL accuracy through four loops. 
In section~\ref{sec:Infrared} we contrast our results with infrared factorisation, which allows us to extract the NNLL contribution to the four-loop soft anomalous dimension and the finite part of QCD and $\mathcal{N}=4$ SYM amplitudes. 
In section~\ref{sec:SAD_Regge} we investigate the parametrisation of the soft anomalous 
dimension~\cite{Becher:2019avh} and derive the high-energy limit of the kinematic functions it contains.

\section{The Regge limit of \texorpdfstring{$2 \to 2$}{2->2} scattering: introduction }
\label{sec:formalism}
\subsection{Partonic scattering in the high-energy limit}
\label{subsec:Regge2to2}

\begin{figure}[htb]
\centering
\includegraphics[scale=0.7]{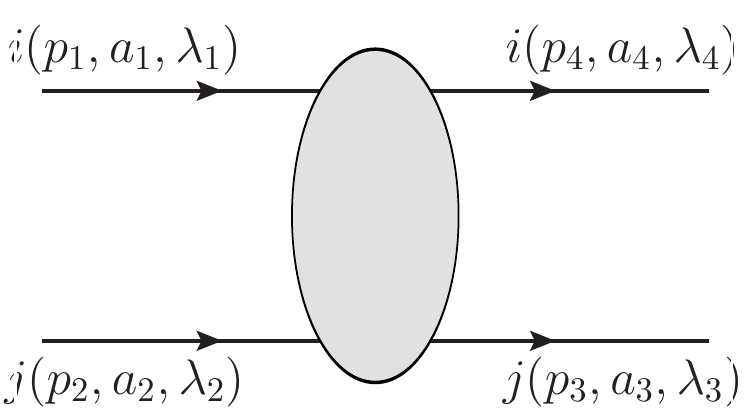}
\caption{2 to 2 particle scattering corresponding to eq.~\eqref{2to2}. The two partons~$i$ and~$j$ can be quarks or gluons, and the arrows indicate the direction of the momenta. The indices~$a_k$ are colour indices, while the labels $\lambda_k$ represent the particle helicities. At leading power in the high-energy limit the amplitude is helicity conserving, as indicated at tree level in eq.~\eqref{eq:treelevel}. }
\label{fig:2to2d}
\end{figure}

Let us consider two-parton scattering 
\beqa\label{2to2}
i(p_1, a_1,\lambda_1) + j(p_2,a_2,\lambda_2) \to j(p_3,a_3,\lambda_3) + i(p_4,a_4,\lambda_4),
\eeqa
where the partons $i$, $j$ can  each be quarks or gluons. Labelling the momenta of the particles as in figure \ref{fig:2to2d}, with $p_1$, $p_2$ incoming, and $p_3$, $p_4$ outgoing, the process is described in terms of the Mandelstam variables
\beq
s=(p_1+p_2)^2>0 \,,\hspace{1cm}\,t=(p_1-p_4)^2<0\,,\hspace{1cm}\,u=(p_1-p_3)^2<0.
\eeq
The high-energy limit is defined by the condition 
\beq
s \gg -t,
\eeq
i.e., the centre-of-mass energy $s$ becomes much larger than the momentum transfer $-t$. 

In this regime, the amplitude can be seen as an expansion in the small parameter $(-t)/s$. We focus in this paper on the leading power term in this expansion, which is dominated by a $t$-channel exchange, and is therefore proportional to $[(-t)/s]^{-1}$.
This leading-power scattering amplitude is further restricted to be helicity-conserving, namely, the partons $i$ and $j$ retain their helicities through the scattering process.

The amplitude develops large logarithms of the ratio $s/(-t)$ and our general aim is to determine these log-enhanced terms. The leading logarithmic (LL) contribution is summed to all orders~\cite{Lipatov:1976zz,Kuraev:1976ge} by dressing the tree-level $t$-channel gluon exchange by the exponential factor
\begin{equation}
\label{LLresum}
\mathcal{M}^{\rm{LL}}_{ij\to ij} = \left(\frac{s}{-t}\right)^{C_A\,\alpha_g(t,\mu^2)}\,\mathcal{M}_{ij\to ij}^{\rm tree},
\end{equation}
where the tree-level amplitude reads
\beq
\label{eq:treelevel}
{\cal M}^{\rm tree}_{ij\to ij}= g_s^2\, \frac{2s}{t} \, (T_i^b)_{a_1 a_4}(T_j^b)_{a_2 a_3} \, 
\delta_{\lambda_1 \lambda_4} \delta_{\lambda_2 \lambda_3},
\eeq
where the factor $\delta_{\lambda_1 \lambda_4} \delta_{\lambda_2 \lambda_3}$ represents helicity conservation, and where the colour matrices depend on the representations of the projectile ($i$) and target ($j$) according to
\hbox{$(T_i^b)_{a_1a_4} = t^b_{a_1 a_4}$} for quarks, \hbox{$(T_i^b)_{a_1a_4} = - t^b_{a_4 a_1}$} for antiquarks, and \hbox{$(T_i^b)_{a_1a_4} = i f^{a_1 b a_4}$} for gluons. 
The exponent $\alpha_g(t,\mu^2)$ in eq.~(\ref{LLresum}) is the famous \emph{gluon Regge trajectory}, whose perturbative expansion reads 
\beq
\label{al_g}
\al_g (t,\mu^2)
= \sum_n \bigg(\frac{\al_s(\mu^2)}{\pi}\bigg)^n \al_g^{(n)}(t,\mu^2).
\eeq
The one-loop coefficient in dimensional regularisation, with $D = 4-2\eps$,
is given by 
\beq
\label{r_gamma}
\al_g^{(1)}(t,\mu^2) = \frac{r_\Gamma}{2\eps}\left(\frac{\mu^2}{-t}\right)^{\epsilon}\,;
\qquad\qquad 
r_\Gamma = e^{\eps\gamma_E}
\frac{\Gamma^2(1-\eps)\Gamma(1+\eps)}{\Gamma(1-2\eps)}.
\eeq
In the following we will often evaluate amplitudes at the renormalisation scale $\mu^2 = -t$. In this case we will omit the scale argument: $\al_g (t) \equiv \al_g (t,-t)$.

The simple exponentiation property in eq.~\eqref{LLresum}, with the characteristic colour charge $C_A$, the quadratic Casimir in the adjoint representation, can be understood as due to the exchange of a single \emph{Reggeized gluon} (dubbed \emph{Reggeon}), which admits a trivial evolution equation in rapidity.\footnote{The concept of rapidity evolution will be important in what follows, and we shall return to it in section~\ref{BFKL_abridged}. We refer the reader to refs.~\cite{Caron-Huot:2013fea,Caron-Huot:2017fxr} for a more complete presentation of the formalism.} The exchange of a single Reggeon can be related to so-called \emph{Regge poles} in the complex angular momentum plane.

As we discuss in section~\ref{sec:subleading_logs} below, beyond the LL approximation, the amplitude develops a more complex analytic structure, which can be associated to the emergence of \emph{Regge cuts} in the complex angular momentum plane~\cite{Mandelstam:1963cw,Eden:1966dnq,Collins:1977jy}. These in turn are related to the exchange of compound states of multiple Reggeons. 
The Balitsky, Fadin, Kuraev and Lipatov (BFKL) equation \cite{Fadin:1975cb,Kuraev:1976ge,Kuraev:1977fs,Balitsky:1978ic} describes the evolution of 
two-Reggeon states. 
Different approaches have been developed to go beyond these cases, e.g. in terms of the Bartels, Kwiecinski  and Praszalowicz (BKP) equation~\cite{Bartels:1980pe,Kwiecinski:1980wb}. 
The evolution of states consisting of any number of Reggeons, as well as mixing between such states, can be described in full generality in terms of the non-linear Balitsky-JIMWLK equation~\cite{Caron-Huot:2013fea,Caron-Huot:2017fxr}. 
In the case of a two-Reggeon exchange, this equation reduces to the BFKL equation, which has been solved in a variety of circumstances, see e.g.~\cite{Lipatov:1985uk,Caron-Huot:2013fea,Caron-Huot:2017zfo,Caron-Huot:2020grv}.  
Going beyond two-Reggeon exchange, 
an exact solution can be found in the planar limit, where the BKP equation becomes integrable~\cite{Lipatov:1993yb,Faddeev:1994zg,Janik:1998xj,Wosiek:1996bf,Derkachov:2001yn,Derkachov:2002wz}.
The non-linear Balitsky-JIMWLK equation is much harder to solve, but restricting it to a given logarithmic accuracy -- and thus restricting the number of Reggeons being exchanged in the $t$ channel -- an iterative solution can be obtained even in the general non-planar, off-forward case~\cite{Caron-Huot:2017fxr,Falcioni:2020lvv}.
We shall review the essentials of this approach in section~\ref{BFKL_abridged} and then use it for the calculation in subsequent sections.     

In the high-energy limit $u \simeq -s$, the amplitude acquires an additional symmetry under the exchange $s \leftrightarrow u$, which is known as the \emph{signature} symmetry. It is thus convenient to split the amplitude into its even and odd components under $s\leftrightarrow u$: 
\begin{equation}
\label{eq:Msig}
 {\cal M}^{(\pm)}(s,t) = \tfrac12\Big( {\cal M}(s,t) \pm {\cal M}(-s-t,t) \Big), 
\end{equation}
where ${\cal M}^{(+)}$ and ${\cal M}^{(-)}$
are referred to, respectively, as the \emph{even}
and \emph{odd} amplitudes. As shown in ref.~\cite{Caron-Huot:2017fxr},
upon using the signature-even combination of logarithms,
\beq
\label{eq:siglog}
L\equiv \log\left(\frac{s}{-t}\right)-\frac{i\pi}{2}=\frac12\left[\log\left(\frac{-s- i0}{-t}\right)+\log\left(\frac{-u-i0 }{-t}\right)\right],
\eeq
and expanding the amplitudes ${\cal M}_{\pm}$ according to 
\begin{equation}\label{eq:expansionDef}
{\mm}^{(\pm)}_{ij\to ij} = \sum_{n=0}^\infty \left(\frac{\al_s}{\pi}\right)^n \sum_{m=0}^nL^m {\mm}^{(\pm,n,m)}_{ij\to ij},
\end{equation}
with ${\mm}^{(-,0,0)}_{ij\to ij}\equiv {\cal M}_{ij\to ij}^{\rm tree}$, the odd amplitude coefficients $\mm^{(-,n,m)}_{ij\to ij}$ are purely \emph{real}, while the even ones $\mm^{(+,n,m)}_{ij\to ij}$ are purely \emph{imaginary}. These two components also have independent, and rather different factorisation properties in the high-energy limit, which we shall review in section~\ref{sec:subleading_logs}. Naturally, they also feature distinct colour components, as we now discuss. 

\subsection{Colour structure of \texorpdfstring{$2\to 2$}{2->2} scattering}
\label{sec:colour_decomposition}

In order to express the dependence of the scattering amplitude on the colour degrees of freedom of the external particles, we write it as a vector in colour-flow space: 
\beq\label{AmpColor}
{\cal M}_{ij\to ij} =\sum_k \, c_{ij}^{[k]} \, {\cal M}_{ij\to ij}^{[k]},
\eeq
where the tensors $c_{ij}^{[k]}$ represent the elements of a colour basis, and ${\cal M}_{ij\to ij}^{[k]}$ are the corresponding amplitude coefficients. A convenient basis in the high-energy limit is given by the $t$-channel exchange orthonormal basis, as defined in ref.~\cite{DelDuca:2014cya} and appendix B of ref.~\cite{Caron-Huot:2017fxr}, to which we refer for further details. 

An orthonormal colour basis in the $t$-channel can be obtained by decomposing the direct product of the colour representations associated to the incoming and outgoing particles
1 and 4 (the top line in figure~\ref{fig:2to2d}) into a direct sum, and taking those representations which have non-zero overlap with the equivalent decomposition obtained for particles 2 and 3 (the bottom line in figure~\ref{fig:2to2d}). In the cases of gluon-gluon, quark-gluon and quark-quark scattering one has\footnote{Notice that the decomposition in \eqref{ColorDecompose} is valid for SU($N_c$), although the representation labels refer to their dimension for $N_c = 3$.} 
\begin{subequations}
\label{ColorDecompose}
\beqa
(8\otimes 8)_{gg\to gg} \quad &=& \quad  1 \oplus 8_s \oplus 8_a \oplus (10 \oplus \overline{10}) \oplus 27 \oplus 0, \label{ColorDecompose_gg}\\
(8\otimes 8)_{qg \to qg} \quad &=& \quad  1 \oplus 8_s \oplus 8_a, \label{ColorDecompose_qg}\\ 
(3\otimes \bar 3)_{qq \to qq} \quad &=& \quad  1 \oplus 8\,. \label{ColorDecompose_qq}
\eeqa
\end{subequations}
This basis is particularly convenient, because it describes the colour flow in the $t$-channel: at tree-level the two-parton amplitude in the high-energy limit is always given by the exchange of a gluon in the $t$-channel, which has the quantum number of an antisymmetric octet, thus the tree-level amplitude in the $t$-channel orthonormal basis is always given by a single component, namely 
\beq
\label{eq:treelevel_using_8a}
{\cal M}^{\rm tree}_{ij\to ij} = c_{ij}^{[8_a]} \,
{\cal M}^{{\rm tree},[8_a]}_{ij\to ij}\,.
\eeq
As is clear from eq.~\eqref{LLresum}, the same is true to all orders for the LL component of the amplitude. This can also be seen as a manifestation of the fact that a single Reggeon carries the quantum numbers of a gluon. One has thus 
\beq \label{LLs_explicit_t_channel_basis}
{\cal M}_{ij\to ij}(s,t)|_{\rm LL} = c_{ij}^{[8_a]} \, 
{\cal M}^{[8_a]}_{ij\to ij}(s,t)|_{\rm LL}\,.
\eeq
Owing to multi-Reggeon exchange, beyond LLs all colour components in the decomposition in eq.~\eqref{ColorDecompose} contribute.  The exchange of an odd number of Reggeons contributes to ${\cal M}^{(-)}$, while an even number of Reggeons contributes to ${\cal M}^{(+)}$~\cite{Caron-Huot:2013fea,Caron-Huot:2017fxr}.

In the cases of gluon-gluon and quark-gluon scattering, Bose symmetry links the kinematic dependence to that of colour, therefore, ${\cal M}_{gg\to gg}^{(\pm)}$ and ${\cal M}_{qg\to qg}^{(\pm)}$ receive contributions from $t$-channel exchange of colour representations which are even and odd under $1\leftrightarrow 4$ (or $2\leftrightarrow 3$) interchange for the even $(+)$ and odd $(-)$ components, respectively. 
Specifically, the odd amplitude ${\cal M}_{gg\to gg}^{(-)}$ consists of $t$-channel exchange of the antisymmetric octet $8_a$ and the $(10 \oplus \overline{10})$ representations, while ${\cal M}_{gg\to gg}^{(+)}$ involves the remaining components in eq.~(\ref{ColorDecompose_gg}). Similarly,  ${\cal M}_{qg\to qg}^{(-)}$ is governed (exclusively!) by the antisymmetric octet $8_a$ representation, while ${\cal M}_{qg\to qg}^{(+)}$ by the singlet and the symmetric octet $8_s$ in eq.~(\ref{ColorDecompose_qg}).
Bose symmetry does not apply to quark-quark scattering, and therefore there is no direct correspondence between ${\cal M}_{qq\to qq}^{(\pm)}$ and the colour components~${\cal M}^{[1]}_{qq\to qq}$,~${\cal M}^{[8]}_{qq\to qq}$. For instance, the exchange of two Reggeons, which contributes to the even amplitude ${\cal M}^{(+)}$ starting at NLL, affects both ${\cal M}^{[1]}_{qq\to qq}$~and~${\cal M}^{[8]}_{qq\to qq}$.

Our main interest in this paper is in the odd-signature amplitude. In this case, the relevant orthonormal basis elements for $qq$, $gg$ and $qg$ scattering are:
\begin{subequations}\label{colourTensors}
\begin{align}
    c_{qq}^{[8]}=\,&\frac{2}{\sqrt{N_c^2-1}}(t^b)^{a_4}{}_{a_1}(t^b)^{a_3}{}_{a_2},\\
    c_{qq}^{[1]}=\,& \frac{1}{N_c}\delta^{a_4}{}_{a_1}\delta^{a_3}{}_{a_2},\\
    c_{gg}^{[8_a]}=\,&\frac{1}{N_c}\frac{1}{\sqrt{N_c^2-1}}f^{a_1a_4b}f^{a_2a_3}{}_b, \\
    c_{gg}^{[10+\bar{10}]}=\,&\sqrt{\frac{2}{(N_c^2-4)(N_c^2-1)}}\left[\frac{1}{2}(\delta^{a_1}{}_{a_2}\delta^{a_3}{}_{a_4}-\delta^{a_3}{}_{a_1}\delta^{a_4}{}_{a_2})-\frac{1}{N_c}f^{a_1a_4b}f^{a_2a_3}{}_b\right],\\
    c_{qg}^{[8_a]} =\,&\sqrt{\frac{2}{N_c(N_c^2-1)}}(t^b)^{a_4}{}_{a_1}if^{a_2a_3b}.
\end{align}
\end{subequations}

While the $t$-channel colour basis is convenient for analysing individual $2\to2$ scattering processes in the high-energy limit, understanding the process-independent features, such as the relation between the exponentiation of high-energy logarithms and that of infrared singularities (see section~\ref{sec:infared_Regge_limit}), is best done in terms of colour operators. 
Indeed, following refs.~\cite{DelDuca:2011wkl,DelDuca:2011ae,Caron-Huot:2013fea,Henn:2016jdu,Caron-Huot:2017fxr,Caron-Huot:2017zfo,Gardi:2019pmk,Caron-Huot:2020grv,Falcioni:2020lvv,Naculich:2020clm}, our analysis of $2\to2$ scattering amplitudes in this paper will be largely based on such operators. 

To this end we use the colour-space formalism introduced in refs.~\cite{Bassetto:1983mvz,Catani:1996jh,Catani:1996vz,Catani:1998bh}: a colour operator $\T_k$ corresponds to the colour generator associated with the $k$-th parton in the scattering amplitude, which acts as an SU($N_c$) matrix on the colour indices of that parton. For instance $\mathbf{T}_1^c$ acting on the tree-level colour structure in eq.~(\ref{eq:treelevel}) can be written, with explicit indices, as
\begin{equation}
    \mathbf{T}_1^c(\mathbf{T}_i^b)_{a_1a_4}(\mathbf{T}_j^b)_{a_2a_3} = (\mathbf{T}_1^c)_{a_1a'_1} (\mathbf{T}_i^b)_{a'_1a_4}(\mathbf{T}_j^b)_{a_2a_3}.
    \label{T_1_action}
\end{equation}
For later use, let us note that generators associated with different particles trivially commute, $\T_k \cdot \T_l = \T_l \cdot \T_k$ for $k\neq l$, while $\T_k^2 = C_k$ where $C_k$ is the quadratic Casimir operator of the corresponding colour representation, i.e.~$C_q = C_F=(N_c^2-1)/(2N_c)$ for quarks and~$C_g = C_A = N_c$ for gluons. 

In the high-energy limit it proves useful to express such colour factors using the basis of 
Casimirs corresponding to colour flow through the three channels~\cite{Dokshitzer:2005ig,DelDuca:2011ae}:
\begin{align}
\label{TtTsTu}
\begin{split}
\T_s &= \T_1+\T_2=-\T_3-\T_4, \\ 
\T_u &= \T_1+\T_3=-\T_2-\T_4, \\
\T_t  &= \T_1+\T_4=-\T_2-\T_3\,,
\end{split}
\end{align}
subject to the colour-conservation identity 
\beq
\label{eq:colcons}
\left(\T_1+\T_2+\T_3+\T_4\right)\mathcal{M}=0\,.
\eeq 
The latter identity also implies
\beq
\label{eq:stucc}
\T_s^2 +\T_t^2+\T_u^2=\sum_{i=1}^4 C_i,
\eeq
which can be used to simplify the colour structure of the amplitude. 

As we have seen above when working with explicit colour tensors, the $t$ channel has a special status in the high-energy limit: in this basis
the tree level (and the tower of LLs in eq.~(\ref{LLs_explicit_t_channel_basis})) is given by a single component, the antisymmetric octet exchange.  Similarly, when working with channel colour operators, the operator $\tts$ has a special status in the high-energy limit: $\tts$ measures the colour charge flowing in the $t$ channel and thus upon acting on the tree-level amplitude~(\ref{eq:treelevel}) it simply yields 
\begin{equation}
\label{ttsToCA}
\tts {\cal M}^{\rm tree}_{ij\to ij} =C_A  {\cal M}^{\rm tree}_{ij\to ij}\,.
\end{equation}
To see this we may choose to expand $\tts$ using $\mathbf{T}_1+\mathbf{T}_4$
 in eq.~(\ref{TtTsTu}), getting
\begin{align}
\begin{split}
  \tts (\mathbf{T}_i^b)_{a_1a_4}(\mathbf{T}_j^b)_{a_2a_3}
  &=
  (\mathbf{T}_1+\mathbf{T}_4)^2
  (\mathbf{T}_i^b)_{a_1a_4}(\mathbf{T}_j^b)_{a_2a_3} 
  \\
  &=
  \Big(\mathbf{T}_1\cdot \mathbf{T}_1 
  +\mathbf{T}_4\cdot \mathbf{T}_4
  +2 \mathbf{T}_1^c \mathbf{T}_4^c\Big)
  (\mathbf{T}_i^b)_{a_1a_4}(\mathbf{T}_j^b)_{a_2a_3}
  \\
 & =
  2C_i  (\mathbf{T}_i^b)_{a_1a_4}(\mathbf{T}_j^b)_{a_2a_3}
  -2 (\mathbf{T}_i^c)_{a_1a'_1} (\mathbf{T}_i^b)_{a'_1a'_4}(\mathbf{T}_i^c)_{a'_4a_4} (\mathbf{T}_j^b)_{a_2a_3}
  \\
  &= C_A
  (\mathbf{T}_i^b)_{a_1a_4}(\mathbf{T}_j^b)_{a_2a_3}\,,
  \end{split}
\end{align}
where after expanding the square we used the rule in eq.~(\ref{T_1_action}) for both $\mathbf{T}_1$ and $\mathbf{T}_4$. Keeping in mind that $\mathbf{T}_k$ (with $k=1$ through 4) acts inwards towards the $t$-channel exchange (see figure~\ref{fig:2to2d}) we observe that in the second line $\mathbf{T}_1^c=\mathbf{T}_i^c$ acts on the original tree-level exchange from the left, while  $\mathbf{T}_4^c=-\mathbf{T}_i^c$ acts on it from the right. The relative minus sign reflects the fact that parton 4 is outgoing, and thus $\mathbf{T}_4^c$ is opposite to the direction of $\mathbf{T}_i^c$. In the final step we use the colour algebra on the $\mathbf{T}_i$ generators, noting the cancellation of the quadratic Casimir $C_i$, and thus recovering eq.~(\ref{ttsToCA}). 

In contrast to $\tts$, the action of $\T_s^2$ or $\T_u^2$ on the tree-level gives rise to other colour tensors.
Given the symmetry properties of (Bose-symmetric) signature eigenstate amplitudes ${\cal M}^{(\pm)}$, it is convenient to introduce a colour operator that is \emph{odd} under $s\leftrightarrow u$ 
interchange:
\beq
\label{eq:Tsudef}
\tsu\equiv \tfrac12 \left(\T_s^2-\T_u^2\right).
\eeq
We will see in section~\ref{subsec:colour} and what follows that a natural basis of operators in the Regge limit can be constructed using $\T_t^2$ and $\tsu$ along with their commutators (see also previous work~\cite{Caron-Huot:2017fxr,Caron-Huot:2017zfo,Gardi:2019pmk,Caron-Huot:2020grv,Falcioni:2020lvv,Naculich:2020clm}). Signature symmetry is readily encoded in such a basis, since signature odd (even) operators must necessarily feature $\tsu$ an odd (even) number of times. Having set up the colour operator formalism, we are ready to review the correspondence between the exponentiation of high-energy logarithms and that of infrared singularities. 

\subsection{Infrared divergences in the Regge limit of \texorpdfstring{$2\to 2$}{2->2} scattering}
\label{sec:infared_Regge_limit}

A salient feature of gauge-theory amplitudes is the presence of long-distance singularities. 
It is well known that these singularities factorise and exponentiate, as described for example in refs.~\cite{Catani:1998bh,Sterman:2002qn,Aybat:2006mz,Aybat:2006wq,Gardi:2009qi,Gardi:2009zv,Becher:2009cu,Becher:2019avh,Almelid:2015jia}.
Specifically, considering the ultraviolet-renomormalised massless scattering amplitude $\mathcal{M}_{ij\to ij}$ in general kinematics, one has
\begin{align}
\label{ZH_intro}
    \mathcal{M}_{ij\to ij}\left(s,t;\mu,\al_s(\mu^2),\eps\right) =& \,\mathbf{Z}_{ij\to ij}\left(s,t;\mu,\al_s(\mu^2),\eps\right)
    \mathcal{H}_{ij\to ij}\left(s,t;\mu,\al_s(\mu^2),\eps\right)\,,
\end{align}
where the so-called infrared renormalisation factor $\mathbf{Z}_{ij\to ij}$ captures all the singularities 
in~$\epsilon$, while the hard function $\mathcal{H}_{ij\to ij}$ on which it acts, is finite for $\epsilon\to 0$. The 
$\mathbf{Z}_{ij\to ij}$ factor exponentiates according to 
\begin{align}
        \mathbf{Z}_{ij\to ij}\left(s,t;\mu,\al_s(\mu^2),\eps\right) =&\, \mathcal{P}\exp{\left\{-\frac{1}{2}\int_0^{\mu^2}\frac{d\lambda^2}{\lambda^2}\bm{\Gamma}_{ij\to ij}(s,t;\lambda,\al_s(\lambda^2))\right\}},\label{eq:ZfactorIntro}
\end{align}
where $\bm{\Gamma}_{ij\to ij}$ is 
the soft anomalous dimension, which is a finite function of the $d=4-2\epsilon$ dimensional coupling
\begin{subequations}
 \begin{align}
     \frac{d\al_s}{d\log\mu} =& -2\eps\al_s - \frac{\al_s^2}{2\pi}\sum_{n=0}^{\infty}b_n\left(\frac{\al_s}{\pi}\right)^n,\\
     \al_s(\lambda^2) =&\, \al_s(\mu^2)\left(\frac{\lambda^2}{\mu^2}\right)^{-\eps}\left[1-\frac{b_0}{4\pi\eps}\left(1-\left(\frac{\lambda^2}{\mu^2}\right)^{-\eps}\right)\al_s(\mu^2)+\mathcal{O}\!\left(\al_s^2\right)\right], \label{runnign_coupling_D_dim}
 \end{align}
 \end{subequations}
where $b_0=(11C_A-4T_Rn_f)/3$ in QCD; in ${\cal N}=4$ SYM the beta function coefficients are identically zero, so the square brackets in eq.~(\ref{runnign_coupling_D_dim}) should be replaced by $1$. Infrared singularities in the exponent of eq.~(\ref{eq:ZfactorIntro}) are generated by integrating over the coupling down to $\lambda\to 0$.

Within the Regge limit, long-distance singularities  are particularly conspicuous as the leading-order gluon Regge trajectory in eq.~(\ref{r_gamma}) is itself proportional to $1/\epsilon$. Thus, eq.~(\ref{LLresum}) implies that the LLs, 
$L^m {\mm}^{(\pm,m,m)}_{ij\to ij}$ in eq.~(\ref{eq:expansionDef}), are proportional to $1/\epsilon^m$. Naturally, similar infrared singularities occur along with subleading high-energy logarithms, $L^m {\mm}^{(\pm,n,m)}_{ij\to ij}$ with $n>m$,  and thus studying factorisation and exponentiation in the Regge limit is directly connected to studying the infrared singularity structure. The intimate interplay between the two has been observed and used in multiple occasions, e.g.~in~refs.~\cite{Korchemskaya:1994qp,Korchemskaya:1996je,DelDuca:2011wkl,DelDuca:2011ae,Caron-Huot:2013fea,Henn:2016jdu,Caron-Huot:2017fxr,Caron-Huot:2017zfo,Almelid:2017qju,Gardi:2019pmk,Caron-Huot:2020grv,Falcioni:2020lvv,Naculich:2020clm}. 
A primary avenue for investigating it, is based on specialising the analysis of infrared singularities in terms of the soft anomalous dimension (according to eqs.~(\ref{ZH_intro}) and~(\ref{eq:ZfactorIntro})) to the Regge limit, obtaining\footnote{In section~\ref{sec:SAD_Regge} we will review the structure of the soft anomalous dimension in general kinematics, and examine how it simplifies in the high-energy limit, obtaining eq.~(\ref{eq:gammaFactorIntro}). Here we only need some of its salient features, which we define below. }
\begin{align}
    \mathbf{\Gamma}_{ij\to ij}\left(\alpha_s,L,\frac{-t}{\lambda^2}\right)=&\,\frac{1}{2}\gamma_{K}(\alpha_s)\left[L\tts + i\pi\tsu\right] +   \Gamma_i\left(\alpha_s,\frac{-t}{\lambda^2}\right)+\Gamma_j\left(\alpha_s,\frac{-t}{\lambda^2}\right)\nn\\&+\dd (L,\alpha_s)\,,\label{eq:gammaFactorIntro}
\end{align}
where the first terms arise from the dipole formula~\cite{Catani:1998bh,Sterman:2002qn,Aybat:2006mz,Aybat:2006wq,Gardi:2009qi,Gardi:2009zv,Becher:2009cu}: $\gamma_K$ is the coefficient of the quadratic Casimir within the lightlike cusp anomalous dimension (see eq.~(\ref{cuspAD}) below) while $\Gamma_i$ and $\Gamma_j$ capture collinear singularities, which depend on the scattered partons (both their spin and their colour representation), but are independent of high-energy logarithms and are proportional to the unit matrix in colour space. In turn,
\begin{equation}
\label{Delta_intro}
\dd (L,\alpha_s) = \sum_{\ell=3}^\infty \left(\frac{\alpha_s}{\pi}\right)^\ell \sum^{\ell-1}_{m=0} L^m \dd^{(\ell,m)},
\end{equation}
represents corrections to the dipole formula, and has a complex structure in colour space. It arises owing to non-planar multi-parton correlations starting from three loops and to quartic (and higher) corrections to the cusp anomalous dimension, starting from four loops. 

The cusp anomalous dimension may be defined through the renormalisation of cusped Wilson loops~\cite{Korchemsky:1987wg} and it corresponds to the coefficient of the leading ultraviolet divergence of this object in the lightlike limit. This translates~\cite{Korchemsky:1985xj} into the leading infrared behaviour of massless amplitudes, where the colour representation of the Wilson line is that of the parton. It may be expanded in Casimirs according to~\cite{Korchemsky:1987wg,Korchemsky:1985xj,Gardi:2009qi,Gardi:2009zv,Becher:2009cu,Becher:2019avh,Boels:2017ftb,Boels:2017skl,Moch:2017uml,Grozin:2017css,Henn:2019swt,Huber:2019fxe,vonManteuffel:2020vjv,Agarwal:2021zft}
\beq
\label{cuspAD}
\Gamma^{\rm{cusp}}_i(\alpha_s(\lambda^2)) = \frac{1}{2} \gamma_K(\alpha_s(\lambda^2))C_i+ \sum_R g_R(\alpha_s(\lambda^2)) \frac{d_{RR_i}}{N_{R_i}} + {\cal}O(\as^5),
\eeq
where $\gamma_K(\alpha_s)$ multiplies the quadratic Casimir $C_i$ in the representation of parton~$i$, while $g_R(\alpha_s)$, starting only at four loops, multiplies the quartic Casimir (defined in eq.~(\ref{eq:drri}) below).
Both $\gamma_K$ and $g_R$ are independent of the representation $R_i$ and are known to four loops in QCD~\cite{Boels:2017ftb,Boels:2017skl,Moch:2017uml,Grozin:2017css,Henn:2019swt,Huber:2019fxe,vonManteuffel:2020vjv,Agarwal:2021zft}. 
For convenience we collect the expansions in eqs.~(\ref{gammaK}) 
and~(\ref{g_R_values}).

The anomalous dimensions $\Gamma_i$ and $\Gamma_j$, associated to the projectile~$i$ and the target~$j$, respectively, 
are the only sources of double poles in $\epsilon$ upon integration in eq.~(\ref{eq:ZfactorIntro}). They take the form
\begin{align}
\label{collAD}
 \Gamma_i\bigg(\alpha_s(\lambda^2),\frac{-t}{\lambda^2}\bigg)&= 2\gamma_i(\alpha_s(\lambda^2))+\Gamma^{\rm{cusp}}_i(\alpha_s(\lambda^2))\log \frac{-t}{\lambda^2},
\end{align} 
where $\Gamma^{\rm{cusp}}_i$ is given in \eqref{cuspAD} and the function $\gamma_i(\alpha_s)$ is the collinear anomalous dimension~\cite{FormFactors,DelDuca:2014cya,Falcioni:2019nxk,Dixon:2017nat} corresponding to the parton~$i$. It has recently been computed to four loops in QCD~\cite{Agarwal:2021zft}. We provide its expansion through two-loop in eqs.~\eqref{gammaq} and \eqref{gammag} for quarks and gluons, respectively.

Having defined the relevant components of the soft anomalous dimension, let us see what infrared singularities they generate at leading and subleading logarithmic accuracy in the Regge limit. 
Upon integrating over eq.~(\ref{eq:gammaFactorIntro}) in eq.~(\ref{eq:ZfactorIntro}), the terms involving $\gamma_K$ result in the following two integrals~\cite{DelDuca:2011wkl,DelDuca:2011ae}:
\begin{subequations}
\begin{align}
    \label{eq:Kdef}
     K(\al_s(\mu^2))&\equiv -\frac{1}{4}\int_0^{\mu^2}\frac{d\lambda^2}{\lambda^2}\gamma_K(\al_s(\lambda^2)) =
     \sum_{n = 0}^{\infty} 
\left( \frac{\al_s(\mu^2)}{\pi} \right)^n K^{(n)}
=
     \frac{1}{2\epsilon}
     \frac{\alpha_s(\mu^2)}{\pi}+\ldots ,\\
     K_D(\al_s(\mu^2))&\equiv -\frac{1}{4}\int_0^{\mu^2}\frac{d\lambda^2}{\lambda^2}\gamma_K(\al_s(\lambda^2)) \ln\left(\frac{\mu^2}{\lambda^2}\right)
     = -\frac{1}{2\epsilon^2}
     \frac{\alpha_s(\mu^2)}{\pi}+\ldots,
\end{align}
\end{subequations}
where the ellipsis stand for higher-order terms in the coupling, generated by the expansion of $\gamma_K$ in eqs.~(\ref{gammaK}), and by the running coupling~(\ref{runnign_coupling_D_dim}).
Of these two integrals, the former, involving a \emph{single pole} in $\epsilon$, contributes $K(\al_s) (L\tts+i\pi\tsu)$ to the exponent while the latter, involving a \emph{double pole} in~$\epsilon$ enters only through $\Gamma_i$ and $\Gamma_j$,  multiplying a unit operator in colour space. 
At this point the comparison with the Regge-pole factorisation picture is immediate: of all the terms in the exponent in eq.~(\ref{eq:ZfactorIntro}) the only one entering at leading logarithmic accuracy is 
$\tts K^{(1)} L {\alpha_s(-t)}/{\pi}$, originating in the leading-order cusp anomalous dimension. 
To identify the colour structure of the gluon Regge trajectory note that $\tts$ in~the exponent of ~${\mathbf{Z}}_{ij\to ij}$ will be acting on $\mathcal{H}_{ij\to ij}\left(s,t;\eps\right)$ in eq.~(\ref{ZH_intro}), which coincides with $\mathcal{M}_{ij\to ij}^{\text{tree}}$ at leading order. Using eq.~(\ref{ttsToCA}) one then recovers the $C_A$ colour factor of eq.~(\ref{LLresum}), and one finds
\begin{equation}
\label{K_alpha_s_relation}
\al_g^{(1)}(t) =  K^{(1)}  +\mathcal{O}(\eps^0)\,,
\end{equation}
where aside from the singular term, the left-hand side features finite terms along with $\epsilon$-suppressed terms summarised by eq.~(\ref{r_gamma}).

We thus see explicitly how the two exponentiation pictures, that of high-energy logarithms and that of infrared singularities are compatible, as they must be. 
One wonders how this extrapolates to higher orders. Naively\footnote{Recall that so far we only discussed the validity of eq.~(\ref{LLresum}) to leading logarithmic accuracy.} comparing the 
infrared-singular terms which are \emph{linear} in the high-energy logarithm in the exponent in 
eq.~(\ref{LLresum}) to those of eq.~(\ref{ZH_intro}) ($K(\al_s) L\tts$, to three loops) one would require
\begin{equation}
\label{questioned_eqality_of_sing_part_of_alpha_g_and_K}
\al_g(t)  \overset{?}{=}  K +\mathcal{O}(\eps^0)\,.
\end{equation}
This rather remarkable connection between the singularities of the Regge trajectory and cusp singularities of Wilson lines was proposed already in~ref.~\cite{Korchemskaya:1994qp,Korchemskaya:1996je}, where it was established analysing quark-quark scattering in terms of Wilson lines, and shown to hold at two loops. Although eq.~(\ref{questioned_eqality_of_sing_part_of_alpha_g_and_K}) has been regarded as general (see e.g.~\cite{DelDuca:2014cya}, where it was used at three loops) its realisation beyond two loops is complicated by the presence of Regge cuts (see e.g.~\cite{Caron-Huot:2017fxr}) and has not been fully understood. 
In the present paper we explain how it is realised at three loops, and conjecturally, beyond this order.

Generally speaking, the comparison between the two exponentiation pictures at subleading logarithmic orders is rather complex. This is both because corrections to the soft anomalous dimension in eq.~(\ref{eq:gammaFactorIntro}) at three loops and beyond may have a complex colour structure, and because Regge-pole factorisation is violated by Regge-cut contributions associated with multiple Reggeon exchange, which is the topic of section~\ref{sec:subleading_logs} below.
As already mentioned, this comparison is highly insightful~\cite{Korchemskaya:1994qp,Korchemskaya:1996je,DelDuca:2011wkl,DelDuca:2011ae,DelDuca:2013ara,DelDuca:2013dsa,DelDuca:2014cya,Caron-Huot:2013fea,Henn:2016jdu,Caron-Huot:2017fxr,Caron-Huot:2017zfo,Almelid:2017qju,Gardi:2019pmk,Caron-Huot:2020grv,Falcioni:2020lvv,Naculich:2020clm}, and provides an important impetus for the current study.

Let us now make some brief comments about NLLs and beyond based on eq.~(\ref{eq:gammaFactorIntro}), leaving the more detailed analysis to sections~\ref{sec:Infrared} and~\ref{sec:SAD_Regge},
where we will infer information regarding the soft anomalous dimension at four loops from the calculations in the Regge limit. While at LL accuracy the soft anomalous dimension is real (thus signature even), starting from NLL it contains both real and imaginary contributions. 
Real NLL contributions arise only from the ${\cal O}(\alpha_s)$ term in $\Gamma_i$ and $\Gamma_j$ and the ${\cal O}(\alpha_s^2L)$ term in the expansion of $K(\al_s) L\tts$. 
Thus, the signature-even part of the NLL anomalous dimension is two-loop exact.\footnote{The absence of real $\dd^{(\ell,\ell-1)}$ contributions can also be seen as a manifestation of Regge-pole factorisation of the odd amplitude -- indeed multiple Reggeon exchange cannot yet arise at this logarithmic order. See section~\ref{sec:subleading_logs} below.}
In turn, signature odd terms in eq.~(\ref{eq:gammaFactorIntro}) have a more complex structure: the leading-order contribution is the ${\cal O}(\alpha_s)$ term in $K(\al_s) i\pi\tsu $. Subsequent corrections in this NLL tower start at four loops (see appendix~\ref{sec:SAD3loopa}). They have been computed to all orders, thus fixing the coefficients $\dd^{(\ell,\ell-1)}$ in eq.~(\ref{Delta_intro}) for any $\ell$ (see eq.~(4.24) in ref.~\cite{Caron-Huot:2017zfo}).
Our interest in the present paper is in the real tower of NNLL contributions, ${\rm Re}\,\dd^{(\ell,\ell-2)}$ in eq.~(\ref{Delta_intro}), which, as we shall see, also start at four loops. Knowing that these NNLL contributions to eq.~(\ref{eq:gammaFactorIntro}) are associated (see details in section~\ref{sec:SAD_Regge}) with multi-leg correlations in the soft anomalous dimension, rather than with the cusp anomalous dimension, it is clear at the outset that they must be non-planar to any loop order.\footnote{In the planar limit, the soft anomalous dimension reduces to a sum over two-particle interactions, where the only kinematic dependence is linear in the (high-energy) logarithm, and thus necessarily beyond NNLL from four loops.} This is an important prediction we will be able to verify from the Regge perspective in what follows.

\subsection{Subleading logarithms, Regge pole and Regge cuts}
\label{sec:subleading_logs}

The LL contribution to the scattering amplitude in eq.~\eqref{LLresum} is generated by a single Reggeon exchange, which appears as a Regge pole in the complex angular momentum plane. A Regge pole corresponds to $s$ raised to a power ($C_A\alpha_g(t)$), which is strictly independent of~$s$. 
Beyond LL accuracy one needs to take into account multiple Reggeon exchanges, which give rise to Regge cuts \cite{DelDuca:2001gu,DelDuca:2013ara,DelDuca:2013dsa,DelDuca:2014cya,Caron-Huot:2013fea,Fadin:2016wso,Caron-Huot:2017fxr,Fadin:2017nka,Falcioni:2020lvv,Fadin:2021csi}. 
In refs.~\cite{Mandelstam:1963cw,Eden:1966dnq,Collins:1977jy} it was shown that Regge cuts arise from non-planar diagrams. This has profound implications on the structure of subleading logarithmic corrections to the amplitude.

In the majority of the present work we split the contributions to the $2\to 2$ amplitude into single-Reggeon states (SRS) and multiple-Reggeon states (MRS). We will see later that the latter contain non-planar as well as planar contributions. 
We will write this separation as
\begin{subequations}
\label{polecut}
\begin{align}
{\cal M}^{(-)} &= {\cal M}^{(-),\,\rm SRS} + {\cal M}^{(-),\,\rm MRS}\,,
\label{polecut_minus}
\\
{\cal M}^{(+)} &=  {\cal M}^{(+),\,\rm MRS}\,
\end{align}
\end{subequations}
where the even amplitude, ${\cal M}^{(+)}$, starts at NLL with the exchange of two Reggeised gluons, and in the odd amplitude, ${\cal M}^{(-)}$, the first term refers to the $t$-channel exchange of a single Reggeon state (SRS), while the second involves the $t$-channel propagation of multiple Reggeon states (MRS). The latter starts at NNLL at two loops, with the exchange of three Reggeised gluons. Starting from three loops it also involves mixing between three Reggeons and one Reggeon~\cite{Caron-Huot:2017fxr}. 
To understand precisely what MRS corresponds to, we refer the reader to figures~\ref{fig:h33_2} through~\ref{fig:h13} below, which we will later on compute using an iterative solution of the Balitsky-JIMWLK rapidity evolution equations~\cite{Balitsky:1995ub,Kovchegov:1999yj,JalilianMarian:1996xn,JalilianMarian:1997gr,Iancu:2001ad} following the methodology of refs.~\cite{Caron-Huot:2013fea,Caron-Huot:2017fxr} (briefly reviewed in section~\ref{BFKL_abridged}). In each of these figures the propagation in the $t$ channel is mediated \emph{at some stage} by a compound state of three Reggeons; this is enough to classify it as a MRS in eq.~(\ref{polecut_minus}). Conversely, a SRS refers to the situation where the entire propagation in the $t$ channel from the projectile to the target is via a single Reggeon. 

At LL accuracy ${\cal M}^{(-),\,\rm SRS}$ is given by eq.~\eqref{LLresum} to all orders in $\alpha_s$. Beyond LL, this contribution retains the same exponential dependence on $s$, but higher-order corrections to the gluon Regge trajectory $\alpha_g(t)$ of eq.~(\ref{al_g}) are required, and in addition \emph{impact factors} must be introduced, which can be seen as effective couplings of the Reggeon with the external scattered particles~\cite{Fadin:2006bj,Fadin:2015zea}. Thus, the SRS contribution can always be written as
\begin{equation}
\label{Regge-Pole-General}
\mm^{(-),\, \rm SRS}_{ij\to ij}  =
 e^{C_A \alpha_g(t) L} \,
C_i(t) \, C_j(t) \,
{\cal M}^{\rm tree}_{ij\to ij}\,, 
\end{equation}
generalising eq.~(\ref{LLresum}). 
The impact factors $C_{i/j}(t)$, much like the trajectory $\alpha_g(t)$, are strictly independent of the centre-of-mass energy $s$.
Importantly, eq.~(\ref{Regge-Pole-General}) also retains the colour tensor structure of the tree-level amplitude in~eq.~(\ref{eq:treelevel}), which is equivalent to stating that the $t$-channel exchange remains an antisymmetric octet exchange.

Contributions at fixed logarithmic accuracy are obtained upon expanding the Regge trajectory and the impact factors in powers of the strong coupling. In particular, the NLL contribution is given by taking the Regge trajectory at ${\cal O}(\as^2)$ in the exponent, and the impact factors at ${\cal O}(\as)$. Both the Regge trajectory and $C_{i/j}(t)$ are theory-dependent. The two-loop Regge trajectory in QCD is~\cite{Fadin:1995xg,Fadin:1996tb,Fadin:1995km,Blumlein:1998ib}
\beq \label{eq:regge2QCD}
\alpha^{(2)}_g(t)|_{\rm QCD} = -\frac{b_0}{16\eps^2} +\frac{1}{8\eps}
\left[\left(\frac{67}{18}-\zeta_2\right)\CA -\frac{10T_R\nf}{9}\right]
+ \CA\left(\frac{101}{108}-\frac{\zeta_3}{8}\right) 
- \frac{7\, T_R\nf}{27} + {\cal O}(\eps)\,.
\eeq
The corresponding expression for ${\cal N}=4$ SYM can be obtained, according to the principle of ``maximum transcendentality''~\cite{Kotikov:2001sc,Kotikov:2002ab,Kotikov:2004er,Kotikov:2007cy}, by selecting the terms in eq.~\eqref{eq:regge2QCD} with the highest transcendental weight, i.e. 
\beq \label{eq:regge2SYM}
\alpha^{(2)}_g(t)|_{\rm{SYM}} = - \frac{C_A}{8}
\bigg(\frac{\zeta_2}{\epsilon}+\zeta_3  \bigg) + {\cal O}(\eps)\,. 
\eeq

An amplitude that can be brought to the form of eq.~(\ref{Regge-Pole-General}) is said to admit \emph{Regge-pole factorisation}.
At NLL, the real part of the amplitude ${\cal M}^{(-)}_{ij\to ij}$ admits Regge-pole factorisation, namely it is fully described by the exchange of a SRS as in eq.~(\ref{Regge-Pole-General}). A well-known consequence of this is the following simple relation between quark-gluon, gluon-gluon and quark-quark scattering~\cite{DelDuca:2001gu}:
\begin{equation}
\label{eq:pole-factorisation}
    \left(\frac{\mathcal{M}^{(-)}_{qg\to qg}|_{\text{NLL}}}{\mathcal{M}_{qg\to qg}^{\text{tree}}}\right)^2=\frac{\mathcal{M}^{(-)}_{gg\to gg}|_{\text{NLL}}}{\mathcal{M}_{gg\to gg}^{\text{tree}}}\cdot\frac{\mathcal{M}^{(-)}_{qq\to qq}|_{\text{NLL}}}{\mathcal{M}^{\text{tree}}_{qq\to qq}}\,,
\end{equation}
which automatically follows from the fact that each of the separate processes computed at NLL admits eq.~(\ref{Regge-Pole-General}), and that the gluon Regge trajectory is independent on the nature of the scattered partons.

Considering odd partonic amplitudes beyond NLL, the above relation is broken by Regge-cut contributions that first arise in the signature-odd amplitudes at NNLL at two loops~\cite{DelDuca:2001gu}. 
These contributions, commonly referred to as Regge-pole factorisation breaking terms, have been interpreted as being due to three-Reggeon exchange -- namely contributions to ${\cal M}^{(-),\,\rm MRS}$ in eq.~(\ref{polecut_minus}) -- in refs.~\cite{Caron-Huot:2017fxr,Fadin:2016wso,Fadin:2017nka}, where the three-loop NNLL amplitude was determined. 
In this paper we take a further step, by fully characterizing the tower of NNLL corrections to the odd amplitude, and extending the explicit computations of these to four loops. 
We will also show that contributions that do not conform with factorisation according to eq.~(\ref{Regge-Pole-General}) (thus nor with a relation of the form of eq.~(\ref{eq:pole-factorisation})) arise only from non-planar diagrams. 
Indeed, planar contributions do conform with the factorisation of eq.~(\ref{Regge-Pole-General}), even if they arise from
multiple Reggeon exchanges, ${\cal M}^{(-),\,\rm MRS}$ in eq.~(\ref{polecut_minus}). 
Our findings are therefore consistent with the general considerations showing that Regge-pole-factorisation breaking is associated with non-planar diagrams~\cite{Mandelstam:1963cw,Eden:1966dnq,Collins:1977jy}.

We conclude that to describe the high-energy behaviour of a signature-odd amplitude at NNLL and beyond one must take into account both terms in eq.~(\ref{polecut}). Consequently, $C_{i/j}(t)$ and $\al_g(t)$ start to depend on how the separation in eq.~(\ref{polecut}) is precisely defined; this will be referred to as a scheme choice. As explained following eq.~(\ref{polecut}), to perform explicit computations we shall use \emph{the MRS scheme}, 
where the entire contribution due to MRS will be explicitly computed using the Balitsky-JIMWLK rapidity evolution equation, while the remaining SRS contribution, which is strictly driven by the evolution of a single Reggeon, will be determined by matching 
\begin{equation}
\label{Schemes_MRS}
{\cal M}_{ij\to ij}^{(-)} =  e^{C_A \alpha_g(t) L} \,
C_i(t) \, C_j(t) \,
{\cal M}^{\rm tree}_{ij\to ij} 
\,+ \,{\cal M}_{ij\to ij}^{(-),\,\rm MRS}\,,
\end{equation}
to the known fixed-order signature-odd amplitude. With the computed ${\cal M}^{(-),\,\rm MRS}$ at hand, and knowing that the pure SRS contribution admits Regge-pole factorisation, we would be able to uniquely define and compute the impact factors $C_i(t)$ and $C_j(t)$ and trajectory $\alpha_g(t)$ in this scheme, as done in ref.~\cite{Caron-Huot:2017fxr}.

Given that planar MRS contributions can be factorised according to  eq.~(\ref{Regge-Pole-General}), one may be inclined to consider another scheme for separating the odd amplitude, \emph{the Regge-cut scheme}, where
\begin{align}
\label{Schemes_Cut}
\begin{split}
{\cal M}_{ij\to ij}^{(-)} \,&=
\, {\cal M}_{ij\to ij}^{(-),\,\rm pole}
\,+ \, {\cal M}_{ij\to ij}^{(-),\,\rm cut}\\
\,&=
e^{C_A \tilde{\alpha}_g(t) L} \,
\tilde{C}_i(t) \, \tilde{C}_j(t) \,
{\cal M}^{\rm tree}_{ij\to ij} 
\,+ \, {\cal M}_{ij\to ij}^{(-),\,\rm cut}\,.
\end{split}
\end{align}
Here the \emph{non-planar part} of ${\cal M}^{(-),\,\rm MRS}$ is represented by ${\cal M}^{(-),\,\rm cut}$, while the planar part of ${\cal M}^{(-),\,\rm MRS}$ is absorbed into the first term, the Regge pole. The latter can be factorised just as in eq.~(\ref{Regge-Pole-General}), but given the presence of planar MRS at NNLL (and beyond), one would obtain different values for the impact factor coefficients starting at two loops, and the gluon Regge trajectory, starting at three loops, as compared to the MRS scheme in eq.~(\ref{Schemes_MRS}).

While the MRS scheme of eq.~(\ref{Schemes_MRS}) is most natural from the calculation perspective, the Regge-cut scheme of eq.~(\ref{Schemes_Cut}) aims to capture the high-energy behaviour of the amplitude along with its behaviour in the large-$N_c$ limit. 
The Regge-pole term dominates in the large-$N_c$ limit, while, as we will see, the cut term at NNLL accuracy is characterised by a highly non-trivial colour structure. 
Furthermore, we will see that it is in this scheme that the aforementioned remarkable relation (\ref{questioned_eqality_of_sing_part_of_alpha_g_and_K}) between the singularities of the gluon Regge trajectory and the cusp anomalous dimension~\cite{Korchemskaya:1994qp,Korchemskaya:1996je}, namely
\begin{equation}
\label{affirmed_eqality_of_sing_part_of_tilde_alpha_g_and_K}
\tilde{\al}_g(t)  =  K +\mathcal{O}(\eps^0)\,,
\end{equation} 
holds at three loops. Its generalisation at four loops is discussed in section \ref{SADgenrep}. We will return to the Regge-cut scheme in section~\ref{sec:pole_cut_explicit_rep} and then discuss in detail the interplay between the exponentiation of infrared singularities and high-energy logarithms in sections~\ref{sec:Infrared} and \ref{sec:SAD_Regge}. At this point we return to the MRS scheme and focus on characterising the NNLL tower of corrections using rapidity evolution.

\subsection{The reduced amplitude}
\label{sec:reduced_ampl}

As a final step of preparation for the study of the NNLL tower of MRS corrections ${\cal M}^{(-),\,\rm MRS}$ using rapidity evolution equations, we briefly review the concept of a \emph{reduced} amplitude~\cite{Caron-Huot:2017fxr,Caron-Huot:2017zfo,Gardi:2019pmk,Caron-Huot:2020grv,Falcioni:2020lvv}, defined by stripping off of the amplitude the single Reggeon evolution and parton-dependent singularities associated with the impact factors. 

To this end we further factorise the impact factors as
\beq\label{eq:impactIRfact}
C_{i/j}(t) = Z_{i/j}(t) \, D_{i/j}(t),
\eeq
where~\cite{Gardi:2009qi,Becher:2009qa,Gardi:2009zv,DelDuca:2011ae,DelDuca:2011wkl}
\begin{equation}
Z_i(t)=\exp\left\{-\frac{1}{2}\int_0^{\mu^2}\frac{d\lambda^2}{\lambda^2}\Gamma_i\left(\al_s(\lambda^2),\frac{-t}{\lambda^2}\right)\right\}\,,
\label{Zi}
\end{equation}
with $\Gamma_i$ defined in eq.~(\ref{collAD}).
Following the convention introduced for the Regge trajectory in eq.~\eqref{al_g}, we expand the impact factors in powers of $\as/\pi$:
\begin{subequations}
\begin{align}
\label{impact_factors_def}
C_i(t) &= \sum_{n = 0}^{\infty} 
\left( \frac{\al_s(-t)}{\pi} \right)^n C^{(n)}_i,\\
Z_i(t) &= \sum_{n = 0}^{\infty} 
\left( \frac{\al_s(-t)}{\pi} \right)^n Z^{(n)}_i,\\
D_i(t) &= \sum_{n = 0}^{\infty} 
\left( \frac{\al_s(-t)}{\pi} \right)^n D^{(n)}_i\,,
\end{align}
\end{subequations}
where $C_i^{(0)}=Z_{i}^{(0)}=D_{i}^{(0)}=1$.

To directly access the multiple Reggeon exchange, we introduce a reduced amplitude ${\cal \hat M}$, as follows~\cite{Caron-Huot:2017fxr}: 
\beq\label{Mreduced}
\Mreduced_{ij\to ij} \equiv 
\left(Z_i  Z_j \right)^{-1} 
\,e^{-\,\T_t^2\, \alpha_g(t) \, L} \, \MM_{ij\to ij}\,,
\eeq
where the effect of the exponential is to remove the evolution of a single Reggeon and where $Z_i$ and $Z_j$, defined in \eqref{Zi}, remove the collinear divergences.

At tree-level ${\cal \hat M}^{\rm tree}_{ij\to ij} = {\cal M}^{\rm tree}_{ij\to ij}$. In general, comparing eq.~\eqref{Regge-Pole-General} with eq.~\eqref{Mreduced}, the structure of the reduced signature-odd amplitude beyond NLL reads
\begin{equation}
\label{Regge-Pole-and-Cut}
\Mreduced^{(-)}_{ij\to ij}  =
D_i(t) D_j(t){\cal M}^{\rm tree}_{ij\to ij}
+ {\hat{{\cal M}}}^{(-),\,\rm MRS}_{ij\to ij}\,,
\end{equation}
where ${\hat{{\cal M}}}^{(-),\,\rm MRS}_{ij\to ij}$ represents
the contribution due to multiple Reggeon exchanges. In section~\ref{BFKL_abridged} we are going to review how these terms can be calculated by solving iteratively the Balitsky-JIMWLK equation, \cite{Caron-Huot:2013fea,Caron-Huot:2017zfo,Caron-Huot:2020grv}. The resulting amplitude takes the form given in eq.~\eqref{eq:expansionDef}, and once the renormalisation scale is set as $\mu^2=-t$, the amplitude coefficients depend only on $\epsilon$ and on colour operators acting on the tree amplitude. 

We point out that the definition in eq.~(\ref{Mreduced}) applies also to the signature-even amplitude.  The first few coefficients of the reduced even amplitude at NLL, as calculated in ref.~\cite{Caron-Huot:2020grv}, 
read:
\begin{subequations}\label{eq:evenAmps}
\begin{align} \label{eq:m1finite} 
\Mreduced^{(+,1,0)} &= 
i\pi \, r_\Gamma \, \bigg\{ \frac{1}{2\eps} \bigg\} \tsu \MMtree, \\  \label{eq:m2finite}
\Mreduced^{(+,2,1)} &= 
i\pi \, \frac{r_\Gamma^2}{2} \,\bigg\{ -\frac{1}{4 \eps^2} \bigg\} 
[\tts,\tsu]\MMtree, \\   \label{eq:m3finite}
\Mreduced^{(+,3,2)} &= 
i\pi \,\frac{r_\Gamma^3}{3!}\, 
\bigg\{\frac{1}{8 \eps^3} 
- \frac{11\zeta_3}{4} \bigg\} [\tts,[\tts,\tsu]]\MMtree, \\  
\Mreduced^{(+,4,3)} &= 
i\pi \,\frac{r_\Gamma^4}{4!}\,
\bigg\{ -\left( \frac{\zeta_3}{8 \eps} 
+ \frac{3\zeta_4}{16}\right) [\tts,[\tts,\tsu]]\tts
-\frac{1}{16 \eps^4} [\tts,[\tts,[\tts,\tsu]]] 
\bigg\} \MMtree,
\end{align}
\end{subequations}
where we suppressed corrections that vanish at $\epsilon\to 0$. 
Note that the commutators in eq.~(\ref{eq:evenAmps}) involve an odd number of $\tsu$ operators, consistently with the signature symmetry (recall that $\MMtree$ is signature-odd). It can be shown (see eq.~\eqref{eq:non-planar-arg} in section \ref{sec:col4}) that colour operators involving commutators correspond to non-planar diagrams, i.e., to subleading contributions in the large-$N_c$ limit. This allows one to conclude that, starting from two-loops, the two-Reggeon MRS contribution to the NLL amplitude is non-planar and, thus, also consistent with the known diagrammatic origin of Regge cuts~\cite{Mandelstam:1963cw,Eden:1966dnq,Collins:1977jy}.

\section{From rapidity evolution to the NNLL odd amplitude} \label{sec:JIMWALK_to_NNLL}

The calculation of the odd-signature amplitude at NNLL accuracy is particularly interesting, because at this logarithmic accuracy one faces for the first time the contribution of a Regge cut in the real part of the amplitude \cite{DelDuca:2001gu,DelDuca:2013ara,DelDuca:2014cya,Caron-Huot:2013fea,Fadin:2016wso,Fadin:2017nka}. This phenomenon is associated with the exchange of three Reggeons, as shown by direct calculations to three loops \cite{Caron-Huot:2017fxr,Fadin:2016wso}. By using the Balitsky-JIMWLK evolution equation which dictates the evolution of one- and three-Reggeon states, we determine here the general form of the entire NNLL tower, ${\mm}^{(-,n,n-2)}$.

\subsection{Amplitudes via Balitsky-JIMWLK evolution}
\label{BFKL_abridged}

As discussed in refs.~\cite{Caron-Huot:2013fea,Caron-Huot:2017fxr}, two-parton scattering in the high-energy limit is conveniently described within the shockwave formalism: fast particles moving in the $(+)$ lightcone direction, to which we refer to as the \emph{projectile}, scatter against fast particles moving in the $(-)$ lightcone direction, referred to as the \emph{target}. The fast projectile $\ket{\psi_i}$ appears as a set of infinite Wilson lines \cite{Korchemskaya:1994qp,Korchemskaya:1996je} at transverse position $z_k \equiv x_{k\perp}$, crossing the target $\bra{\psi_j}$ (seen as a ``shockwave'') at $x_- = 0$: one has  
\beq
\ket{\psi_i} \sim U(z_1) \otimes \cdots \otimes U(z_n),
\eeq
where $\otimes$ denote convolution, and a Wilson line at transverse position $z$ reads 
\beq
\label{Udef}
U(z) = \mathcal{P}\exp\!\left[ig_s \mathbf{T}^a\!\!\int_{-\infty}^{+\infty}\!\!\!dx^+ A^a_+(x^+,x^-\!=\!0,z) \right].
\eeq
In perturbation theory the unitary matrices $U(z)$ are close to the identity, therefore they can be parameterised in terms of a colour-adjoint field $W^a$:
\beq
U(z) = e^{ig_s\,T^aW^a(z)}\,, \label{Uparam}
\eeq
which is identified as a source for Reggeised gluons. The projectile and target are thus expanded in Reggeon fields:
\beq
\label{sum_of_n_Reggeons}
\ket{\psi_i} =\sum_{n=1}^{\infty}\ket{\psi_{i,n}},
\eeq
where $\ket{\psi_{i,n}}$ represents a state of $n$ Reggeons. In turn, the latter has an expansion in powers of the strong coupling: 
\begin{align}
  \begin{split}
  \label{eq:ketexp}
 \ket{\psi_{i,n}} =\ket{\psi_{i,n}}^{\text{LO}}
 +\sum_{k=1}^{\infty}
 \ket{\psi_{i,n}}^{\text{N}^k\text{LO}}\,,
  \end{split}
\end{align}
which accounts for the non-trivial dynamics of the projectile. The leading-order projectile is given in terms of a single Wilson line: in momentum space one has  
\beq \label{simple_eikonal}
\ket{\psi_i}^{\LO} = U_i(p), 
\eeq
where $U_i(p)$ is defined by means of Fourier transformation: 
\beq\label{Utransform}
U_i(p) = \int [dz] \,e^{-ipz}\,U_i(z), \qquad
U_i(z) = \int [\dbar p] \,e^{ipz}\, U_i(p),
\eeq
where $U_i(z)$ is given by eq.~(\ref{Udef}) with the generator $\mathbf{T}$ identified as $\mathbf{T}_i$, and we have further defined the transverse-space integration measures
\beq
[dz]\equiv d^{2{-}2\eps}z, \qquad {\rm and }
\qquad 
[\dbar p] \equiv \frac{d^{2-2\eps} p}{(2\pi)^{2-2\eps}}.
\eeq 
Upon expansion at fixed order in the strong coupling, this gives states with 
a definite number of Reggeons, which, according to eq.~\eqref{eq:ketexp} read \cite{Caron-Huot:2017fxr}:
\begin{subequations}
\label{123Reggeons} 
\beqa 
 \ket{\psi_{i,1}}^{\text{LO}} &=& ig_s\,\T_i^a \, W^a(p), \\
 \ket{\psi_{i,2}}^{\text{LO}} &=& - \frac{g_s^2}{2} \T_i^a\T_i^b \int [\dbar q] \, W^a(q)W^b(p{-}q), \\ 
 \ket{\psi_{i,3}}^{\text{LO}} &=& -\, \frac{ig_s^3}{6} \T_i^a\T_i^b\T_i^c 
\int [\dbar q_1] [\dbar q_2] \, W^a(q_1)W^b(q_2)W^c(p{-}q_1{-}q_2),
\eeqa
\end{subequations}
where the Reggeon fields $W^a(p)$ in momentum space are defined as
\beq\label{Wtransform}
W^a(p) = \int [dz] \,e^{-ipz}\,W^a(z), \qquad
W^a(z) = \int [\dbar p] \,e^{ipz}\, W^a(p).
\eeq
The non-trivial dependence on the transverse coordinates 
is parameterised in terms of impact factors: at NLO one has for instance
\begin{subequations}
\beqa \label{NLO_wavefunction} 
\ket{\psi_{i,1}}^{\text{NLO}} &=& ig_s\,\T_i^a \, W^a(p)\,
 \frac{\as}{\pi} D_i^{(1)}(p), \\
\ket{\psi_{i,2}}^{\text{NLO}} &=& 
- \frac{g_s^2}{2} \T_i^a\T_i^b \int [\dbar q] 
\, \frac{\as}{\pi} \psi^{(1)}_i(p,q) \, W^a(q)W^b(p{-}q),
\eeqa
\end{subequations}
while at NNLO the single-Reggeon wavefunction is given by
\beq \label{NNLO_wavefunction}
\ket{\psi_{i,1}}^{\text{NNLO}} = ig_s\, \T_i^a \, W^a(p)\, 
\left(\frac{\as}{\pi}\right)^2 D_i^{(2)}(p). 
\eeq
In these equations, $D_i^{(n)}(p)$ defines the single-Reggeon impact factors at $n$-th order in perturbation theory, while $\psi^{(n)}_i(p,q)$ represent a two-Reggeon impact factor. On a technical note, let us remark that the wavefunctions $\ket{\psi_{i}}$ defined above have the meaning of ``collinear subtracted'' wavefunctions, i.e. they describe the effective coupling for the emission of a given number of Reggeons, in which collinear 
divergences have been subtracted according to $\ket{\psi_{i}} = Z_{i}^{-1} \ket{\chi_{i}}$,
where $\ket{\chi_{i}}$ represents the full effective projectile-Reggeon(s) coupling 
with unsubtracted collinear singularities. For this reason, the single-Reggeon impact factors $D^{(n)}_i$ in eqs.~(\ref{NLO_wavefunction}) and~(\ref{NNLO_wavefunction}) coincide with the collinear-subtracted impact factors introduced in eq.~\eqref{Regge-Pole-General}. 
We shall determine their values by matching matrix elements involving $\ket{\psi_{i,1}}^{\rm{NLO}}$ and $\ket{\psi_{i,1}}^{\rm{NNLO}}$, defined in eqs.~(\ref{NLO_wavefunction}) and~(\ref{NNLO_wavefunction}), to the full amplitude at one and at two loops, respectively.

The $n$-Reggeon states $\ket{\psi_{i,n}}$ depend on the transverse momenta of the Reggeons, but not on the centre-of-mass energy. In this formulation the energy dependence enters through the fact that infinite Wilson lines develop rapidity divergences. These may be regulated by introducing a cutoff, which can be identified with $L$ of eq.~(\ref{eq:siglog})~\cite{Caron-Huot:2013fea}. The dependence on~$L$ is encoded in a rapidity evolution equation for the projectile (and the target)
\beq \label{rapidity_evolution}
-\frac{d}{dL}\,\ket{\psi_i} = H\, \ket{\psi_i},
\eeq
where $H$ is the Balitsky-JIMWLK Hamiltonian \cite{Balitsky:1995ub,Kovchegov:1999yj,JalilianMarian:1996xn,JalilianMarian:1997gr,Iancu:2001ad}. A key feature of eq.~\eqref{rapidity_evolution} is the non-linearity of $H$: evolution of the full projectile $\ket{\psi_i} \sim U(z_1) \otimes \cdots \otimes U(z_n)$ generates an increasing number of Wilson lines $U(z_j)$, eventually leading to the phenomenon of gluon saturation. However, in applications to partonic scattering processes as considered in this paper, it is appropriate to take the limit of dilute projectile and target~\cite{Caron-Huot:2013fea,Caron-Huot:2017fxr}, in which case the Balitsky-JIMWLK Hamiltonian acts on states with a given number of Reggeon fields~$W$. In this perturbative regime, $H$ is diagonal to leading order in $g_s^2$; the non-linearity of $H$ manifests itself at higher orders in the coupling, producing transitions between states with different number of Reggeon fields: a transition $k \to k+2n$ is of order $g_s^{2(1+n)}$. Also, note that, as a consequence of the signature symmetry, only transitions of the type $k \to k \pm 2$ are allowed; transitions of the type $k \to k \pm 1$ are forbidden, because they would induce mixing between the even and odd parts of the amplitude. In short, the r.h.s of eq.~\eqref{rapidity_evolution} has the form
\beqa \label{Hamiltonian_schematic_form}
H  \left( 
\begin{array}{c}
  W      \\  WW     \\  WWW  \\  \cdots
\end{array}
\right) &\equiv&
\left(
\begin{array}{cccc}
 H_{1{\to}1} & 0  & H_{3{\to}1} & \ldots \\
 0 & H_{2{\to}2}  & 0  &  \ldots \\
 H_{1{\to}3} & 0  & H_{3{\to}3} & \ldots\\
 \cdots & \cdots  & \cdots & \cdots\\
\end{array}
\right)\left(
\begin{array}{c}
  W      \\  WW     \\  WWW  \\  \cdots\end{array}
\right)
\nn\\&\simeq&
\left(
\begin{array}{cccc}
 g_s^2 & 0  & g_s^4 & \ldots \\
 0 & g_s^2  & 0  &  \ldots \\
 g_s^4 & 0  & g_s^2 & \ldots\\
 \cdots & \cdots  & \cdots & \cdots\\
\end{array}
\right) \left(
\begin{array}{c}
  W      \\  WW     \\  WWW  \\  \cdots\end{array}
\right) 
\eeqa
where the non-vanishing entries in the second line display the perturbative order at which each transition Hamiltonian $H_{k\to l}$ starts to contribute.
We will see in section~\ref{sec:equation12} that for the calculation of the odd amplitude at NNLL accuracy, only the $k \to k$, $1\to 3$, and $3 \to 1$ transitions Hamiltonians are needed. 
They have been  calculated \cite{Caron-Huot:2017fxr}: the diagonal $k \to k$ transitions, with $k >1$ are entirely given in terms of the leading-order BFKL equation, while the 
$1\to 3$, and $3 \to 1$ transitions are extracted from the leading-order Balitsky-JIMWLK equation. In momentum space the $k \to k$ Hamiltonian reads
\beq \label{ktok_momentum_space} 
H_{k{\to}k} = A_{k\to k} + B_{k\to k},
\eeq
where 
\beqa \label{ktok_momentum_space11} 
A_{k\to k} &=& -\int [dp] \, C_A \, \alpha_g(p^2,\mu^2) \,
W^a(p)\frac{\delta}{\delta W^a(p)}, \\[0.2cm] \nn
\label{ktok_momentum_space33}
B_{k\to k} &=& \as(\mu^2)\int [\dbar q][dp_1][dp_2]\, 
H_{22}(q;p_1,p_2) \,W^x(p_1{+}q)W^y(p_2{-}q) \\ 
&&\hspace{4.0cm}\cdot \, 
(F^xF^y)^{ab}\frac{\delta}{\delta W^a(p_1)}
\frac{\delta}{\delta W^b(p_2)},
\eeqa
with $(F^{x})^{ab}\equiv if^{axb}$ and the kernel $H_{22}(q;p_1,p_2)$ reads
\beq \label{Ham22}
H_{22}(q;p_1,p_2) = \frac{(p_1+p_2)^2}{p_1^2p_2^2}
-\frac{(p_1+q)^2}{p_1^2q^2}-\frac{(p_2 - q)^2}{q^2p_2^2}. 
\eeq
Notice that the form of $H_{22}$ implies 
that $B_{k{\to}k}$ is scale-invariant in exactly two transverse dimensions. 
Indeed, this is a consequence of the fact that the evolution 
of multiple Reggeon states up to NNLL accuracy is completely 
determined by the LO Balitsky-JIMWLK Hamiltonian, which is 
itself invariant under conformal transformations 
\cite{Caron-Huot:2013fea}. This simple consideration 
allows us to anticipate that also the other terms 
in eq.~\eqref{Hamiltonian_schematic_form}, such as 
$H_{1\to 3}$ and $H_{3\to 1}$, are scale invariant.
While we can rely on this property for the calculation 
of the odd amplitude at NNLL accuracy, let us remark that 
the same will not be true for the even amplitude at 
NNLL accuracy, which requires the Balitsky-JIMWLK 
Hamiltonian at NLO,~\cite{Balitsky:2013fea,Kovner:2013ona,Kovner:2014xia,Kovner:2014lca,Caron-Huot:2015bja,Lublinsky:2016meo}, for which 
scale-invariance is broken.

In momentum space the $1\to 3$ Hamiltonian reads 
\beqa \label{Ham13} \nn
H_{1{\to}3} &=& \as^2(\mu^2)\int [\dbar p_1][\dbar p_2][dp] \,
{\rm Tr}[F^aF^bF^cF^d] \,  \\
&&\hspace{2.0cm}\cdot \,W^b(p_1)W^c(p_2)W^d(p_3) 
\, H_{13}(p_1,p_2,p_3)\, \frac{\delta}{\delta W^a(p)},
\eeqa
where $p_3=p-p_1-p_2$ and the kernel is
\beqa \label{H13kin} \nn
H_{13}(p_1,p_2,p_3) &=&
\frac{2\pi}{3}S_\eps(\mu^2)\int [\dbar q]\bigg[
\frac{(p_1{+}p_2)^2}{q^2(p_1{+}p_2{-}q)^2} 
+\frac{(p_2{+}p_3)^2}{q^2(p_2{+}p_3-q)^2} \\
&&\hspace{3.0cm}-\, 
\frac{(p_1{+}p_2{+}p_3)^2}{q^2(p_1{+}p_2{+}p_3{-}q)^2}
-\frac{p_2^2}{q^2 (p_2{-}q)^2} \bigg] \\ \nn
&=& \frac{r_\Gamma}{3\eps} \left[
\left(\frac{\mu^2}{(p_1{+}p_2{+}p_3)^2}\right)^{\eps}
+\left(\frac{\mu^2}{p_2^2}\right)^{\eps} 
 -\left(\frac{\mu^2}{(p_1{+}p_2)^2}\right)^{\eps}
-\left(\frac{\mu^2}{(p_2{+}p_3)^2}\right)^{\eps} 
\right],
\eeqa
where $S_\eps(\mu^2)=\left(\frac{\mu^2}{4\pi e^{-\gamma_E}}\right)^{\eps}$.
The $3 \to 1$ transition can be obtained by symmetry, by requiring that the two matrix elements 
$$ 
\big(\bra{WWW}H\big)\ket{W}=\bra{WWW}\big(H\ket{W}\big),
$$
are equal, yielding~\cite{Caron-Huot:2017fxr}
\beqa\label{Ham31}\nn
H_{3{\to}1} &=& \as^2(\mu^2)\int [dp_1][dp_2][dp_3] \,
{\rm Tr}[F^aF^bF^cF^d]\, W^d(p_1{+}p_2{+}p_3)\,  \\ 
&&\hspace{1.0cm}\times 
(-1)\frac{(p_1{+}p_2{+}p_3)^2}{p_1^2p_2^2p_3^2}
\, H_{13}(p_1,p_2,p_3) \, \frac{\delta}{\delta W^a(p_1)}\frac{\delta}{\delta W^b(p_2)}
\frac{\delta}{\delta W^c(p_3)}. 
\eeqa

To summarise, a scattering amplitude in the high-energy limit is given as an expectation value between states with fixed number of Reggeon fields $W$ evolved to equal rapidity, by means of eq.~\eqref{rapidity_evolution}:
\beq\label{eq:ampEqualsPsi}
\frac{i(Z_iZ_j)^{-1}}{2s}\mathcal{M}_{ij\to ij} = \braket{\psi_j|e^{-HL}|\psi_i}.
\eeq
Note that this equation takes into account the fact that collinear divergences are removed from the projectile and the target, as discussed following eq.~\eqref{NNLO_wavefunction}. Considering the reduced amplitude introduced in eq.~(\ref{Mreduced}), rapidity evolution is governed by a suitably-modified \emph{reduced} Hamiltonian. Starting from eq.~(\ref{Mreduced}) and using the 
amplitude definition in eq.~\eqref{eq:ampEqualsPsi} we define
\beq \label{eq:MhatDef}
\frac{i}{2s}\Mreduced = e^{-\,\T_t^2\, \alpha_g(t) \, L}
\braket{\psi_j|e^{-H L}|\psi_i} \equiv 
\braket{\psi_j|e^{-\hat H L}|\psi_i}.
\eeq
The explicit form of $\hat H$ is found by noticing that the Hamiltonian $H$ in eq.~\eqref{eq:MhatDef} is given by a matrix operator in the space labelled by the Reggeon number, as defined in eq.~\eqref{Hamiltonian_schematic_form}. In this space, the subtraction term $\T_t^2\, \alpha_g $ must be interpreted as a diagonal matrix, such that one has \beq\label{eq:HhatH}
{\hat{H}}_{k\to k+2n} = H_{k\to k+2n} +\delta_{n0}\tts\al_g(t).
\eeq

After evolution has been performed, one is left with an inner product between states with a fixed number of Reggeon fields at equal rapidity. Such matrix elements must be evaluated in full QCD, as discussed in section 2.3 of ref.~\cite{Caron-Huot:2017fxr}. Following refs.~\cite{Caron-Huot:2013fea,Caron-Huot:2017fxr} we \emph{define} the Reggeon propagator as 
\beq \label{inner_product_oneW}
G_{11'} \equiv \langle W_{1} | W_{1'}\rangle 
= i\,\frac{\delta^{a_1a_1'}}{p_1^2}\,
\delta^{(2-2\eps)}(p_1-p_1')\,,
\eeq
with no ${\cal O}(g_s^2)$ corrections, along with a vanishing overlap between states with different numbers of Reggeon at equal rapidity, namely, e.g.,
\beqa\label{3WtoW-Wto3W}
\langle W_1W_2W_3 | W_4 \rangle =
\langle W_4 | W_1W_2W_3 \rangle  = 0\,. 
\eeqa    
While other definitions of the inner product are possible, including higher-order corrections in eqs.~(\ref{inner_product_oneW}) and~(\ref{3WtoW-Wto3W}) would also result in a modification of the impact factors. Given that the latter are determined by matching matrix elements involving eqs.~(\ref{NLO_wavefunction}) and~(\ref{NNLO_wavefunction}) to the full amplitude, the final result would not be altered by such alternative definitions. 

Having fixed the inner product at equal rapidity, multi-Reggeon correlators are simply obtained by Wick contractions:
\beqa \label{inner_product}\nn
\langle W_{1}W_{2} | W_{1'}W_{2'}\rangle  &=& 
G_{11'}G_{22'}+G_{12'}G_{21'} + {\cal O}(g_s^2), \\ \nn
\langle  W_{1}W_{2}W_3 | W_{1'}W_{2'}W_{3'}\rangle 
&=& G_{11'}G_{22'}G_{33'}+\mbox{(5 permutations)}
+{\cal O}(g_s^2), \\ 
&&\mbox{etc.}
\eeqa
We now have all the definitions necessary to derive an all-order expression for the odd amplitude at NNLL accuracy, to be discussed in the next section.

\subsection{The odd reduced amplitude at \texorpdfstring{$n$}{n} loops}
\label{sec:equation12}

We are now ready to derive the general form of the all-order NNLL reduced amplitude by expanding eq.~(\ref{eq:MhatDef}). First of all, we notice that, as discussed after eq.~\eqref{rapidity_evolution}, the even and odd amplitudes are orthogonal under the action of $\hat H$, as a consequence of signature symmetry. Therefore eq.~(\ref{eq:MhatDef}) splits as 
\beq \label{Regge-odd-Even-Amplitude}
\frac{i}{2s}\Mreduced_{ij\to ij} \xrightarrow{{\rm Regge}}
\frac{i}{2s}\left(\Mreduced^{(+)}_{ij\to ij} + 
\Mreduced^{(-)}_{ij\to ij} \right) 
\equiv \bra{\psi^{(+)}_j}e^{-\hat H L}\ket{\psi^{(+)}_i}
+\bra{\psi^{(-)}_j}e^{-\hat H L}\ket{\psi^{(-)}_i},
\eeq
where the notation $\ket{\psi^{(+)}_i}$, $\ket{\psi^{(-)}_i}$ indicates a restriction to states with an even or odd number of Reggeons $W$, respectively, see eq.~\eqref{123Reggeons}. In this paper we focus on the odd contribution to the amplitude at NNLL, i.e. we consider directly the second term in \eqref{Regge-odd-Even-Amplitude}, which involves only states with an odd number of Reggeons. It is useful to expand this term explicitly for the first few orders. At leading order in the strong coupling one obviously has
\beq
\label{3loop_odd_bfkl-0}
\frac{i}{2s} \Mreduced_{ij\to ij}^{(-)\, \textrm{tree}} =
\langle \psi_{j,1} | \psi_{i,1}\rangle^{\rm LO},
\eeq
where the superscript LO indicates that the target and projectile are taken at leading order, as defined in eq.~\eqref{123Reggeons}. Next, the expansion at first and second order gives 
\begin{subequations}
\label{3loop_odd_bfkl-12}
\begin{align}
\label{3loop_odd_bfkl-1}
\frac{i}{2s} \Mreduced_{ij\to ij}^{(-)\, \textrm{1-loop}}\, =& \,
- L  \bra{\psi_{j,1}}\hat H_{1{\to1}}\ket{\psi_{i,1}}^{\LO}
+\langle \psi_{j,1} | \psi_{i,1}\rangle^{\rm NLO}, \\[0.3cm] \nn
\label{3loop_odd_bfkl-2}
\frac{i}{2s} \Mreduced_{ij\to ij}^{(-)\, \textrm{2-loops}} \, =& \,
\frac{L^2}{2} \bra{\psi_{j,1}}(\hat H_{1{\to}1})^2\ket{\psi_{i,1}}^{\rm LO}
- L \bra{\psi_{j,1}}\hat H_{1{\to1}}\ket{\psi_{i,1}}^{\rm NLO} \\ 
&+\, \langle \psi_{j,3}| \psi_{i,3} \rangle^{\rm LO} 
+\langle \psi_{j,1} | \psi_{i,1}\rangle^{\rm NNLO}\,,
\end{align}
\end{subequations}
where the superscripts NLO and NNLO indicate that one needs to take expectation values at NLO and NNLO, respectively, using the expressions for the target and projectile wavefunction provided in eqs.~\eqref{NLO_wavefunction} and~\eqref{NNLO_wavefunction}.
Here, for illustrative purposes, we have listed all terms that follows from the formal expansion of the second term in eq.~\eqref{Regge-odd-Even-Amplitude}, taking into account eq.~\eqref{3WtoW-Wto3W}, but without any specific assumption about $\hat H$. However, we recall that the $1\to 1$ transition is given, according to eqs.~\eqref{ktok_momentum_space} and~\eqref{ktok_momentum_space11}, by the Regge trajectory $H_{1{\to}1} = - C_A \, \alpha_{g}(t)$. Inspecting eq.~\eqref{eq:HhatH} we see that 
$\hat H_{1{\to}1}= \left(\tts - C_A \right)\, \alpha_{g}(t)$
and then we deduce
\beq
\label{Hhat-LL}
\bra{\psi_{j,1}}\hat H_{1{\to}1}O\ket{\psi_{i,n}}=\bra{\psi_{j,n}}O\hat H_{1{\to}1}\ket{\psi_{i,1}} = 0,
\eeq
where $O$ is a general operator mediating between single and $n$-Reggeon states. We use the fact that any transition between these states is proportional (in colour space) to the tree-level (which we explain in section~\ref{sec:col3}) and that we can commute $\tts$ through the bracket and apply eq.~(\ref{ttsToCA}). Thus, eq.~\eqref{3loop_odd_bfkl-12} simplifies to
\begin{subequations}
\label{3loop_odd_bfkl-12_simp}
\begin{align}
\label{3loop_odd_bfkl-1_simp}
\frac{i}{2s} \Mreduced_{ij\to ij}^{(-)\, \textrm{1-loop}}\, =& \,
\langle \psi_{j,1} | \psi_{i,1}\rangle^{\rm NLO}, \\[0.3cm] 
\label{3loop_odd_bfkl-2_simp}
\frac{i}{2s} \Mreduced_{ij\to ij}^{(-)\, \textrm{2-loops}} \, =& \,
 \langle \psi_{j,3}| \psi_{i,3} \rangle^{\rm LO} 
+\langle \psi_{j,1} | \psi_{i,1}\rangle^{\rm NNLO}.
\end{align}
\end{subequations}
Taking now into account eq.~\eqref{3WtoW-Wto3W} as well as eq.~\eqref{Hhat-LL}, expansion at third order gives 
\beqa \nn
\label{3loop_odd_bfkl-3}
\frac{i}{2s} \Mreduced_{ij\to ij}^{(-)\, \textrm{3-loops}} \, &=& \,
-L \Big[\bra{\psi_{j,3}} \hat H_{3{\to}3}\ket{\psi_{i,3}}
+\bra{\psi_{j,3}} \hat H_{1{\to}3}\ket{\psi_{i,1}}  
+\bra{\psi_{j,1}} \hat H_{3{\to}1}\ket{\psi_{i,3}} \Big]^{\rm LO} \\[0.1cm] 
&&\hspace{1.0cm} +\, \langle \psi_{j,3} | \psi_{i,3}\rangle^{\rm NLO}
+\langle \psi_{j,1} | \psi_{i,1}\rangle^{\rm (N^3LO)}.
\eeqa

Here we are interested to determine the NNLL contribution to $\Mreduced_{ij\to ij}^{(-)}$ to all orders. These terms are of order $\as^n L^{n-2}$, thus their contribution starts at two loops. Inspecting eq.~\eqref{3loop_odd_bfkl-12_simp} and~\eqref{3loop_odd_bfkl-3} we see that the NNLL terms read 
\begin{subequations}
\label{3loop_odd_bfkl-nnll}
\begin{align}
\label{3loop_odd_bfkl-nnll-2}
\frac{i}{2s} \Mreduced_{ij\to ij}^{(-,2,0)} \, =& \,
 \langle \psi_{j,3}| \psi_{i,3} \rangle^{\rm LO} 
+\langle \psi_{j,1} | \psi_{i,1}\rangle^{\rm NNLO}, \\[0.2cm]
\label{3loop_odd_bfkl-nnll-3}
\frac{i}{2s} \Mreduced_{ij\to ij}^{(-,3,1)} \, =& \,
- \Big[\bra{\psi_{j,3}} \hat H_{3{\to}3}\ket{\psi_{i,3}} 
+\bra{\psi_{j,3}} \hat H_{1{\to}3}\ket{\psi_{i,1}}  
+\bra{\psi_{j,1}} \hat H_{3{\to}1}\ket{\psi_{i,3}} \Big]^{\rm LO},
\end{align}
\end{subequations}
where we used the notation for the amplitude coefficients introduced in eq.~\eqref{eq:expansionDef}. In general, at $n$-loop, the NNLL contribution to the odd amplitude is proportional to $L^{n-2}$, which is obtained from $n-2$ repeated actions of the Hamiltonian $H$ in eq.~(\ref{eq:MhatDef}). The diagonal transitions $H_{k\to k}$ in eq.~(\ref{ktok_momentum_space}) are $\mathcal{O}(\al_s)$ while next-to-diagonal transitions $H_{k\to k\pm 2n}$ with $n=1$ in eq.~(\ref{Ham13}) are $\mathcal{O}(\al_s^2)$. Noting that $\ket{\psi_{i,n}}$ is $\mathcal{O}(g_s^{n-1})$ we immediately find four different types of contributions of the form $\mathcal{O}(\al_s^{n}L^{n-2})$
\begin{align}\label{braket1331}
\begin{array}{cc}
\braket{\psi_{j,3}| \hat H_{3\to3}^k|\psi_{i,3}}, & 
\qquad\qquad 
\braket{\psi_{j,1}| \hat H_{3\to1}\, \hat H_{3\to3}^{k-2}\, 
\hat H_{1\to3}|\psi_{i,1}}, \\[0.2cm]
\braket{\psi_{j,1}| \hat  H_{3\to1}\, \hat  H_{3\to3}^{k-1}|\psi_{i,3}}, & 
\qquad\qquad 
\braket{\psi_{j,3}| \hat H_{3\to3}^{k-1} \, \hat H_{1\to3}|\psi_{i,1}}.
\end{array}
\end{align}
These constitute the NNLL tower. The two transitions in the second line of eq.~(\ref{braket1331}) are related by target-projectile symmetry. 

In order to express the contributions due to eq.~\eqref{braket1331} in a compact form, we find useful to make the power-counting in the strong coupling manifest, by introducing normalised Reggeon states $\ket{i_n}$ and a normalised reduced Hamiltonian $\tilde{H}$, defined as 
\begin{align}\label{eq:newNormalisations}
\begin{split}
(r_\Gamma\al_s)^{(n-1)/2} \,\ket{i_n} \, &\equiv\ket{\psi_{i,n}},
\\
\left(\frac{\al_s}{\pi} r_\Gamma\right)^{1+n} \,\pi^n
\, {\tilde H}_{k\to k+2n}  
\, & \equiv 
\, {\hat{H}}_{k\to k+2n} = H_{k\to k+2n} +\delta_{n0}\tts\al_g(t) \,,
\end{split}
\end{align}
where we introduce rescalings by the constant $r_\Gamma$ defined in eq.~(\ref{r_gamma}) and we recall the relation between $\hat{H}$ and $H$ according to eq.~(\ref{eq:HhatH}).
In terms of these quantities the all-order contribution to the odd amplitude at NNLL accuracy reads  
\begin{align}
\begin{split}
\label{eq:mhatNNLL}
\frac{i}{2s}\hat{\mm}^{(-),\text{NNLL}}_{ij\to ij} =& \,
\pi^2 \Bigg[\sum_{\ell=0}^{\infty}\frac{(-L)^\ell}{\ell!}
\left(\frac{\al_s}{\pi}r_\Gamma \right)^{\ell+2}
\braket{j_{3}|\tilde{H}_{3\to3}^\ell|i_{3}}  \\ 
&\hspace{-1.6cm}+\, \sum_{\ell=1}^{\infty}
\frac{(-L)^{\ell}}{\ell!}\left(\frac{\al_s}{\pi}
r_\Gamma\right)^{\ell+2} \left[
\braket{j_{1}|\tilde{H}_{3\to1}\tilde{H}_{3\to3}^{\ell-1}|i_{3}}
+\braket{j_{3}|\tilde{H}_{3\to3}^{\ell-1}\tilde{H}_{1\to3}|i_{1}}\right] \\ 
&\hspace{-1.6cm} +\, \sum_{\ell=2}^{\infty}
\frac{(-L)^{\ell}}{\ell!}\left(\frac{\al_s}{\pi} r_\Gamma\right)^{\ell+2}
\braket{j_{1}|\tilde{H}_{3\to1}\tilde{H}_{3\to3}^{\ell-2}\tilde{H}_{1\to3}|i_{1}}
\Bigg]^{\text{LO}}\!\!\!\!
+\left(\frac{\alpha_s}{\pi}\right)^2
\braket{j_{1}|i_{1}}^{\text{NNLO}}\!\!.
\end{split}    
\end{align}
One can check that this expression gives eqs.~\eqref{3loop_odd_bfkl-nnll-2} and~\eqref{3loop_odd_bfkl-nnll-3}, at two- and three-loops. It is interesting to notice that the whole NNLL tower is described in terms of the leading-order formalism, with the sole exception of the two-loop contribution $\braket{j_{1}|i_{1}}^{\text{NNLO}}$. The latter has been extracted from two-loop amplitudes in ref.~\cite{Caron-Huot:2017fxr}, and does not enter the calculation of reduced amplitudes to higher orders. All remaining $\mathrm{NNLL}$ terms are found by expanding the projectile and the target in states with either one or three Reggeons, according to eqs.~(\ref{sum_of_n_Reggeons}) and (\ref{123Reggeons}), and applying the Hamiltonians $\tilde{H}_{3\to3}$, $\tilde{H}_{3\to1}$ and $\tilde{H}_{1\to3}$, given in  eq.~(\ref{eq:newNormalisations}) and with $H_{k\to k+2n}$ of section~\ref{BFKL_abridged}.
Finally, one is left with the inner product of Reggeon fields at leading order, which is evaluated as in eq.~(\ref{inner_product_oneW}). Apart from $\braket{j_{1}|i_{1}}^{\text{NNLO}}$, eq.~(\ref{eq:mhatNNLL}) is therefore universal, in that it applies in any gauge theory, fully governed by the \emph{leading order} Balitsky-JIMWLK evolution of infinite Wilson lines in eq.~(\ref{rapidity_evolution}).

Eq.~\eqref{eq:mhatNNLL} has characteristic analytic properties, that are interesting to highlight at this point. First of all, notice that the rescaling adopted in eq.~\eqref{eq:newNormalisations} is particularly useful, because it manifests the fact that (with the exception of the two-loop single-Reggeon contribution) the NNLL amplitude is proportional to a factor of $\pi^2$ to all orders in perturbation theory. Furthermore, owing to the form of the Hamiltonian presented in 
section~\ref{BFKL_abridged}, the amplitude has maximal, uniform transcendental weight, when the dimensional regularisation parameter $\epsilon$ is counted as having transcendental weight $-1$. This will become apparent in sections \ref{subsec:integrals} and~\ref{sec:results}, where explicit results for the loop integrals and matrix elements involved in the calculation of eq.~\eqref{eq:mhatNNLL} at fixed order will be presented.

\section{Computation of the NNLL transition amplitudes}
\label{sec:computation}
In this section we expand eq.~(\ref{eq:mhatNNLL}) up to four loops and compute the transition amplitudes explicitly. Each contribution to the transition amplitudes has a transparent diagrammatic interpretation, presented in sections~\ref{subsubsec:Mhat2} through \ref{subsubsec:Mhat4} for two, three and four loops, respectively. Subsequently, we will discuss the calculation of the colour structure and of the integrals associated to each diagram, respectively in section~\ref{subsec:colour} and in section~\ref{subsec:integrals}. 

\subsection{Diagrammatic interpretation of the transition amplitudes\label{subsec:dia}}

\subsubsection{Two loops\label{subsubsec:Mhat2}}

The two-loop transition amplitude is given in eq.~\eqref{3loop_odd_bfkl-2}. Here we focus on the $3\to 3$ transition, which is obtained by contracting the Reggeons emitted by target and projectile, without any action of the Hamiltonian operator
\begin{equation}
\label{eq:amp33_2loops}
\langle \psi_{j,3}  | \psi_{i,3} \rangle = i\,g_s^6\,\mathbf{C}_{33}^{(2)}\,S_\epsilon^2(\mu^2)\,\int\,\frac{[\dbar k_1\dbar k_2\dbar k_3]_\delta}{k_1^2 k_2^2 k_3^2},
\end{equation}
where an $\overline{{\rm MS}}$ scale factor $S_\eps(\mu^2)=\left(\frac{\mu^2}{4\pi e^{-\gamma_E}}\right)^{\eps}$ is introduced for each loop.
The coupling constant $g_s$ is evaluated in the $\overline{\rm MS}$ scheme at the renormalisation scale $\mu^2$ and in this section, whenever not specified, both $g_s$ and $\alpha_s$ are assumed to be defined at~$\mu^2$. In eq.~(\ref{eq:amp33_2loops}) we define the integration measure
\begin{equation}
    [\dbar k] = \frac{d^{d-2}k}{(2\pi)^{d-2}},\qquad [\dbar k_1\dbar k_2\dbar k_3]_\delta = [\dbar k_1][\dbar k_2][\dbar k_3]\,(2\pi)^{d-2}\,\delta^{(d-2)}(p-k_1-k_2-k_3).
\end{equation}
The colour factor of eq.~(\ref{eq:amp33_2loops}) is
\begin{equation}
\label{eq:colA332}
\mathbf{C}_{33}^{(2)}=\frac{1}{6}\left(\mathbf{T}^{\{a,b,c\}_+}\right)_i\left(\mathbf{T}^{\{a,b,c\}_+}\right)_j,
\end{equation}
where we define the shorthand notation $\mathbf{T}^{a_1}_i\dots\mathbf{T}^{a_n}_i = \left(\mathbf{T}^{a_1,\dots, a_n}\right)_i$ and
\begin{equation}\label{eq:Tsym}
    \left(\mathbf{T}^{\{a_1,\dots,a_n\}_+}\right)_i =\{\mathbf{T}^{a_1}_i,\mathbf{T}^{a_2}_i,\dots,\mathbf{T}^{a_n}_i\}_+= \frac{1}{n!}\,\sum_{\sigma\in S_n}\,\mathbf{T}_i^{a_{\sigma(1)}}\dots\mathbf{T}_i^{a_{\sigma(n)}}.
\end{equation}
\begin{figure}[htbp]
    \centering
    \includegraphics[scale=1.2]{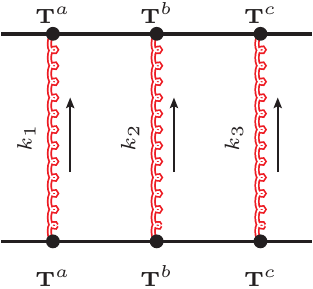}
    \caption{A two-loop diagram representing the leading contribution to the amplitude due to triple Reggeon exchange. Note that the diagram drawn is just one of the six distinct orderings in eq.~(\ref{eq:colA332}).  }
    \label{fig:h33_2}
\end{figure}
The integrand appearing in eq.~(\ref{eq:amp33_2loops}) is constructed with three free Reggeon propagators, with a fixed total transverse momentum $p$. Therefore, the transition amplitude in eq.~(\ref{eq:amp33_2loops}) is represented diagrammatically in figure~\ref{fig:h33_2} by exchanges of three Reggeons. These need to be taken in all possible orderings. Finally, in the notation of 
eq.~(\ref{eq:mhatNNLL}) we have 
\begin{equation}
    \langle j_3|i_3\rangle = \frac{1}{\alpha_s^2r_\Gamma^2}\, \langle \psi_{j,3}|\psi_{i,3}\rangle\,,
\end{equation}
which fixes the leading-order ($\ell=0$) contribution to this tower of logarithms. 
\subsubsection{Three loops\label{subsubsec:Mhat3}}

According to eq.~(\ref{eq:mhatNNLL}), three-loop NNLL transitions either involve the evolution of three-Reggeon states $\langle j_3|$ and $|i_3\rangle$, by means of a single application of the Hamiltonian $\tilde{H}_{3\to3}$, or transitions between single- and triple-Reggeon states, governed by $\tilde{H}_{1\to3}$ 
and~$\tilde{H}_{3\to1}$. Let us begin by considering the former contributions, for which we get
\begin{equation}
\label{eq:Hhat33_3loop}
    \langle j_{3} | \tilde{H}_{3\to3} | i_{3} \rangle = \frac{\pi}{\alpha_s^3 r_\Gamma^3}\,\Big[\langle \psi_{j,3} | H_{3\to3} | \psi_{i,3} \rangle + \alpha_g(t,\mu^2) \langle \psi_{j,3} | \tts | \psi_{i,3} \rangle\Big],
\end{equation}
where the normalisation factor follows from the rescaling in eq.~(\ref{eq:newNormalisations}) and 
$H_{3\to 3}$ is defined according to eqs.~(\ref{ktok_momentum_space}), (\ref{ktok_momentum_space11}) and (\ref{ktok_momentum_space33}), where
$\alpha_g(t,\mu^2)$ is written in eq.~(\ref{al_g}).
The last term in eq.~(\ref{eq:Hhat33_3loop}) is immediately proportional to the two-loop result
\begin{equation}
\label{eq:agtts}
    \alpha_g(t,\mu^2)\langle \psi_{j,3} |\tts| \psi_{i,3}\rangle = \alpha_g(t,\mu^2)\, \tts\,\langle \psi_{j,3} | \psi_{i,3} \rangle,
\end{equation}
where $\langle \psi_{j,3} | \psi_{i,3}\rangle$ is given in eq.~(\ref{eq:amp33_2loops}). This relation stems from the fact that $\tts$ may be interpreted as acting on either the projectile or the target, rather than on simultaneously on both. 
The first term in eq.~(\ref{eq:Hhat33_3loop}) involves evolution of the three-Reggeon state
\begin{align}
\begin{split}
\label{eq:H33_3loop}
\langle \psi_{j,3} | H_{3\to3} | \psi_{i,3} \rangle &=-3i\,g_s^6\,S_\epsilon^2(\mu^2)\,C_A\,\mathbf{C}_{33}^{(2)}\,\int\,\frac{[\dbar k_1\dbar k_2\dbar k_3]_\delta}{k_1^2 k_2^2 k_3^2}\,\alpha_g(k_1^2,\mu^2)\\
&
+i\,\alpha_s\,g_s^6\,S_\epsilon^3(\mu^2)\,\mathbf{C}_{33}^{(3)}\,\int\,[\dbar q]\,\frac{[\dbar k_1\dbar k_2\dbar k_3]_\delta}{k_1^2 k_2^2 k_3^2}H_{22}(q;k_1-q,k_2+q)\,,
\end{split}
\end{align}
\begin{figure}[htbp]
\centering
\subfloat[\label{fig:h33_3a}]{\includegraphics[scale=1.2]{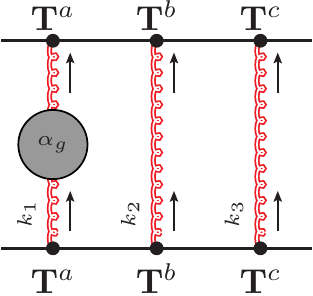}}
\hspace{60pt}
\subfloat[\label{fig:h33_3}]{\includegraphics[scale=1.2]{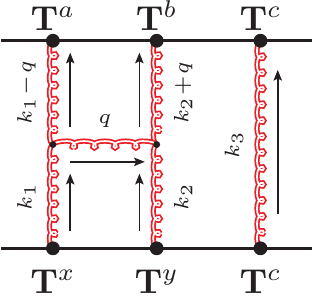}}
   \caption{Diagrammatic representation of the first term (a) and of the second term (b) in eq.~(\ref{eq:H33_3loop}).}
\label{fig:33-threeloops}
\end{figure}
where the two terms correspond respectively to the two components, eqs.~(\ref{ktok_momentum_space11}) and (\ref{ktok_momentum_space33}), in the $H_{3\to3}$ Hamiltonian, and the new three-loop colour factor arising in the latter, $\mathbf{C}_{33}^{(3)}$, is
\begin{align}
     \label{eq:colA333}
     \mathbf{C}_{33}^{(3)}&=\left(F^xF^y\right)^{ab}\left(\mathbf{T}^{\{a,b,c\}_+}\right)_i\left(\mathbf{T}^{\{x,y,c\}_+}\right)_j.
\end{align}
A diagrammatic representation of both the colour structure and the momentum flow of the loop integrals of each of the two contributions in eq.~(\ref{eq:H33_3loop}) is shown respectively in 
figures~\ref{fig:h33_3a} and~\ref{fig:h33_3}. The former differs from the two-loop transition amplitude of figure \ref{fig:h33_2} only by the insertion of the Regge trajectory $\alpha_g(k_1^2,\mu^2)$ in the integrand. In turn, figure \ref{fig:h33_3} depicts the action of a gluon exchange between the two Reggeons, where the emission and absorption of the gluon manifests in eq.~(\ref{eq:colA333}) as an adjoint generator, $F^x$ and $F^y$, respectively.

As mentioned in the beginning of this section, at this loop order, there are also contributions involving transitions between one and three Reggeon states, given by the second line of eq.~(\ref{eq:mhatNNLL}). These read
\begin{equation}
    \langle j_1 | \tilde{H}_{3\to 1} | i_3\rangle + \langle j_3 | \tilde{H}_{1\to 3} | i_1\rangle = \frac{\pi}{\alpha_s^3r_\Gamma^3}\Big[\langle \psi_{j,1} | H_{3\to 1} | \psi_{i,3} \rangle+\langle \psi_{j,3} | H_{1\to 3} | \psi_{i,1} \rangle\Big].
\end{equation}
By using the expressions of the Hamiltonian $H_{3\to1}$ of eq.~(\ref{Ham31}), we find that the transition amplitude $\langle \psi_{j,1} | H_{3\to 1} | \psi_{i,3} \rangle$ is
\begin{align}
    \begin{split}
    \label{eq:H31_3loop}
        \langle \psi_{j,1} | H_{3\to 1} | \psi_{i,3} \rangle &= -i\alpha_s^2\,g_s^4\,S_\epsilon^3\left(\mu^2\right)\,\mathbf{C}_{13}^{(3)}\,\int\frac{[\dbar k_1\dbar k_2\dbar k_3]_\delta}{k_1^2 k_2^2 k_3^2}\,H_{13}\left(k_1,k_2,k_3\right),
    \end{split}
\end{align}
where
\begin{align}
    \begin{split}
    \label{eq:C13_3}
    \mathbf{C}_{13}^{(3)} &=\text{Tr}\left[F^aF^bF^cF^d\right]\,\left(\mathbf{T}^{\{a,b,c\}_+}\right)_i\left(\mathbf{T}^d\right)_j.
    \end{split}
\end{align}
\begin{figure}
    \centering
    \subfloat[\label{fig:h31-3}]{\includegraphics[scale=.6]{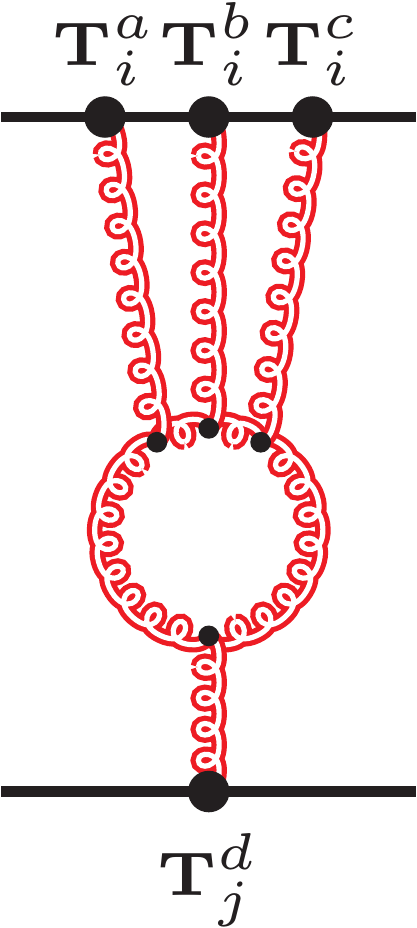}}
    \hspace{120pt}
    \subfloat[\label{fig:h13-3}]{\includegraphics[scale=.6]{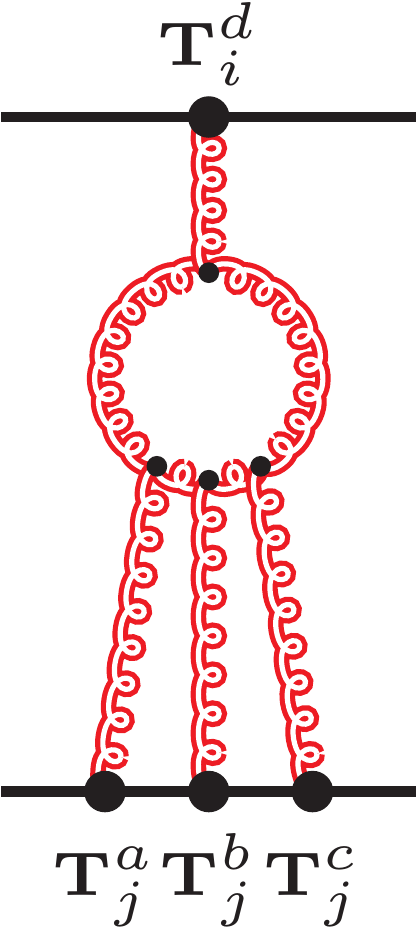}}
    \caption{Diagrammatic representation of eq.~(\ref{eq:H31_3loop}). Note that the colour factor in eq.~(\ref{eq:C13_3}) involves a trace of four generators in the adjoint representation, symmetrically summed over six different orderings.}
\end{figure}
\noindent 
The diagrammatic representation of eq.~(\ref{eq:H31_3loop}) is given in 
figure~\ref{fig:h31-3}. We obtain the symmetric contribution $\langle \psi_{j,3}|H_{1\to3}| \psi_{i,1}\rangle$ by replacing $i\leftrightarrow j$ in eq.~(\ref{eq:C13_3}), which corresponds diagrammatically to applying target-projectile symmetry on figure~\ref{fig:h31-3}, giving figure~\ref{fig:h13-3}. Therefore, summing $\langle \psi_{j,1}|H_{3\to 1}| \psi_{i,3}\rangle$ and $\langle \psi_{j,3}|H_{1\to 3}| \psi_{i,1}\rangle$ we finally obtain
\begin{align}
\begin{split}
\label{eq:H13loop}
        \langle \psi_{j,1} | H_{3\to 1} | \psi_{i,3} \rangle &+ \langle \psi_{j,3} | H_{1\to 3} | \psi_{i,1} \rangle =\\ 
&-i\alpha_s^2\,g_s^4\,S_\epsilon^3(\mu^2)\,\left(\mathbf{C}_{13}^{(3)}+\mathbf{C}_{31}^{(3)}\right)\,\int\frac{[\dbar k_1\dbar k_2\dbar k_3]_\delta}{k_1^2 k_2^2 k_3^2}\,H_{13}\left(k_1,k_2,k_3\right),
\end{split}
\end{align}
with
\begin{align}
    \begin{split}
    \label{eq:C31_3}
    \mathbf{C}_{31}^{(3)} &=\text{Tr}\left[F^aF^bF^cF^d\right]\,\left(\mathbf{T}^{\{a,b,c\}_+}\right)_j\left(\mathbf{T}^d\right)_i.
    \end{split}
\end{align}

\subsubsection{Four loops\label{subsubsec:Mhat4}}

\noindent
As shown in the square bracket of eq.~(\ref{eq:mhatNNLL}), the NNLL reduced amplitude at four loops includes (i) $3\to 3$ transitions, which connect three-Reggeon states in the target $|\psi_{i,3}\rangle$ and in the projectile $\langle \psi_{j,3}|$ involving two insertions of the Hamiltonian $\tilde{H}_{3\to 3}$;
(ii) $3\to 1$ and $1\to 3$ transitions, connecting $|\psi_{i,3}\rangle$ with $\langle \psi_{j,1}|$ and $| \psi_{i,1}\rangle$ with $\langle \psi_{j,3}|$, respectively; and 
(iii) $1\to 1$ transitions from $|\psi_{i,1}\rangle$ to $\langle \psi_{j,1}|$, which are mediated by three-Reggeon states. 
Below, we discuss the interpretation of each of these in turn in terms of diagrams, integrals and colour factors.

\vspace{1em}
\noindent
\textbf{The $\mathbf{3\to 3}$ transition} at four loops is generated by iterating the Hamiltonian $\tilde{H}_{3\to3}$ twice. Applying the rescaling of eq.~\eqref{eq:newNormalisations} and the subtraction defined in eq.~\eqref{eq:MhatDef} one has 
\begin{align}
\langle j_{3} | \tilde{H}_{3\to3}^2 | i_{3} \rangle =& \frac{\pi^2}{\left(\alpha_s r_\Gamma\right)^4}\Big[\langle \psi_{j,3} | H_{3\to3}^2 | \psi_{i,3} \rangle + \alpha_g(t,\mu^2)\langle \psi_{j,3} | \left(\tts\,H_{3\to3}+H_{3\to3}\tts\right) | \psi_{i,3} \rangle \nonumber\\*
& \qquad\qquad+ \alpha_g^2(t,\mu^2)\,\langle \psi_{j,3}|\left(\tts\right)^2| \psi_{i,3} \rangle\Big]\, .\label{eq:Hhat33_4loop}
\end{align}
The first term of eq.~(\ref{eq:Hhat33_4loop}) is the most complicated contribution: by writing the Hamiltonian $H_{3\to3}$ according to eq.~\eqref{ktok_momentum_space} we find
\begin{equation}
\label{eq:H33_4loop}
    \langle \psi_{j,3} | H_{3\to3}^2 | \psi_{i,3} \rangle = \langle \psi_{j,3} | B_{3\to3}^2 | \psi_{i,3} \rangle + \langle \psi_{j,3} | B_{3\to3} A_{3\to3} + A_{3\to3} B_{3\to 3} | \psi_{i,3} \rangle + \langle \psi_{j,3} | A_{3\to3}^2 | \psi_{i,3} \rangle,
\end{equation}
where we distinguish between terms involving interactions between Reggeons, generated by $B_{3\to3}$ of eq.~\eqref{ktok_momentum_space33}, from insertions of the one-loop Regge trajectory $\alpha_g$ on a single Reggeon, generated by $A_{3\to3}$ of eq.~\eqref{ktok_momentum_space11}. Diagrammatically, as we have already seen at three loops,
the action of $B_{3\to3}$ is represented by the addition of a rung in the triple-Reggeon ladders, as in figure~\ref{fig:h33_3}, while $A_{3\to 3}$ corresponds to an insertion of the trajectory $\alpha_g$ on one of the Reggeons, as in figure~\ref{fig:h33_3a}. Eq.~(\ref{eq:H33_4loop}) represents two such insertions in any combination.
\begin{figure}[htbp]
    \centering
    \subfloat[\label{fig:h33DL}]{\includegraphics[scale=1.2]{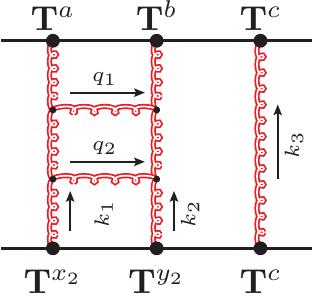}}\hspace{60pt}
    \subfloat[\label{fig:h33ML}]{\includegraphics[scale=1.2]{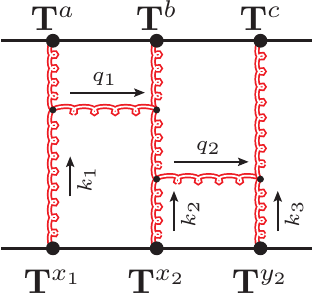}}
    \caption{Diagrammatic representation of a) the double ladder in eq.~(\ref{eq:H33_DL}) and b) the mixed ladder in eq.~(\ref{eq:H33_ML}).}
    \label{fig:my_label}
\end{figure}

The first term in eq.~(\ref{eq:H33_4loop}) reads 
\begin{equation}
\label{eq:Htild332}
\langle \psi_{j,3} | B_{3\to3}^2 | \psi_{i,3} \rangle = \langle \psi_{j,3} | B_{3\to3}^2 | \psi_{i,3} \rangle_{\text{DL}} 
+ \langle \psi_{j,3} | B_{3\to3}^2 | \psi_{i,3} \rangle_{\text{ML}},  
\end{equation}
and includes two distinct contributions 
\begin{subequations}
\begin{align}
        \langle \psi_{j,3} | B_{3\to3}^2 | \psi_{i,3} \rangle_{\text{DL}}&= 2i\,\alpha_s^2\,g_s^6\,S_\epsilon^4(\mu^2)\,\mathbf{C}_{\text{DL}}\int\,[\dbar q_1\dbar q_2]\frac{[\dbar k_1\dbar k_2\dbar k_3]_\delta}{k_1^2 k_2^2 k_3^2}H_{22}(q_2;k_1-q_2,k_2+q_2)\nonumber\\*
        &\hspace{3.8cm}\times H_{22}(q_1;k_1-q_1-q_2,k_2+q_1+q_2),\label{eq:H33_DL}
    \\ \nonumber\\
        \langle \psi_{j,3} | B_{3\to3}^2 | \psi_{i,3} \rangle_{\text{ML}}&= 4i\,\alpha_s^2\,g_s^6\,S_\epsilon^4(\mu^2)\,\mathbf{C}_{\text{ML}}\int\,[\dbar q_1\dbar q_2]\frac{[\dbar k_1\dbar k_2\dbar k_3]_\delta}{k_1^2 k_2^2 k_3^2}H_{22}(q_2;k_2-q_2,k_3+q_2)\nonumber\\*
        &\hspace{3.8cm}\times H_{22}(q_1;k_1-q_1,k_2+q_1-q_2),\label{eq:H33_ML}
\end{align}
\end{subequations}
which correspond respectively to the two possible ways of adding a second rung to the ladder, depicted in figure~\ref{fig:h33DL} and figure~\ref{fig:h33ML}.

The colour structures $\mathbf{C}_{\text{DL}}$ and $\mathbf{C}_{\text{ML}}$, associated to the {\textit{double ladder}} diagram in figure~\ref{fig:h33DL} and to the {\textit{mixed ladder}} diagram in figure~\ref{fig:h33ML} respectively, are
\begin{subequations}
\begin{align}
\label{def:CDL}
\mathbf{C}_{\text{DL}}&=\left(F^{x_1}F^{y_1}\right)^{ab}\left(F^{x_2}F^{y_2}\right)^{x_1y_1}\,\left(\mathbf{T}^{\{a,b,c\}_+}\right)_i\left(\mathbf{T}^{\{x_2,y_2,c\}_+}\right)_j,\\
\label{def:CML}
\mathbf{C}_{\text{ML}}&=\left(F^{x_1}F^{y_1}\right)^{ab}\left(F^{x_2}F^{y_2}\right)^{y_1c}\,\left(\mathbf{T}^{\{a,b,c\}_+}\right)_i\left(\mathbf{T}^{\{x_2,y_2,x_1\}_+}\right)_j.
\end{align}
\end{subequations}
The remaining terms in eq.~(\ref{eq:H33_4loop}) are
\begin{subequations}
\begin{align}
\begin{split}
\label{eq:H33Atilde}
&\langle \psi_{j,3} | B_{3\to3} A_{3\to3} + A_{3\to3} B_{3\to 3} | \psi_{i,3} \rangle  = -2i\,g_s^6\alpha_s\,S_\epsilon^3(\mu^2)\,C_A\mathbf{C}^{(3)}_{33}\,\int[\dbar q]\frac{[\dbar k_1\dbar k_2\dbar k_3]_\delta}{k_1^2 k_2^2 k_3^2}\\
&\hspace{1.0cm}\times \, 
\Big\{ H_{22}(q;k_1-q,k_2+q)\Big[\alpha_g\left(k_1^2,\mu^2\right)+\alpha_g\left(k_2^2,\mu^2\right)+\alpha_g\left(k_3^2,\mu^2\right)\Big]\Big\},
\end{split}
\\
\begin{split}
\label{eq:AtildeAtilde}
&\langle \psi_{j,3} | A_{3\to3}^2 | \psi_{i,3} \rangle = 3i\,g_s^6\,S_\epsilon^2(\mu^2)\,C_A^2\mathbf{C}_{33}^{(2)}\,\int\frac{[\dbar k_1\dbar k_2\dbar k_3]_\delta}{k_1^2 k_2^2 k_3^2}\,\Big[\left(\alpha_g\left(k_1^2,\mu^2\right)\right)^2
\\&\hspace{8.0cm}
+\,2\alpha_g\left(k_1^2,\mu^2\right)\alpha_g\left(k_2^2,\mu^2\right)\Big].
\end{split}
\end{align}
\end{subequations}
We compute $\langle \psi_{j,3} | H_{3\to3}^2 |\psi_{i,3}\rangle$ by adding eqs.~(\ref{eq:H33_DL}), (\ref{eq:H33_ML}), (\ref{eq:H33Atilde}) and (\ref{eq:AtildeAtilde}).
In order to calculate the reduced amplitude $\langle \psi_{j,3} | \tilde{H}_{3\to3}^2 |\psi_{i,3}\rangle$, according to eq.~(\ref{eq:Hhat33_4loop}), we add 
\begin{subequations}
\begin{equation}
\label{eq:H33ag}
\alpha_g(t,\mu^2)\langle \psi_{j,3} | \left(\tts\,H_{3\to3}+H_{3\to3}\tts\right) | \psi_{i,3} \rangle = 2\alpha_g(t,\mu^2) \,\tts\,\langle \psi_{j,3} | H_{3\to3} |\psi_{i,3} \rangle,
\end{equation}
\begin{equation}
\label{eq:agag}
    \alpha_g^2(t,\mu^2)\,\langle \psi_{j,3} | \left(\tts\right)^2| \psi_{i,3} \rangle = \alpha_g^2(t,\mu^2)\left(\tts\right)^2\,\langle \psi_{j,3} | \psi_{i,3} \rangle,
\end{equation}
\end{subequations}
where $\langle \psi_{j,3}| H_{3\to3} | \psi_{i,3}\rangle$ and $\langle \psi_{j,3}|\psi_{i,3}\rangle$ are given in eq.~(\ref{eq:amp33_2loops}) and in eq.~(\ref{eq:H33_3loop}), respectively. In eqs.~(\ref{eq:H33ag}) and (\ref{eq:agag}) we notice that $\tts$ commutes with the application of $H_{3\to3}$ and with the contraction of target and projectile, as in eq.~(\ref{eq:agtts}).

\vspace{1em}
\noindent
\textbf{The $\mathbf{3\to1}$ and $\mathbf{1\to3}$ transitions} at four loops read 
\begin{subequations}
\begin{align}
\label{eq:defH31H33}
    \langle j_{1} | \tilde{H}_{3\to1}\tilde{H}_{3\to3} | i_{3} \rangle &= \frac{\pi^2}{\alpha_s^4 r_\Gamma^4}\,\langle \psi_{j,1} | H_{3\to1} \left(H_{3\to3}-H_{1\to 1}\right) | \psi_{i,3}\rangle,\\
    \langle j_{3} | \tilde{H}_{3\to3}\tilde{H}_{1\to3} | i_{1} \rangle &= \frac{\pi^2}{\alpha_s^4 r_\Gamma^4}\,\langle \psi_{j,3} | \left(H_{3\to3}-H_{1\to1}\right) H_{1\to3} | \psi_{i,1}\rangle.
\end{align}
\end{subequations}
We decompose each transition amplitude into three contributions
\begin{subequations}
\begin{align}
\begin{split}
\label{eq:Hhat3133}
    \langle \psi_{j,1} | H_{3\to1}\,\left(H_{3\to3}-H_{1\to 1}\right) | \psi_{i,3} \rangle &= \langle \psi_{j,1} | H_{3\to1}\,B_{3\to3} | \psi_{i,3} \rangle + \langle \psi_{j,1} | H_{3\to1}\, A_{3\to3} | \psi_{i,3} \rangle \\
    &+ \alpha_g(t,\mu^2) \langle \psi_{j,1} | H_{3\to1} \, \tts| \psi_{i,3} \rangle,
\end{split}\\
\begin{split}
\label{eq:Hhat3313}
    \langle \psi_{j,3} | \left(H_{3\to3}-H_{1\to 1}\right)\,H_{1\to3} | \psi_{i,1} \rangle &= \langle \psi_{j,3} | B_{3\to3}\,H_{1\to3} | \psi_{i,1} \rangle + \langle \psi_{j,3} | A_{3\to3}\,H_{1\to3} | \psi_{i,1} \rangle \\
    &+ \alpha_g(t,\mu^2) \langle \psi_{j,3} | \tts\,H_{1\to3} | \psi_{i,1} \rangle.
    \end{split}
\end{align}
\end{subequations}
\begin{figure}[htbp]
  \centering
  \subfloat[\label{fig:h13-4}]{\includegraphics[scale=.6]{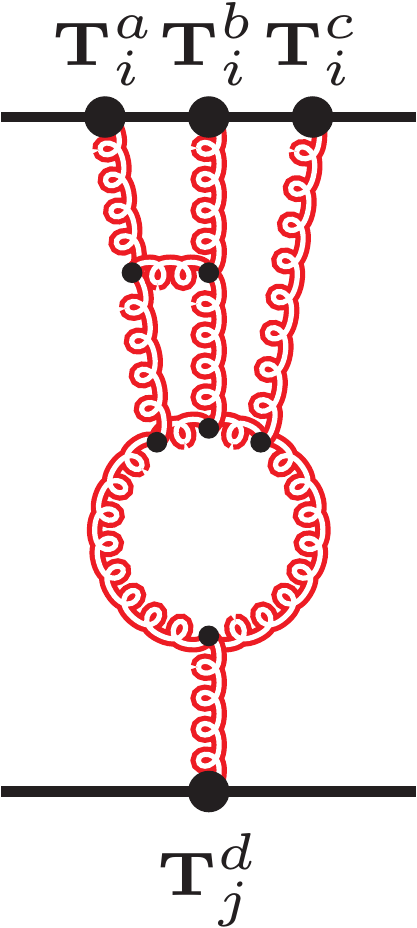}}\hspace{60pt}
  \subfloat[\label{fig:h13-4b}]{\includegraphics[scale=.6]{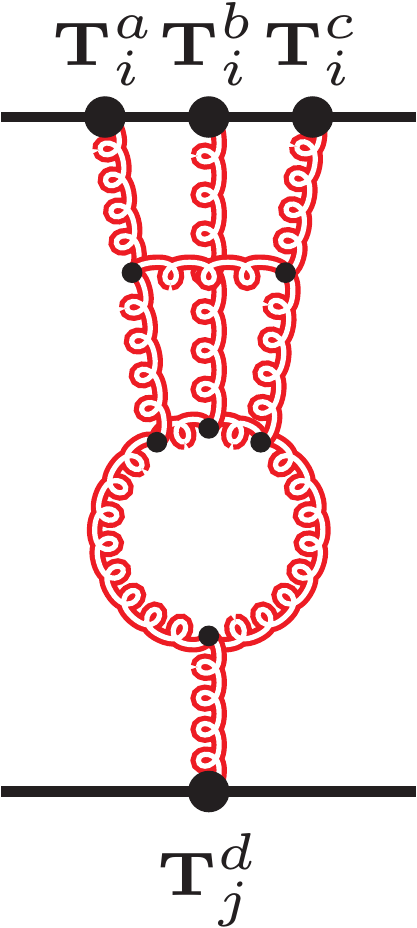}}
  \caption{Diagrammatic representation of the first (a) and second (b) terms of eq.~(\ref{eq:Hhat31Htilde33}).}
\end{figure}
The amplitudes in eqs.~(\ref{eq:Hhat3133}) and (\ref{eq:Hhat3313}) are related by target-projectile symmetry. Below we discuss the former equation. The first term in eq.~(\ref{eq:Hhat3133}) gives
\begin{align}
    \begin{split}
    \label{eq:Hhat31Htilde33}
        \langle \psi_{j,1} | H_{3\to1} B_{3\to3} | \psi_{i,3} \rangle =& -2i\alpha_s^3\,g_s^4\,S_\epsilon^4(\mu^2)\,\mathbf{C}_{13}^{(4),A}\,\int[\dbar q]\frac{[\dbar k_1\dbar k_2\dbar k_3]_\delta}{k_1^2 k_2^2 k_3^2}\,H_{13}\left(k_3,k_1,k_2\right)\\
        &\phantom{-2i\alpha_s^3\,g^4\,S_\epsilon^4(\mu^2)\,\mathbf{C}_{13}^{(4),A}\,}\times\,H_{22}\left(q;k_1-q,k_3+q\right)\\
        &-i\alpha_s^3\,g_s^4\,S_\epsilon^4(\mu^2)\,\mathbf{C}_{13}^{(4),B}\,\int[\dbar q]\frac{[\dbar k_1\dbar k_2\dbar k_3]_\delta}{k_1^2 k_2^2 k_3^2}\,H_{13}\left(k_1,k_2,k_3\right)\\
        &\phantom{-i\alpha_s^3\,g_s^4\,S_\epsilon^4(\mu^2)\,\mathbf{C}_{13}^{(4),B}\,}\times\,H_{22}\left(q;k_1-q,k_3+q\right),
    \end{split}
\end{align}
where the colour factors $\mathbf{C}_{13}^{(4),A}$ and $\mathbf{C}_{13}^{(4),B}$ are
\begin{subequations}
\begin{align}
\begin{split}
\label{def:C13_4}
    \mathbf{C}_{13}^{(4),A} &= \left(F^xF^y\right)^{ac}\,\text{Tr}\left[F^xF^y\left(F^bF^d+F^dF^b\right)\right]\,\left(\mathbf{T}^{\{a,b,c\}_+}\right)_i\left(\mathbf{T}^d\right)_j,
\end{split}\\
\begin{split}
\label{eq:h13_4b}
    \mathbf{C}_{13}^{(4),B}&=\left(F^xF^y\right)^{ac}\,\left(\text{Tr}\left[F^xF^bF^yF^d\right]+\text{Tr}\left[F^xF^dF^yF^b\right]\right)\,\left(\mathbf{T}^{\{a,b,c\}_+}\right)_i\,\left(\mathbf{T}^d\right)_j.
\end{split}
\end{align}
\end{subequations}
Diagrammatically, the two terms on the right-hand side of eq.~(\ref{eq:Hhat31Htilde33}) are represented in figures~\ref{fig:h13-4} and~\ref{fig:h13-4b}, respectively. These diagrams are obtained by dressing the three-loop transition amplitude $\langle \psi_{j,1} |H_{3\to1} |\psi_{i,3}\rangle$, depicted in figure~\ref{fig:h31-3}, with one additional rung. 

The second term in eq.~(\ref{eq:Hhat3133}) is 
\begin{align}
    \begin{split}
    \label{eq:Hhat13A}
        \langle \psi_{j,1} | H_{1\to 3}\, A_{3\to3} | \psi_{i,3} \rangle &= i\alpha_s g_s^6\,S_\epsilon^3(\mu^2)\,C_A\mathbf{C}_{13}^{(3)}\,\int\frac{[\dbar k_1\dbar k_2\dbar k_3]}{k_1^2 k_2^2 k_3^2}\,H_{13}\left(k_1,k_2,k_3\right)\\
        &\phantom{i\alpha_s g_s^6\,S_\epsilon^3(\mu^2)\,C_A\mathbf{C}_{13}^{(3)}\,\int\frac{[\dbar k_1\dbar k_2\dbar k_3]}{k_1^2 k_2^2 k_3^2}}\times\left[2\alpha_g\!\left(k_1^2,\mu^2\right)+\alpha_g\!\left(k_2^2,\mu^2\right)\right],
    \end{split}
\end{align}
where we used the symmetry of $H_{13}\left(k_1,k_2,k_3\right)$ of eq.~(\ref{H13kin}) under $k_1\leftrightarrow k_3$.
Note that eq.~(\ref{eq:Hhat13A}) features the same colour factor $\mathbf{C}_{13}^{(3)}$ as the three-loop diagram, multiplied by~$C_A$.
The last term in eq.~(\ref{eq:Hhat3133}) is proportional to the three-loop transition amplitude
\begin{equation}
\label{eq:agH13}
\alpha_g(t,\mu^2) \langle \psi_{j,1} | \tts\,H_{3\to1} | \psi_{i,3} \rangle = \alpha_g(t,\mu^2)\, \tts\, \langle \psi_{j,1} | H_{1\to3} | \psi_{i,3} \rangle,
\end{equation}
where $\langle \psi_{j,1} |H_{3\to1} | \psi_{i,3} \rangle = \frac{\left(\alpha_s r_\Gamma\right)^3}{\pi}\langle j_1 |\tilde{H}_{3\to1} | i_3 \rangle$ is obtained from eq.~(\ref{eq:H31_3loop}). 

Target-projectile symmetry implies that the amplitudes in eq.~(\ref{eq:Hhat3313}), namely the terms $\langle \psi_{j,3} | B_{3\to3}H_{1\to3} | \psi_{i,1} \rangle$,
$\langle \psi_{j,3}| A_{3\to3} H_{1\to3} | \psi_{i,1} \rangle$ and $\langle \psi_{j,3} | \tts H_{1\to3} | \psi_{i,1} \rangle$, are obtained from eqs.~(\ref{eq:Hhat31Htilde33}), (\ref{eq:Hhat13A}) and (\ref{eq:agH13}), respectively, by replacing colour factors $\mathbf{C}_{13}^{(3)}\to\mathbf{C}_{31}^{(3)}$, defined in eq.~(\ref{eq:C31_3}), and  $\mathbf{C}_{13}^{(4),A}\to\mathbf{C}_{31}^{(4),A}$, $\mathbf{C}_{13}^{(4),B}\to\mathbf{C}_{31}^{(4),B}$, with
\begin{subequations}
\begin{align}
\begin{split}
\label{def:C31_4}
    \mathbf{C}_{31}^{(4),A} &= \left(F^xF^y\right)^{ac}\,\text{Tr}\left[F^xF^y\left(F^bF^d+F^dF^b\right)\right]\,\left(\mathbf{T}^{\{a,b,c\}_+}\right)_j\left(\mathbf{T}^d\right)_i,
\end{split}\\    
\begin{split}
\label{def:tildeC31_4}
    \mathbf{C}_{31}^{(4),B} &= \left(F^xF^y\right)^{ac}\,\left(\text{Tr}\left[F^xF^bF^yF^d\right]+\text{Tr}\left[F^xF^dF^yF^b\right]\right)\,\left(\mathbf{T}^{\{a,b,c\}_+}\right)_j\,\left(\mathbf{T}^d\right)_i.
\end{split}
\end{align}
\end{subequations}

\begin{figure}[tb]
  \centering
  \subfloat[\label{fig:h3113a}]{\includegraphics[scale=.5]{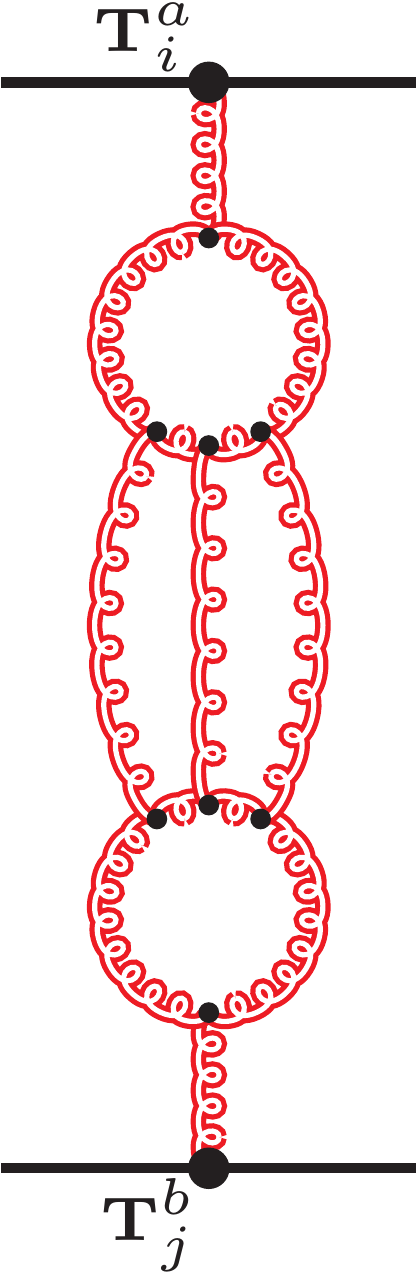}}\hspace{60pt}
  \subfloat[\label{fig:h3113b}]{\includegraphics[scale=.5]{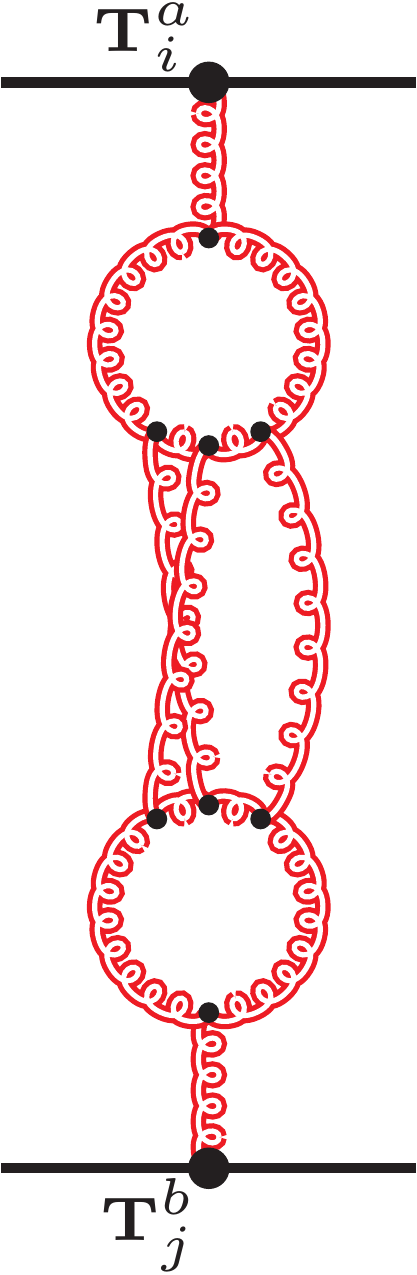}}
  \caption{Diagrammatic representation of a) $\langle \psi_{j,1} | H_{3\to1}\,H_{1\to3} | \psi_{i,1} \rangle_D$ in eq.~(\ref{eq:defH31H13D}) and b) $\langle \psi_{j,1} | H_{3\to1}\,H_{1\to3} | \psi_{i,1} \rangle_X$ in eq.~(\ref{eq:defH31H13X}).}
\end{figure}

\vspace{1em}
\noindent
\textbf{The $\mathbf{1\to 1}$ transition mediated by three Reggeon states} arises for the first time at four loops where it takes the form
\begin{equation}
\label{eq:Hhat31Hhat13}
    \langle j_1 | \tilde{H}_{3\to1}\tilde{H}_{1\to3} | i_1 \rangle =\frac{\pi^2}{\alpha_s^4\,r_\Gamma^4} \Big[\langle \psi_{j,1} | H_{3\to1}\,H_{1\to3} | \psi_{i,1} \rangle_D + \langle \psi_{j,1} | H_{3\to1}\,H_{1\to3} | \psi_{i,1} \rangle_X\Big],
\end{equation}
where we have distinguished two contributions characterised by different colour structures and integrals. These are defined as
\begin{subequations}
\begin{align}
    \begin{split}
    \label{eq:defH31H13D}
        \langle \psi_{j,1} | H_{3\to1}\,H_{1\to3} | \psi_{i,1} \rangle_D &= i\alpha_s^4g_s^2\,S_\epsilon^4(\mu^2)\,\mathbf{C}_{3113}^{(4),D}\int\frac{[\dbar k_1\dbar k_2\dbar k_3]_\delta}{k_1^2 k_2^2 k_3^2}\,\Big[H_{13}\left(k_1,k_2,k_3\right)\Big]^2,
    \end{split}\\
    \begin{split}
    \label{eq:defH31H13X}
        \langle \psi_{j,1} | H_{3\to1}\,H_{1\to3} | \psi_{i,1} \rangle_X &= i\alpha_s^4g_s^2\,S_\epsilon^4(\mu^2)\,\mathbf{C}_{3113}^{(4),X}\\
        &\times\int\frac{[\dbar k_1\dbar k_2\dbar k_3]_\delta}{k_1^2 k_2^2 k_3^2}\,\Bigg[H_{13}\left(k_1,k_2,k_3\right)\,H_{13}\left(k_3,k_1,k_2\right)\Bigg],
    \end{split}
\end{align}
\end{subequations}
with colour factors
\begin{subequations}
\begin{align}
\begin{split}
\label{def:C3113_D}
    \mathbf{C}_{3113}^{(4),D} &= \text{Tr}\left[F^aF^cF^dF^e\right]\text{Tr}\left[F^bF^cF^dF^e+F^bF^eF^dF^c\right]\,\Big(\mathbf{T}^a\Big)_i\Big(\mathbf{T}^b\Big)_j,
\end{split}\\
\begin{split}
\label{def:C3113_X}
        \mathbf{C}_{3113}^{(4),X} &= \text{Tr}\left[F^bF^cF^eF^d+F^bF^eF^cF^d+F^bF^dF^eF^c+F^bF^dF^cF^e\right]\\
        &\times\text{Tr}\left[F^aF^cF^dF^e\right]\Big(\mathbf{T}^a\Big)_i\Big(\mathbf{T}^b\Big)_j.
\end{split}
\end{align}
\end{subequations}
$\mathbf{C}_{3113}^{(4),D}$ and $\mathbf{C}_{3113}^{(4),X}$ correspond to the diagrams in figures~\ref{fig:h3113a} and \ref{fig:h3113b}, respectively.

All the transition amplitudes contributing to the NNLL reduced amplitude through four loops are written in terms of a handful of colour structures and of integrals involving $H_{22}$, $H_{13}$ and $\alpha_g$ as kernels. In sections~\ref{subsec:colour} and~\ref{subsec:integrals} we will describe the techniques we used to compute colour structures and integrals, respectively.

\subsection{Colour structure\label{subsec:colour}}

The definitions of the colour factors in section~\ref{subsec:dia} are independent of the representations of external particles, therefore they equally apply to quark-quark scattering, to gluon-gluon scattering and to quark-gluon scattering. Each colour structure can be easily evaluated by specialising the generators $\mathbf{T}_i$ and $\mathbf{T}_j$ to the representation of the projectile and the target, respectively, and by picking a colour basis for each process, such as the $t$-channel bases given in ref.~\cite{DelDuca:2014cya}. 
However, an explicit basis would obscure the universal features of the amplitudes, such as the structure of their infrared divergences and the universality of the Regge limit. Therefore, we need specialised techniques to compute colour factors for general representations of the external particles. In sections~\ref{sec:col2} and \ref{sec:col3}, we will briefly review the fundamental technique introduced in ref.~\cite{Caron-Huot:2017fxr} and compute the colour structures of the two-loop and three-loop reduced amplitudes. We will see, however, that one must generalise this method in order to compute the four-loop colour factors introduced in section~\ref{subsubsec:Mhat4}. To this end, in section~\ref{sec:col4} we derive and apply new identities that allow us to provide explicit expressions for all colour factors through four loops without specifying the representation of the scattering particles.
Eventually, every colour structure will be written in terms of operators $\tts$ and $\tsu$, along with quartic Casimir invariants, acting on the colour structure of the tree-level amplitude, $\mathbf{T}^a_i\mathbf{T}^a_j$.

\subsubsection{Colour structure at two loops\label{sec:col2}}

At two loops, the reduced amplitude is characterised by the colour structure $\mathbf{C}_{33}^{(2)}$ defined in eq.~(\ref{eq:colA332}). In that expression, the contractions of colour generators may be evaluated in a straightforward way, for any representation of the projectile and the target. Indeed, contractions of outermost generators on the top and on the bottom line may be written in terms of the colour channel operators $\tts$ and $\tsu$ defined in eqs.~(\ref{TtTsTu}) and (\ref{eq:Tsudef}). 
As described in ref.~\cite{Caron-Huot:2017fxr}, we have
\begin{subequations}\label{eq:relTsuPMTt/2}
\begin{align}
   \begin{split}
   \label{eq:relTsu-Tt/2}
        &\left(\mathbf{T}^a_i\,\mathbf{T}^{b_1}_i\mathbf{T}^{b_2}_i\dots\right)\,\left(\mathbf{T}^{a}_j\,\mathbf{T}^{c_1}_j\mathbf{T}^{c_2}_j\dots\right)=\left(\mathbf{T}^{b_1}_i\mathbf{T}^{b_2}_i\dots\mathbf{T}^a_i\right)\,\left(\mathbf{T}^{c_1}_j\mathbf{T}^{c_2}_j\dots\mathbf{T}^{a}_j\right)\\
        &\phantom{\left(\mathbf{T}^a_i\,\mathbf{T}^{b_1}_i\mathbf{T}^{b_2}_i\dots\right)\,\left(\mathbf{T}^{a}_j\,\mathbf{T}^{c_1}_j\mathbf{T}^{c_2}_j\dots\right)}=\frac{1}{2}\Big[\mathbf{T}^2_{s-u}-\frac{\mathbf{T}_t^2}{2}\Big]\left(\mathbf{T}^{b_1}_i\mathbf{T}^{b_2}_i\dots\right)\left(\mathbf{T}^{c_1}_j\mathbf{T}^{c_2}_j\dots\right),
    \end{split}\\
    \begin{split}
    \label{eq:relTsu+Tt/2}
        &\left(\mathbf{T}^a_i\,\mathbf{T}^{b_1}_i\mathbf{T}^{b_2}_i\dots\right)\,\left(\mathbf{T}^{c_1}_j\mathbf{T}^{c_2}_j\dots\mathbf{T}^{a}_j\right)=\left(\mathbf{T}^{b_1}_i\mathbf{T}^{b_2}_i\dots \mathbf{T}^a_i\right)\,\left(\mathbf{T}^{a}_j\mathbf{T}^{c_1}_j\mathbf{T}^{c_2}_j\dots\right)\\
        &\phantom{\left(\mathbf{T}^a_i\,\mathbf{T}^{b_1}_i\mathbf{T}^{b_2}_i\dots\right)\,\left(\mathbf{T}^{c_1}_j\mathbf{T}^{c_2}_j\dots\mathbf{T}^{a}_j\right)}=\frac{1}{2}\Big[\mathbf{T}^2_{s-u}+\frac{\mathbf{T}_t^2}{2}\Big]\left(\mathbf{T}^{b_1}_i\mathbf{T}^{b_2}_i\dots\right)\left(\mathbf{T}^{c_1}_j\mathbf{T}^{c_2}_j\dots\right).
    \end{split}
\end{align}
\end{subequations}
By applying these identities to eq.~(\ref{eq:colA332}) we get
\begin{equation}
    \label{res:colA332}
    \mathbf{C}^{(2)}_{33}=\frac{1}{24}\left[\left(\tsu\right)^2-\frac{C_A^2}{12}\right]\,\mathbf{T}^a_i\mathbf{T}^a_j.
\end{equation}
Here, we isolated a universal operator, written in terms of $\tts$ and $\tsu$, which acts upon the tree-level amplitude, $\mathbf{T}^a_i\mathbf{T}^a_j$. 

\subsubsection{Colour structures at three loops\label{sec:col3}}

Let us apply the method described in section~\ref{sec:col2} to compute the colour structure $\mathbf{C}_{33}^{(3)}$, defined in eq.~(\ref{eq:colA333}). We write the structure constants $\left(F^xF^y\right)^{ab}$ in terms of commutators of colour generators, using the Lie algebra relation
\begin{equation}
\label{eq:LieAlgebra}
    \mathbf{T}^a_i\,\left(F^x\right)^{ab} = [\mathbf{T}^x_i,\mathbf{T}^b_i] \equiv \mathbf{T}^{[x,b]}_i.
\end{equation}
There is a unique way of applying eq.~(\ref{eq:LieAlgebra}) to eq.~(\ref{eq:colA333}) such that each generator $\mathbf{T}_i$, in the representation of the target, is contracted with a generator $\mathbf{T}_j$, in the representation of the projectile, preserving manifest symmetry under exchange of target and projectile, and it leads to
\begin{equation}
    \label{eq:colA333b}
    \mathbf{C}^{(3)}_{33}=\left(\mathbf{T}^{\{[a,x],b,c\}_+}\right)_i\,\left(\mathbf{T}^{\{[b,x],a,c\}_+}\right)_j.
\end{equation}
Target-projectile symmetry is immediately verified by exchanging representations of the generators $i\leftrightarrow j$. By applying eqs.~(\ref{eq:relTsu-Tt/2}) and (\ref{eq:relTsu+Tt/2}) to eq.~(\ref{eq:colA333b}) we get
\begin{equation}
\label{res:colA333}
    \mathbf{C}^{(3)}_{33}=\left(3C_A-\tts\right)\,\mathbf{C}^{(2)}_{33}.
\end{equation}
\begin{figure}[htbp]
  \centering
  \includegraphics[scale=.6]{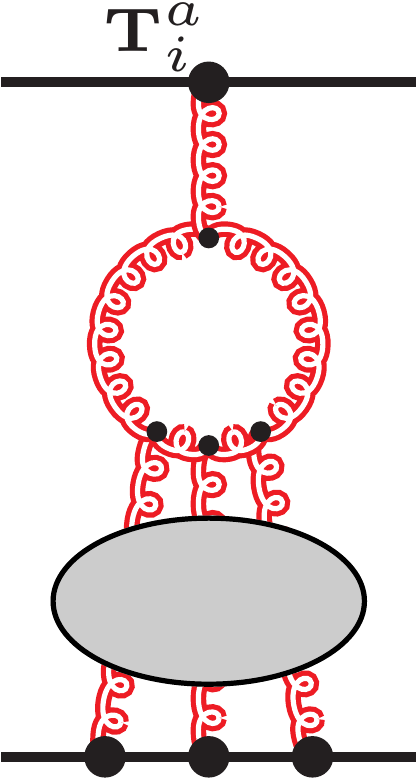}
  \caption{Diagrammatic representation of a generic transition amplitude connecting one- and three-Reggeon states. \label{fig:h13}}
\end{figure}

The remaining colour structures at three loops are $\mathbf{C}_{13}^{(3)}$ and $\mathbf{C}_{31}^{(3)}$, defined in eqs.~(\ref{eq:C13_3}) and (\ref{eq:C31_3}), respectively. As noticed in eq.~(\ref{eq:H13loop}), the reduced amplitude features their sum,  $\mathbf{C}_{13}^{(3)}+\mathbf{C}_{31}^{(3)}$. This colour factor was computed in ref.~\cite{Caron-Huot:2017fxr} by applying the identities in eqs.~(\ref{eq:relTsu-Tt/2}), (\ref{eq:relTsu+Tt/2}) and~(\ref{eq:LieAlgebra}), getting
\begin{align}
    \label{res:col133a}
    \mathbf{C}_{13}^{(3)}+\mathbf{C}_{31}^{(3)} = \left\{\frac{\tts}{4}\left[\left(\tsu\right)^2+\frac{C_A^2}{12}\right]+\frac{3}{4}\tsu\left[\tsu,\tts\right]\right\}\,\mathbf{T}^a_i\mathbf{T}^a_j\,.
\end{align}
However, the result above can be simplified further. Both in $\mathbf{C}_{13}^{(3)}$ and in $\mathbf{C}_{31}^{(3)}$, colour flows through a single generator, either on the target or on the projectile side. This single generator acts\footnote{We thank Simon Caron-Huot for raising this point.} as a projection operator~\cite{Cvitanovic:2008zz} on the adjoint representation in the $t$-channel, independently of the representations of the target and of the projectile. 
This in turn implies that, to all orders in perturbation theory, the colour structure of transition amplitudes connecting one- and three-Reggeon states, depicted in figure~\ref{fig:h13}, is proportional to the tree-level amplitude. This observation implies
\begin{subequations}
\label{eq:Simon_trick12}
\begin{align}
\label{eq:Simon_trick1}
    &\mathbf{C}_{13}^{(3)} \equiv \text{Tr}\left[F^aF^bF^cF^d\right]\,\left(\mathbf{T}^{\{a,b,c\}_+}\right)_i\left(\mathbf{T}^d\right)_j = C_{13}^{(3)}\,\mathbf{T}_i^a\,\mathbf{T}_j^a,\\
\label{eq:Simon_trick2}
    &\mathbf{C}_{31}^{(3)} \equiv \text{Tr}\left[F^aF^bF^cF^d\right]\,\left(\mathbf{T}^{\{a,b,c\}_+}\right)_j\left(\mathbf{T}^d\right)_i = C_{31}^{(3)}\,\mathbf{T}_i^a\,\mathbf{T}_j^a,
\end{align}
\end{subequations}
where $C_{13}^{(3)}$ and $C_{31}^{(3)}$ are scalar functions of the Casimir invariants of the projectile and the target, respectively.
In order to determine $C_{13}^{(3)}$, we consider the case where the target is in the same representation as the projectile. Upon setting $\mathbf{T}^d_j = \mathbf{T}^d_i$ on the left-hand side of eq.~(\ref{eq:Simon_trick1}) and $\mathbf{T}^a_j = \mathbf{T}^a_i$  on the right-hand side, we contract all indices, getting
\begin{equation}
\label{eq:FFFFTTT}
    \text{Tr}\left[F^aF^bF^cF^d\right]\,\text{Tr}_{R_i}\left[\mathbf{T}^{\{a,b,c\}_+}_i\,\mathbf{T}_i^d\right] = C_{13}^{(3)}\,\text{Tr}_{R_i}\left[\mathbf{T}_i^a\,\mathbf{T}_i^a\right],
\end{equation}
where the subscript $R_i$ specifies the representation of the colour generators inside the trace.
After computing the traces we find
\begin{equation}
\label{eq:resC13_3}
    C_{13}^{(3)}=\frac{d_{AR_i}}{N_{R_i}}\frac{1}{C_i}.
\end{equation}
Here $d_{AR_i}$ is the quartic Casimir operator
\begin{equation}
\label{eq:dARi}
  d_{AR_i}=\frac{1}{4!}\sum_{\sigma\in\mathcal{S}_4}\mathrm{Tr}\left[F^{\sigma(a)}F^{\sigma(b)}F^{\sigma(c)}F^{\sigma(d)}\right]\mathrm{Tr}\left[\mathbf{T}^a_i\mathbf{T}^{b}_i\mathbf{T}^{c}_i\mathbf{T}^{d}_i\right],
\end{equation}
and explicitly for a $\mathrm{SU}(N_c)$ gauge group, $d_{AR_i}$ is
\begin{equation}
\label{eq:dAA}
  \frac{d_{AA}}{N_A}=\frac{N_c^2(N_c^2+36)}{24},\qquad\quad \frac{d_{AF}}{N_F}=\frac{(N_c^2+6)(N_c^2-1)}{48},
\end{equation}
where $N_A=N_c^2-1$ and $N_F=N_c$ are, respectively, the dimensions of the fundamental and of the adjoint representations of $\mathrm{SU}(N_c)$. Following the same steps leading to eq.~(\ref{eq:resC13_3}), we determine
\begin{equation}
\label{eq:resC31_3}
    C_{31}^{(3)}=\frac{d_{AR_j}}{N_{R_j}}\frac{1}{C_j}.
\end{equation}
Hence, the colour structure in eq.~(\ref{res:col133a}) becomes
\begin{align}
    \label{res:col133b}
    \mathbf{C}_{13}^{(3)}+\mathbf{C}_{31}^{(3)} = \left(\frac{d_{AR_i}}{N_{R_i}}\frac{1}{C_i}+\frac{d_{AR_j}}{N_{R_j}}\frac{1}{C_j}\right)\,\mathbf{T}^a_i\mathbf{T}^a_j.
\end{align}
The result above implies the following non-trivial identity involving colour operators $\tts$ and $\tsu$
\begin{equation}
\label{eq:Simon_trick}
    \left\{\frac{\tts}{4}\left[\left(\tsu\right)^2+\frac{C_A^2}{12}\right]+\frac{3}{4}\tsu\left[\tsu,\tts\right]\right\}\,\mathbf{T}^a_i\mathbf{T}^a_j = \left(\frac{d_{AR_i}}{N_{R_i}}\frac{1}{C_i}+\frac{d_{AR_j}}{N_{R_j}}\frac{1}{C_j}\right)\,\mathbf{T}^a_i\mathbf{T}^a_j,
\end{equation}
for any representation $R_i$ and $R_j$ of the projectile and the target. This three-loop quartic Casimir colour identity will be useful also at four loops.
We stress that the basic argument leading to eqs.~(\ref{eq:Simon_trick12}) holds to all orders in perturbation theory: every transition amplitude involving a single Reggeon exchange, either from the target or from the projectile, has the same colour structure as the tree-level amplitude up to a multiplicative factor depending on Casimir operators. We therefore expect to be able to derive identities analogous to eq.~(\ref{eq:Simon_trick}) at every perturbative order by applying the above argument. In particular, at four loops we shall derive one more relation of this kind, given in eq.~(\ref{eq:zerorel}) below. 

\subsubsection{Colour structure at four loops\label{sec:col4}}

\begin{figure}
    \centering
    \includegraphics{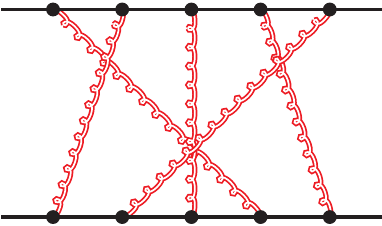}
    \caption{An example of an entangled colour configuration. This specific one is the first term of $d_1$ in eq.~(\ref{def:d1}).}
    \label{fig:5-ent}
\end{figure}

\textbf{Colour factors appearing in $3\to 3$ transitions.} 
We encounter two new colour structures in the transition amplitude $\langle j_3| \tilde{H}_{3\to3}^2 | i_3 \rangle$. These are $\mathbf{C}_{\text{DL}}$ and $\mathbf{C}_{\text{ML}}$, defined in eqs.~(\ref{def:CDL}) and~(\ref{def:CML}) and depicted in figure~\ref{fig:h33DL} and~\ref{fig:h33ML}, respectively. As discussed above eq.~(\ref{eq:colA333b}), all the structure constants may be  written as commutators of the 
gauge-group generators by means of the Lie algebra, so eqs.~(\ref{def:CDL}) and~(\ref{def:CML}) may be expressed as
\begin{subequations}
\begin{align}
    \begin{split}
    \label{eq:CDLt}
        \mathbf{C}_{\text{DL}}&=\left(\mathbf{T}^{\left\{a,b,[[c,d],e]\right\}_+}\right)_i\,\left(\mathbf{T}^{\left\{a,c,[[b,e],d]\right\}_+}\right)_j\,,
    \end{split}\\
    \begin{split}
    \label{eq:CMLt}
        \mathbf{C}_{\text{ML}}&=\left(\mathbf{T}^{\left\{a,[b,c],[d,e]\right\}_+}\right)_i\,\left(\mathbf{T}^{\left\{e,[d,c],[b,a]\right\}_+}\right)_j\,,
    \end{split}
\end{align}
\end{subequations}
where we contract each generator $\mathbf{T}_i^a$, in the representation of the projectile, with one generator $\mathbf{T}^a_j$ associated to the target and we preserve manifest target-projectile symmetry. 
As we will clarify below, both features are crucial to evaluate $\mathbf{C}_{\text{DL}}$ and $\mathbf{C}_{\text{ML}}$ in general representations. To this end, we repeatedly  apply identities~(\ref{eq:relTsu-Tt/2}) and (\ref{eq:relTsu+Tt/2}) to eqs.~(\ref{eq:CDLt}) and (\ref{eq:CMLt}), in order to write contractions of the outermost generators on each line in terms of $\tts$ and $\tsu$. However, these identities are applicable only to some of the terms.
While for two- and three-loop colour factors one could identify at every stage a gluon whose both ends are external, thus fitting the pattern of either~(\ref{eq:relTsu-Tt/2}) or (\ref{eq:relTsu+Tt/2}), starting from four loops this is no more the case. 
Specifically, both $\mathbf{C}_{\text{DL}}$ and $\mathbf{C}_{\text{ML}}$ involve {\textit{entangled}} configurations, where all gluons whose emission point is external on line~$i$, have its absorption point internal (that is, appearing in between other generators) on line $j$, and vice versa. In this case, the reduction into a sequence of $\tts$ and $\tsu$ operators faces an obstruction.
An example of such an entangled configuration is shown in figure~\ref{fig:5-ent}.

In total we identified 8 sets, $d_1\dots d_8$, of entangled colour structures, characterised by target-projectile symmetry, that enter in $\mathbf{C}_{\text{DL}}$ and $\mathbf{C}_{\text{ML}}$. We present these contributions in Appendix~\ref{sec:colourappendix}, where we also prove a set of identities that express $d_1\dots d_8$ in terms of operators $\tts$, $\tsu$ and of the quartic Casimir invariant $d_{AA}$. The derivation of these identities, given in eqs.~(\ref{eq:resd1})-(\ref{eq:resd8}), relies on target-projectile symmetry and a  
representation of the colour structures as multiple pairwise interactions where 
each generator on the target line is contracted with one on the projectile line, similar to $\mathbf{C}_{\text{DL}}$ and $\mathbf{C}_{\text{ML}}$ in eqs.~(\ref{eq:CDLt}) and (\ref{eq:CMLt}).  After applying eqs.~(\ref{eq:resd1})-(\ref{eq:resd8}), as well as eqs.~(\ref{eq:relTsu-Tt/2}) and (\ref{eq:relTsu+Tt/2}), to eqs.~(\ref{eq:CDLt}) and (\ref{eq:CMLt}) we obtain
\begin{subequations}
\begin{align}
\begin{split}
\label{eq:rescol0DL}
    \mathbf{C}_{\text{DL}} &= \left\{-\frac{1}{36}\left(\frac{d_{AA}}{N_A}+\frac{C_A^4}{12}\right)+\frac{7\,C_A^2}{48}(\tsu)^2+\frac{C_A}{48}\left(\frac{1}{3}\tts(\tsu)^2-7\tsu\tts\tsu\right)\right.\\
    &\left.+\frac{1}{24}\left(\frac{3}{2}\tsu(\tts)^2\tsu-\tts\tsu\tts\tsu+\frac{1}{3}(\tts)^2(\tsu)^2\right)
    \right\}\mathbf{T}^a_i\mathbf{T}^a_j,
\end{split}\\
\begin{split}
\label{eq:rescol0ML}
    \mathbf{C}_{\text{ML}} &= 
    \left\{\frac{1}{72}\left(\frac{d_{AA}}{N_A}-\frac{C_A^4}{6}\right)+\frac{7\,C_A^2}{192}(\tsu)^2+\frac{11\,C_A}{48}\left(\frac{1}{4}\tsu\tts\tsu-\frac{1}{3}\tts(\tsu)^2\right)\right.\\
    &\left.+\frac{1}{96}\left(\frac{7}{2}\tts\tsu\tts\tsu-3\tsu(\tts)^2\tsu-\frac{1}{6}(\tts)^2(\tsu)^2\right)    
    \right\}\mathbf{T}^a_i\mathbf{T}^a_j.
\end{split}
\end{align}
\end{subequations}

For later use, we derive alternative expressions for $\mathbf{C}_{\text{DL}}$ and $\mathbf{C}_{\text{ML}}$, which make it easy to identify terms that are leading in the large-$N_c$ limit.
To this end we need a new four-loop colour identity 
\begin{equation}
\label{eq:zerorel}
    \bigg\{3\bigg[\tsu,\Big[\tts,\big[\tts,\tsu\big]\Big]\bigg]+\left(\tts-3C_A\right)\left[\tts,\left(\tsu\right)^2\right]\bigg\}\,\mathbf{T}^a_i\mathbf{T}^a_j = 0,
\end{equation}
which will be derived in the second part of the present section, following similar steps to those leading to eq.~(\ref{eq:Simon_trick}).
By applying the three- and four-loop  identities in eqs.~(\ref{eq:Simon_trick}) and (\ref{eq:zerorel}) to eqs.~(\ref{eq:rescol0DL}) and~(\ref{eq:rescol0ML}), we obtain
\begin{subequations}
\begin{align}
\begin{split}
\label{eq:rescolDL}
\mathbf{C}_{\text{DL}}&=\left\{-\frac{1}{36}\left[\frac{d_{AA}}{N_A}-6C_A\left(\frac{d_{AR_i}}{N_{R_i}C_i}+\frac{d_{AR_j}}{N_{R_j}C_j}\right)+\frac{5}{24}C_A^4\right]+\frac{1}{16}\big[\tsu,\tts\big]\tts\tsu\right.\\
&\phantom{=\Bigg\{}\left.-\frac{1}{48}\tts\left[\left(\tsu\right)^2,\tts\right]\right\}\mathbf{T}^a_i\mathbf{T}^a_j\,,
\end{split}\\
\begin{split}
\label{eq:rescolML}
\mathbf{C}_{\text{ML}}&=\left\{\frac{1}{72}\left[\frac{d_{AA}}{N_A}+6C_A\left(\frac{d_{AR_i}}{N_{R_i}C_i}+\frac{d_{AR_j}}{N_{R_j}C_j}\right)-\frac{7}{24}C_A^4\right]-\frac{3}{32}\Big[\tsu,\big[\tsu,\tts\big]\Big]\tts\right.\\
&\left.-\frac{1}{32}\big[\tsu,\tts\big]\tts\tsu+\frac{1}{16}\tsu\big[\tsu,\tts\big]\tts\right\}\mathbf{T}^a_i\mathbf{T}^a_j.
\end{split}
\end{align}
\end{subequations}
In eqs.~(\ref{eq:rescolDL}) and (\ref{eq:rescolML}), we collected in square brackets the leading contributions in the large-$N_c$ limit of $\mathbf{C}_{\text{DL}}$ and $\mathbf{C}_{\text{ML}}$, respectively. These terms are proportional to the unit matrix in colour space. Subleading contributions at large $N_c$ have non-trivial colour structure, characterised by the action of commutators, such as $[\tts,\tsu]$ on the tree-level colour structure. Indeed, the occurrence of such commutators is automatically associated to non-planar colour diagrams. To see this, consider the action of the commutator $[\mathbf{T}^2_s,\tts]$ on a general colour structure 
\begin{align}
\label{eq:non-planar-arg}
\begin{split}
    \raisebox{27pt}{$\displaystyle\frac{1}{4}\big[\mathbf{T}^2_s,\tts\big]$}\includegraphics[scale=0.9]{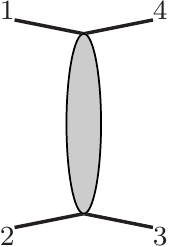}&\raisebox{27pt}{\quad$=$\quad}\,\raisebox{27pt}{$\big[\mathbf{T}_1^a,\mathbf{T}_1^b\big]\,\mathbf{T}_2^a\mathbf{T}_4^b$}\includegraphics[scale=0.9]{Figures/hard-crop.pdf}\\
    & \raisebox{27pt}{\quad$=$\quad}\includegraphics[scale=.9]{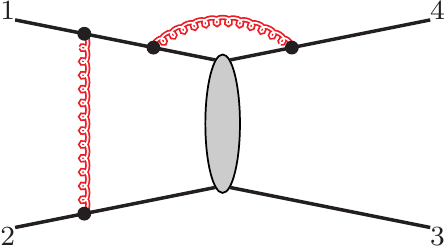} \raisebox{27pt}{$-$}\includegraphics[scale=.9]{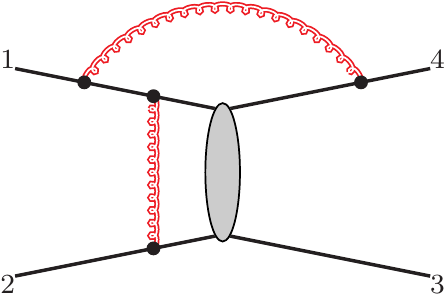}\\
    &\\
    &\raisebox{27pt}{\quad$=$\quad}\,\includegraphics[scale=.9]{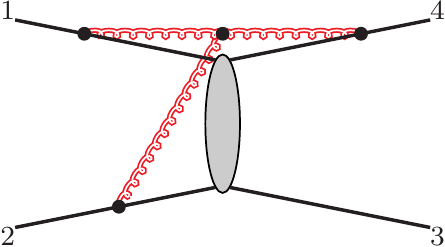}\,.\\
\end{split}
\end{align}

\noindent
Here the grey blob represents a general colour tensor with four indices. In the final line of eq.~(\ref{eq:non-planar-arg}), we used the commutation relation of the Lie algebra. Having done that, three of the four external lines are connected together giving rise to a non-planar diagram. Similarly, one shows that $\big[\mathbf{T}_u^2,\tts\big]$ is non-planar, hence $\big[\tsu,\tts\big]$ is subleading in the large-$N_c$ limit. Thus, by organising the colour structure in terms of commutators, we can immediately distinguish between purely non-planar terms and those that have a planar contribution.  For instance, we observe that all signature-even amplitudes in eqs.~(\ref{eq:evenAmps}) at two-loops and beyond are expressible in terms of (nested) commutators and are hence subleading at large $N_c$. It also becomes clear now that the only terms that contribute in the large-$N_c$ limit in eqs.~(\ref{eq:rescolDL}) and (\ref{eq:rescolML}) are those in the square brackets.

\vskip 10pt

\noindent
\textbf{Colour factors appearing in $3\to 1$ and $1\to 3$ transitions.}
We now proceed to compute the structures $\mathbf{C}_{13}^{(4),A}$ and $\mathbf{C}_{31}^{(4),A}$, defined in eqs.~(\ref{def:C13_4}) and (\ref{def:C31_4}), respectively.
Following the observation at the end of section~\ref{sec:col3}, these colour tensors are proportional to $\mathbf{T}^a_i\mathbf{T}^a_j$. The proportionality factors may be found by repeating the steps that lead from eqs.~(\ref{eq:Simon_trick1}) to (\ref{eq:resC13_3}). For instance, in the case of $\mathbf{C}_{13}^{(4),A}$ we have
\begin{align}
\label{ex:C134A}
 \mathbf{C}_{13}^{(4),A} &= \left(F^xF^y\right)^{ac}\,\text{Tr}\left[F^xF^y\left(F^bF^d+F^dF^b\right)\right]\,\left(\mathbf{T}^{\{a,b,c\}_+}\right)_i\left(\mathbf{T}^d\right)_j = C_{13}^{(4),A}\,\mathbf{T}^a_i\cdot\mathbf{T}^a_j,
\end{align}
where $C_{13}^{(4),A}$, which is a scalar function of the Casimir invariants of the representation $R_i$ of the projectile, can be determined by choosing $\mathbf{T}^d_j=\mathbf{T}^d_i$ and $\mathbf{T}^a_j=\mathbf{T}^a_i$ in eq.~(\ref{ex:C134A}) and contracting all indices. This yields
\begin{align}
\label{ex:C134Astep1}
    C_{13}^{(4),A} &= \frac{1}{N_{R_i}C_i}\,\text{Tr}_{R_i}\left[\mathbf{T}^{\{a,b,c\}_+,d}_i\right]\,\left(F^xF^y\right)^{ac}\,\text{Tr}\left[F^xF^y\left(F^bF^d+F^dF^b\right)\right].
\end{align}
Similar expressions for the target-projectile symmetric colour factor $\mathbf{C}_{31}^{(4),A}$ are simply obtained from of eqs.~(\ref{ex:C134A}) and (\ref{ex:C134Astep1}), by replacing the representation of the projectile $R_i$ with the one of the target, $R_j$.
After computing traces with the codes \verb+color+~\cite{vanRitbergen:1998pn} and \verb+ColorMath+~\cite{Sjodahl:2012nk}, we find
\begin{subequations}
\begin{align}
\label{eq:resC13_4}
  \mathbf{C}^{(4),A}_{13}&=C_A\,\frac{d_{AR_i}}{N_{R_i}C_i}\,\mathbf{T}^a_i\mathbf{T}^a_j,\\
\label{eq:resC31_4}    
  \mathbf{C}^{(4),A}_{31}&=C_A\,\frac{d_{AR_j}}{N_{R_j}C_j}\,\mathbf{T}^a_i\mathbf{T}^a_j.
\end{align}
\end{subequations}
By target-projectile symmetry, both $\mathbf{C}^{(4),A}_{13}$ and $\mathbf{C}^{(4),A}_{31}$ multiply the same integrals, defined in the first line of in eqs.~(\ref{eq:Hhat31Htilde33}). Therefore, the transition amplitude involves the colour factor 
\begin{equation}
\label{eq:resC13C31_4}
    \mathbf{C}^{(4),A}_{13}+\mathbf{C}^{(4),A}_{31} = C_A\,\left( \frac{d_{AR_i}}{N_{R_i}C_i} + \frac{d_{AR_j}}{N_{R_j}C_j} \right)\,\mathbf{T}^a_i\mathbf{T}^a_j,
\end{equation}
which simply corresponds to the three-loop colour structure in eq.~(\ref{res:col133b}) multiplied by~$C_A$. 

We compute $\mathbf{C}_{13}^{(4),B}$, associated to the diagram in figure~\ref{fig:h13-4b}, and its symmetric partner under exchange of target and projectile, $\mathbf{C}_{31}^{(4),B}$, by applying the same procedure. These colour tensors are defined in eqs.~(\ref{eq:h13_4b}) and (\ref{def:tildeC31_4}). Furthermore, they admit
\begin{equation}
    \mathbf{C}_{13}^{(4),B}=C_{13}^{(4),B}\,\mathbf{T}^a_i\,\mathbf{T}^a_j,\qquad\mathbf{C}_{31}^{(4),B}=C_{31}^{(4),B}\,\mathbf{T}^a_i\,\mathbf{T}^a_j.
\end{equation}
We determine the scalar coefficients $C_{13}^{(4),B}$ and $C_{31}^{(4),B}$ by following the steps described in eqs.~(\ref{ex:C134Astep1}) and~(\ref{eq:resC13_4}), getting 
\begin{subequations}
\begin{align}
\begin{split}
\label{eq:restildeC13}
    C_{13}^{(4),B}&=\frac{1}{N_{R_i}C_i}\,\text{Tr}\left[\mathbf{T}_i^{\{a,b,c\}_+,d}\right]\,\left(F^xF^y\right)^{ac}\,\left(\text{Tr}\left[F^xF^bF^yF^d\right]+\text{Tr}\left[F^xF^dF^yF^b\right]\right)
    = 0,
\end{split}\\
\begin{split}
\label{eq:restildeC31}
    C_{31}^{(4),B}&=\frac{1}{N_{R_j}C_j}\,\text{Tr}\left[\mathbf{T}_j^{\{a,b,c\}_+,d}\right]\,\left(F^xF^y\right)^{ac}\,\left(\text{Tr}\left[F^xF^bF^yF^d\right]+\text{Tr}\left[F^xF^dF^yF^b\right]\right)
    = 0,
\end{split}
\end{align}
\end{subequations}
where we used the code \texttt{color} \cite{vanRitbergen:1998pn} to compute the traces in eqs.~(\ref{eq:restildeC13}) and (\ref{eq:restildeC31}). This holds for general representations in any semi-simple gauge group. 
The vanishing of $\mathbf{C}_{13}^{(4),B}$ and $\mathbf{C}_{31}^{(4),B}$ simplifies the calculation of the $3\to 1$ transition at four loops, defined in eq.~(\ref{eq:Hhat31Htilde33}), and its symmetric contribution under target-projectile exchange, the $1\to3$ transition. 

Using the calculation of $\mathbf{C}_{13}^{(4),B}$ and $\mathbf{C}_{31}^{(4),B}$ above, we now derive the identity quoted in eq.~(\ref{eq:zerorel}) above, which was applied to simplify the four-loop colour structures in eqs.~(\ref{eq:rescolDL}) and (\ref{eq:rescolML}). To this end we compute $\mathbf{C}_{13}^{(4),B}$ and $\mathbf{C}_{31}^{(4),B}$ again, starting with  eqs.~(\ref{eq:h13_4b}) and (\ref{def:tildeC31_4}) and applying the identities in appendix \ref{sec:colourappendix}, along with eqs.~(\ref{eq:relTsu-Tt/2}) and (\ref{eq:relTsu+Tt/2}),  obtaining
\begin{equation}
\label{eq:c134+c314}
\mathbf{C}_{13}^{(4),B} + \mathbf{C}_{31}^{(4),B} = \left\{3\bigg[\tsu,\Big[\tts,\big[\tts,\tsu\big]\Big]\bigg]+\left(\tts-3C_A\right)\left[\tts,\left(\tsu\right)^2\right]\right\}\,
\mathbf{T}^a_i\mathbf{T}^a_j.
\end{equation}
By comparing the sum of eqs.~(\ref{eq:restildeC13}) and (\ref{eq:restildeC31}) to eq.~(\ref{eq:c134+c314}), we obtain eq.~(\ref{eq:zerorel}). As a cross-check we re-compute eq.~(\ref{eq:resC13C31_4}) by applying eqs.~(\ref{eq:relTsu-Tt/2}) and (\ref{eq:relTsu+Tt/2}), as well as identities in appendix~\ref{sec:colourappendix}, to the definitions of $\mathbf{C}_{13}^{(4),A}$ and $\mathbf{C}_{31}^{(4),A}$, respectively in eq.~(\ref{def:C13_4}) and (\ref{def:C31_4}) and we get
\begin{align}
\begin{split}
\label{eq:relC13_4}
\mathbf{C}_{13}^{(4),A}+\mathbf{C}_{31}^{(4),A} &= C_A\,\left\{\frac{\tts}{4}\left[\left(\tsu\right)^2+\frac{C_A^2}{12}\right]+\frac{3}{4}\tsu\big[\tsu,\tts\big]\right\}\,\mathbf{T}^a_i\mathbf{T}^a_j\\
&-\frac{5}{48}\,\left\{3\bigg[\tsu,\Big[\tts,\big[\tts,\tsu\big]\Big]\bigg]+\left(\tts-3C_A\right)\left[\tts,\left(\tsu\right)^2\right]\right\}\,\mathbf{T}^a_i\mathbf{T}^a_j.
\end{split}
\end{align}
Here, the second line vanishes by eq.~(\ref{eq:zerorel}), while the first line, reproduces the expected result in eq.~(\ref{eq:resC13C31_4}) upon using eq.~(\ref{eq:Simon_trick}).

Finally, we compute the colour structures $\mathbf{C}_{3113}^{(4),D}$ and $\mathbf{C}_{3113}^{(4),X}$, which were defined in eqs.~(\ref{def:C3113_D}) and (\ref{def:C3113_X}), respectively. By a similar argument to that leading to eq.~(\ref{eq:Simon_trick12}) we have
\begin{subequations}
\begin{align}
    \mathbf{C}_{3113}^{(4),D} &= C_{3113}^{(4),D}\,\mathbf{T}^a_i\,\mathbf{T}^a_j,\\
    \mathbf{C}_{3113}^{(4),X} &= C_{3113}^{(4),X}\,\mathbf{T}^a_i\,\mathbf{T}^a_j,
\end{align}
\end{subequations}
where the scalar functions $C_{3113}^{(4),D}$ and $C_{3113}^{(4),X}$ are
\begin{subequations}
\begin{eqnarray}
C_{3113}^{(4),D}&=&\frac{1}{N_A}\,\text{Tr}\left[F^a F^c F^d F^e\right]\,\text{Tr}\left[F^a F^c F^d F^e+ F^a F^e F^d F^c\right]\nn \\
&=&\left(2\frac{d_{AA}}{N_A}+\frac{C_A^4}{6}\right),\label{eq:resC3113_D}\\
C_{3113}^{(4),X} &=& \frac{1}{N_A}\,\text{Tr}\left[F^aF^cF^dF^e\right]\,\text{Tr}\left[F^a F^c F^e F^d+ F^a F^e F^c F^d + F^a F^d F^e F^c + F^a F^d F^c F^e\right]\nn
\\
&=&\left(4\frac{d_{AA}}{N_A}-\frac{C_A^4}{6}\right), \label{eq:resC3113_X}
\end{eqnarray}
\end{subequations}
where we computed the traces using the code of ref.~\cite{vanRitbergen:1998pn}.
It is easy to check using eq.~(\ref{eq:dAA}) that $C_{3113}^{(4),X}$ is subleading in the large-$N_c$ limit, while $C_{3113}^{(4),X}$ has a leading ${\cal O}(N_c^4)$ component. This can be expected based on the non-planar and planar nature of figures~\ref{fig:h3113b} and~\ref{fig:h3113a}, respectively.  

In conclusion, we expressed all the colour structures that contribute to the reduced amplitudes through four loops in terms of colour channel operators $\tts$ and $\tsu$. In addition, they contain quartic Casimir invariants 
associated with the colour representations of the scattered particles, $d_{AR_i}$ and $d_{AR_j}$, starting at three loops, and purely adjoint ones,~$d_{AA}$, starting at four loops.
These arise in both the $3\to3$ transitions of eqs.~(\ref{eq:rescolDL}) 
and~(\ref{eq:rescolML}) and the $1\to 3\to 1$ transition of eqs.~(\ref{eq:resC3113_D}) and~(\ref{eq:resC3113_X}).
Note that the $d_{AA}$ terms are driven entirely by gluon loops, thus, they are entirely independent of the representations of the scattered particles and on the matter content of the theory.

\subsection{Integrals\label{subsec:integrals}}

The transition amplitudes defined in section~\ref{subsec:dia} involve a restricted set of integrals in $d=2-2\epsilon$ dimensional Euclidean space, transverse to the lightcone of the colliding particles. The associated integrands are constructed in term of massless propagators of the three exchanged Reggeons, with momenta $k_1$, $k_2$ and $k_3 = p-k_1-k_2$, and integration kernels $H_{22}$, $H_{13}$ and $\alpha_g$. It is convenient to use the notation of ref.~\cite{Caron-Huot:2017fxr},
\begin{equation}
\label{def:Iint}
I[N] \equiv \left(\frac{4\pi S_\eps(p^2)}{r_\Gamma}\right)^2\int\frac{d^{2-2\eps}p_1}{(2\pi)^{2-2\eps}}\int\frac{d^{2-2\eps}p_2}{(2\pi)^{2-2\eps}}\frac{p^2}{p_1^2p_2^2(p-p_1-p_2)^2}N,
\end{equation}
where $N$ indicates a general integrand depnding on $p_1$ and $p_2$. All integrals appearing in transition amplitudes through four loops may be expressed in terms of the following set
\begin{align}
\label{def:Jint}
 J_1 = I\left[\frac{1}{\eps^2}\right], &&
 J_2 = I\left[\frac{1}{\eps^2}\ratio{1}\right], &&
 J_3 = I\left[\frac{1}{\eps^2}\ratio{12}\right], \nonumber\\
 J_4 = I\left[\frac{1}{\eps^2}\left(\ratio{1}\right)^{\!2}\right], &&
 J_5 = I\left[\frac{1}{\eps^2}\left(\ratio{12}\right)^{\!2}\right], &&
 J_6= I\left[\frac{1}{\eps^2}\ratio{1}\ratio{2}\right], \nonumber\\
 J_7 = I\left[\frac{1}{\eps^2}\ratio{1}\ratio{12}\right], &&
 J_8 = I\left[\frac{1}{\eps^2}\ratio{1}\ratio{23}\right], &&
 J_9 = I\left[\frac{1}{\eps^2}\ratio{12}\ratio{13}\right],\nonumber\\
\end{align}
where $p_{ij}\equiv p_i+p_j$.
We emphasise that despite describing $2\to 2$ amplitudes, the integrals above have a single kinematic scale, namely the squared transferred momentum of the process $p^2=-t$, because the dependence on $s$ is completely determined by Balitsky-JIMWLK evolution.

With exception of $J_9$, all the integrals may be computed by repeatedly applying the one-loop propagator formula \cite{Tkachov:1981wb,Kazakov:1984bw}
\begin{align}
\label{def:bub}
\int\frac{[\dbar k]}{\big(k^2\big)^\alpha\big(\left(p-k\right)^2\big)^\beta}&=\frac{B_{\alpha,\beta}}{\left(4\pi\right)^{1-\epsilon}}\,\left(p^2\right)^{1-\epsilon-\alpha-\beta},
\end{align}
where as above, $[\dbar k] = \frac{d^{d-2}k}{(2\pi)^{d-2}}$, and the function $B_{\alpha,\beta}$ is defined by
\begin{align}
\label{def:Bfunction}
B_{\alpha,\beta}&=\frac{\Gamma(1-\epsilon-\alpha)\Gamma(1-\epsilon-\beta)\Gamma(\alpha+\beta-1+\epsilon)}{\Gamma(\alpha)\Gamma(\beta)\Gamma(2-2\epsilon-\alpha-\beta)}.
\end{align}
Using the notation above, the integrals $J_1\dots J_8$ evaluate to
\begin{subequations}
\begin{align}
\label{eq:resj1}
  J_1&=\frac{4 B_{1,\eps+1}}{\eps^4 B_{1,1}}=\frac{3}{\eps^4}-\frac{18 \zeta_{3}}{\eps}-27 \zeta_{4}+ {\cal{O}}\left(\epsilon\right)\\
\label{eq:resj2} 
  J_2&=\frac{4 B_{\eps+1,\eps+1}}{\eps^4 B_{1,1}}=\frac{2}{\eps^4}-\frac{44 \zeta_{3}}{\eps}-66 \zeta_{4}+ {\cal{O}}\left(\epsilon\right)\\
\label{eq:resj3}  
  J_3&=\frac{4 B_{1,2 \eps+1}}{\eps^4 B_{1,1}}=\frac{8}{3 \eps^4}-\frac{128 \zeta_{3}}{3 \eps}-64 \zeta_{4}+ {\cal{O}}\left(\epsilon\right)\\
\label{eq:resj4}  
  J_4&=J_8=\frac{4 B_{2 \eps+1,\eps+1}}{\eps^4 B_{1,1}}=\frac{5}{3 \eps^4}-\frac{230 \zeta_{3}}{3 \eps}-115 \zeta_{4}+ {\cal{O}}\left(\epsilon\right)\\
\label{eq:resj5}  
  J_5&=\frac{4 B_{1,3 \eps+1}}{\eps^4 B_{1,1}}=\frac{5}{2 \eps^4}-\frac{75 \zeta_{3}}{\eps}-\frac{225 \zeta_{4}}{2}+ {\cal{O}}\left(\epsilon\right)\\
\label{eq:resj6}  
  J_6&=\frac{4 B_{1,\eps+1} B_{\eps+1,2 \eps+1}}{\eps^4 B_{1,1}^2}=\frac{5}{4 \eps^4}-\frac{65 \zeta_{3}}{\eps}-\frac{195 \zeta_{4}}{2}+ {\cal{O}}\left(\epsilon\right)\\
\label{eq:resj7}  
  J_7&=\frac{4 B_{1,3 \eps+1} B_{\eps+1,1}}{\eps^4 B_{1,1}^2}=\frac{15}{8 \eps^4}-\frac{135 \zeta_{3}}{2 \eps}-\frac{405 \zeta_{4}}{4}+ {\cal{O}}\left(\epsilon\right)
\end{align}
\end{subequations}
Note the absence of poles of order $\epsilon^{-3}$ and $\epsilon^{-2}$, the latter being associated with the absence of $\zeta_2$. This will reflect in the final results of the amplitudes. 

\begin{figure}[htbp]
    \centering
    \includegraphics{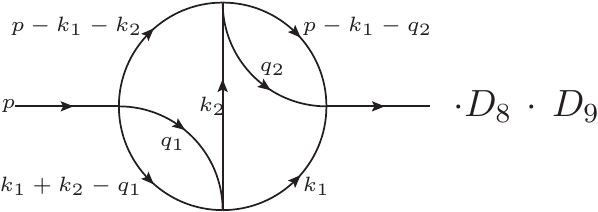}
    \caption{The integral $T^{(2-2\epsilon)}_{1,1,1,1,1,1,1,-1,-1,0,0,0,0,0}$ which is related to the required 
    integral~$J_9$ through eq.~(\ref{eq:j9tot9}). Arrows on the scalar propagators are added to keep track of the momentum flow. The factors $D_8$ and $D_9$ represent numerator insertions.}
    \label{fig:j9}
\end{figure}

In order to compute the remaining integral $J_9$, we define an integral family generated by the following inverse scalar propagators 
\begin{equation}
\label{def:topo}
\begin{array}{llll}
D_1 = k_1^2 & D_2=k_2^2 & D_3=(p-k_1-k_2)^2\quad & D_4=q_1^2\\
D_5 = (k_1+k_2-q_1)^2\quad & D_6=q_2^2\qquad & D_7=(p-k_1-q_2)^2\quad & D_8=(k_1+k_2)^2\\
D_9 = (p-k_1)^2 & D_{10} = (k_1+q_1)^2 \quad& D_{11} = (k_1+q_2)^2 & D_{12} = (k_2+q_2)^2\\
D_{13} =(q_1+q_2)^2 & D_{14} = (p-q_1)^2 & 
\end{array}    
\end{equation}
Denoting integrals in this topology as
\begin{equation}
\label{def:t9}
    T^{(d)}_{n_1,\dots,n_{14}} \equiv \left(p^2\,e^{\gamma_E}\right)^{4\epsilon}\int\frac{d^dk_1\,d^dk_2\,d^dq_1\,d^dq_2}{\left(\pi^{{d}/{2}}\right)^4}\,\frac{p^2}{D_1^{n_1}\dots D_{14}^{n_{14}}},
\end{equation}
we consider $T^{(d)}_{1,1,1,1,1,1,1,-1,-1,0,0,0,0,0}$, associated to the diagram in figure~\ref{fig:j9}, which is related to the integral $J_9$ by
\begin{equation}
\label{eq:j9tot9}
    J_9 =\left(\frac{e^{-\epsilon\gamma_E}}{\epsilon\,B_{1,1}\,r_\Gamma}\right)^2\,T^{(2-2\epsilon)}_{1,1,1,1,1,1,1,-1,-1,0,0,0,0,0}\,,
\end{equation}
where $r_\Gamma$ is defined in eq.~(\ref{r_gamma}) and $B_{1,1}$ in eq.~(\ref{def:Bfunction}). 

We use the code \texttt{LiteRed} \cite{Lee:2013mka} to apply raising dimensional recurrence relations \cite{Tarasov:1996br} and express the integral $T^{(2-2\epsilon)}_{1,1,1,1,1,1,1,-1,-1,0,0,0,0,0}$ in terms of $2$-point integrals in $4-2\epsilon$ dimensions, $T^{(4-2\epsilon)}_{n_1,n_2,n_3,n_4,n_5,n_6,n_7,n_8,n_9,0,0,0,0,0}$. The latter are computed with the program \texttt{Forcer} \cite{Ruijl:2017cxj} and expanded\footnote{Note that the topology defined in eq.~(\ref{def:topo}) is absent in \texttt{Forcer}. Instead, the topology \texttt{t1star24} is defined in terms of $D_1$, $D_2$, $D_3$, $D_8$ and $D_9$, where the latter two denominators are allowed to assume non-integer powers. The conversion of integrals $T^{(4-2\epsilon)}_{n_1,n_2,n_3,n_4,n_5,n_6,n_7,n_8,n_9,0,0,0,0,0}$ in \texttt{Forcer} notation is immediately obtained by integrating the one-loop sub-integrals over $q_1$ and $q_2$.} in $\epsilon$, giving
\begin{align}
    \label{eq:rest9}
    T^{(2-2\epsilon)}_{1,1,1,1,1,1,1,-1,-1,0,0,0,0,0} &= \frac{28}{3\epsilon^4}-\frac{56\zeta_2}{3\epsilon^2}-\frac{3352\zeta_3}{9\epsilon}-\frac{1508\zeta_4}{3}+\epsilon\left(\frac{6704}{9}\zeta_2\zeta_3-\frac{73432}{15}\zeta_5\right)\nn\\
    &+ \epsilon^2\left(\frac{230264}{27}\zeta_3^2-9458\zeta_6\right)+{\cal{O}}\left(\epsilon^3\right)
   \end{align}
and 
\begin{equation}
\label{eq:resj9}
    J_9=\frac{7}{3 \eps^4}-\frac{214 \zeta_{3}}{3 \eps}-107\,\zeta_{4} -1166\,\zeta_5\,\epsilon + \left(\frac{4094}{3}\zeta_3^2-\frac{8210}{3}\zeta_6\right)\eps^2 + {\cal{O}}\left(\epsilon^3\right).
\end{equation}
We checked the result in eq.~(\ref{eq:resj9}) by verifying up to ${\cal{O}}(\epsilon^5)$ that it yields 
uniform-weight zeta numbers\footnote{In contrast to the integrals $J_1\dots J_8$, which all yield only single zeta values, $J_9$ features multiple zeta values, such as $\zeta_{5,3}$.}, by comparing it 
with ref.~\cite{Caron-Huot:2013fea}, which provides $J_9$ to ${\cal{O}}(\epsilon)$, and by comparing with the numerical result, obtained up to ${\cal{O}}(\epsilon)$ with \texttt{pySecDec} \cite{Borowka:2017idc}.

The integrals $J_1\dots J_9$ are all pure transcendental numbers of uniform weight 4, when $\epsilon$ is assigned weight $-1$. We note that the factor $r_\Gamma$ in the denominator of eq.~(\ref{def:Iint}) removes all the occurrences of $\zeta_2$ from eqs.~(\ref{eq:resj1})-(\ref{eq:resj7}) and (\ref{eq:resj9}). Furthermore, all occurrences of even zeta values may be removed by re-defining the odd zeta numbers, 
as in refs.~\cite{Baikov:2018gap,Baikov:2018wgs,Baikov:2019zmy,Kotikov:2019bqo}. This can be immediately verified by absorbing $\zeta_4$ and $\zeta_6$ in eqs.~(\ref{eq:resj1})-(\ref{eq:resj7}) and (\ref{eq:resj9}) using the definition
\begin{equation}
\label{def:hatz}
    \hat{\zeta}_3 = \zeta_3 + \frac{3}{2}\epsilon\zeta_4 - \frac{5}{2}\epsilon^3\zeta_6+ {\cal{O}}\left(\epsilon^5\right),\quad\hat{\zeta}_5 = \zeta_5 + \frac{5}{2}\epsilon \zeta_6 + {\cal{O}}\left(\epsilon^3\right)
\end{equation}
The appropriate definitions of hatted representation of the zeta values is derived in ref.~\cite{Baikov:2019zmy,Kotikov:2019bqo} including higher orders\footnote{The systematic construction of hatted single zeta values has been developed for all orders in $\epsilon$ in ref.~\cite{Kotikov:2019bqo}.} in $\epsilon$.

\section{The NNLL amplitude through four loops: results}
\label{sec:results}
We now provide results for the reduced amplitude $\hat{\mathcal{M}}^{(-),\text{NNLL}}_{ij\to ij}$, eq.~(\ref{eq:mhatNNLL}), through four loops. The relevant expressions for the transition amplitudes are defined in section~\ref{subsec:dia}. Here we evaluate them in terms of the colour factors computed in section~\ref{subsec:colour} and of the basis of integrals introduced in section~\ref{subsec:integrals}. At each perturbative order, we discuss the planar limit of the reduced amplitude, by separating the dominant terms at large $N_c$.

In section~\ref{sec:pole_cut_explicit_rep} we construct the complete amplitudes, according to eq.~(\ref{Mreduced}), and we disentangle the contributions of the Regge pole from those associated to the Regge cut, as specified in eq.~(\ref{Schemes_Cut}). We provide explicit results for quark-quark, quark-gluon and gluon-gluon scattering amplitudes. Finally, using this \textit{Regge-cut scheme}, we extract two-loop quark and gluon impact factors, $\tilde{C}_q^{(2)}$ and $\tilde{C}_g^{(2)}$, in QCD, as well as the three-loop Regge trajectory in $\mathcal{N}=4$ SYM.

\subsection{Two loops}\label{sec:twoLoopsComp}

The transitions that contribute to two-loop reduced amplitudes can be read off the expansion of the general formula eq.~(\ref{eq:mhatNNLL}) to ${\cal{O}}\left(\alpha_s^2\right)$
\begin{equation}
\label{reduced_ampl_two_loop}
    \frac{i}{2s}\hat{\mathcal{M}}^{(-,2,0)}_{ij\to ij} = \langle j_1|i_1\rangle^{\text{NNLO}} + \pi^2r_\Gamma^2\,\langle j_3|i_3\rangle.
\end{equation}
Here the first term describes the contribution of single-Reggeon exchange to the two-loop reduced amplitude. It may be obtained by substituting the NNLO impact factors introduced in eq.~(\ref{impact_factors_def}) into the first term in eq.~(\ref{Regge-Pole-and-Cut}) and expanding it to two loops.
This fixes the first term in eq.~(\ref{reduced_ampl_two_loop}) to be
\begin{equation}
\label{eq:resJ1I1NNLO}
\langle j_1 | i_1\rangle^{\text{NNLO}} = \left[D_i^{(2)}+D_j^{(2)}+D_i^{(1)}D_j^{(1)}\right]\langle j_1 | i_1\rangle,
\end{equation}
where the impact-factor coefficients $D_i^{(1)}$ and $D_i^{(2)}$ were defined in eq.~(\ref{NLO_wavefunction}) and (\ref{NNLO_wavefunction}), respectively. These coefficients have been determined in QCD in ref.~\cite{Caron-Huot:2017fxr}, by matching the expression above with explicit results of two-loop helicity amplitudes~\cite{Bern:2002tk,Bern:2003ck,DeFreitas:2004kmi}. They are quoted here in 
appendix~\ref{ImpactFactors_Appendix}.
Of course, this term is proportional to the tree-level amplitude,
\begin{equation}
\label{eq:resTREE}
    \langle j_1 | i_1\rangle = \frac{i}{2s}\hat{\mathcal{M}}^{\text{tree}}_{ij\to ij}=i\,\frac{g_s^2}{t}\,\mathbf{T}^a_i\mathbf{T}^a_j,
\end{equation}
a purely $t$-channel octet exchange.

\vspace{1em}
\noindent
{\textbf{The ${\mathbf{3\to 3}}$ transition}}, $\langle j_3|i_3\rangle$ entering eq.~(\ref{reduced_ampl_two_loop}), provides the first contribution of three-Reggeon states to scattering amplitudes. Substituting eq.~(\ref{eq:amp33_2loops}) into 
\begin{equation}
\label{eq:j3i3}
    \langle j_3 | i_3\rangle = \frac{\langle \psi_{j,3} |\psi_{i,3}\rangle}{\left(\alpha_sr_\Gamma\right)^2}
\end{equation}
and evaluating the integral, we obtain
\begin{align}
\label{eq:resJ3I3}
    \langle j_3 | i_3\rangle &=-\frac{i g_s^2}{t}\,\epsilon^2\,J_1\,\mathbf{C}_{33}^{(2)}\\&\nonumber =-\frac{1}{8}\left(\frac{1}{\epsilon^2}-6\epsilon\,\hat{\zeta}_3+{\cal{O}}\left(\epsilon^3\right)\right)\,\left[\left(\tsu\right)^2-\frac{C_A^2}{12}\right]\langle j_1 | i_1\rangle,
\end{align}
where we expressed the colour factor $\mathbf{C}^{(2)}_{33}$ as in eq.~(\ref{res:colA332}) and we set the scale of the coupling $\mu^2=-t$ for simplicity. The value of $\hat{\zeta}_3$ is given in eq.~(\ref{def:hatz}). In this section, we will consistently suppress the occurrence of logarithms of $\frac{\mu^2}{-t}$ by making this scale choice.

Because of the action of $(\tsu)^2$, the colour structure of eq.~(\ref{eq:resJ3I3}) is strikingly different to single-Reggeon exchange $\langle j_1|i_1\rangle$, breaking Regge-pole factorisation \cite{DelDuca:2011ae,DelDuca:2011wkl,DelDuca:2013ara,DelDuca:2013dsa,DelDuca:2014cya}. 
Importantly, the leading term in the large-$N_c$ expansion of $(\tsu)^2\langle j_1|i_1\rangle$ is always proportional to $\langle j_1|i_1\rangle$, and furthermore, the proportionality constant does not depend on the representation of the scattered particles.  
This can be understood by rewriting eq.~(\ref{eq:Simon_trick}) as 
\begin{align}
\label{eq:TsuNc}
\frac{N_c(\tsu)^2}{4}\mathbf{T}^a_i\mathbf{T}^a_j &= \left(\frac{d_{AR_i}}{N_{R_i}C_i} + \frac{d_{AR_j}}{N_{R_j}C_j} - \frac{N_c^3}{48} \right)\mathbf{T}^a_i\mathbf{T}^a_j \\
&-\frac{3}{4}\tsu\,\big[\tsu,\tts\big]\,\mathbf{T}^a_i\mathbf{T}^a_j-\frac{1}{4}\left[\tts,\left(\tsu\right)^2\right]\,\mathbf{T}^a_i\mathbf{T}^a_j.\nonumber
\end{align}
Here, the second line is manifestly non-planar, being written in terms of commutators (see eq.~(\ref{eq:non-planar-arg})). In order to determine the large-$N_c$ limit of the first line, we notice that the invariant $\frac{d_{AR_i}}{N_{R_i}C_i}$ takes a simple form, when the representation $R_i$ is either the fundamental or the adjoint. From eq.~(\ref{eq:dAA}) we find
\begin{align}
    \label{eq:dAAlargeNc}
    \frac{d_{AR_i}}{N_{R_i}C_i} &= \frac{N_c^3}{24} + N_c\,\left\{
    \begin{array}{l}
         \frac{1}{4}  \qquad\text{if } R_i=F\\
         \\
         \frac{3}{2}  \qquad\text{if } R_i=A
    \end{array}
    \right. \,\nonumber\\
    &\equiv\, \frac{N_c^3}{24} + N_c\,\kappa(R_i).
\end{align}
After plugging the expression above into eq.~(\ref{eq:TsuNc}) and multiplying by $i\,\frac{g_s^2}{t}$ we get 
\begin{align}
\label{eq:TsuNc2}
 (\tsu)^2\langle j_1|i_1\rangle &= \left[\frac{N_c^2}{4}+4\,\kappa(R_i) + 4\,\kappa(R_j)\right]\langle j_1|i_1\rangle \\
 &-\frac{3}{N_c}\tsu\big[\tsu,\tts\big]\langle j_1|i_1\rangle-\frac{1}{N_c}\left[\tts,\left(\tsu\right)^2\right]\langle j_1|i_1\rangle.\nonumber
\end{align}
In the large-$N_c$ limit, the first term dominates. It features the colour factor of the tree-level amplitude $\langle j_1|i_1\rangle$, corresponding to the exchange of a colour octet, multiplied by the factor $N_c^2/4$, irrespective of whether the projectile and the target are quarks or gluons. Subleading terms in $N_c$ are separated into two parts. In the first line, terms proportional to the function $\kappa(R)$ are still characterised by colour octet structure. By changing the representation, either fundamental or adjoint, of the scattered partons, these terms are easily constructed as the symmetric sum of one contribution associated to the projectile and one to the target. The terms in the second line of eq.~(\ref{eq:TsuNc2}) have non-trivial colour structure and, to the best of our knowledge, their expressions for definite representations of the external particles do not obey simple relations.

Finally, plugging eqs.~(\ref{eq:resJ1I1NNLO}) and (\ref{eq:resJ3I3}) into eq.~(\ref{reduced_ampl_two_loop}) we recover the result of two-loop amplitudes~\cite{Caron-Huot:2017fxr}:
\begin{equation}
    \mExpM{2}{0} =\left[D_i^{(2)}+D_j^{(2)}+D_i^{(1)}D_j^{(1)}+\pi^2 r_\Gamma^2\,S^{(2)}(\epsilon)\left((\tsu)^2-\frac{1}{12}C_A^2\right)\right]\mTree\label{eq:M20},
\end{equation}
where we define\footnote{We slightly depart from the notation of ref.~\cite{Caron-Huot:2017fxr}, where the expansion of the reduced amplitude involved $R^{(2)}(\epsilon)=r_\Gamma^2\,S^{(2)}(\epsilon)$. Similarly, at three loops we will introduce the quantities $S^{(3)}_A(\epsilon)$, $S^{(3)}_B(\epsilon)$ and $S^{(3)}_C(\epsilon)$, which are related to corresponding expressions in \cite{Caron-Huot:2017fxr}, $R^{(3)}_i(\epsilon) = r_\Gamma^3\,S^{(3)}_i(\epsilon)$ with $i=A,B,C$.}
\begin{equation}\label{eq:R2Res}
    S^{(2)}(\epsilon)=-\frac{\epsilon^2}{24}J_1 = -\frac{1}{8\eps^2} +\frac{3}{4}\eps\zeta_3 + \frac{9}{8}\eps^2\zeta_4 +{\cal O}(\eps^3).
\end{equation}
We recall that the entire dependence on the type of the scattered partons is encoded in the impact factors $D_i$ and $D_j$, while the three-Reggeon contribution in operator form is entirely process-independent\footnote{Note the distinction we make between process-independence of the operator itself, versus the process dependence of its projection on the various components in an explicit colour basis, which we discuss in section~\ref{sec:pole_cut_explicit_rep} and appendix \ref{subsec:ExplicitRep} below (where we choose the $t$-channel basis).}.
Likewise the former depends on the specific gauge theory considered, while the latter is entirely independent of the matter content and is thus universal. As commented already in section~\ref{sec:equation12} these features (scattering-process independence in operator form and matter-contents independence) apply to the entire tower of MRS contributions to the reduced odd amplitude at NNLL to all loops. In appendix~\ref{subsec:ExplicitRep}, we compute explicitly the reduced amplitude in eq.~(\ref{eq:M20}), by picking orthonormal colour bases in the $t$-channel for quark-quark, gluon-gluon and quark-gluon scattering processes. We obtain the colour octet and colour singlet contributions to the quark-quark amplitude, respectively in eqs.~(\ref{eq:M20qqOctet}) and (\ref{eq:M20qqSinglet}), (antisymmetric) octet and decuplet in gluon-gluon scattering, respectively in eqs.~(\ref{eq:M20ggOctet}) and (\ref{eq:M20ggDecuplet}), and an (antisymmetric) octet in quark-gluon scattering, in eq.~(\ref{eq:M20qgOctet}).

In the planar limit, eq.~(\ref{eq:M20}) becomes
\begin{equation}
   \left. \mExpM{2}{0}\right|_{\text{planar}} =
   \left[D_i^{(2)}+D_j^{(2)}+D_i^{(1)}D_j^{(1)}+\pi^2r_\Gamma^2\,\frac{N_c^2}{6}S^{(2)}(\epsilon)\right]\mTree,\label{eq:M20planar}
\end{equation}
where we apply the identity in eq.~(\ref{eq:TsuNc2}) and we retain only the terms leading in $N_c$ in $D_i^{(1)}$, $D_j^{(1)}$ and $D_i^{(2)}$, $D_j^{(2)}$. 

\subsection{Three loops\label{subsec:res3}}

The expansion of eq.~(\ref{eq:mhatNNLL}) to three loops gives 
\begin{equation}
\label{eq:expaMhatnnll3}
    \frac{i}{2s}\hat{\mathcal{M}}^{(-,3,1)}_{ij\to ij}= -\pi^2 r_\Gamma^3\Big[\langle j_3| \tilde{H}_{3\to3} | i_3\rangle\,+\,\left(\langle j_1 |\tilde{H}_{3\to1}|i_3\rangle + \langle j_3 |\tilde{H}_{1\to3}|i_1\rangle\right)\Big].
\end{equation}
Here, the first term is associated to $3\to 3$ transitions, while the two terms in brackets represent $3\to1$ and $1\to 3$ transitions. Each contribution is described in turn below.

\vspace{1em}
\noindent
{\textbf{The $\mathbf{3\to 3}$ transition}} is defined in eq.~(\ref{eq:Hhat33_3loop}) in terms of the integrals in eqs.~(\ref{eq:agtts}) and (\ref{eq:H33_3loop}). These are evaluated at $\mu^2=-t$ in terms of the basis of integrals $J_1\dots J_8$ of section~\ref{subsec:integrals}, getting
\begin{align}
\begin{split}
    \langle j_3 | \tilde{H}_{3\to3} | i_3\rangle &= \frac{ig_s^2}{t}
    \frac{\epsilon}{2}
    \Bigg[(J_3-2J_2)\,\mathbf{C}^{(3)}_{33}+{3\,J_2}\,C_A\,\mathbf{C}^{(2)}_{33}-{J_1}\tts\,\mathbf{C}^{(2)}_{33}\Bigg].
\end{split}    
\end{align}
Here, the first two terms correspond to eq.~(\ref{eq:H33_3loop}). The last term, with the colour structure $\tts\,\mathbf{C}^{(2)}_{33}$, arises from eq.~(\ref{eq:agtts}). Using the expressions for the colour factors $\mathbf{C}^{(2)}_{33}$ and $\mathbf{C}^{(3)}_{33}$ given in eqs.~(\ref{res:colA332}) and (\ref{res:colA333}), respectively, we get 
\begin{align}
\begin{split}
    \label{eq:resJ3H33I3}
    \langle j_3 | \tilde{H}_{3\to3} | i_3\rangle &=-24\,S_C^{(3)}(\epsilon)\,\left(\frac{d_{AR_i}}{N_{R_i}C_i}+\frac{d_{AR_j}}{N_{R_j}C_j}\right)\langle j_1 | i_1\rangle\\
    &+\Big[S_A^{(3)}(\epsilon)\tsu\big[\tsu,\tts\big]+S_B^{(3)}(\epsilon)\big[\tsu,\tts\big]\tsu+S_C^{(3)}(\epsilon)C_A^3\Big]\,\langle j_1| i_1\rangle,
\end{split}    
\end{align}
where
\begin{subequations}
\label{RABC}
\begin{align}
\label{eq:defRA3}
S_A^{(3)}(\epsilon) &\equiv \frac{\epsilon}{16}(J_1-J_3)=
\frac{1}{48\epsilon^3}+\frac{37\,\hat{\zeta}_3}{24}\,+\,{\cal{O}}\left(\epsilon^2\right)\,,\\
\label{eq:defRB3}
S_B^{(3)}(\epsilon) &\equiv  -\frac{\epsilon}{16}(J_2-J_3) =
\frac{1}{24\epsilon^3} +\frac{\hat{\zeta}_3}{12} \,+\,{\cal{O}} \left(\epsilon^2\right)\,,\\
\label{eq:defRC3}
S_C^{(3)}(\epsilon) &\equiv \frac{\epsilon}{288}\left(J_1+J_2-2J_3\right) =
-\frac{1}{432}\left(\frac{1}{2\epsilon^3}-35\hat{\zeta}_3 + {\cal{O}}\left(\epsilon^2\right)\right)\,, 
\end{align}
\end{subequations}
where we expanded each of the combinations of integrals in $\epsilon$ and where $\hat{\zeta}_3$ is given in eq.~(\ref{def:hatz}).
In order to obtain the form in eq.~(\ref{eq:resJ3H33I3}) we applied the three-loop colour identity of eq.~(\ref{eq:Simon_trick}), separating the contribution proportional to the tree-level amplitude $\langle j_1|i_1\rangle$ with the proportionality factor involving the quartic Casimir of the scattered particles, from the remaining colour structures generated by the operators $\tsu[\tsu,\tts]$ and $[\tsu,\tts]\tsu$ and a pure $C_A^3$ term. 

\vspace{1em}
\noindent
\textbf{The $\mathbf{1\to 3}$ and $\mathbf{3\to 1}$ transitions} are defined in eq.~(\ref{eq:H13loop}). By replacing the colour factor with the expression in eq.~(\ref{res:col133b}) and performing the integrals, we get
\begin{align}
\begin{split}
\label{eq:resJ1H31I3}
    \langle j_3 | \tilde{H}_{1\to3} | i_3\rangle + \langle j_1 |\tilde{H}_{3\to1}|i_3\rangle &=\frac{\epsilon}{12}\,\left(J_1+J_2-2J_3\right)\,\left(\frac{d_{AR_i}}{N_{R_i}C_i}+\frac{d_{AR_j}}{N_{R_j}C_j}\right)\,\langle j_1|i_1\rangle,\\
    &=24\,S_C^{(3)}(\epsilon)\,\left(\frac{d_{AR_i}}{N_{R_i}C_i}+\frac{d_{AR_j}}{N_{R_j}C_j}\right)\,\langle j_1|i_1\rangle\,,
\end{split}
\end{align}
where $S_C^{(3)}(\epsilon)$ is given in eq.~(\ref{eq:defRC3}). 
We notice that the result is just the negative of the quartic-Casimir-dependent part of the $3\to 3$ transition amplitude in eq.~(\ref{eq:resJ3H33I3}), so these terms will cancel exactly in the reduced amplitude.

\vspace{1em}
\noindent
\textbf{The three-loop reduced amplitude}, eq.~(\ref{eq:expaMhatnnll3}), is computed by summing eqs.~(\ref{eq:resJ3H33I3}) and (\ref{eq:resJ1H31I3}). As noted above, quartic Casimir invariants cancel between the two transition amplitudes and we obtain
\begin{equation}
    \mExpM{3}{1} =-\pi^2r_\Gamma^3\left[S_A^{(3)}(\epsilon)\tsu\left[\tsu,\tts\right]+S_B^{(3)}(\epsilon)\left[\tsu,\tts\right]\tsu+S_C^{(3)}(\epsilon)C_A^3\right]\mTree\label{eq:M31},
\end{equation}
where $S_A^{(3)}(\epsilon)$, $S_B^{(3)}(\epsilon)$ and $S_C^{(3)}(\epsilon)$ are defined in eqs.~(\ref{eq:defRA3}), (\ref{eq:defRB3}) and (\ref{eq:defRC3}), respectively. 
In appendix~\ref{subsec:ExplicitRep}, eq.~(\ref{eq:M31}) is expanded in the dimensional regulator $\epsilon$ and evaluated in $t$-channel bases. Results are presented for the octet and for the singlet in quark-quark scattering, respectively in eqs.~(\ref{eq:M31qqOctet}) and (\ref{eq:M31qqSinglet}), for the antisymmetric octet and the decuplet in gluon-gluon scattering, respectively in eqs.~(\ref{eq:M31ggOctet}) and (\ref{eq:M31ggDecuplet}), and for the antisymmetric octet in quark-gluon scattering, eq.~(\ref{eq:M31qgOctet}).

In the planar limit the last term in eq.~(\ref{eq:M31}) dominates, while the commutators are subleading, and eq.~(\ref{eq:M31}) then becomes
\begin{equation}
\label{eq:M31planar}
   \left. \mExpM{3}{1}\right|_{\text{planar}} 
   = -\pi^2 r_\Gamma^3 \, S_C^{(3)}(\epsilon) C_A^3\,\mTree\,,
\end{equation}
which is simply proportional to the tree-level amplitude $\langle j_1|i_1\rangle$, times a factor that is independent of the scattered partons, similarly to the planar contribution to the two-loop amplitude, eq.~(\ref{eq:M20planar}).

\subsection{Four loops}\label{sec:fourLoopComp}

By expanding eq.~(\ref{eq:mhatNNLL}) to four loops, we find the reduced amplitude
\begin{align}
\begin{split}
\label{eq:expaMhatnnll4}
\frac{i}{2s}\hat{\mathcal{M}}^{(-,4,2)}_{ij\to ij} &= \frac{\pi^2r_\Gamma^4}{2}\;\Big[\langle j_3|\tilde{H}_{3\to3}^2|i_3\rangle + \left(\langle j_1| \tilde{H}_{3\to1}\tilde{H}_{3\to3}|i_3\rangle + \langle j_3 |\tilde{H}_{3\to3}\tilde{H}_{1\to3}|i_1\rangle \right) \\
&\hspace*{2cm}+ \langle j_1|\tilde{H}_{3\to1}\tilde{H}_{1\to3}|i_1\rangle \Big],
\end{split}
\end{align}
where the terms above describe respectively $3\to 3$, the sum of $3\to1$ and $1\to3$ and finally $1\to1$ transitions, mediated by an intermediate three-Reggeon state. Each term is evaluated in turn using the results of sections~\ref{subsec:dia}, \ref{subsec:colour} and \ref{subsec:integrals}, as discussed below.

\vspace{1em}
\noindent
\textbf{The $\mathbf{3\to 3}$ transition} involves three contributions, given on the right-hand side of \\eq.~(\ref{eq:Hhat33_4loop}). The first term yields
\begin{align}
\begin{split}
\label{eq:reshat33hat33_p1}
 \frac{\pi^2}{\alpha_s^4r_\Gamma^4}\langle \psi_{j,3} | H_{3\to3}^2 | \psi_{i,3}\rangle &=\frac{ig^2_s}{t}\Bigg[ \left(2J_7-\frac{J_5}{2}-J_6-J_8\right)\mathbf{C}_{\text{DL}}-\left(3J_6-2J_7-J_8+J_9\right)\mathbf{C}_{\text{ML}}\\
 &+\left({J_4+2J_6-J_7-\frac{J_8}{2}}\right)\,C_A\,\mathbf{C}^{(3)}_{33}-\frac{3}{4}\left(J_4+2J_6\right)\,C_A^2\,\mathbf{C}^{(2)}_{33}\Bigg],
\end{split}    
\end{align}
where the factors multiplying~$\mathbf{C}_{\text{DL}}$ and $\mathbf{C}_{\text{ML}}$ in the first line arise from the integrals in eqs.~(\ref{eq:H33_DL}) and (\ref{eq:H33_ML}), respectively; the contribution proportional to $\mathbf{C}^{(3)}_{33}$, in the second line, is obtained by integrating eq.~(\ref{eq:H33Atilde}) and the last term in eq.~(\ref{eq:reshat33hat33_p1}) corresponds to eq.~(\ref{eq:AtildeAtilde}). The remaining terms in eq.~(\ref{eq:Hhat33_4loop}), defined in eqs.~(\ref{eq:H33ag}) and (\ref{eq:agag}), yield
\begin{subequations}
\begin{align}
\label{eq:reshat33hat33_p2}
\frac{\pi^2}{\alpha_s^4r_\Gamma^4}\Bigg[ 2\alpha_g(t) \,\tts\,\langle \psi_{j,3} | H_{3\to3} |\psi_{i,3} \rangle\Bigg]&=\frac{ig^2_s}{t}\Bigg[\left(J_3-2J_2\right)\,\mathbf{T}^2_t\mathbf{C}^{(3)}_{33}+\frac{3J_2}{2}C_A\,\mathbf{C}^{(3)}_{33}\Bigg],\\
\label{eq:reshat33hat33_p3}
    \frac{\pi^2}{\alpha_s^4r_\Gamma^4}\,\alpha_g^2(t)\,\left(\mathbf{T}^2_t\right)^2\,\langle\psi_{j,3} | \psi_{i,3}\rangle &= -\frac{ig_s^2}{t}\,\frac{J_1}{4}\left(\mathbf{T}^2_t\right)^2\mathbf{C}^{(2)}_{33}.
\end{align}
\end{subequations}
The $3\to 3$ transition amplitude is obtained by combining eqs.~(\ref{eq:reshat33hat33_p1}), (\ref{eq:reshat33hat33_p2}) and (\ref{eq:reshat33hat33_p3}). The relevant colour factors are given in eqs.~(\ref{res:colA332}), (\ref{res:colA333}), (\ref{eq:rescolDL}) and (\ref{eq:rescolML}). Furthermore, we make use of the three- and four-loop identities in eq.~(\ref{eq:Simon_trick}) and (\ref{eq:zerorel}) to simplify the colour structures and to isolate the leading contributions in the planar limit. We find
\begin{align}
\begin{split}
    \label{eq:reshat33hat33simb}
\langle j_3|\tilde{H}_{3\to3}^2| i_3\rangle &=\Bigg[\frac{d_{AA}}{72\,N_A}\Big(3J_4+J_5-J_6-2J_7-J_9\Big)\\
&+\frac{C_A^4}{1728}\Big(3J_1+6J_2-12J_3+5J_5+J_6-10J_7+7J_9\Big)\\
&-\frac{1}{32}{\Big[\tsu,\big[\tsu,\tts\big]\Big]\tts}\,\Big(4J_4-3J_6+2J_7-3J_9\Big)\\
&+\frac{1}{96}{\tts\left[\left(\tsu\right)^2,\tts\right]}\,\Big(J_1-4J_2+2J_3+2J_4+J_5+2J_6-4J_7\Big)\\
&-\frac{1}{32}{\big[\tsu,\tts\big]\tts\tsu}\,\Big(3J_4+J_5-J_6-2J_7-J_9\Big)\\
&+\frac{1}{32}{\tsu\big[\tsu,\tts\big]\tts}\Big(J_1+2J_2-4J_3+3J_4-4J_6+4J_7-2J_9\Big)\\
&-\frac{C_A}{24}\left(\frac{d_{AR_i}}{N_{R_i}C_i}+\frac{d_{AR_j}}{N_{R_j}C_j}\right)\Big(J_1+2J_2-4J_3+J_4+2J_5-4J_7+2J_9\Big)\Bigg]\,\langle j_1|i_1\rangle.
\end{split}    
\end{align}
Using section~\ref{subsec:integrals} we substitute the results for the integrals $J_1\dots J_9$, expanded in powers of $\epsilon$, into eq.~(\ref{eq:reshat33hat33simb}), getting
\begin{align}
\label{eq:reshat33hat33}
    \begin{split}
\langle j_3|\tilde{H}_{3\to3}^2| i_3\rangle &=\Bigg[\frac{\mathbf{C}^{(4,-4)}}{\epsilon^4}+\frac{\hat{\zeta}_3}{\epsilon}\,\mathbf{C}^{(4,-1)}+{\cal{O}}\left(\epsilon\right)\Bigg]\,\langle j_1|i_1\rangle,
    \end{split}
\end{align}
where we have defined the following operators in colour space
\begin{subequations}
\begin{align}
\begin{split}
\label{eq:resC44}
&\mathbf{C}^{(4,-4)}=\frac{1}{432}\left[\frac{d_{AA}}{N_A}-3C_A\left(\frac{d_{AR_i}}{N_{R_i}C_i}+\frac{d_{AR_j}}{N_{R_j}C_j}\right)+\frac{C_A^4}{12}\right]-\frac{1}{192}[\tsu,\tts]\tts\tsu\\
&\phantom{\mathbf{C}^{(4,-4)}=}+\frac{1}{96}\Big[\tsu,\big[\tsu,\tts\big]\Big]\tts+\frac{7}{576}\tts\left[\left(\tsu\right)^2,\tts\right]-\frac{5}{192}\tsu\big[\tsu,\tts\big]\tts,
\end{split}    
\end{align}
\begin{align}
\begin{split}
\label{eq:resC41}
&\mathbf{C}^{(4,-1)}= -\frac{101}{216}\left[\frac{d_{AA}}{N_A}-\frac{312}{101}C_A\left(\frac{d_{AR_i}}{N_{R_i}C_i}+\frac{d_{AR_j}}{N_{R_j}C_j}\right)+\frac{211C_A^4}{2424}\right]+\frac{101}{96}\big[\tsu,\tts\big]\tts\tsu\\
&\phantom{\mathbf{C}^{(4,-1)}=}+\frac{49}{48}\Big[\tsu,\big[\tsu,\tts\big]\Big]\tts-\frac{47}{288}\tts\,\left[\left(\tsu\right)^2,\tts\right]-\frac{49}{48}\tsu\left[\tsu,\tts\right]\tts.
\end{split}
\end{align}
\end{subequations}
The terms in square brackets in eqs.~(\ref{eq:resC44}) and (\ref{eq:resC41}) are leading in the planar limit, while all other terms, involving commutators of $\tsu$ and $\tts$, are suppressed for large $N_c$. 

\vspace{1em}
\noindent
\textbf{The $\mathbf{3\to1}$ and $\mathbf{1\to3}$ transitions}, defined in eq.~(\ref{eq:defH31H33}), are computed by summing eqs.~(\ref{eq:Hhat31Htilde33}), (\ref{eq:Hhat13A}) and (\ref{eq:agH13}), together with their symmetric contributions under exchange of the target and the projectile. The latter are obtained from the former by modifying the colour factors, as explained below eq.~(\ref{eq:agH13}). Therefore, summing each expression in eqs.~(\ref{eq:Hhat31Htilde33}), (\ref{eq:Hhat13A}) and (\ref{eq:agH13}), with its target-projectile symmetric partner, we obtain, respectively, the following compact results:
\begin{subequations}
\begin{align}
\begin{split}
\label{eq:resHhat31Htilde33}
\frac{\pi^2}{\alpha_s^4\,r_\Gamma^4}\Big[\langle \psi_{j,3} | B_{3\to3}H_{1\to3} |\psi_{i,1}\rangle + \langle \psi_{j,1} | H_{3\to1}B_{3\to3} |\psi_{i,3}\rangle\Big] 
& \\
&\hspace{-6.0cm}=i\frac{g_s^2}{t}\,\frac{\mathbf{C}^{(4),A}_{31}+\mathbf{C}^{(4),A}_{13}}{12}\Big[2J_2-J_3+J_5+J_6-4J_7+J_9\Big],
\end{split}    
\end{align}
\begin{align}
\begin{split}
\label{eq:resHhat13A}
\frac{\pi^2}{\alpha_s^4r_\Gamma^4}\left[\langle\psi_{j,1}|H_{3\to 1}\,A_{3\to3}|\psi_{i,3}\rangle + \langle\psi_{j,3}| A_{3\to3}\,H_{1\to 3}|\psi_{i,1}\rangle\right]& \\
&\hspace{-6.0cm}
=-\frac{ig_s^2}{t}\frac{C_A\left(\mathbf{C}^{(3)}_{13}+\mathbf{C}^{(3)}_{31}\right)}{24}\Big[3J_2+J_4+
2J_6-4J_7-2J_8\Big],
\end{split}    
\end{align}
\begin{align}
\begin{split}
\label{eq:resagH13}
\frac{\pi^2\alpha_g(t)}{\alpha_s^4r_\Gamma^4}\tts\left[\langle\psi_{j,1}|H_{3\to 1}|\psi_{i,3}\rangle + \langle\psi_{j,3}|H_{1\to 3}|\psi_{i,1}\rangle\right]& \\
&\hspace{-3.0cm}=\frac{ig_s^2}{t}\frac{\tts\,\left(\mathbf{C}^{(3)}_{13}+\mathbf{C}^{(3)}_{31}\right)}{24}\Big[J_1+J_2-2J_3\Big]\,.
\end{split}    
\end{align}
\end{subequations}
Next, summing eqs.~(\ref{eq:resHhat31Htilde33}), (\ref{eq:resHhat13A}) and (\ref{eq:resagH13}), and using the expressions in eqs.~(\ref{res:col133b}) and (\ref{eq:resC13C31_4}) for the colour factors, we find
\begin{align}
\label{eq:resJ1H31H33I3simb}        
\begin{split}
\langle j_1 |\tilde{H}_{3\to1}\tilde{H}_{3\to3}|i_3\rangle &+ \langle j_3 |\tilde{H}_{3\to3}\tilde{H}_{1\to3}|i_1\rangle =\\
&=\frac{C_A}{24}\left(\frac{d_{AR_i}}{N_{R_i}C_i}+\frac{d_{AR_j}}{N_{R_j}C_j}\right)(J_1+2J_2-4J_3+J_4+2J_5-4J_7+2J_9)\,
\\ &=\frac{C_A}{144}\left(\frac{d_{AR_i}}{N_{R_i}C_i}+\frac{d_{AR_j}}{N_{R_j}C_j}\right)\left(\frac{1}{\epsilon^4}-\frac{208\hat{\zeta}_3}{\epsilon}+{\cal{O}}\left(\epsilon\right)\right)\langle j_1|i_1\rangle\,,
\end{split}
\end{align}
where in the second line we expanded the integrals in $\epsilon$ using the results in section~\ref{subsec:integrals}.
We notice that the first line of eq.~(\ref{eq:resJ1H31H33I3simb}) is just the negative of the terms proportional to quartic Casimirs $d_{AR_i}$ and $d_{AR_j}$ in the $3\to 3$ transition, eq.~(\ref{eq:reshat33hat33simb}). Therefore, in the four-loop reduced amplitude, eq.~(\ref{eq:expaMhatnnll4}), these contributions will cancel exactly to all orders in $\epsilon$, as in the case of the three-loop transition amplitudes, eq.~(\ref{eq:resJ3H33I3}) and (\ref{eq:resJ1H31I3}). 

\vspace{1em}
\noindent
\textbf{The $\mathbf{1\to1}$ transition, mediated by three-Reggeon states}, defined in eq.~(\ref{eq:Hhat31Hhat13}), involves two contributions, given in eqs.~(\ref{eq:defH31H13D}) and (\ref{eq:defH31H13X}). After performing the relevant integrals of the kernel $H_{13}$, we obtain
\begin{subequations}
\begin{equation}
\frac{\pi^2}{\alpha_s^4r_\Gamma^4}\langle\psi_{j,1}| H_{3\to1}H_{1\to3}|\psi_{i,1}\rangle_D=-\frac{ig_s^2}{t}\frac{\mathbf{C}^{(4),D}_{3113}}{144}\,\left[J_1+2J_2-4J_3+J_4+2\left(J_5-2J_7+J_9\right)\right]
\end{equation}    
\begin{equation}
\frac{\pi^2}{\alpha_s^4r_\Gamma^4}\langle\psi_{j,1}| H_{3\to1}H_{1\to3}|\psi_{i,1}\rangle_X=-\frac{ig_s^2}{t}\frac{\mathbf{C}^{(4),X}_{3113}}{144}\left[J_1+2J_2-4J_3+J_5+J_6-2\left(J_7+J_8\right)+3J_9\right]
\end{equation}
\end{subequations}
By substituting the colour factors $\mathbf{C}^{(4),D}_{3113}$ and $\mathbf{C}^{(4),X}_{3113}$ given in eqs.~(\ref{eq:resC3113_D}) and (\ref{eq:resC3113_X}), respectively, we get
\begin{align}
\begin{split}
\label{eq:resJ1H31H13I1simb}
\langle j_1| \tilde{H}_{3\to1}\tilde{H}_{1\to3}|i_1\rangle&=-\frac{1}{72}\Bigg[\frac{d_{AA}}{N_A}(3J_1+6J_2-12J_3-3J_4+4J_5+2J_6-8J_7+8J_9)\\
&\phantom{-\frac{1}{72}\Bigg[}+\frac{C_A^4}{12}(3J_4+J_5-J_6-2J_7-J_9)\Bigg]\langle j_1|i_1\rangle.   
\end{split}
\end{align}
We expand the integrals $J_1\dots J_9$, finding
\begin{align}
\begin{split}
\label{eq:resJ1H31H13I1}
\langle j_1| \tilde{H}_{3\to1}\tilde{H}_{1\to3}|i_1\rangle&=\Bigg[-\frac{1}{432\epsilon^4}\left(\frac{d_{AA}}{N_A}+\frac{C_A^4}{12}\right)+\frac{55\,\hat{\zeta}_3}{108\epsilon}\left(\frac{d_{AA}}{N_A}+\frac{101C_A^4}{1320}\right)\Bigg]\langle j_1|i_1\rangle.
\end{split}    
\end{align}
The $1\to 1$ transition amplitude mediated by three-Reggeon states in eq.~(\ref{eq:resJ1H31H13I1}) has the same colour structure of the tree-level amplitude, $\langle j_1|i_1\rangle$. Furthermore, the result is independent on the representation of the colliding particles and it depends only on the degrees of freedom of the exchanged Reggeized gluons, through the Casimir invariants $C_A$ and $d_{AA}$.

\vspace{1em}
\noindent
\textbf{The four-loop reduced amplitude} is given by the sum in eq.~(\ref{eq:expaMhatnnll4}) of the $3\to 3$, $1\to 3$ and $3\to1$, and $1\to 3\to 1$ transition amplitudes. Using the corresponding expressions in eqs.~(\ref{eq:reshat33hat33simb}), (\ref{eq:resJ1H31H33I3simb}) and (\ref{eq:resJ1H31H13I1}), respectively, we get
\begin{align}
\label{eq:M42simb}
\hat{\mathcal{M}}^{(-,4,2)} &= \frac{\pi^2r_\Gamma^4}{2}\Bigg\{-\frac{1}{24}\left(\frac{d_{AA}}{N_A}-\frac{C_A^4}{24}\right)(J_1+2J_2-4J_3-2J_4+J_5+J_6-2J_7+3J_9)  \nn \\ \nn
&-\frac{1}{32}{\Big[\tsu,\big[\tsu,\tts\big]\Big]\tts}\,\Big(4J_4-3J_6+2J_7-3J_9\Big)\\ \nn
&+\frac{1}{96}{\tts\left[\left(\tsu\right)^2,\tts\right]}\,\Big(J_1-4J_2+2J_3+2J_4+J_5+2J_6-4J_7\Big)\\ \nn
&-\frac{1}{32}{\big[\tsu,\tts\big]\tts\tsu}\,\Big(3J_4+J_5-J_6-2J_7-J_9\Big)\\ 
&+\frac{1}{32}{\tsu\big[\tsu,\tts\big]\tts}\Big(J_1+2J_2-4J_3+3J_4-4J_6+4J_7-2J_9\Big)\Bigg\}\,\langle j_1|i_1\rangle.  
\end{align}
Here we notice some important aspects of the reduced amplitude. As anticipated, all terms arising from $3\to 1$ and $1\to 3$ transitions, eq.~(\ref{eq:resJ1H31H33I3simb}), which are proportional to the quartic Casimir invariants $d_{AR_i}$ and $d_{AR_j}$, cancel in the sum with the $3\to3$ amplitude, eq.~(\ref{eq:reshat33hat33simb}) exactly, to all orders in $\epsilon$. This cancellation has already been observed in the three-loop reduced amplitude, in eq.~(\ref{eq:resJ1H31I3}). Based on this observation, we conjecture that transition amplitudes connecting three-Reggeon states with a single Reggeon cancel in the reduced amplitude $\hat{\mathcal{M}}^{(-),\text{NNLL}}$ to all perturbative orders. More precisely, we expect a cancellation in eq.~(\ref{eq:mhatNNLL}) between the quartic Casimir contributions associated with the projectile and the target inside $\langle j_3|\tilde{H}_{3\to3}^l|i_3\rangle$ and the entire contribution due to $\langle j_1 | \tilde{H}_{3\to 1} \tilde{H}_{3\to3}^l| i_3 \rangle$ and $\langle j_3 |\tilde{H}_{3\to3}^l  \tilde{H}_{1\to 3} | i_1 \rangle$, for all $l$.

The first line in eq.~(\ref{eq:M42simb}) is proportional to the colour factor $\frac{d_{AA}}{N_A}-\frac{C_A^4}{24}$, which emerges from the sum of eqs.~(\ref{eq:reshat33hat33simb}) and (\ref{eq:resJ1H31H13I1}). These are indeed the only contributions that involve $d_{AA}$ and $C_A^4$. Interestingly, this combination is non-planar, as it can be easily checked by using the expression of $d_{AA}$ in $\mathrm{SU}(N_c)$, given in eq.~(\ref{eq:dAA})
\begin{equation}\label{eq:dAAminusCa4}
    \frac{d_{AA}}{N_A}-\frac{C_A^4}{24} = 0\cdot N_c^4+\frac{3}{2}N_c^2.
\end{equation}
All remaining terms in eq.~(\ref{eq:M42simb}) are manifestly non-planar, because they are written in terms of commutators of $\tsu$ and $\tts$. This guarantees that the four-loop reduced amplitude is non-planar as a whole. Notably, all planar contributions in the $3\to3$ transition, which are proportional to the colour factors \{$d_{AR_i}$, $d_{AR_j}$\} and \{$d_{AA}$, $C_A^4$\}, are cancelled exactly by the $1\to3$ and $3\to1$ transitions, and by the $1\to 3\to1$ transition, respectively, to all orders in $\epsilon$. This mechanism is in place starting at four loops, where all transitions amplitudes are available. At two and at three loops, there remain planar terms in the reduced amplitudes, which are associated to the $3\to3$ transitions in eq.~(\ref{eq:resJ3I3}) and (\ref{eq:resJ3H33I3}), respectively.

We expand the result in eq.~(\ref{eq:M42simb}) using the $J_i$ integrals in section~\ref{subsec:integrals} and obtain
\begin{equation}
    \label{eq:M42}
    \mExpM{4}{2}=\frac{\pi^2\,r^4_\Gamma}{2}\Bigg[\frac{1}{\epsilon^4} \mathbf{K}^{(4)} +
    \left( \frac{1}{\epsilon} \zeta_3 + \frac{3}{2}\zeta_4 \right)
     \mathbf{K}^{(1)}+ {\cal O}(\epsilon) \Bigg]\mTree,
\end{equation}
where we replaced $\hat{\zeta}_3$ using eq.~(\ref{def:hatz}) to explicitly display the resulting ${\cal O}(\epsilon^0)$ contribution. This result is exact in as far as divergent and finite contributions to $\mExpM{4}{2}$ are concerned. Higher orders in the $\epsilon$ expansion, which we suppressed here, can be readily obtained from eq.~(\ref{eq:M42simb}) using the integrals in section~\ref{subsec:integrals}. The colour structures in eq.~(\ref{eq:M42}) are defined by
\begin{subequations}
\begin{align}
\begin{split}
\label{eq:K4}
\mathbf{K}^{(4)}&=\frac{1}{96}\Big[\tsu,\big[\tsu,\tts\big]\Big]\tts+\frac{7}{576}\tts\left[\left(\tsu\right)^2,\tts\right]
\\&
-\frac{1}{192}\left[\tsu,\tts\right]\tts\tsu
-\frac{5}{192}\tsu\left[\tsu,\tts\right]\tts,
\end{split}\\
\begin{split}
\label{eq:K1pole}
\mathbf{K}^{(1)}&=\frac{49}{48}\Big[\tsu,\big[\tsu,\tts\big]\Big]\tts-\frac{47}{288}\tts\left[\left(\tsu\right)^2,\tts\right]\\
&+\frac{101}{96}\big[\tsu,\tts\big]\tts\tsu-\frac{49}{48}\tsu\big[\tsu,\tts\big]\tts + \frac{1}{24}\left(\frac{d_{AA}}{N_A}-\frac{C_A^4}{24}\right)\,. 
\end{split}
\end{align}
\end{subequations}
The colour tensors in eqs.~(\ref{eq:K4}) and (\ref{eq:K1pole}) have a transparent interpretation in the large-$N_c$ limit. However, different choices of colour bases can also be considered. For instance, ref.~\cite{Naculich:2020clm} proposes a basis of nested commutators, which allows one to uncover all-orders relations between the Regge limits of $\mathcal{N}=4$ SYM and $\mathcal{N}=8$ supergravity amplitudes, and verifies that such basis is suitable to write the three-loop $\mathcal{N}=4$ amplitude of ref.~\cite{Henn:2016jdu}. We checked that also the four-loop amplitude in eq.~(\ref{eq:M42}) can be expressed in the basis of ref.~\cite{Naculich:2020clm}, by computing explicitly eqs.~(\ref{eq:K4}) and (\ref{eq:K1pole}) in the adjoint representation.

Explicit results of the four-loop reduced amplitude, eq.~(\ref{eq:M42}), expanded in $\epsilon$ and evaluated in $t$-channel colour basis are presented in appendix~\ref{subsec:ExplicitRep}. Specifically, the octet and the singlet components of the quark-quark reduced amplitudes are presented, respectively, in eqs.~(\ref{eq:M42qqOctet}) and~(\ref{eq:M42qqSinglet}). The antisymmetric octet and the decuplet of the gluon-gluon amplitude are given in eqs.~(\ref{eq:M42ggOctet}) and~(\ref{eq:M42ggDecuplet}), respectively. The antisymmetric octet component of the quark-gluon amplitude is reported in eq.~(\ref{eq:M42qgOctet}).

\subsection{Disentangling the pole and cut contributions}
\label{sec:pole_cut_explicit_rep}

We have been computing amplitudes in the MRS scheme of eq.~(\ref{Schemes_MRS}), where we have separated the contributions from transitions mediated purely by single-Reggeon states from those involving multiple-Reggeon states, which we explicitly computed to four loops. In this section we will show that by redefining the two-loop impact factors $C_i^{(2)}$ and the three-loop Regge trajectory $\al_g^{(3)}$, we can absorb the entire set of planar corrections at two and three loops arising in the multi-Reggeon exchanges of eqs.~(\ref{eq:M20}) and~(\ref{eq:M31}) respectively, into a Regge-pole factorising term, thus rendering the remaining term, which is associated with the Regge cut, non-planar. This implements the Regge-cut scheme of eq.~(\ref{Schemes_Cut}).

The fact that this transformation can be done is important and non-trivial. It requires in particular that the planar MRS corrections would themselves be process-independent. In fact, we already know that this is the case, see eqs.~(\ref{eq:M20planar}) and (\ref{eq:M31planar}) above.
To expose the different contributions explicitly, we will compute the full NNLL odd amplitude at two loops, three loops and four loops for $qq$, $gg$ and $qg$ scattering. We begin with the  decomposition in eq.~(\ref{Schemes_MRS}), which we expand in the coupling and the high-energy logarithm as follows:
\begin{align}\label{eq:Mdecomp}
\begin{split}
    \M^{(-)}_{ij\to ij} =& \, e^{C_A \alpha_g(t) L} \,
C_i(t) \, C_j(t) \,
{\cal M}^{\rm tree}_{ij\to ij} + \M^{(-),\text{MRS}}_{ij\to ij}\\
\equiv& \sum_{m, n=0}^\infty\left(\frac{\al_s}{\pi}\right)^mL^n\left[p_{ij}^{(m,n)}c_{ij}^{[8_a]} + (i\pi)^2(r_\Gamma)^mf_{ij}^{(m,n)}\right]\M_{ij\to ij}^{\text{tree}, [8_a]},
\end{split}
\end{align}
where we define the functions $p_{ij}^{(m,n)}$ and $f_{ij}^{(m,n)}$ for the SRS and MRS parts respectively, and where the tree amplitude with the octet colour tensor $c_{ij}^{[8_a]}$ stripped off, $\M_{ij\to ij}^{\text{tree}, [8_a]}$, is defined in eq.~(\ref{eq:treelevel_using_8a}). Note that $f_{ij}^{(m,n)}$ is a linear combination of colour tensors (depending on the representations of the scattered partons $i$ and~$j$) while from $p_{ij}^{(m,n)}$ we have extracted the colour tensor structure, so as to emphasise the fact that the Regge pole contributes only to the $c_{ij}^{[8_a]}$ component and does not involve other colour components. 
The impact factors $C_{i/j}(t)$ are assumed to be evaluated at the renormalisation scale $\mu^2=-t$. We utilise the orthornormal $t$-channel basis of refs.~\cite{DelDuca:2014cya,Caron-Huot:2017fxr}, given in eq.~(\ref{colourTensors}) for the relevant colour tensors.

The function $p_{ij}^{(m,n)}$ can be found from expanding the exponential in eq.~(\ref{eq:Mdecomp})
where the quark and gluon impact factors and gluon Regge trajectory in QCD are given in appendix~\ref{ImpactFactors_Appendix},
while $f_{ij}^{(m,n)}$ can be found from the reduced amplitudes given in sections~\ref{sec:twoLoopsComp} through~\ref{sec:fourLoopComp}. Using eqs.~(\ref{Mreduced}),~(\ref{Regge-Pole-and-Cut}) and~(\ref{eq:Mdecomp}) we have 
\begin{align}
\label{eq:fIJExtract}
\begin{split}
    \M^{(-),\text{MRS}}_{ij\to ij}&=\,(i\pi)^2
    \sum_{m=2}^\infty\left(\frac{\al_s}{\pi} r_\Gamma\right)^m \sum_{n=0}^{m-2} L^n f_{ij}^{(m,n)}\M_{ij\to ij}^{\text{tree},[8_a]}
    \\
    &= Z_i(t)Z_j(t)e^{\tts\al_g(t)L}\left(\hat{\M}^{(-)}_{ij\to ij} - D_i(t)D_j(t)\M^{\text{tree}}_{ij\to ij}\right),
    \end{split}
\end{align}
where we indicated that the expansion starts at two loops and that the highest logarithms are NNLL, with $n=m-2$. The combination in parentheses in the second line is the $\hat\MM_{ij\to ij}^{(-),\,\rm MRS}$ of eq.~(\ref{Regge-Pole-and-Cut}). We now evaluate $p_{ij}$ and~$f_{ij}$ at NNLL accuracy at two, three and four loops in the MRS scheme. Subsequently, we will use these values to compute the corresponding coefficients in the cut scheme of eq.~(\ref{Schemes_Cut}):
\begin{align}
\label{eq:Mdecomp_cut_scheme}
\begin{split}
    \M^{(-)}_{ij\to ij} =& \, e^{C_A \tilde{\alpha}_g(t) L} \,
\tilde{C}_i(t) \, \tilde{C}_j(t) \,
{\cal M}^{\rm tree}_{ij\to ij} + \M^{(-),\text{cut}}_{ij\to ij}\\
\equiv& \sum_{m, n=0}^\infty\left(\frac{\al_s}{\pi}\right)^mL^n\left[\tilde{p}_{ij}^{(m,n)}c_{ij}^{[8_a]} + (i\pi)^2(r_\Gamma)^m\tilde{f}_{ij}^{(m,n)}\right]\M_{ij\to ij}^{\text{tree}, [8_a]}\,,
\end{split}
\end{align}
where $\tilde{f}_{ij}$ is related to ${f}_{ij}$ by simply removing the planar $8_a$ component (there is no change of the remaining components). Of course, the sum of terms in the square brackets in eq.~(\ref{eq:Mdecomp_cut_scheme}) must exactly match that in eq.~(\ref{eq:Mdecomp}), that is, the relation between the NNLL coefficients in the MRS and cut schemes is consistent with
\begin{equation}
\label{Cut_MRS_transformation}
 (i\pi)^2(r_\Gamma)^m\Big(
 \tilde{f}_{ij}^{(m,m-2)}- 
{f}_{ij}^{(m,m-2)}\Big) =-\Big( \tilde{p}_{ij}^{(m,m-2)}-
{p}_{ij}^{(m,m-2)}\Big)c_{ij}^{[8_a]}  \,.
\end{equation}
In the following we shall see that the freedom to make such shifts exists only for two and three loops ($m=2,3$) corresponding respectively to fixing the impact factor $\tilde{C}_i^{(2)}$ and the three-loop Regge trajectory $\tilde{\al}_g^{(3)}$. At four loop and beyond there are no parameters that could be tuned at NNLL, and given that (on general grounds~\cite{Mandelstam:1963cw,Eden:1966dnq,Collins:1977jy}) Regge cuts must be associated with non-planar diagrams, one should expect  
${f}_{ij}^{(m,m-2)}$ to be non-planar at four loops and beyond, and of course then $\tilde{f}_{ij}^{(m,m-2)}={f}_{ij}^{(m,m-2)}$ for all $m\geq 4$. We shall see explicitly that this expectation is realised 
at four loop: result of the calculation of ${f}_{ij}^{(4,2)}$ is strictly non-planar.

One might also contemplate alternative schemes in which the difference in eq.~(\ref{Cut_MRS_transformation}) involves non-planar contributions beyond the planar ones. In what follows we shall see that at two loops there is indeed some freedom to do that, and absorb further non-planar corrections from the MRS component into the impact factors of the Regge pole. 
This directly reproduces the scheme for the Regge cut proposed by Fadin and Lipatov in ref.~\cite{Fadin:2017nka}.
At three loops, in turn, there is no such freedom, since non-planar terms would be unnatural in $\tilde{\al}_g^{(3)}$, which is expected to be computable as a correlator of cusped (lightlike) Wilson lines.\footnote{At two loops both the singular and finite parts of ${\alpha}_g(t)$ (known from refs.~\cite{Fadin:1995xg,Fadin:1996tb,Fadin:1995km,Blumlein:1998ib,DelDuca:2001gu}) have been observed (see equation (1.6) of ref.~\cite{Falcioni:2019nxk}) to be related to the simple wedge configuration of two semi-infinite lightlike Wilson lines meeting at a cusp, which was computed to this order in refs.~\cite{Erdogan:2011yc,Falcioni:2019nxk}. Note that the finite part differs by a term proportional to $b_0\pi^2$, which is yet to be understood. Three-loop computations of this Wilson-line correlator are still lacking.} The non-abelian exponentiation theorem~\cite{Gatheral1983ExponentiationOE,Frenkel1984NonabelianEE} then implies a simple, maximally non-Abelian colour structure from any gluonic diagrams, with $(N_c\alpha_s)^m$ for $m=2,3$, respectively.

\subsubsection{Two loops}\label{subsec:pole_cut_sep_two}

At two loops we find the NNLL functions $p_{ij}^{(2,0)}$ and $f_{ij}^{(2,0)}$ as follows: $p_{ij}^{(2,0)}$ is obtained from eq.~(\ref{eq:Mdecomp}) using the expansion of the impact factors defined in eq.~(\ref{impact_factors_def}),
\begin{align}
\label{p2result}
    p_{ij}^{(2,0)}=&\,C_i^{(2)}+ C_j^{(2)} +C_i^{(1)}C_j^{(1)}\,.
\end{align}
In turn, $f_{ij}^{(2,0)}$ is obtained from eqs.~(\ref{eq:fIJExtract}) and~(\ref{eq:M20}):
\begin{align}
    (i\pi)^2(r_\Gamma)^2f_{ij}^{(2,0)}\M_{ij\to ij}^{\text{tree},[8_a]}=&\mExpM{2}{0}-
    \Big(D_i^{(2)}+D_j^{(2)}+D_i^{(1)}D_j^{(1)}\Big)\, \M_{ij\to ij}^{\text{tree}}
    \nn\\
    =&\,\pi^2(r_\Gamma)^2 S^{(2)}(\epsilon)\left((\tsu)^2-\frac{1}{12}C_A^2\right)\M_{ij\to ij}^{\text{tree}}\,,\label{eq:f20All}
\end{align}
where $S^{(2)}(\epsilon)$ is given in eq.~(\ref{eq:R2Res}). 
The impact-factor coefficients $C_i^{(n)}$ in eq.~(\ref{p2result}) are theory-dependent and can be found in QCD from the expressions for $D_i^{(n)}$ and $Z_i^{(n)}$ given in Appendix~\ref{ImpactFactors_Appendix}. Here, $f_{ij}^{(2,0)}$ is driven solely by the three-Reggeon exchange in eq.~(\ref{eq:resJ3I3}). Evaluating this function in the $t$-channel basis for different partonic processes, we have
\begin{subequations}\label{eq:f20}
\begin{align}
    f^{(2,0)}_{qq}=&\,- S^{(2)}(\epsilon) \bigg[\left(\frac{N_c^2}{6}-1+\frac{3}{N_c^2}\right) c_{qq}^{[8]}+\sqrt{N_c^2-1} \left(\frac{1}{2}-\frac{2}{ N_c^2}\right)c_{qq}^{[1]}\bigg], \\
    f^{(2,0)}_{gg}=&\,- S^{(2)}(\epsilon)\left[\left(\frac{N_c^2}{6}+6\right) c_{gg}^{[8_a]}+3\sqrt{\frac{N_c^2-4}{2}}c_{gg}^{[10+\bar{10}]}\right],\\
    f^{(2,0)}_{qg}=&\,- S^{(2)}(\epsilon) \left(\frac{N_c^2}{6}+1\right) c_{qg}^{[8_a]}\,,
\end{align}
\end{subequations}
 Now, it is evident that the leading-colour coefficient of $c_{ij}^{[8_a]}$ is the same for the different processes:
\begin{equation}
\label{f20planar}
   \left. f^{(2,0)}_{ij}\right|_{\text{planar}} = -\frac{N_c^2}{6} S^{(2)}(\epsilon)c_{ij}^{[8_a]} =
   \frac{N_c^2}{6} \left(\frac{1}{8\eps^2}-\frac{3}{4}\zeta_3\eps-\frac{9}{8}\zeta_4\eps^2+\mathcal{O}(\eps^3)\right)c_{ij}^{[8_a]}\,,
\end{equation}
which is in agreement with the universal result of $\left. \mExpM{2}{0}\right|_{\text{planar}}$ in eq.~(\ref{eq:M20planar}), as it must be. We then shift to the Regge-cut basis according to
\begin{align}
\begin{split}
\label{eq:ftilde20}
    (i\pi)^2(r_\Gamma)^2\tilde{f}^{(2,0)}_{ij} \equiv&\,    (i\pi)^2(r_\Gamma)^2\left[f_{ij}^{(2,0)}+\frac{N_c^2}{6}
    S^{(2)}(\epsilon)
    c^{[8_a]}_{ij} \right]\\
=&
    (i\pi)^2(r_\Gamma)^2f_{ij}^{(2,0)}+\left(p_{ij}^{(2,0)}-\tilde{p}_{ij}^{(2,0)}\right)c^{[8_a]}_{ij}\,,
    \end{split}
\end{align}
where in the first line we defined $\tilde{f}^{(2,0)}_{ij}$
by subtracting the planar part of $f_{ij}^{(2,0)}$ using eq.~(\ref{f20planar}) and in the second we
required a corresponding shift of $p_{ij}^{(2,0)}$ using eq.~(\ref
{Cut_MRS_transformation}).
To implement the latter we now recall eq.~(\ref{p2result}) and similarly expand the pole term in eq.~(\ref{eq:Mdecomp_cut_scheme}) to two loops getting
\begin{align}
    \tilde{p}_{ij}^{(2,0)}=&\,\tilde{C}_i^{(2)}+ \tilde{C}_j^{(2)} +C_i^{(1)}C_j^{(1)} \,.
    \end{align}
    Note that since $\M^{(-), \text{cut}}_{ij\to ij}$ vanishes at NLL, the pole term is uniquely defined at that order and we have $\tilde{C}_{i/j}^{(1)}=C_{i/j}^{(1)}$. The required shift of the impact factor coefficient is 
 \begin{align}
 \begin{split}
    \tilde{C}_{i/j}^{(2)} 
    \, = &\, C_{i/j}^{(2)} + N_c^2(r_\Gamma)^2\,\frac{\pi^2}{6}\frac12 S^{(2)}(\epsilon)
    \\
    \, = &\, C_{i/j}^{(2)} - N_c^2(r_\Gamma)^2\,\frac{\pi^2}{6}\frac12 \left(\frac{1}{8 \eps^2}-\frac{3}{4}\zeta_3\eps-\frac{9}{8}   \zeta_4\eps^2+\mathcal{O}(\eps^3)\right),
    \end{split}
    \label{eq:Ctilde}
\end{align}
where we retain the target-projectile symmetry by shifting $C_{i}^{(2)}$ and $C_{j}^{(2)}$ equally.

Our two-loop amplitude is then written in the Regge-cut scheme from eq.~(\ref{eq:Mdecomp_cut_scheme}) as
\begin{align}
\begin{split}
    \M^{(-,2,0)}_{ij\to ij} =&\left[ 
    \tilde{p}_{ij}^{(2,0)} c_{ij}^{[8_a]} 
    + (i\pi)^2(r_\Gamma)^2 \tilde{f}^{(2,0)}_{ij}\right] \M_{ij\to ij}^{\text{tree}, [8_a]}\\
    =&\left[\tilde{p}_{ij}^{(2,0)} + \pi^2(r_\Gamma)^2S^{(2)}(\eps)\left((\tsu)^2-\frac{C_A^2}{4}\right)\right]\M_{ij\to ij}^{\text{tree}}
    \end{split}
\end{align}
where we can then write
\begin{align}
\label{M20cut}
    \M^{(-,2,0),\,\text{cut}}_{ij\to ij} = \pi^2(r_\Gamma)^2S^{(2)}(\eps)\left((\tsu)^2-\frac{C_A^2}{4}\right)\, \M_{ij\to ij}^{\text{tree}}\,,
\end{align}
which can be easily verified to be non-planar by the identity in eq.~(\ref{eq:TsuNc2}). By evaluating the contribution to the octet of $\M^{(-,2,0),\,\text{cut}}_{ij\to ij}$ for the scattering particles in the fundamental or the adjoint representations, we find agreement with the Regge-pole factorisation breaking term $R_{ij}^{(2),0,[8]}$, which was defined in refs.~\cite{DelDuca:2013ara,DelDuca:2014cya} based on considerations about the infrared singularities of the amplitudes rather than direct calculation.

Another proposed separation is that of Fadin and Lipatov in ref.~\cite{Fadin:2017nka}, which we will name the FL scheme. It can be achieved by not only absorbing the leading-colour component of $f_{ij}^{(2,0)}$ into the Regge-pole term, but also the $\kappa(R_i)$ terms from the decomposition of $(\tsu)^2$ in eq.~(\ref{eq:TsuNc2}). The impact factors are shifted from the MRS scheme by
\begin{align}
    C_{i}^{\text{FL},(2)} 
    \, &= \, C_{i}^{(2)} + (r_\Gamma)^2\,\pi^2\left(\frac{N_c^2}{12}+4\,\kappa(R_i)\right)S^{(2)}(\epsilon)\label{FLimpactFactor}
\end{align}
and $C_{j}^{\text{FL},(2)}$ can be found by replacing $C_{i}^{(2)}\to C_{j}^{(2)}$ and $\kappa(R_i) \to \kappa(R_j)$ above. Then the Regge-cut term in the FL scheme can be found to be
\begin{align}
\label{FLM20cut}
    \M^{(-,2,0),\,\text{FL-cut}}_{ij\to ij} = \pi^2(r_\Gamma)^2S^{(2)}(\eps)\left((\tsu)^2-\frac{C_A^2}{4}-4\Big(\kappa(R_i)+\kappa(R_j)\Big)\right)\, \M_{ij\to ij}^{\text{tree}}\,.
\end{align}
The colour structure of eq.~(\ref{FLM20cut}) for the different processes aligns completely with the $C_{ij}^C$ coefficients of ref.~\cite{Fadin:2017nka}. This can be checked using eq.~(\ref{eq:f20}) and the definition of the $\kappa(R_i)$ in eq.~(\ref{eq:dAAlargeNc}).
Upon using eq.~(\ref{eq:TsuNc2}) to substitute for $(\tsu)^2$ in eq.~(\ref{FLM20cut}) we see that the Regge-cut term in the FL scheme at two loops may be written directly in terms of commutators of $\tts$ and~$\tsu$.

\subsubsection{Three loops}\label{subsec:pole_cut_sep_three}

At three loops, the SRS terms obtained from the expansion of eq.~(\ref{eq:Mdecomp}) read:
\begin{align}\label{p31}
    p_{ij}^{(3,1)} =&\, N_c \left[\al_g^{(3)}(t)+\al_g^{(2)}(t) \left(C_i^{(1)}+C_j^{(1)}\right)+\al_g^{(1)}(t) \left(C_i^{(1)} C_j^{(1)}+C_i^{(2)}+C_j^{(2)}\right)\right],
\end{align}
while the MRS contribution is obtained from eq.~(\ref{eq:fIJExtract}) and using the expression from the first line of eq.~(\ref{eq:f20All}) we have
\begin{align}
    (i\pi)^2(r_\Gamma)^3f_{ij}^{(3,1)}\M_{ij\to ij}^{\text{tree},[8_a]} =& \, \mExpM{3}{1} + \al_g^{(1)}(t)\tts (i\pi)^2(r_\Gamma)^2f_{ij}^{(2,0)}\M_{ij\to ij}^{\text{tree},[8_a]},
\end{align}
where $\mExpM{3}{1}$ is given by eq.~(\ref{eq:M31}).
Taking the planar part of the above equation and using eq.~(\ref{eq:M31planar}) we obtain (upon dropping an overall factor of $(i\pi)^2$):
\begin{align}
\label{planar_f3}
\begin{split}
    \left.r_\Gamma^3f_{ij}^{(3,1)} \right|_{\text{planar}}
    =& \, r_\Gamma^3 \, S_C^{(3)}(\epsilon) N_c^3\,c_{ij}^{[8_a]}\, +
    \left.\al_g^{(1)}(t) N_c (r_\Gamma)^2f_{ij}^{(2,0)}\right|_{\text{planar}}\,
    \\
    =& \,  r_\Gamma^3 N_c^3\left[
     \, S_C^{(3)}(\epsilon) \, - \frac{1}{12\epsilon} S^{(2)}(\epsilon)
    \right]c_{ij}^{[8_a]}
    \\
    =& \,  r_\Gamma^3 N_c^3\left[\frac{1}{108 \eps^3}+\frac{\zeta_{3}}{54}+\frac{\zeta_4}{36}\eps+\mathcal{O}(\eps^2)
    \right]c_{ij}^{[8_a]}\,,
    \end{split}
\end{align}
where in the second line we used eq.~(\ref{f20planar}) and in the third we substituted the explicit expansions of $S^{(2)}(\epsilon)$ and $S_C^{(3)}(\epsilon)$ from eqs.~(\ref{eq:M20}) and~(\ref{eq:defRC3}), respectively. 

As a check, let us evaluate $f^{(3,1)}_{ij}$ for the different partonic processes,
\begin{subequations}
\begin{align}
    f^{(3,1)}_{qq}=&\bigg[N_c^3 \left(\frac{1}{108 \eps^3}+\frac{\zeta_{3}}{54}+\frac{\zeta_4}{36}\eps\right)+N_c \left(-\frac{1}{12 \eps^3}+\frac{11 \zeta_{3}}{6}+\frac{11\zeta_4}{6}\eps\right)\nonumber\\&\hskip4.4cm+\frac{1}{N_c}\left(\frac{5}{24 \eps^3}-\frac{31 \zeta_{3}}{12}-\frac{31\zeta_4}{8}\eps\right) + \mathcal{O}(\eps^2)\bigg]c_{qq}^{[8]}\nonumber\\&\qquad
    +\sqrt{N_c^2-1} \frac{N_c^2-4}{N_c}\bigg[\left(\frac{1}{48 \eps^3}+\frac{\zeta_{3}}{24}+\frac{\zeta_4}{16}\eps\right)+ \mathcal{O}(\eps^2)\bigg]c_{qq}^{[1]}, \\
    f^{(3,1)}_{gg}=&\, \bigg[N_c^3 \left(\frac{1}{108 \eps^3}+\frac{\zeta_{3}}{54}+\frac{\zeta_4}{36}\eps \right)+N_c \left(\frac{1}{3 \eps^3}+\frac{2 \zeta_{3}}{3}+\zeta_4\eps\right) + \mathcal{O}(\eps^2)\bigg]c_{gg}^{[8_a]}\nonumber\\&
    \qquad\qquad\qquad+\sqrt{\frac{N_c^2-4}{2}} N_c \left(\frac{7}{24 \eps^3}-\frac{65 \zeta_{3}}{12}-\frac{65\zeta_4}{8}\eps  + \mathcal{O}(\eps^2)\right)c_{gg}^{[10+\bar{10}]},\\
    f^{(3,1)}_{qg}=&\,\bigg[N_c^3 \left(\frac{1}{108 \eps^3}+\frac{\zeta_{3}}{54}+\frac{\zeta_4}{36}\eps\right)+N_c \left(\frac{1}{48 \eps^3}+\frac{61 \zeta_{3}}{24}+\frac{61\zeta_4}{16}\eps\right) + \mathcal{O}(\eps^2)\bigg]c_{qg}^{[8_a]}.
\end{align}
\end{subequations}
It is then clear that the leading-colour component of $c_{ij}^{[8_a]}$ is the same, consistently with the general result of eq.~(\ref{planar_f3}). Now, as before, we change to the Regge-cut scheme by expanding the pole term in eq.~(\ref{eq:Mdecomp_cut_scheme}) to three loops
\begin{align}\label{p31Tilde}
    \tilde{p}_{ij}^{(3,1)} =&\, N_c \left[\tilde{\al}_g^{(3)}(t)+\al_g^{(2)}(t) \left(C_i^{(1)}+C_j^{(1)}\right)+\al_g^{(1)}(t) \left(C_i^{(1)} C_j^{(1)}+\tilde{C}_i^{(2)}+\tilde{C}_j^{(2)}\right)\right].
\end{align}
Note that again, since the pole term at NLL is uniquely defined, we have $\tilde\al_g^{(2)}(t)=\al_g^{(2)}(t)$. The $\tilde{f}_{ij}^{(3,1)}$ is found by subtracting the planar piece of eq.~(\ref{planar_f3})
\begin{align}
\label{eq:ftilde31}
    \tilde{f}_{ij}^{(3,1)} =& \,f_{ij}^{(3,1)} -  \left.f_{ij}^{(3,1)} \right|_{\text{planar}}\,.
\end{align}
Using eq.~(\ref{Cut_MRS_transformation}) we then have
\begin{align}
\begin{split}
    (i\pi)^2r_\Gamma^3\left.f_{ij}^{(3,1)} \right|_{\text{planar}}=& -\left(p_{ij}^{(3,1)}-\tilde{p}_{ij}^{(3,1)}\right)c^{[8_a]}_{ij}\\
    =&-N_c\left[\al_g^{(3)}(t)-\tilde\al_g^{(3)}(t)+\al_g^{(1)}(t)\left(C_i^{(2)}-\tilde C_i^{(2)}+C_j^{(2)}-\tilde C_j^{(2)}\right)\right]c_{ij}^{[8_a]}\,,
    \end{split} 
\end{align}
where we used the expressions for $p_{ij}^{(3,1)}$ and $\tilde p_{ij}^{(3,1)}$ in eqs.~(\ref{p31}) and~(\ref{p31Tilde}), respectively. Substituting the value of the difference of the impact factors in eq.~(\ref{eq:Ctilde}) we can see that the ensuing term directly cancels the  $S^{(2)}(\eps)$ term of $\left.f_{ij}^{(3,1)} \right|_{\text{planar}}$ in eq.~(\ref{planar_f3}). Thus, 
upon rearranging for $\tilde{\al}_g^{(3)}(t)$, we find 
\begin{equation}
    \label{eq:reggeTilde}
    \tilde{\al}_g^{(3)}(t) =\, \al_g^{(3)}(t) - (r_\Gamma)^3N_c^2\pi^2S_C^{(3)}(\eps).
\end{equation}

We can evaluate $\tilde{\al}_g^{(3)}(t)$ explicitly in $\mathcal{N}=4$ SYM, as $\al_g^{(3)}(t)$ in the MRS scheme was extracted in ref.~\cite{Caron-Huot:2017fxr} upon comparison to the three-loop four-gluon amplitude of ref.~\cite{Henn:2016jdu}. There it is found that
\begin{equation}\label{alg3:2017}
    \al_g^{(3)}(t)|_{\text{SYM}}=N_c^2\left(-\frac{\zeta_2}{144}\frac{1}{\eps^3}+\frac{49\zeta_4}{192}\frac{1}{\eps}+\frac{107}{144}\zeta_2\zeta_3+\frac{\zeta_5}{4} + \mathcal{O}(\eps) \right) .
\end{equation}
Then, using eq.~(\ref{eq:reggeTilde}) and the value for $S_C^{(3)}(\eps)$ in eq.~(\ref{eq:defRC3}), we conclude that
\begin{equation}
\label{eq:alphaTilde3}
    \tilde{\al}_g^{(3)}(t)\rvert_{\text{SYM}}=N_c^2\left(\frac{11}{48}\zeta_4 \,\frac{1}{ \eps}\,+\, \frac{5}{24}\zeta_2 \zeta_3+\frac{\zeta_5}{4}\, +\, \mathcal{O}(\eps)\right).
\end{equation}
As expected from the discussion around eq.~(\ref{affirmed_eqality_of_sing_part_of_tilde_alpha_g_and_K}), this new definition of the Regge trajectory in the Regge-cut scheme has the same singularity structure as $K^{(3)}$, the integral of the three-loop coefficient of the lightlike cusp anomalous dimension. This can be verified when one substitutes the maximal transcendental part of $\gamma_K^{(3)}=\frac{11}{4}\zeta_4$ from eq.~(\ref{gammaK}) into $K^{(3)}$ of eq.~(\ref{Kexplicit}), remembering that the beta function coefficients vanishes in $\mathcal{N}=4$ SYM. 
It is in complete agreement with the result from the planar theory~\cite{Drummond:2007aua,Naculich:2007ub,DelDuca:2008jg}, in line with the expectation that it is an eikonal quantity, featuring only maximally non-Abelian colour structures.

Let us also express the three-loop amplitude in the Regge-cut scheme as
\begin{align}
    \M^{(-,3,1)}_{ij\to ij} =&\left[ 
    \tilde{p}_{ij}^{(3,1)} c_{ij}^{[8_a]} 
    + (i\pi)^2(r_\Gamma)^3 \tilde{f}^{(3,1)}_{ij}\right] \M_{ij\to ij}^{\text{tree}, [8_a]}
\end{align}
where, upon comparison to eq.~(\ref{eq:Mdecomp_cut_scheme}), we have after substituting $\tilde{f}_{ij}^{(3,1)}$ from eq.~(\ref{eq:ftilde31}) and using the $\M^{(-,2,0),\,\text{cut}}_{ij\to ij}$ of eq.~(\ref{M20cut})
\begin{align}
    \M^{(-,3,1),\,\text{cut}}_{ij\to ij} =&\, (i\pi)^2r_\Gamma^3\bigg[S_A^{(3)}(\eps)\tsu[\tsu,\tts]+S_B^{(3)}(\eps)[\tsu,\tts]\tsu\bigg]\M_{ij\to ij}^{\text{tree}}\nn\\&+\al_g^{(1)}\tts\M^{(-,2,0),\,\text{cut}}_{ij\to ij}\label{M31cut}.
\end{align}
We verify that the contribution of $\M^{(-,3,1),\,\text{cut}}_{ij\to ij}$ to the colour octet matches the factorisation breaking term in the infrared singularities of the three-loop amplitudes, defined as the function $R^{(3),1,[8]}_{ij}$ in refs.~\cite{DelDuca:2013ara,DelDuca:2014cya}.

In the FL scheme, 
the three-loop Regge trajectory is still found according to the definition in eq.~(\ref{eq:reggeTilde}).
Therefore, the expression for the FL cut will be eq.~(\ref{M31cut}) with the replacement of~$\M^{(-,2,0),\,\text{cut}}_{ij\to ij}$ by $\M^{(-,2,0),\,\text{FL-cut}}_{ij\to ij}$ of eq.~(\ref{FLM20cut}) and it agrees with the results of refs.~\cite{Fadin:2016wso,Fadin:2017nka}.

\subsubsection{Four loops}
At four loops the expansions of eqs.~(\ref{eq:Mdecomp}) and (\ref{eq:fIJExtract}) in the MRS scheme give
\begin{align}
    p_{ij}^{(4,2)} =&\, N_c^2\bigg[ \al_g^{(1)}(t)\al_g^{(3)}(t)+\frac{1}{2}\left(\al_g^{(2)}(t)\right)^2+\al_g^{(1)}(t)\al_g^{(2)}(t) \left(C_i^{(1)}+C_j^{(1)}\right)\nonumber\\&\hskip4.5cm+\frac{1}{2}\left(\al_g^{(1)}(t)\right)^2 \left(C_i^{(1)} C_j^{(1)}+C_i^{(2)}+C_j^{(2)}\right)\bigg],
\end{align}
and
\begin{align}
\label{f43_MRS}
    (i\pi)^2(r_\Gamma)^4f_{ij}^{(4,2)}\M_{ij\to ij}^{\text{tree},[8_a]} =&\,\, \mExpM{4}{2} + \al_g^{(1)}\tts\mExpM{3}{1} \nonumber\\&\qquad\qquad+ \frac{1}{2}\left(\al_g^{(1)}\right)^2\left(\tts\right)^2(i\pi)^2(r_\Gamma)^2f_{ij}^{(2,0)}\M_{ij\to ij}^{\text{tree},[8_a]}.
\end{align}
As discussed following eq.~(\ref{eq:M42}) the MRS result for $\mExpM{4}{2}$ is non-planar. Therefore, the only planar corrections to $f_{ij}^{(4,2)}$ in eq.~(\ref{f43_MRS}) come from the other terms, which, as discussed previously, are universal. Evaluating $f_{ij}^{(4,2)}$ for the different processes we find
\begin{subequations}\label{eq:f42Explicit}
\begin{align}
    f^{(4,2)}_{qq}=&\,\bigg[N_c^4\left(\frac{7}{3456 \eps^4}+\frac{43 {\zeta}_3}{1728 \eps}+\frac{43\zeta_4}{1152}\right)-N_c^2 \left(\frac{3}{128 \eps^4}-\frac{9 {\zeta}_3}{32 \eps}-\frac{27\zeta_4}{64}\right)\nonumber\\&+\frac{7}{128 \eps^4}-\frac{{\zeta}_3}{2 \eps}-\frac{3\zeta_4}{4}+\mathcal{O}(\eps)\bigg]c_{qq}^{[8]}
    +\sqrt{N_c^2-1}\frac{N_c^2-4}{256}\left(\frac{1}{\eps^4}-\frac{2\zeta_3}{ \eps}-3\zeta_4+\mathcal{O}(\eps)\right)c_{qq}^{[1]} , \\
    f^{(4,2)}_{gg}=&\, \bigg[N_c^4\left(\frac{7}{3456 \eps^4}+\frac{43 {\zeta}_3}{1728 \eps}+\frac{43\zeta_4}{1152}\right)+N_c^2 \left(\frac{17}{192 \eps^4}-\frac{217 {\zeta}_3}{96 \eps}-\frac{217\zeta_4}{64}\right)\nonumber\\&\hskip6.5cm+\frac{1}{16 \eps^4}-\frac{101 {\zeta}_3}{8 \eps}-\frac{303\zeta_4}{16}+\mathcal{O}(\eps)\bigg]c_{gg}^{[8_a]}\nonumber\\&
    +\sqrt{\frac{N_c^2-4}{512}}\bigg[N_c^2 \left(\frac{15}{8 \eps^4}-\frac{305 {\zeta}_3}{4 \eps}-\frac{915\zeta_4}{8}\right)+\frac{1}{2\eps^4}-\frac{101 {\zeta}_3}{ \eps}-\frac{303\zeta_4}{2}+\mathcal{O}(\eps)\bigg]c_{gg}^{[10+\bar{10}]},\\
    f^{(4,2)}_{qg}=&\,\bigg[N_c^4\left(\frac{7}{3456 \eps^4}+\frac{43{\zeta}_3}{1728 \eps}+\frac{43\zeta_4}{1152}\right)+N_c^2 \left(\frac{5 {\zeta}_3}{16 \eps}+\frac{15\zeta_4}{32}\right)+\mathcal{O}(\eps)\bigg]c_{qg}^{[8_a]}.
\end{align}
\end{subequations}
In the Regge-cut scheme we have instead
\begin{align}
    \tilde{p}_{ij}^{(4,2)} =&\, N_c^2\bigg[ \al_g^{(1)}(t)\tilde{\al}_g^{(3)}(t)+\frac{1}{2}(\al_g^{(2)}(t))^2+\al_g^{(1)}(t)\al_g^{(2)}(t) \left(C_i^{(1)}+C_j^{(1)}\right)\nonumber\\&\qquad\qquad\qquad\qquad\qquad\qquad\qquad+\frac{1}{2}(\al_g^{(1)}(t))^2 \left(C_i^{(1)} C_j^{(1)}+\tilde{C}_i^{(2)}+\tilde{C}_j^{(2)}\right)\bigg],
\end{align}
for the pole term in the amplitude of eq.~(\ref{eq:Mdecomp_cut_scheme}). 
Thus, the resulting shift is 
\begin{align}
    p_{ij}^{(4,2)}-\tilde{p}_{ij}^{(4,2)}
    &=N_c^2\left[
    \al_g^{(1)}(t)\Big(\al_g^{(3)}(t)- \tilde{\al}_g^{(3)}(t)\Big)
    +\frac12 \left(\al_g^{(1)}(t)\right)^2 
    \left(C_i^{(2)}- \tilde{C}_i^{(2)}+
    C_j^{(2)}- \tilde{C}_j^{(2)}
    \right)
    \right]\nn
    \\
    &=N_c^4(r_\Gamma)^4 \pi^2\frac{1}{2\eps}\left[ 
    S_C^{(3)}(\epsilon) -
    \frac{1}{24\eps}S^{(2)}(\epsilon)
    \right]\,,\label{eq:p42Minusp42}
\end{align}
where we have substituted eqs.~(\ref{eq:reggeTilde}) and~(\ref{eq:Ctilde}).
Then, using eq.~(\ref{Cut_MRS_transformation}) we have
\begin{align}
\label{eq:ftilde42}
    (i\pi)^2(r_\Gamma)^4\tilde{f}_{ij}^{(4,2)} =& (i\pi)^2(r_\Gamma)^4f_{ij}^{(4,2)}+\left(p_{ij}^{(4,2)}-\tilde{p}_{ij}^{(4,2)}\right)c^{[8_a]}_{ij} \\
    =& (i\pi)^2(r_\Gamma)^4\left[f_{ij}^{(4,2)}-N_c^4\left(\frac{7}{3456 \eps^4}+\frac{43 \zeta_3}{1728 \eps}+\frac{43 \zeta_4}{1152}+\mathcal{O}(\eps)\right)c^{[8_a]}_{ij}\right]\,.\nn
\end{align}
Comparing eq.~(\ref{eq:ftilde42}) with the explicit results of the different processes in eq.~(\ref{eq:f42Explicit}) it is clear that $\tilde f_{ij}^{(4,2)}$ is non-planar in the octet, as expected. This cut contribution in eq.~(\ref{eq:ftilde42}), after some algebra using eq.~(\ref{eq:p42Minusp42}) and the two and three-loop cut amplitudes in eqs.~(\ref{M20cut}) and~(\ref{M31cut}) respectively, can be written as
\begin{align}
\label{eq:M42cut}
    \M^{(-,4,2),\,\text{cut}}_{ij\to ij}=\mExpM{4}{2} + \al_g^{(1)}(t)\tts\M^{(-,3,1),\,\text{cut}}_{ij\to ij}-\frac{1}{2}(\al_g^{(1)}(t))^2(\tts)^2\M^{(-,2,0),\,\text{cut}}_{ij\to ij}.
\end{align}

We have shifted all planar contributions arising from the three-Reggeon exchange and its mixing with a single Reggeon into the redefined two-loop impact factors $\tilde{C}_i^{(2)}$ and the three-loop Regge trajectory $\tilde{\al}_g^{(3)}(t)$, rendering the cut contribution non-planar. This will remain true to all loop orders,
\begin{align}\label{eq:McutAllOrder}
    \M^{(-,n,n-2),\,\text{cut}}_{ij\to ij}&=\sum_{m=0}^{n-4}\frac{1}{m!}\left(\al_g^{(1)}(t)\tts\right)^m\mExpM{n-m}{n-m-2}\\&\hskip-14pt+\frac{1}{(n-3)!}\left(\al_g^{(1)}(t)\tts\right)^{n-3}\M^{(-,3,1),\,\text{cut}}_{ij\to ij}-\frac{n-3}{(n-2)!}\left(\al_g^{(1)}(t)\tts\right)^{n-2}\M^{(-,2,0),\,\text{cut}}_{ij\to ij}.\nn
\end{align}
With the expectation that the reduced amplitude at four loops and beyond will be non-planar (which, at least at four loops, is realised through cancellation of the planar contributions in the pure $3\to3$ transitions with the entire contribution due to the $1\to 3$ and $3\to 1$ mixing), eq.~(\ref{eq:McutAllOrder}) shows that the NNLL cut contribution to the amplitude in the cut scheme will also be non-planar. We stress that the freedom we used to define this scheme at two and three loops is no more there at four loops and beyond, so for $\M^{(-,n,n-2),\,\text{cut}}_{ij\to ij}$ to comply with the general expectation~\cite{Mandelstam:1963cw,Eden:1966dnq,Collins:1977jy} that the cut is non-planar, the reduced amplitude at four loops and beyond must indeed be non-planar.

\section{Intersection of Regge and infrared factorisation}
\label{sec:Infrared} 
In addition to the high-energy limit of scattering amplitudes, where Regge factorisation and exponentiation are observed,
there is a complementary picture of factorisation and exponentiation which is valid in general kinematics, namely infrared factorisation in fixed-angle scattering. The latter is governed by the soft anomalous dimension.  
In this section we will explore the intersection of these two, namely the soft anomalous dimension in the asymptotic high-energy limit. We will use the four-loop NNLL reduced amplitude computed in eq.~(\ref{eq:M42}) to study infrared divergences at the same logarithmic accuracy.

\subsection{Infrared factorisation in the high-energy limit}\label{sec:factorHE}

It is well known that long-distance singularities factorise from generic massless $n$-particle  scattering amplitudes $\mathcal{M}_n$ (which we assume to have been 
ultraviolet-renormalised) in the following way \cite{Catani:1998bh,Sterman:2002qn,Aybat:2006mz,Aybat:2006wq,Ma:2019hjq,Feige:2014wja}
\begin{equation}
    \mathcal{M}_n\left(\{s_{ij}\},\mu, \alpha_s(\mu^2),\eps\right) = \mathbf{Z}_n\left(\{s_{ij}\},\mu, \alpha_s(\mu^2),\eps\right)\cdot\mathcal{H}_n\left(\{s_{ij}\},\mu, \alpha_s(\mu^2),\eps\right),\label{eq:factorFormula}
\end{equation}
where $s_{ij}$ are the kinematic invariants and $\eps$ is the dimensional regulator. Eq.~(\ref{eq:factorFormula}) is the $n$-particle generalisation of eq.~(\ref{ZH_intro}). The divergences of the amplitude $\mathcal{M}_n$ as $\eps\to 0$ are captured in an infrared-renormalisation factor we denote by $\mathbf{Z}_n$, leaving a finite, so-called \emph{hard} function $\mathcal{H}$. The soft anomalous dimension $\bm{\Gamma}_n$ is defined by the evolution equation~\cite{Catani:1998bh,Sterman:2002qn,Aybat:2006mz,Aybat:2006wq,Gardi:2009qi,Gardi:2009zv,Dixon:2009ur,Becher:2009cu,Becher:2019avh,Almelid:2015jia,Almelid:2017qju,Magnea:2021fvy} 
\begin{equation}
\label{eq:Gamma_n_def}
    \mu\frac{d}{d\mu}\mathbf{Z}_n\left(\{s_{ij}\},\mu, \alpha_s(\mu^2),\eps\right) = -\mathbf{Z}_n\left(\{s_{ij}\},\mu, \alpha_s(\mu^2),\eps\right)\bm{\Gamma}_n(\{s_{ij}\},\mu, \alpha_s(\mu^2))\,,
\end{equation}
whose solution is the exponential 
\begin{equation}
    \mathbf{Z}_n\left(\{s_{ij}\},\mu, \alpha_s(\mu^2),\eps\right) = \mathcal{P}\exp{\left\{-\frac{1}{2}\int_0^{\mu^2}\frac{d\lambda^2}{\lambda^2}\bm{\Gamma}_n(\{s_{ij}\},\lambda, \alpha_s(\lambda^2))\right\}}.\label{eq:Zfactor}
\end{equation}
Here $\mathcal{P}$ stands for ``path'' ordering of the colour matrices appearing in the expansion of the exponential in correspondence with the scale $\lambda$, such that harder instances of $\bm{\Gamma}_n$ appear closer to the hard function ${\cal H}$. The anomalous dimension is expanded in terms of the $d=4-2\epsilon$ dimensional coupling (\ref{runnign_coupling_D_dim}), such that integration over this coupling in the exponent of eq.~(\ref{eq:Zfactor}) generates the infrared poles in $\eps$ as $\lambda\to 0$. 
 The soft anomalous dimension is known through three loops~\cite{Almelid:2015jia,Almelid:2017qju} and we present an ansatz for it at four loops in section~\ref{subsec:introSAD} following ref.~\cite{Becher:2019avh}, which we then constrain using the high-energy limit.

 In the high-energy limit of $2\to 2$ scattering, the soft anomalous dimension takes the form of eq.~(\ref{eq:gammaFactorIntro}) which we repeat here for clarity~\cite{DelDuca:2011wkl,DelDuca:2011ae,Caron-Huot:2017fxr}
\begin{align}
 \begin{split}
\label{eq:softADdef6}
\mathbf{\Gamma}_{ij\to ij}\left(\alpha_s(\lambda^2),L,\frac{-t}{\lambda^2}\right)=&\,\frac{1}{2}\gamma_{K}(\alpha_s)\left[L\tts + i\pi\tsu\right] +   \Gamma_i\left(\alpha_s,\frac{-t}{\lambda^2}\right)+\Gamma_j\left(\alpha_s,\frac{-t}{\lambda^2}\right)\\&+\sum_{\ell=3}^\infty \left(\frac{\alpha_s}{\pi}\right)^\ell \sum^{\ell-1}_{m=0} L^m \dd^{(\ell,m)},
\end{split}
\end{align} 
 where $\Gamma_i$ is the anomalous dimension generating collinear singularities, defined in eq.~\eqref{collAD}, $\gamma_K$ is the component of the cusp anomalous dimension~(\ref{collAD}) proportional to the quadratic Casimir, $L$ the signature-even large logarithm of eq.~\eqref{eq:siglog}, $i$ corresponds to the projectile and $j$ to the target. The first line of eq.~(\ref{eq:softADdef6}) is the so-called dipole formula in the high-energy limit. Through two loops there are only dipole correlations, as there are no tripole terms in massless scattering~\cite{Aybat:2006mz,Aybat:2006wq,Gardi:2009qi,Becher:2009cu}. All terms that do not conform to this structure are collected in $\dd$, which starts at three loops (see appendix \ref{sec:SAD3loopa}).

Instead of using the full amplitude computed in the Regge limit to compare to the factorisation formula in eq.~(\ref{eq:factorFormula}), we will work here directly with the reduced amplitude in eq.~(\ref{Mreduced}). Using the latter along with eq.~(\ref{eq:factorFormula}) we can write
\begin{align}
\label{eq:modifiedZ}
\begin{split}
    \hat{\mathcal{M}} =&\, e^{-\tts\al_g(t) L}Z_i^{-1}Z_j^{-1}\mathbf{Z}\mathcal{H}\\
    \equiv&\, e^{-\tts\al_g(t) L}\,\tilde{\mathbf{Z}}\,\mathcal{H}
    \end{split}
\end{align}
where $Z_i$ is determined by $\Gamma_i$ according to eq.~(\ref{Zi}) and we defined a modified infrared-renormalisation factor $\tilde{\mathbf{Z}}$, which generates no collinear singularities. This will itself have a modified soft anomalous dimension 
\begin{align}
\label{eq:modifiedGamma}
\tilde{\mathbf{\Gamma}}\left(L,\alpha_s\right) \equiv\,&\, \mathbf{\Gamma}_{ij\to ij}\left(\alpha_s,L,\frac{-t}{\lambda^2}\right) - \Gamma_i\left(\alpha_s,\frac{-t}{\lambda^2}\right) -\Gamma_j\left(\alpha_s,\frac{-t}{\lambda^2}\right)\nn\\
=&\,\frac{1}{2}\gamma_{K}(\alpha_s)\left[L\tts + i\pi\tsu\right] +\sum_{\ell=3}^\infty \left(\frac{\alpha_s}{\pi}\right)^\ell \sum^{\ell-1}_{m=0} L^m \dd^{(\ell,m)}\,,
\end{align}
such that the corresponding infrared-renormalisation factor we defined in eq.~(\ref{eq:modifiedZ}) is obtained, similarly to eq.~(\ref{eq:Zfactor}), by
\begin{equation}
\label{eq:Ztildefactor}
    \tilde{\mathbf{Z}}\left(L,\alpha_s(\mu^2),\eps\right) = \mathcal{P}\exp{\left\{-\frac{1}{2}\int_0^{\mu^2}\frac{d\lambda^2}{\lambda^2}{\tilde{\mathbf{\Gamma}}}(L,\alpha_s(\lambda^2))
    \right\}}.
\end{equation}
Just as the amplitude can be separated by signature, so can $\tilde{\mathbf{\Gamma}}$. Since $\tilde{\mathbf{\Gamma}}$ appears as the coefficient of the $\frac{1}{\epsilon}$ poles multiplying the tree-level amplitude in eq.~\eqref{eq:treelevel}, which itself is odd,  $\tilde{\mathbf{\Gamma}}^{(\pm)}$ corresponds to $\hat{\mathcal{M}}^{(\mp)}$.
We have $\tilde{\mathbf{\Gamma}}=\tilde{\mathbf{\Gamma}}^{(+)}+\tilde{\mathbf{\Gamma}}^{(-)}$ with
\begin{subequations}
\begin{align}
\label{eq:modifiedGammaPlusMinus}
\tilde{\mathbf{\Gamma}}^{(+)}\left(L,\alpha_s\right) 
=&\,\frac{1}{2}\gamma_{K}(\alpha_s) L\tts  +\sum_{\ell=3}^\infty \left(\frac{\alpha_s}{\pi}\right)^\ell \sum^{\ell-1}_{m=0} L^m \dd^{(+,\ell,m)}\,,
\\
\tilde{\mathbf{\Gamma}}^{(-)}\left(L,\alpha_s\right) 
=&\,\frac{1}{2}\gamma_{K}(\alpha_s) i\pi\tsu  +\sum_{\ell=3}^\infty \left(\frac{\alpha_s}{\pi}\right)^\ell \sum^{\ell-1}_{m=0} L^m \dd^{(-,\ell,m)}\,,
\end{align}
\end{subequations}
where we note that $\tilde{\mathbf{\Gamma}}^{(+)}$ is even under $s\leftrightarrow u$ whereas $\tilde{\mathbf{\Gamma}}^{(-)}$ is odd.

Upon inverting eq.~(\ref{eq:modifiedZ}) we have
\begin{align}
    \mathcal{H} = \tilde{\mathbf{Z}}^{-1}e^{\tts\al_g(t) L}
    \hat{\mathcal{M}}.\label{eq:hardFunction}
\end{align}
By demanding $\mathcal{H}$ to be finite, we can use our knowledge of $\hat{\mathcal{M}}$ to determine~$\tilde{\mathbf{Z}}$. Specifically, we can find the single-pole behaviour of $\tilde{\mathbf{Z}}$, corresponding to the soft anomalous dimension coefficients $\dd^{(\ell,m)}$ in eq.~(\ref{eq:modifiedGamma}). Furthermore, the expansion provides a highly non-trivial consistency check on the 
higher-order poles of the high-energy amplitude through four loops.  

\subsection{Expanding the factorisation formula in the high-energy limit}\label{sec:factorExpand}

We now proceed to expand eq.~(\ref{eq:hardFunction}) to four loops and NNLL accuracy. The expansion of $\tilde{\mathbf{Z}}$ is not only complicated by it being a matrix exponential but also the path ordering present in eq.~(\ref{eq:Ztildefactor}) dictates the order of different instances of $\tilde{\mathbf{\Gamma}}$ evaluated at different scales. It is clear that the latter difficulty only starts being relevant at four loops, where there is a term from the interference of the one-loop $\gamma_K^{(1)}$ and the three-loop $\dd^{(3)}$ in the expansion of the exponential with $\tilde{\mathbf{\Gamma}}$ of eq.~(\ref{eq:modifiedGamma}) as the exponent. The term contributing to the odd amplitude at NNLL is proportional to
\begin{align}
    \int_0^{\mu^2}\frac{d\lambda^2}{\lambda^2}\int_0^{\lambda^2}&\frac{d\xi^2}{\xi^2}\bigg\{\gamma_K^{(1)}(\al_s(\xi^2))\left[\tts\dd^{(+,3,1)}(\al_s(\lambda^2))+i\pi\tsu\dd^{(-,3,2)}(\al_s(\lambda^2))\right] \nn\\&\hskip18pt+ \left[\dd^{(+,3,1)}(\al_s(\xi^2))\tts+i\pi\dd^{(-,3,2)}(\al_s(\xi^2))\tsu\right]\gamma_K^{(1)}(\al_s(\lambda^2))\bigg\}.\label{eq:RealPartZ}
\end{align}
By explicit calculation of $\dd^{(3)}$ \cite{Almelid:2015jia,Caron-Huot:2017fxr} we have
\begin{equation}
\label{Vanishing_Delta_components_at_three_loops}
\dd^{(-,3,2)}=\dd^{(+,3,1)}=0\,,
\end{equation}
see appendix~\ref{sec:SAD3loopa}. As such (\ref{eq:RealPartZ}) vanishes, so path-ordering does not in fact complicate the expansion relevant for our NNLL analysis of the odd amplitude. We mention though that path-ordering does need to be taken into account if we were to examine the NNLL even amplitude or the N$^3$LL odd amplitude, since the non-zero $\dd^{(-,3,1)}$ and $\dd^{(+,3,0)}$  will contribute.

To proceed we recall the definition of $K$ in eq.~(\ref{eq:Kdef}) and define the integral of $\dd$, both of which appear in the exponent in eq.~(\ref{eq:Ztildefactor}),
\begin{align}\label{eq:KandQDef}
K(\al_s(\mu^2))= -\frac{1}{4}\int_0^{\mu^2}\frac{d\lambda^2}{\lambda^2}\gamma_K(\al_s(\lambda^2))\,, \qquad Q_\dd(\al_s(\mu^2)) = -\frac{1}{2}\int_0^{\mu^2}\frac{d\lambda^2}{\lambda^2}\dd(\al_s(\lambda^2))\,.
\end{align}
Explicit results for $\gamma_K(\al_s)$ are provided in eq.~(\ref{gammaK}). Expressions for $\dd(\al_s)$ at three loops are summarised in appendix \ref{sec:SAD3loopa} and corresponding ones at four loops will be determined in the present section.

We may now expand the exponentials in eq.~(\ref{eq:modifiedZ}) using the Zassenhaus formula~\cite{Magnus_Wilhelm}, as follows:
\begin{subequations}\begin{align}
  \tilde{\mathbf{Z}} =& \exp\left[K(L\tts + i\pi\tsu) + Q_\dd\right] \\[0.1cm]
=& \,\exp\left[K(L\tts + i\pi\tsu)\right]\, e^{Q_\dd}+\dots\label{eq:ExpTogether}\\
  =& \,e^{KX}e^{KY}e^{-\frac{K^2}{2}[X,Y]}\exp\left[\frac{K^3}{6}
  \Big(2[Y,[X,Y]]+[X,[X,Y]]\Big)\right]\nonumber\\&\times\exp\left[-\frac{K^4}{4!}\Big([[[X,Y],X],X] + 3[[[X,Y],X],Y] + 3[[[X,Y],Y],Y]\Big)\right]e^{Q_\dd}+\dots\label{eq:ExpExpanded}
\end{align}\end{subequations}
where for simplicity we defined $X\equiv L\tts$ and $Y\equiv i\pi\tsu$. The ellipses in eqs.~(\ref{eq:ExpTogether}) and~(\ref{eq:ExpExpanded}) denote terms that do not contribute to our required accuracy. The inverse of eq.~(\ref{eq:ExpExpanded}) can then be easily found to be
\begin{align}
 \tilde{\mathbf{Z}}^{-1} &= \,(1-Q_\dd)\left(1+\frac{K^4}{4!}\Big([[[X,Y],X],X] + 3[[[X,Y],X],Y] + 3[[[X,Y],Y],Y]\Big)\right)\times\nonumber\\&\hskip-15pt\left(1-\frac{K^3}{3!}
 \Big(2[Y,[X,Y]]+[X,[X,Y]]\Big)\right)\left(1+\frac{K^2}{2}[X,Y]+\frac{1}{2}\left(\frac{K^2}{2}[X,Y]\right)^2\right)e^{-KY}e^{-KX},
\end{align}
again, up to corrections irrelevant for NNLL accuracy through four loops. Then, following ref. \cite{Caron-Huot:2017fxr}, we define a cusp-subtracted Regge trajectory 
\begin{equation} \label{alphagtildeDef}
    \hat{\al}_g \equiv \al_g(t) - K\left(\as(-t)\right).
\end{equation} 
This definition is inspired by eq.~(\ref{questioned_eqality_of_sing_part_of_alpha_g_and_K}), where we have seen that $\hat{\al}_g^{(1)}$ and $\hat{\al}_g^{(2)}$ are finite.
It is important to keep in mind though, that $\ag{3}$ (in the MRS scheme\footnote{
We have seen in section~\ref{subsec:pole_cut_sep_three}, that in the MRS scheme $\ag{3}$ is divergent, while in the Regge-cut scheme 
$\hat{\tilde{\al}}_g^{(3)} =\tilde{\al}_g^{(3)}-K^{(3)}$ is finite
(see~(\ref{hat_tilde_alpha_g})).}, in which we work here) is divergent. Using the definition of $\hat\al_g$ in eq.~(\ref{alphagtildeDef}) the hard function in eq.~(\ref{eq:hardFunction}) admits
\begin{align}
\label{eq:hardFunctionExpand}
\begin{split}
\mathcal{H}&=(1-Q_\dd)\left(1+\frac{K^4}{4!}
\Big([[[X,Y],X],X] + 3[[[X,Y],X],Y] + 3[[[X,Y],Y],Y]\Big)\right)\\&\hskip-10pt\times\left(1-\frac{K^3}{3!}\Big(2[Y,[X,Y]]+[X,[X,Y]]\Big)\right)\left(1+\frac{K^2}{2}[X,Y]+\frac{K^4}{8}[X,Y]^2\right)e^{-KY}e^{\hat{\al}_g X}\hat{\mathcal{M}}.
\end{split}
\end{align}
We will now expand the left and right hand side of eq.~(\ref{eq:hardFunctionExpand}) in terms of $\al_s/\pi$ and $L$. As $Y$ is logarithmically suppressed compared to $X$, we notice that $Q_\dd$, which is a pure pole in $\epsilon$, cannot receive contributions at LL accuracy. Furthermore, the leading-logarithmic hard function $\mathcal{H}^{(n,n)}$ comes from only the first term in the expansion of the last exponential in eq.~(\ref{eq:hardFunctionExpand}). It is given solely by the one-loop $\hat{\al}_g^{(1)}$
\begin{equation}
\label{Hnn}
    \mathcal{H}^{(-,n,n)} = \frac{1}{n!}\left(\hat{\al}_g^{(1)}\tts\right)^n\mTree,
\end{equation}
and $\mathcal{H}^{(+,n,n)}=0$. Without knowing the result for $\ag{1}$, eq.~(\ref{Hnn}) immediately implies that it must be finite. Indeed, using the explicit quantities, namely $\gamma_K^{(1)}$ of eq.~(\ref{gammaK}) in (\ref{eq:KandQDef}) along with eq.~(\ref{r_gamma}), we have
\begin{equation}
K^{(1)}=\frac{1}{2\eps};\qquad \ag{1} = \al_g^{(1)}(t) - K^{(1)} = \frac{r_\Gamma}{2\eps}-\frac{1}{2\eps}=\mathcal{O}(\eps)\,,\label{eq:ag1Explicit}
\end{equation}
which we have seen from eq.~(\ref{K_alpha_s_relation}).

\subsection{Extracting information from lower-loop orders}

If we were to proceed directly to the four-loop NNLL hard function, $\mathcal{H}^{(4,2)}$ in the expansion of eq.~(\ref{eq:hardFunctionExpand}) using the results of 
section~\ref{sec:results} for~$\hat{\mathcal{M}}$,
we would find that it depends on $\al_g$ and~$D_i$. A priori these quantities are theory-dependent and are found after matching to the full scattering amplitude calculation. However, using lower-loop results we will see that the divergent terms of these quantities are universal (to the required logarithmic accuracy) and do not depend on the underlying theory. The derivation of the formulae presented in this section, up to three loops, follows closely section 4 of ref.~\cite{Caron-Huot:2017fxr}.

\subsubsection{One loop}
\label{subsec:oneloopHard}
At one loop the expansion of eq.~(\ref{eq:hardFunctionExpand}) at NLL accuracy gives for the hard function
\begin{align}
  \mathcal{H}^{(1,0)} &= \hat{\mathcal{M}}^{(1,0)} - i\pi K^{(1)}\tsu\mTree.\label{eq:H10}
\end{align}

Taking the real and imaginary parts of eq.~(\ref{eq:H10}) we gain access to the odd and even parts of the amplitude, respectively, 
\begin{subequations}\begin{align}
  \mathcal{H}^{(-,1,0)} &= \hat{\mathcal{M}}^{(-,1,0)}&&=\left(D_i^{(1)} + D_j^{(1)}\right)\mTree\\
  \mathcal{H}^{(+,1,0)} &= \hat{\mathcal{M}}^{(+,1,0)} - i\pi K^{(1)}\tsu\mTree &&= i\pi\left(\frac{r_\Gamma}{2\eps} - K^{(1)}\right)\tsu\mTree
\end{align}\end{subequations}
where we have used the results for $\mExpP{1}{0}$ and $\mExpM{1}{0}$ in eqs.~(\ref{eq:m1finite}) and the one-loop~(\ref{Regge-Pole-and-Cut}), respectively. We can then deduce that the one-loop impact factors need to be finite. After matching to the one-loop amplitude, they are given by \cite{Caron-Huot:2017fxr}
\begin{subequations}\label{eq:D1}
  \begin{align}
    D_g^{(1)} =& -N_c\left(\frac{67}{72}-\zeta_2\right)+\frac{5}{36}n_f + \mathcal{O}(\eps), \\
    D_q^{(1)} =& N_c\left(\frac{13}{72}+\frac{7}{8}\zeta_2\right) + \frac{1}{N_c}\left(1-\frac{1}{8}\zeta_2\right)-\frac{5}{36}n_f + \mathcal{O}(\eps).
\end{align}
\end{subequations}
Higher-order terms in $\epsilon$ can be found in appendix~\ref{ImpactFactors_Appendix}.

\subsubsection{Two loops}
\label{subsec:twoloopHard}

The expansion of eq.~(\ref{eq:hardFunctionExpand}) at two loops gives the following expressions for the hard function at NLL and NNLL,
\begin{subequations}\begin{align}
  \mathcal{H}^{(2,1)} &= \hat{\mathcal{M}}^{(2,1)}+\hat{\al}_g^{(1)}\tts\hat{\mathcal{M}}^{(1,0)}+\hat{\al}_g^{(2)}\tts\mTree \nonumber\\&\qquad\qquad\qquad\qquad\qquad+ i\pi K^{(1)}\left(\frac{1}{2}K^{(1)}[\tts,\tsu] - \hat{\al}_g^{(1)}\tsu\tts\right)\mTree,\label{eq:H21}\\
  \mathcal{H}^{(2,0)} &= \hat{\mathcal{M}}^{(2,0)}-\frac{\pi^2}{2}( K^{(1)}\tsu)^2\mTree - i\pi\left(K^{(2)}\tsu\mTree + K^{(1)}\tsu\hat{\mathcal{M}}^{(1,0)}\right).\label{eq:H20}
\end{align}\end{subequations}
We now take the odd part of eq.~(\ref{eq:H21}) and using the result for the reduced amplitudes $\mExpM{1}{0}$ in eq.~(\ref{Regge-Pole-and-Cut}) and that $\mExpM{2}{1}=0$ we have
\begin{align}
  \mathcal{H}^{(-,2,1)} &= \left(\ag{2}+\ag{1}\left(D_i^{(1)}+D_j^{(1)}\right)\right)\tts\mTree.\label{eq:reH21}
\end{align}
As $\ag{1}$ and $D_i^{(1)}$ are finite in eq.~(\ref{eq:reH21}) then we know that $\ag{2}$ needs to be finite. For the real part of eq.~(\ref{eq:H20}) we have
\begin{align}
\label{eq:ReH20Res}
\begin{split}
  \mathcal{H}^{(-,2,0)} &= \bigg[D_i^{(2)}+D_j^{(2)}+D_i^{(1)}D_j^{(1)}-\frac{\pi^2}{12}(r_\Gamma)^2S^{(2)}(\eps)C_A^2
  \\&\qquad\qquad\,\,\,\,+\pi^2\left((r_\Gamma)^2S^{(2)}(\eps)+\frac{1}{2}(K^{(1)})^2+K^{(1)}\ag{1}\right)(\tsu)^2\bigg]\mTree,
  \end{split}
\end{align}
where we have used the result of $\mExpM{2}{0}$ in eq.~(\ref{eq:M20}). Using eq.~(\ref{eq:ReH20Res}) and the explicit expressions for $S^{(2)}(\eps)$ in eq.~(\ref{eq:R2Res}), and $K^{(1)}$ and $\ag{1}$ in eq.~(\ref{eq:ag1Explicit}), we find the universal divergent
terms of the two-loop impact factors,
\begin{align}
\label{eq:Di2Explicit}
\begin{split}
    D_i^{(2)} + D_j^{(2)} =& -D_i^{(1)}D_j^{(1)} + \frac{\pi^2}{12}(r_\Gamma)^2S^{(2)}(\eps)C_A^2 + \mathcal{O}(\eps^0)\\
    =&-C_A^2\frac{\zeta_2}{16\eps^2}+ \mathcal{O}(\eps^0).
    \end{split}
\end{align}

It may seem surprising that the collinear-subtracted impact factors contain further divergences. We point out that the coefficients in eq.~(\ref{eq:Di2Explicit}) depend on the choice of MRS scheme. In the Regge-cut scheme, we define impact factors $\tilde{C}_i(t)$, which include universal terms that arise from three-Reggeon exchanges according to eq.~(\ref{eq:Ctilde}). The associated collinear-subtracted impact factors are given by
\begin{equation}
    \tilde{C}_{i/j}(t) = Z_{i/j}(t)\,\tilde{D}_{i/j}(t).
\end{equation}
With this scheme choice, we collect contributions factorising in a Regge pole form eq.~(\ref{Schemes_Cut}),
\begin{equation}
\label{eq:colsubpole-cut-scheme}
    \mathcal{M}^{(-),\text{pole}}_{ij\to ij} = Z_i(t)\,\tilde{D}_i(t)\,Z_j(t)\,\tilde{D}_j(t)\,e^{C_A\tilde\alpha_g(t)\,L}\,\mathcal{M}^{\text{tree}}_{ij\to ij}.
\end{equation}
It is customary to rearrange the expression above into the crossing-symmetric sum of exponential factors \cite{Fadin:1993wh}
\begin{equation}
\label{eq:colsubpole-vit-scheme}
    \mathcal{M}^{(-),\text{pole}}_{ij\to ij} = Z_i(t)\,\bar D_i(t)\,Z_j(t)\,\bar D_j(t)\,\left[\left(\frac{-s-i0}{-t}\right)^{C_A \tilde\alpha_g(t)} + \left(\frac{-u-i0}{-t}\right)^{C_A \tilde\alpha_g(t)}\right]\mathcal{M}^{\text{tree}}_{ij\to ij},
\end{equation}
where $\bar D_{i/j}(t) = 1 +{\cal{O}}(\alpha_s)$, are \emph{finite} through two loops when $\epsilon\to0$~\cite{DelDuca:2014cya}. We compare eqs.~(\ref{eq:colsubpole-cut-scheme}) and (\ref{eq:colsubpole-vit-scheme}), finding
\begin{align}
\label{eq:DtildeToN}
    \tilde{D}_{i/j}(t) &= \bar D_{i/j}(t)\,\sqrt{\cos\left(\frac{\pi\, C_A \tilde\alpha_g(t)}{2}\right)}=\bar D_{i/j}(t)\left[1-\left(\frac{\alpha_s}{\pi}\right)^2\frac{\pi^2\,r_\Gamma^2}{64\epsilon^2}N_c^2 + {\cal{O}}(\alpha_s^3)\right],
\end{align}
where we used the one-loop Regge trajectory, eq.~(\ref{r_gamma}) and set the renormalisation scale $\mu^2=-t$. 

Given that ${\bar D}_{i/j}(t)$ is finite through two loops, eq.~(\ref{eq:DtildeToN}) implies that, by defining $\tilde{D}_{i/j}(t)=1+\sum_{n>0}\tilde{D}_{i/j}^{(n)}\,\left(\frac{\alpha_s(-t)}{\pi}\right)^n$, we must have 
\begin{equation}
\label{eq:tildeDi1tildeDi2poles}
\tilde{D}_i^{(1)} = {\cal{O}}(\epsilon^0),\qquad \tilde{D}_i^{(2)} = -\frac{\pi^2\,N_c^2}{64\epsilon^2} + {\cal{O}}(\epsilon^0),
\end{equation}
where the divergence in $\tilde{D}^{(2)}_{i/j}$ is fully fixed by the Regge trajectory according to eq.~(\ref{eq:DtildeToN}). Eq.~(\ref{eq:tildeDi1tildeDi2poles}) is consistent with our results in the MRS scheme for the collinear-subtracted impact factors at one and two loops, $D_{i/j}^{(1)}$ and $D_{i/j}^{(2)}$ in eqs.~(\ref{eq:D1}) and (\ref{eq:Di2Explicit}), respectively. Indeed, by using $\tilde{C}_i^{(1)} = C_i^{(1)}$ and eq.~(\ref{eq:Ctilde}), we find
\begin{equation}
\tilde{C}_i(t) = C_i(t) + \left(\frac{\alpha_s(-t)}{\pi}\right)^2\pi^2r_\Gamma^2\,S^{(2)}(\epsilon)\,\frac{N_c^2}{12} + {\cal{O}}(\alpha_s^3).    
\end{equation}
Upon dividing by~$Z_i$ on both sides and expanding in the strong coupling, we obtain the coefficients of the collinear-subtracted impact factors in the cut scheme
\begin{align}
    \label{eq:checkDtilde1}
    \tilde{D}_i^{(1)} &= D_i^{(1)} = {\cal{O}}(\epsilon^0),\\
    \tilde{D}_i^{(2)} &= D_i^{(2)} + \pi^2r_\Gamma^2\,S^{(2)}(\epsilon)\,\frac{N_c^2}{12} = -\frac{\pi^2\,N_c^2}{64\epsilon^2} + {\cal{O}}(\epsilon^0),
    \label{eq:checkDtilde2}
\end{align}
where we used the expansion of $S^{(2)}(\epsilon)$ in eq.~(\ref{eq:R2Res}). Eq.~(\ref{eq:checkDtilde2}) agrees with eq.~(\ref{eq:DtildeToN}), thus explaining the origin of the divergence in eq.~(\ref{eq:Di2Explicit}).

\subsubsection{Three loops}
\label{subsec:threeloopHard}

At three loops we first see the appearance of the correction to the dipole formula, $\dd$. 
Expanding eq.~(\ref{eq:hardFunctionExpand}) to three loops we will only require the real part of the NNLL amplitude:
\begin{align}
\label{H31_deriv}
\mathcal{H}^{(-,3,1)}&=  \mExpM{3}{1} -Q_{\dd}^{(+,3,1)}\mTree+\ag{2} \tts \mExpM{1}{0} + \ag{1} \tts \mExpM{2}{0}+\ag{3} \tts\mTree\nonumber\\&+ 
   (\kusp{1})^2 \frac{\pi^2}{6}\left\{ \kusp{1} \left( \tsu[\tts,\tsu]+[\tts,(\tsu)^2] \right)- 
    3 \ag{1}(\tsu)^2\tts\right\}\mTree \nonumber\\&+i\pi\kusp{1}\left\{\left(\frac{\kusp{1}}{2} [\tts,\tsu] - \ag{1}\tsu\tts\right)\mExpP{1}{0}-\tsu \mExpP{2}{1}\right\}\nonumber\\
  &=\bigg\{C_A\left(\ag{3}+\ag{2}(D_i^{(1)}+D_j^{(1)})+\ag{1}(D_i^{(2)}+D_j^{(2)}+D_i^{(1)}D_j^{(1)})\right) \nonumber\\&\qquad\qquad
  -C_A^3 \pi^2 \left(r_\Gamma^3 S^{(3)}_C+\frac{1}{12} \ag{1}r_\Gamma^2 S^{(2)} \right)
  -Q_{\dd}^{(+,3,1)}
     \bigg\}\mTree+\mathcal{O}(\eps^0),
\end{align}
where in the final expression we have dropped finite terms and used the results of the odd amplitudes $\mExpM{2}{0}$ in eq.~(\ref{eq:M20}), $\mExpM{3}{1}$ in eq.~(\ref{eq:M31}), and the even NLL amplitudes $\mExpP{1}{0}$ and $\mExpP{2}{1}$ in eqs.~(\ref{eq:m1finite}) and~(\ref{eq:m2finite}). The two terms proportional to $C_A^3$ in the last line of the final expression originate in $\mExpM{3}{1}$ and $\ag{1} \tts \mExpM{2}{0}$, respectively. Using the explicit result that $\dd^{(+,3,1)}=0$ from ref.~\cite{Almelid:2015jia,Caron-Huot:2017fxr}, and the fact that the singular terms in the curly brackets in eq.~(\ref{H31_deriv}) must vanish, we find 
\begin{align}
\begin{split}
&\ag{3}+\ag{2}(D_i^{(1)}+D_j^{(1)})+\ag{1}(D_i^{(2)}+D_j^{(2)}+D_i^{(1)}D_j^{(1)}) =
\\
&\hspace{20pt}
=C_A^2 \pi^2 \left(r_\Gamma^3 S^{(3)}_C+\frac{1}{12} \ag{1}r_\Gamma^2 S^{(2)} \right) + \mathcal{O}(\eps^0)
=
- C_A^2\frac{\pi^2}{864}\left(\frac{1}{\eps^3}-\frac{15\zeta_2}{4\eps}\right) + \mathcal{O}(\eps^0)\,.
\end{split}
\end{align}
Next, using the values of $\ag{1}$ in eq.~(\ref{eq:ag1Explicit}), $D_i^{(2)}$ in eq.~(\ref{eq:Di2Explicit}), and that $D_i^{(1)}$ and $\ag{2}$ are finite we find for the universal divergent part of $\ag{3}$:
\begin{align}
\label{eq:AG3const}
\begin{split}
&\ag{3}= C_A^2 \pi^2 \,r_\Gamma^3 S^{(3)}_C
+O\left(\eps^0\right)
=
-C_A^2\left(\frac{\zeta_2}{144 \eps^3}-\frac{5 \zeta_4}{192}\frac{1}{\eps}\right)+O\left(\eps^0\right).
\end{split}
\end{align}
The universal divergence of eq.~(\ref{eq:AG3const}) is in line with $\al_g^{(3)}$ in $\mathcal{N}=4$ SYM extracted in~ref.~\cite{Caron-Huot:2017fxr} and quoted in eq.~(\ref{alg3:2017}). Here we use the MRS scheme in the calculation of the reduced amplitudes, as in ref.~\cite{Caron-Huot:2017fxr}. In section~\ref{subsec:pole_cut_sep_three} we saw that in the Regge-cut scheme such a divergence is not present, as $\tilde\al_g^{(3)}-K^{(3)}$ is finite (compare 
eq.~(\ref{eq:reggeTilde}) with eq.~(\ref{eq:AG3const})).

\subsection{The four-loop soft anomalous dimension and hard function}
\label{subsec:fourloopHard}

We are now ready to compute the NNLL hard function at four loops in terms of the reduced amplitudes of the previous section. This will facilitate extracting the soft anomalous dimension at this accuracy. The real part of the expansion of eq.~(\ref{eq:hardFunctionExpand}) gives
\begin{align}
\label{Hminus42}
\begin{split}
 \mathcal{H}^{(-,4,2)} &=\sum_{n=2}^4\frac{(\ag{1}\tts)^{4-n}}{(4-n)!}\mExpM{n}{n-2}+ \ag{2} \tts\left(
 \ag{1} \tts \mExpM{1}{0}+ \mExpM{2}{1}\right)  \\& 
 -\bigg\{    \frac{\pi^2}{8}\left( \kusp{1}\right)^4  
 \Big( [\tts, \tsu]^2 + 
       \frac{4}{3}  [\tts, [\tts, \tsu]]\tsu + 
        [[[\tts, \tsu], \tts], \tsu]\Big)\\&
       \,\,\,\,\,\,\,\,\,\,\,\,+\text{Re}\left[Q_\Delta^{(4,2)}\right] + 
   \frac{\pi^2}{4} \left(\ag{1}\kusp{1}\right)^2 (\tsu)^2(\tts)^2 - 
    \frac{1}{2} \Big((\ag{2})^2 + 2 \ag{1} \ag{3}\Big) (\tts)^2\\&
    \qquad -
   \frac{\pi^2}{6} \ag{1} 
   \left(\kusp{1}\right)^3  \Big(2  [\tsu, [\tts, \tsu]]\tts + 
       3 [\tts, \tsu] \tsu  \tts\Big) \bigg\}\mTree\\&
       +i\pi\frac{\kusp{1}}{2}\bigg\{ \Big( \kusp{1} [\tts, \tsu] - 2\ag{1} \tsu  \tts\Big)\mExpP{2}{1} -2 \tsu \mExpP{3}{2}  \\&
       \qquad\qquad+\Big[ \ag{1} \kusp{1}  [\tts, \tsu]\tts - 
   \frac{1}{3}\left(\kusp{1}\right)^2 
   [\tts,[\tts, \tsu]] 
   \\& \qquad\qquad- 
   \left(\ag{1}\right)^2 \tsu ( \tts)^2\Big] \mExpP{1}{0}\bigg\}\,,
   \end{split}
  \end{align}
where we have used $Q_\dd^{(-,3,2)}=0$ and  $Q_\dd^{(+,3,1)}=0$ following eq.~(\ref{Delta3_results}). Using the results of $K^{(1)}$ and $\ag{1}$ in eq.~(\ref{eq:ag1Explicit}), $\,\ag{2}=\mathcal{O}(\eps^0)$, $\,\ag{3}$ in eq.~(\ref{eq:AG3const}), the odd amplitudes \hbox{$\mExpM{2}{1}=0$}, $\,\mExpM{2}{0}$ in eq.~(\ref{eq:M20}), $\,\mExpM{3}{1}$ in eq.~(\ref{eq:M31}), the even amplitudes in eq.~(\ref{eq:evenAmps}) and the newly computed $\mExpM{4}{2}$ in eq.~(\ref{eq:M42}), we find, up to finite terms
\begin{align}
\label{eq:H42}
   \mathcal{H}^{(-,4,2)} =&\bigg\{\frac{\pi ^2 \zeta_3}{48\eps} \left(\frac{d_{AA}}{N_A}-\frac{C_A^4}{24}+\frac{1}{4}\tts[\tts,(\tsu)^2]+\frac{3}{4}[\tsu,\tts]\tts\tsu\right)\nonumber\\&\qquad\qquad\qquad\qquad\qquad\qquad\qquad\qquad\qquad\,\,\,-Q_{\mathbf{\Delta}}^{(+,4,2)} +\mathcal{O}(\eps^0)\bigg\}\mTree.
\end{align}
Using the fact that $\mathcal{H}^{(-,4,2)}$ is finite, this equation directly amounts to determining $Q_{\mathbf{\Delta}}^{(+,4,2)}$. To determine $\dd^{(+,4,2)}$ using eq.~(\ref{eq:KandQDef}) it is sufficient to use the leading-order expression for the running coupling in eq.~(\ref{runnign_coupling_D_dim}), which yields~$\dd^{(+,4,2)}= 8\eps\, Q_\dd^{(+,4,2)} + \dots$, so we obtain\footnote{This result was first reported in ref.~\cite{Falcioni:2020lvv}.}
\begin{align}
\label{eq:delta42}
\begin{split}
   \dd^{(+,4,2)}
  &=\zeta_2\zeta_3\bigg\{\frac{d_{AA}}{N_A}-\frac{C_A^4}{24} + \frac{1}{4}\tts[\tts,(\tsu)^2]+\frac{3}{4}[\tsu,\tts]\tts\tsu\bigg\}
  \\* &\equiv\zeta_2\zeta_3\,C_{\dd}^{(+,4,2)}\,,
  \end{split}
\end{align}
where in the second line we defined the colour structure $C_{\dd}^{(+,4,2)}$. We note that the latter is entirely non-planar, as it is subleading in the large-$N_c$ limit compared to $N_c^4$. This can easily be seen as, according to the argument of eq.~(\ref{eq:non-planar-arg}), commutators are inherently subleading and that the term proportional to the identity behaves in the large-$N_c$ limit as eq.~(\ref{eq:dAAminusCa4}).

The finite term of the hard function in eq.~(\ref{Hminus42}) can be found to be
\begin{align}
    \mathcal{H}^{(-,4,2)}=&\left\{\frac{C_A^2}{2}\left(\ag{2,0}\right)^2+\frac{3}{16}\zeta_4\zeta_2C_{\mathbf{\Delta}}^{(+,4,2)}+\mathcal{O}(\eps)\right\}\mTree,\label{eq:hardFunctionTheoryDependence}
\end{align}
where we encounter again the very same colour structure $C_{\mathbf{\Delta}}^{(+,4,2)}$ defined in eq.~(\ref{eq:delta42}) and $\ag{2,0}$ is the $\mathcal{O}(\eps^0)$ term of the finite $\ag{2}$. Since the term proportional to $C_{\mathbf{\Delta}}^{(+,4,2)}$ is universal, eq.~(\ref{eq:hardFunctionTheoryDependence}) conveniently displays the theory dependence of $\mathcal{H}^{(-,4,2)}$, which is restricted to the two-loop $\ag{2}$. 
One can also interpret eq.~(\ref{eq:hardFunctionTheoryDependence}) as dressing Reggeons with two infrared-renormalised Regge trajectory insertions associated with the Regge pole, plus non-planar corrections stemming from the multi-Reggeon cuts. 

We can find $\ag{2,0}$ for QCD, which is calculated using the explicit two-loop Regge trajectory in eq.~(\ref{eq:regge2QCD}) and the two-loop $K^{(2)}$ of eq.~(\ref{eq:Kdef}) with the values for $\gamma_K$ in eq.~(\ref{gammaK}). It is just the finite parts of $\al_g^{(2)}(t)$,
\begin{align}
    \ag{2,0} =&\, C_A \left(\frac{101}{108} - \frac{\zeta_3}{8}\right) - \frac{7}{27}T_R n_f.
\end{align}
Inserting this into eq.~(\ref{eq:hardFunctionTheoryDependence}) we have
\begin{align}
    \mathcal{H}^{(-,4,2)}_{\text{QCD}}=&\,\bigg\{ C_A^2 T_R^2n_f^2\frac{49}{1458}+C_A^3 T_Rn_f \left(\frac{7 \zeta_3}{216}-\frac{707}{2916}\right)\label{eq:QCDHard}\\&\qquad\qquad\qquad+C_A^4 \left(\frac{\zeta_3^2}{128}-\frac{101 \zeta_3}{864}+\frac{10201}{23328}\right)+\frac{3}{16}\zeta_4\zeta_2C_{\mathbf{\Delta}}^{(+,4,2)}+\mathcal{O}(\eps)\bigg\}\mTree.\nonumber
\end{align}
The $\mathcal{N}=4$ SYM result is
\begin{align}
  \mathcal{H}^{(-,4,2)}_{\text{SYM}}=&\,\left\{\frac{C_A^4}{128}\zeta_3^2+\frac{3}{16}\zeta_4\zeta_2C_{\mathbf{\Delta}}^{(+,4,2)}+\mathcal{O}(\eps)\right\}\mTree,\label{eq:hardFuncRes}
\end{align}
which is precisely the maximal weight terms of the QCD result in eq.~(\ref{eq:QCDHard}), using the well-known relations between quantities in these theories~\cite{Kotikov:2001sc,Kotikov:2002ab,Kotikov:2004er,Kotikov:2007cy}. In the planar limit only the first term survives and matches the well-known BDS ansatz~\cite{Bern:2005iz}.

\subsection{The soft anomalous dimension in the Regge limit -- summary}
\label{subsec:sumSADreg}

 The soft anomalous dimension in the high-energy limit is given in eq.~\eqref{eq:softADdef6}. In this section, we review the corrections to the dipole formula that start at three loops 
 \beq
 \dd= \sum_{\ell=3}^\infty \left(\frac{\alpha_s}{\pi}\right)^\ell \sum^{\ell-1}_{m=0}\dd^{(\ell,m)} L^m,
 \eeq
 where $\ell$ is the loop order and $L$ the signature-even logarithm of eq. \eqref{eq:siglog}.
 
Non-vanishing corrections to the dipole formula at three-loop order in the high-energy limit of $2\to 2$ scattering, begin at NNLL with a purely signature-odd contribution~\cite{Almelid:2015jia, Caron-Huot:2017fxr}, namely
 \begin{align}
 \label{three_loop_Delta}
         \begin{split}
            \dd^{(3)}&=0L^2 +i \pi \Big[\tts,[\tts,\tsu]\Big] \frac{11}{4} \zeta_3 L+O(L^0),
                     \end{split}
     \end{align}
     where $\mathcal{O}(L^0)$ terms are summarised in appendix \ref{sec:SAD3loopa}.
  
  At NLL, there is a tower of signature-odd contributions to the soft anomalous dimension (i.e. signature-even amplitude) emanating from two-Reggeon-exchange ladder diagrams, which have been determined to high orders~\cite{Caron-Huot:2017zfo,Caron-Huot:2020grv} (the singularities at NLL are in fact known to all loops~\cite{Caron-Huot:2017zfo}). These contributions start at four loops with~\cite{Caron-Huot:2013fea}
 \beq
\label{eq:delta43min6}
\dd^{(-,4,3)}=-i\pi\frac{\zeta_3}{24} \Big[\tts,[\tts,\tsu]\Big]\tts=-i\pi\frac{\zeta_3}{24}\bm{C}_{\Delta}^{(-,4,3)}\,.
\eeq  
Note that both eqs.~(\ref{three_loop_Delta}) and (\ref{eq:delta43min6}) contain a single power of $i\pi \tsu$, consistently with the odd signature.

The even signature part of the soft anomalous dimension at NLL is 
two-loop exact~\cite{Lipatov:1976zz,Kuraev:1976ge,Caron-Huot:2017zfo}, that is, the terms $\dd^{(+,\ell,\ell-1)}$ vanish at three loops and above. In particular, at four loops,
\beq
\label{eq:delta43plus6}
\dd^{(+,4,3)}=0.
\eeq 
At NNLL, the odd contribution $\dd^{(-,4,2)}$ is not yet known. The even contribution is given by eq. \eqref{eq:delta42}.
 Using the results of eqs. (\ref{eq:delta43min6}, \ref{eq:delta43plus6}, \ref{eq:delta42}) we can write eq. \eqref{eq:softADdef6} displaying all the explicitly-known contributions 
 \begin{align}
 \begin{split}
\label{eq:softADdef4loopsymD6}
\mathbf{\Gamma}^{(4)}_{ij\to ij}(L) = &- L^3 i \pi \frac{\zeta_3}{24}\Big[\tts,[\tts,\tsu]\Big]\tts+L^2\dd^{(-,4,2)}\\&+L^2 \zeta_2 \zeta_3 \bigg(\frac{d_{AA}}{N_A} -\frac{C_A^4}{24}-\frac{1}{4}\tts[(\tsu)^2,\tts]+\frac{3}{4}[\tsu,\tts]\tts\tsu\bigg)\\& + {\cal O}(L),
\end{split}
\end{align}
  where $\dd^{(-,4,2)}$ and corrections at $O(L)$ and $O(L^0)$ are still to be determined. The fact that the NNLL corrections in $\dd$ are all non-planar is to be expected: only diagrams that correlate up to two partons survive in the anomalous dimension in the planar limit. 

It is important to point out that all the corrections displayed explicitly in eq.~(\ref{eq:softADdef4loopsymD6}) are universal, namely they are independent on the matter content of the theory considered\footnote{In contrast, $\dd^{(-,4,2)}$ and N$^3$LL corrections, are expected to involve non-universal corrections, as can be inferred from the high-energy limit perspective. Theory dependence would enter for example through running-coupling effects, subleading corrections to the Balitsky-JIMWLK Hamiltonian and impact factors.}. Specifically, these corrections are generated by purely gluonic diagrams, and therefore do not involve the matter content. Furthermore, in operator form they are completely independent of the representation of the scattered particles (of course, as we have seen in section~\ref{sec:pole_cut_explicit_rep}, such dependence would show up once acted on the tree-level amplitude and projected onto a colour basis).

\section{Infrared singularities in  \texorpdfstring{$2 \to 2$}{2->2}
scattering in the high-energy limit}
\label{sec:SAD_Regge}
In the previous section we have obtained the soft anomalous dimension in the high-energy limit $s\gg t$, for $2 \to 2$ parton scattering.
In particular, starting from the formula of eq.~\eqref{eq:softADdef6}, we have determined the non-dipole contributions at three and four loops, as summarised in section \ref{subsec:sumSADreg}.
 This result proves useful to constrain the structure of the soft anomalous dimension in general kinematics, too. The latter was determined by direct computation to three loops in ref.~\cite{Almelid:2015jia}. Later, it was shown that it can also be determined by means of a bootstrap approach \cite{Almelid:2017qju}: indeed,
 the functional form of the three-loop soft anomalous dimension in general kinematics was recovered in ref.~\cite{Almelid:2017qju}
 using the colour structure, the analytic properties of the relevant class of integrals, factorisation and symmetry constraints, along with constraints from various kinematic limits. In this perspective, the structure of the soft anomalous dimension in the high-energy limit determined in section \ref{subsec:sumSADreg} provides useful information for a future bootstrap approach at four loops. 
 
 The purpose of this section is to perform a preliminary analysis, determining the implications of the high-energy result for a parametrisation of the soft anomalous dimension recently constructed through four loops in ref.~\cite{Becher:2019avh}.
After a short introduction of the colour and kinematic variables in the general $n$ massless parton case, we focus on $2\to2$ scattering. Specialising further to the high-energy limit, we identify the functions contributing through NNLL accuracy at four-loop order in eq.~\eqref{eq:SADNNLL}.
Subsequently, in section~\ref{sec:factor2to2} the colour structures are expressed in terms of a so-called \emph{Regge-limit basis} of colour operators which we construct using commutators of $\tts$ and $\tsu$, based on the definitions introduced at the end of section \ref{subsec:Regge2to2} and the identities derived in section  \ref{subsec:colour}. Furthermore, 
the (unknown) kinematic functions are specialised to the high-energy limit and are expressed in terms of the signature-even logarithm~$L$ in section~\ref{subsec:Reggelimit}. At this point, in section~\ref{subsec:Constraints} we are able to constrain the form of these kinematic functions by comparing the parametrisation of the four-loop soft anomalous dimension with the expressions we computed through NNLLs, as summarised  in section~\ref{subsec:sumSADreg}.
Finally, in section~\ref{SADgenrep} we discuss the four-loop soft anomalous dimension in the high-energy limit beyond NNLLs. We generalise the relation in eq.~(\ref{affirmed_eqality_of_sing_part_of_tilde_alpha_g_and_K}) between the singularities of the gluon Regge trajectory and the cusp anomalous dimension to four loops, taking into account quartic Casimir contributions, and discuss the implications of this generalisation on the structure of the soft anomalous dimension.

\subsection{Introduction to the soft anomalous dimension}\label{subsec:introSAD}

Given a scattering amplitude $\mathcal{M}$ involving $n$ coloured partons and any number of other (colourless) particles, the soft anomalous dimension ${\bf \Gamma}_n$ defined in eq.~(\ref{eq:Gamma_n_def})~\cite{Catani:1998bh,Sterman:2002qn,Aybat:2006mz,Aybat:2006wq,Gardi:2009qi,Gardi:2009zv,Dixon:2009ur,Becher:2009cu,Becher:2019avh,Almelid:2015jia,Almelid:2017qju,Magnea:2021fvy}  
depends on the kinematic invariants formed between the partons,
\beq
\label{eq:sij}
(-s_{ij})\equiv 2|p_i\cdot p_j| e^{-i \pi \sigma_{ij}},
\eeq 
where $p_i$ represents the momentum of parton $i$, and $\sigma_{ij} = 1$ if both partons $i$ and $j$ are in the initial or final state, otherwise $\sigma_{ij} = 0$. In addition, it depends on the colour generators $\left\{{\bf T}_i\right\}_{i=1}^{n}$ associated to the external partons, which have been defined after eq.~\eqref{eq:treelevel}. 
${\bf \Gamma}_n$ is defined subject to colour conservation 
\beq
\label{eq:colcons_n}
\sum_{i=1}^n\T_i\mathcal{M}=\sum_{i=1}^n\T_i\mathcal{H}=0\, ,
\eeq 
where $\mathcal{H}$ is the hard function of eq.~(\ref{ZH_intro}).

The form of the soft anomalous dimension is strongly constrained. Setting aside collinear singularities (which
are captured by the collinear anomalous dimension $\gamma_i$, and depend on individual scattered partons, namely their spins and colour representations, but involves no kinematic nor colour-flow dependence) all remaining contributions to the soft anomalous dimension follow from the ultraviolet divergence of the corresponding Wilson-line correlator.  Specifically, to describe the singularities of an amplitude of $n$ massless partons, one defines a correlator of a product of $n$ semi-infinite lightlike Wilson lines emanating from a single hard-interaction vertex, following the classical trajectories of the $n$ partons. The ultraviolet divergences of this correlator, associated with the vertex where they meet, map to the soft divergences of the amplitude~\cite{Korchemsky:1985xj,Korchemsky:1987wg,Korchemsky:1988si,Kidonakis:1998nf}. 
This observation can be used to compute the anomalous dimension in a process-independent way~\cite{Gardi:2016ttq,Almelid:2015jia}, rather than extracting it from partonic amplitudes. It also allows one to understand the salient all-order properties~\cite{Becher:2009cu,Gardi:2009qi,Gardi:2009zv,Dixon:2009ur,Almelid:2015jia,Almelid:2017qju,Becher:2019avh} of the soft anomalous dimension, notably,
\begin{enumerate}
\item{} its colour structure is highly constrained by the non-Abelian exponentiation theorem~\cite{Sterman:1981jc,Gatheral1983ExponentiationOE,Frenkel1984NonabelianEE,Gardi:2013ita}: it consists of fully connected diagrams\footnote{Here \emph{connected} refers to the gluon web representing the colour structure, which must be fully connected after the Wilson lines have been removed. See refs.~\cite{Gardi:2010rn,Mitov:2010rp,Gardi:2011yz,Dukes:2013gea,Gardi:2013ita,Agarwal:2020nyc,Agarwal:2021him} for further details on diagrammatic exponentiation in the context of correlators of multiple Wilson lines.};
\item{} it admits Bose symmetry under permutations of any of the Wilson lines, independently of whether these represent quarks or gluons;
\item{} its kinematic dependence is largely constrained by the invariance under scaling of individual Wilson-line velocities.
\end{enumerate}
While the first property dictates the colour structure of the soft anomalous dimension, the third strongly constrains its kinematic dependence. Finally, owing to the second property, colour and kinematics are directly correlated.

The third property, namely rescaling invariance, is violated only through the so-called \emph{collinear anomaly}, namely overlapping soft and collinear divergences, which are responsible for an explicit dependence of $\mathbf{\Gamma}_n\left(\{s_{kl}\},\lambda, \alpha_s(\lambda^2)\right)$ on the scale $\lambda$, beyond its indirect dependence on the scale through the running coupling. 
Stated differently, the factorisation property of soft and collinear singularities, combined with the aforementioned rescaling property of Wilson-line velocities, translates into the following set of all-order constraints~\cite{Becher:2009cu,Becher:2019avh,Gardi:2009zv,Gardi:2009qi} on the soft anomalous dimension
$\mathbf{\Gamma}_n\left(\{s_{kl}\},\lambda, \alpha_s(\lambda^2)\right)$:
\begin{align}
\label{eq:BNdipinh}
\sum_{j\neq i}\frac{d\,
\mathbf{\Gamma}_n\left(\{s_{kl}\},\lambda, \alpha_s(\lambda^2)\right) }{d(l_{ij})}=\Gamma^{\rm{cusp}}_i(\alpha_s(\lambda^2)),
\end{align}
for all $i$, where 
\beq
\label{eq:lij}
l_{ij} \equiv \log\frac{-s_{ij}}{\lambda^2}\,.
\eeq
Here $\Gamma^{\rm{cusp}}_i(\as)$ is the cusp anomalous dimension in the representation of parton~$i$, which can be expanded in Casimirs as in eq.~\eqref{cuspAD}:
\beq
\label{cuspAD_Sec_anom_dim}
\Gamma^{\rm{cusp}}_i(\alpha_s(\lambda^2)) = \frac{1}{2} \gamma_K(\alpha_s(\lambda^2))C_i+ \sum_R g_R(\alpha_s(\lambda^2)) \frac{d_{RR_i}}{N_{R_i}} + {\cal}O(\as^5)\,.
\eeq
The collinear anomaly constraints are key to understanding the structure of higher-order corrections to the soft anomalous dimension.

To understand the implications of eq.~(\ref{eq:BNdipinh}) note first that it is a matrix equation: while the left-hand-side is a non-trivial colour matrix, the right-hand side is proportional to the unit matrix in colour space. 
Upon solving the differential equations in eq.~(\ref{eq:BNdipinh}) one obtains an inhomogeneous solution, which is strictly linear in the logarithms $l_{ij}$ and in the cusp anomalous dimension $\Gamma^{\rm{cusp}}_i(\alpha_s)$ plus a homogeneous solution. The latter contains non-trivial structures in colour space and must be invariant under rescaling of any of the Wilson-line velocities, and is therefore a function of the so-called \emph{conformally invariant cross-ratios} (CICRs) 
\beq
\label{eq:rho}
\rho_{ijkl}=\frac{(-s_{ij})(-s_{kl})}{(-s_{ik})(-s_{jl}) }, 
\eeq
whose logarithm 
\beq
\label{eq:beta}
\beta_{ijkl}=\log \rho_{ijkl}=\log\frac{(-s_{ij})(-s_{kl})}{(-s_{ik})(-s_{jl}) }=l_{ij}+l_{kl}-l_{ik}-l_{jl},
\eeq
satisfies
\beq
\label{eq:dCICR}
\sum_{v \neq u} \frac{d \beta_{ijkl}}{d l_{uv}}=0\,,
\eeq
for any $u\in \{i,j,k,l\}$. 
The logarithm of CICRs has the symmetry properties:
\beq
\label{eq:betasym}
\beta_{ijkl}=\beta_{jilk}=-\beta_{ikjl}=-\beta_{ljki}=\beta_{klij}.
\eeq

Returning to the inhomogeneous component, a particular solution for $n\geq3$ is given by~\cite{Mark}: 
\begin{align}
\begin{split}
\label{eq:mans_main_text}
\Gamma_n'\left(\left\{\frac{s_{ij}}{\lambda^2}\right\}, \alpha_s\right)&=\frac{1}{2(n-1)}\sum_{i=1}^{n} \sum_{j \neq i}^{n} l_{ij}\left(\Gamma^{\rm{cusp}}_i(\alpha_s) +\Gamma^{\rm{cusp}}_j(\alpha_s)-\frac{1}{(n-2)} \sum_{k \neq i, j}^{n} \Gamma_k^{\text{cusp}}(\alpha_s)\right),
\end{split}
\end{align}
implying that the general solution of (\ref{eq:BNdipinh}) takes the form
\begin{align}
\label{eq:BNdipinh_sol}
\mathbf{\Gamma}_n\left(\{s_{ij}\},\lambda, \alpha_s\right) =  \Gamma_n'\left(\left\{\frac{s_{ij}}{\lambda^2}\right\}, \alpha_s\right) + \mathbf{\delta\Gamma}_n \left(\{\beta_{ijkl}\}, \alpha_s\right) \,,
\end{align}
where the second term depends on the kinematics only through a function of the CICRs.
It is clear at the outset, that while $\Gamma_n'$ satisfies eq.~(\ref{eq:BNdipinh}) -- and is therefore useful to understand some of its features -- it cannot on its own be considered a candidate for $\mathbf{\Gamma}_n$, not even at leading order. One way to see this is to note that 
any function arising from Feynman diagram calculation of the above-mentioned Wilson-line correlator has the property that dependence on a kinematic variable associated with a given Wilson line must only appear in terms that feature colour generators of that line. The simplest term that complies with the latter property is a dipole interaction, $\T_i\cdot\T_j\,l_{ij}$, which is the coefficient of the single-pole divergence arising from a single gluon exchange between lines $i$ and $j$.
Indeed, it is well known that the complete result through two loops for the soft anomalous dimension in massless scattering with general kinematics takes the form of a sum over colour dipoles:
\beq\label{Gamma-n-dip}
\mathbf{\Gamma}^{\rm dip.}_n\left(\{s_{ij}\},\lambda, \alpha_s\right)
= -\frac{1}{4} \gamma_{K}(\alpha_s)\sum_{(i,j)}\,\T_i\cdot\T_j\,l_{ij} +\sum_i^n\,\gamma_i(\alpha_s),
\eeq
where the sum goes over all pairs $(i,j)$ that are formed between the $n$ scattered partons. 
Upon substituting $\mathbf{\Gamma}^{\rm dip.}_n$ of eq.~(\ref{Gamma-n-dip}) into the left-hand side of eq.~(\ref{eq:BNdipinh}) and using colour conservation, one obtains the quadratic Casimir component of $\Gamma^{\rm{cusp}}_i(\alpha_s)$, as required. As already mentioned, while $\mathbf{\Gamma}_n$ receives non-dipole corrections from three loops~\cite{Almelid:2015jia}, the anomalous dimensions entering eq.~(\ref{Gamma-n-dip}) are known to four loops in QCD~\cite{Boels:2017ftb,Boels:2017skl,Moch:2017uml,Grozin:2017css,Henn:2019swt,vonManteuffel:2020vjv,Agarwal:2021zft}: $\gamma_i$ is the collinear anomalous dimension~\cite{FormFactors,DelDuca:2014cya,Falcioni:2019nxk,Dixon:2017nat} corresponding to the parton~$i$, while $\gamma_{K}(\alpha_s)$ is the coefficient of the quadratic Casimir in the cusp anomalous dimension of eq.~(\ref{cuspAD_Sec_anom_dim}).

In view of our goal of understanding the structure of the soft anomalous dimension to higher orders, there are several immediate questions raised by the discussion above. One is how to reconcile the sum-over-dipoles formula (\ref{Gamma-n-dip}) with the solution of eqs.~(\ref{eq:BNdipinh_sol}) and (\ref{eq:mans_main_text}).
A related one, noting that the quartic Casimir is absent in eq.~(\ref{Gamma-n-dip}), is how higher Casimir corrections enter there. The latter was addressed at four loops in ref.~\cite{Becher:2019avh}; we connect the two questions in appendix~\ref{sec:cuspappendix}, where we discuss the form of the CICR-dependent function $\mathbf{\delta\Gamma}_n$, which is linear in the logarithms for both the quadratic Casimir and quartic Casimir components. 

A separate aspect, which is salient to the simplicity of eq.~(\ref{Gamma-n-dip}) is the absence of ``tripole'' type contributions connecting three Wilson lines at two loops. Such corrections do appear for non-lightlike lines, but are forbidden by the constraint of (\ref{eq:BNdipinh}) (conformal cross ratios cannot be formed out of three Wilson-line velocities). A very interesting argument\footnote{The argument is based on the colour decomposition of a different Wilson-line correlator associated with multi-parton scattering, which was shown to be related to the soft anomalous dimension via a conformal mapping.} has been given by Vladimirov in ref.~\cite{Vladimirov:2017ksc} stating that only terms consisting of an even number of generators can appear in the soft anomalous dimension at any order, thus excluding for example terms with five generators at four loops and beyond.

At three loops, an explicit computation has been performed~\cite{Almelid:2015jia} fixing the form of the soft anomalous dimension at this order. As was expected, three loop corrections to the dipole formula (\ref{Gamma-n-dip}) depend exclusively on CICRs. The structure of the result is of direct relevance to the functions we shall encounter at four loops, and therefore we shall review it below in the context of the general form the anomalous dimension takes at four loops.

Taking into account the complete set of connected colour structures complying with the non-Abelian exponentiation theorem~\cite{Gatheral1983ExponentiationOE,Frenkel1984NonabelianEE,Gardi:2013ita}, Becher and Neubert wrote down~\cite{Becher:2019avh} a general parametrisation\footnote{Previous work along these lines has been done e.g. in refs.~\cite{Gardi:2009qi,Becher:2009qa,Dixon:2009ur,Ahrens:2012qz,Almelid:2017qju}.} -- with unknown kinematic functions -- which satisfied the aforementioned collinear anomaly constraints of eq.~(\ref{eq:BNdipinh}) along with Bose symmetry. The contributions appearing through four loops can be classified as follows:
\begin{align}
\label{eq:SADintrogamma} 
\begin{split}
\mathbf{\Gamma}_n\left(\{s_{ij}\},\lambda, \alpha_s(\lambda^2)\right) 
& = \mathbf{\Gamma}^{\rm dip.}_n\left(\{s_{ij}\},\lambda, \alpha_s\right) 
+\mathbf{\Gamma}_{n,\rm 4T-3L}\left(\alpha_s\right) 
+\mathbf{\Gamma}_{n,\rm 4T-4L}\left(\{\beta_{ijkl}\}, \alpha_s\right) \\[0.1cm]
&+\mathbf{\Gamma}_{n,\rm Q4T-2,3L}\left(\{s_{ij}\},\lambda, \alpha_s\right)
+\mathbf{\Gamma}_{n,\rm Q4T-4L}\left(\{\beta_{ijkl}\}, \alpha_s\right)\\[0.1cm]
&
+\mathbf{\Gamma}_{n,\rm 5T-4L}\left(\{\beta_{ijkl}\}, \alpha_s\right) +\mathbf{\Gamma}_{n,\rm 5T-5L}\left(\{\beta_{ijkl}\}, \alpha_s\right)
+ {\cal O}(\alpha_s^5)\,,
\end{split}
\end{align}
where the subscript for each term includes, in addition to the number of partons $n$, the following attributes of each (connected) colour factor:$\,{\mathrm Q}$ for a quartic Casimir related contribution; a number followed by ${\mathrm T}$ to indicate the number generators, and a number followed by ${\mathrm L}$ to indicate the number of distinct lines that interact.  

The first term in eq.~(\ref{eq:SADintrogamma}) is the sum-over-dipoles formula, which is the complete result for $\mathbf{\Gamma}_n$ to two loops, while all others start contributing at three (the second and third terms) and four loops (all others). 
Notice that this formula includes terms with explicit dependence on the scale (the first and the fourth) as well as terms that depend exclusively on CICRs (all others). The former can be identified with those that constitute a particular solution satisfying eq.~\eqref{eq:BNdipinh} (through four loops) and are strictly linear in $l_{ij}$, while the latter involve higher transcendental functions of the CICRs. The last two terms, consisting of five generators, are excluded based on the argument of ref.~\cite{Vladimirov:2017ksc}. We retain them here to see, independently of the latter argument, what constraints emerge from the Regge-limit analysis. Let us now introduce explicitly each term of eq.~\eqref{eq:SADintrogamma} in turn, where we adopt much of the notation from ref.~\cite{Becher:2019avh}. 

The second and third terms in (\ref{eq:SADintrogamma}) start at three loops, where they were explicitly computed~\cite{Almelid:2015jia}. These terms involve the colour and kinematic degrees of freedom of subsets of three or four partons, they are non-planar and depend exclusively on CICRs. They read: 
\beqa
\label{Gamman-4T3L}
\mathbf{\Gamma}_{n,\rm 4T-3L}\left(\alpha_s\right) 
&=& f(\alpha_s) \sum_{(i,j,k)}\bm{{\cal T}}_{iijk}, \\ 
\label{Gamman-4T4L}
\mathbf{\Gamma}_{n,\rm 4T-4L}\left(\{\beta_{ijkl}\}, \alpha_s\right) 
&=& \sum_{(i,j,k,l)} \bm{{\cal T}}_{ijkl}\,
{\cal F}(\beta_{ijlk},\beta_{iklj};\alpha_s),
\eeqa
where the summation is over tuples (with no restriction on the relative order of indices). The colour structure involves four generators: 
\beq
\label{col:4T}
\bm{{\cal T}}_{ijkl} \equiv f^{ade} f^{bce} \{\T_i^a, \T_j^b, \T_k^c, \T_l^d \}_+.
\eeq
Notice that the curly brackets represent symmetrisation, defined as
\beq
\{\T_i^{a_1}, \T_j^{a_2}, \dots \T_l^{a_n}\}_+\equiv\frac{1}{n!} \sum_\pi \T_i^{a_{\pi(1)}} \T_j^{a_{\pi(2)}} \dots \T_l^{a_{\pi(n)}},
\eeq
where the sum is over all permutations of the indices. The symmetrisation only acts on generators attached to the same line, as those attached to distinct lines commute. For example, 
\[
\bm{{\cal T}}_{iijk}=f^{ade} f^{bce} \{\T_i^a, \T_i^b\}_+ \T_j^c \T_k^d  
=\frac12 f^{ade} f^{bce}\left(\T_i^a \T_i^b+\T_i^b \T_i^a\right) \T_j^c \T_k^d .
\]
The functions $f(\alpha_s)$ and ${\cal F}(\beta_{ijlk},\beta_{iklj};\alpha_s)$ have a perturbative expansion 
\beq
\label{eq:fexp}
f(\alpha_s)=\left(\frac{\alpha_s}{\pi}\right)^3 f^{(3)} +  \left(\frac{\alpha_s}{\pi}\right)^4 \sum_R f^{(4)}_R+ O(\alpha_s^5),
\eeq
and
\beq
\label{eq:Fexp}
{\cal F}(\beta_{ijlk},\beta_{iklj};\alpha_s)=\left(\frac{\alpha_s}{\pi}\right)^3 {\cal F}^{(3)}(\beta_{ijlk},\beta_{iklj}) + \left(\frac{\alpha_s}{\pi}\right)^4  \sum_R {\cal F}^{(4)}_R(\beta_{ijlk},\beta_{iklj}) + O(\alpha_s^5)\,,
\eeq
where $f^{(\ell)}$ are transcendental constants while ${\cal F}^{(\ell)}$ are transcendental functions of the CICRs defined in eq.~(\ref{eq:rho}). At four loops, these two functions involve a sum over the gauge group representations $R$, which we write explicitly in eqs.~\eqref{eq:fexp} and~\eqref{eq:Fexp}. This is a general feature of the colour structures appearing in the anomalous dimension: there is an implicit sum over the representations, once they are considered at a loop order higher than when they first appear. This is a manifestation of the fact that any structure first appears owing to a purely gluonic diagram, and, as such, it has a universal nature, being entirely independent of the matter contents of the theory.

The functions $f(\alpha_s)$ and ${\cal F}(\beta_{ijlk},\beta_{iklj};\alpha_s)$ have been calculated at three loops~\cite{Almelid:2015jia}. Expressing the CICRs in terms of variables $z_{ijkl}$ and $\bar z_{ijkl}$:
\begin{align}
\label{eq:rhoz}
\rho_{ijkl}&=z_{ijkl}\bar{z}_{ijkl}, \,\,\,\hspace{1cm}
\rho_{ilkj}=(1-z_{ijkl})(1-\bar{z}_{ijkl}),
\end{align}
the function ${\cal F}(\as)$ reads 
\begin{align}
\label{eq:F3}
{\cal F}^{(3)}(\beta_{ijlk},\beta_{iklj})
&=\frac{1}{32}\bigg(F(1-z_{ijlk})-F(z_{ijlk})\bigg),
\end{align}
where in turn $F(z)$ is a function of single-valued harmonic polylogarithms \cite{Brown:2004ugm,Dixon:2012yy,Brown:2013gia,Schnetz:2013hqa}:
\beq
F(z)\equiv {\cal L}_{10101}(z)+ 2\zeta_2 \Big[{\cal L}_{001}(z)+ {\cal L}_{100}(z)\Big]\,,
\eeq
while
\begin{align}
\label{eq:f3}
  f^{(3)}&=\frac{1}{4}\left(\zeta_5 +2\zeta_2\zeta_3\right)\,.
\end{align}

The other terms in eq.~\eqref{eq:SADintrogamma} start at four loops. The quartic term involving 
four generators with attachments to two and three legs can be expressed as~\cite{Becher:2019avh} 
\beq\label{Gamman-4T2LQ}
\mathbf{\Gamma}_{n,\rm Q4T-2,3L}\left(\{s_{ij}\},\lambda, \alpha_s\right) 
= -\frac{1}{2}\sum_R g_R(\alpha_s)\bigg[ \sum_{(i,j)}\,
\big( \bm{{\cal D}}_{iijj}^R + 2 \bm{{\cal D}}_{iiij}^R \big)  l_{ij} 
+ \sum_{(i,j,k)} \bm{{\cal D}}_{ijkk}^R\,l_{ij} \bigg],
\eeq
where the colour operator is defined as
\beq
\label{Dijkl}
\bm{{\cal D}}_{ijkl}^R \equiv  \frac{1}{4!}\sum_{\sigma \in S_4} \operatorname{Tr}_R\left(T^{\sigma(a)}T^{\sigma(b)}T^{\sigma(c)}T^{\sigma(d)}\right)
\,\T_i^a \T_j^b \T_k^c \T_l^d \,.
\eeq
Similarly to the dipole term, $\mathbf{\Gamma}_{n,\rm Q4T-2,3L}$ is part of the inhomogeneous solution of eq.~(\ref{eq:BNdipinh}): upon substituting eq.~(\ref{Gamman-4T2LQ}) into the left-hand side of eq.~(\ref{eq:BNdipinh}) and using colour conservation, one obtains the quartic Casimir component of $\Gamma^{\rm{cusp}}_i(\alpha_s)$, where
\beq
\label{eq:drri}
\bm{\mathcal{D}}^R_{iiii}=\frac{d_{R R_i}}{N_{R_i}}= \frac{1}{4!}\sum_{\sigma\in\mathcal{S}_4}\mathrm{Tr}_R\left[T^{\sigma(a)} T^{\sigma(b)}T^{\sigma(c)}T^{\sigma(d)}\right]\mathbf{T}^a_i\mathbf{T}^b_i\mathbf{T}^c_i\mathbf{T}^d_i\,,
\eeq 
and where $g_R(\alpha_s)$ is the function multiplying $\frac{d_{R R_i}}{N_{R_i}}$ in eq.~\eqref{cuspAD_Sec_anom_dim}.
As discussed in section~\ref{sec:infared_Regge_limit}, the function $g_R(\alpha_s)$ begins at 
four loops and is known at this order~\cite{Henn:2019swt,Huber:2019fxe,vonManteuffel:2020vjv}. The result is quoted in eq.~\eqref{g_R_values}. We provide a more detailed discussion concerning the relation between $\mathbf{\Gamma}_n(\as)$ and $\Gamma^{\rm{cusp}}_i(\as)$ in appendix~\ref{sec:cuspappendix}.

The remaining terms in \eqref{eq:SADintrogamma} are part of the solution to the homogeneous equation associated to eq.~\eqref{eq:BNdipinh}, therefore, the functions appearing in these terms depend exclusively on CICRs.
The term ${\mathrm{Q4T-4L}}$ involves the quartic Casimir operator as well, and reads\footnote{Owing to the complete permutation symmetry of the colour factor $\bm{{\cal D}}_{ijkl}^R$ with respect to $i,j,k$ and $l$, the kinematic function ${\cal G}_R$ admits a similar symmetry. Consequently, ${\cal G}_R$ may be factored out of the sum over the permutations of a given subset of indices. } 
\beq\label{Gamman-4T4LQ}
\mathbf{\Gamma}_{n,\rm Q4T-4L}\left(\{\beta_{ijkl}\}, \alpha_s\right)
= \sum_R \sum_{(i,j,k,l)}\!\bm{{\cal D}}_{ijkl}^R\,{\cal G}_R(\beta_{ijlk},\beta_{iklj};\alpha_s).
\eeq
Finally, there are then two terms involving five colour generators: they are given by
\begin{subequations}
\begin{align}
\label{Gamman-5T4L}
\mathbf{\Gamma}_{n,\rm 5T-4L}\left(\{\beta_{ijkl}\}, \alpha_s\right)  
&= \sum_{(i,j,k,l)}\!\bm{{\cal T}}_{ijkli}\,{\cal H}_1(\beta_{ijlk},\beta_{iklj};\alpha_s), \\ 
\label{Gamman-5T5L}
\mathbf{\Gamma}_{n,\rm 5T-5L}\left(\{\beta_{ijkl}\}, \alpha_s\right) 
&= \sum_{(i,j,k,l,m)}\!\bm{{\cal T}}_{ijklm}\,
{\cal H}_2(\beta_{ijkl},\beta_{ijmk},\beta_{ikmj},\beta_{jiml},\beta_{jlmi};\alpha_s),
\end{align}
\end{subequations}
where the colour structure is defined as
\beq
\label{col:5T}
\bm{{\cal T}}_{ijklm} 
= i f^{adf}f^{bcg}f^{efg} \{ \T_i^a, \T_j^b, \T_k^c, \T_l^d,\T_m^e\}_+.
\eeq

The functions ${\cal G}_R(\as)$, ${\cal H}_1(\as)$ and ${\cal H}_2(\as)$ start at four loops, and have not yet been computed. Similarly, the four-loop contributions to the functions $f(\as)$ and ${\cal F}(\as)$ are to date unknown. In all these cases, the structure of these functions can be constrained by analysing amplitudes in specific kinematic limits, where additional information can be obtained. The collinear limit offers one such instance~\cite{Becher:2009qa,Dixon:2009ur,Almelid:2017qju,Becher:2019avh}, and we briefly summarise below what constraints it provides based on ref.~\cite{Becher:2019avh}, before returning to the Regge limit. 

It is well known that when two particles in either the initial or final state become collinear, the amplitude $\mathcal{M}_n$ factorises into a splitting amplitude $\textbf{Sp}$ and the parent amplitude $\mathcal{M}_{n-1}$ \cite{Berends:1988zn,Mangano_1991,Bern_1995,Kosower_1999,Catani:2011st}, with 
\beq
\label{Coll_fact}
\lim_{p_1||p_2}\mathcal{M}_n \left(\{ p_1,\dots,p_n\},\lambda,\al_s\right)= \textbf{Sp} \left(\{p_1,p_2\},\lambda, \al_s\right) \mathcal{M}_{n-1}\left(\{p_1 +p_2, p_3, \dots p_n\},\lambda,\al_s\right).
\eeq
The splitting amplitude has an anomalous dimension defined as 
\beq
\frac{d}{d \log \lambda} \textbf{Sp}\left( \{p_1, p_2\}, \lambda,\al_s\right)= {\bf \Gamma}_{\text{Sp}}\left( \{p_1, p_2\}, \lambda,\al_s\right) \textbf{Sp}\left(\{p_1,p_2\},\lambda,\al_s\right)
\eeq
which, just like the function $\textbf{Sp}$ itself, is independent of the momenta and colour generators of the particles that are not collinear. Performing infrared factorisation of each of the ingredients in eq.~(\ref{Coll_fact}), one obtains~\cite{Becher:2009qa,Dixon:2009ur}
\beq
\label{eq:sp1}
{\bf\Gamma}_{\bf Sp}\left(\{ p_1,p_2 \}, \lambda ,\al_s\right)=\underset{p_1 || p_2}{\text{lim}}{\bf \Gamma}_n\left(\{ p_1, \dots, p_n\}, \lambda, \al_s\right) - {\bf \Gamma}_{n-1}\left(\{p_1 + p_2, p_3, \dots, p_n\},\lambda, \al_s \right).
\eeq
This provides the non-trivial constraint on ${\bf \Gamma}_n$ itself: the splitting amplitude anomalous dimension on the left-hand side only depends on the two particles that become collinear, hence so must the right-hand side of eq. \eqref{eq:sp1}. This translates into concrete  constraints for the functions in eq.~\eqref{eq:SADintrogamma}. The functions $f(\as)$ and ${\cal F}(\as)$ are related by the condition~\cite{Becher:2009qa,Dixon:2009ur,Almelid:2017qju,Becher:2019avh}  
\beq
\label{eq:Fcol}
\underset{\beta_{12ij}\to -\infty}{\text{lim}}{\cal F}(\beta_{12ij},0;\alpha_s)= \frac{f(\alpha_s)}{2},
\eeq
which, in particular, provide a constraint for the coefficients $f^{(4)}_R$ and ${\cal F}^{(4)}_R$. Similarly, the functions ${\cal G}_R(\as)$ and $g_R(\alpha_s)$ are related by~\cite{Becher:2019avh}  
\beq
\label{eq:Gcol}
\underset{\beta_{12ij}\to -\infty}{\text{lim}}{\cal G}_R(\beta_{12ij},0;\alpha_s)= -\frac{g_R(\alpha_s)}{12}\,\beta_{12ij}.
\eeq
Furthermore, one has~\cite{Becher:2019avh} 
\beq\label{eq:Hcol}
\underset{\beta_{12ij}\to -\infty}{\text{lim}}
{\cal H}_1(\beta_{12ij},0;\alpha_s) = 0\,.
\eeq

Last, we have constraints from the high-energy limit, which is of course the topic of this section. Given our explicit calculation of $2\to 2$ parton scattering in this limit, we are able to determine the four-loop contribution to the functions appearing in eq.~\eqref{eq:SADintrogamma} in this limit. In order to proceed we need first to specialise  
eq.~\eqref{eq:SADintrogamma} to the case of two-parton scattering, and then take the high-energy limit. In this kinematic configuration no constraints can be obtained for  
${\cal H}^{(4)}_2$, which involves at least five partons. However, we are able to obtain constraints for ${\cal F}^{(4)}$ and ${\cal G}_R^{(4)}$, as well as ${\cal H}^{(4)}_1$.

\subsection{The soft anomalous dimension 
in the high-energy limit}\label{sec:factor2to2}
We now take the general form of the soft anomalous dimension as written in eq.~\eqref{eq:SADintrogamma}, and specialise it to the case of $2 \to 2 $ particle scattering in the high-energy limit. In short, the procedure is as follows. 
\begin{itemize}
\item In eq.~\eqref{eq:SADintrogamma} we drop the contributions which only appear for more than four external partons, i.e., we do not consider ${\cal H}_2$.
\item We express the colour operators of eq.~\eqref{eq:SADintrogamma} in what we call \emph{a Regge-limit basis}, i.e., in terms of a minimal\footnote{By using the matrix expression of $\tts$ and $\tsu$, obtained by specialising to projectile and target states either in the fundamental or in the adjoint representation \cite{Caron-Huot:2017fxr}, we verified that there are no linear relations among the colour structures appearing in the reduced amplitude of eq.~(\ref{eq:M42simb}).} set of colour operators made out of $\T_{t}^2$, $\T_{s-u}^2$, their commutators and quartic Casimir operators, as discussed in 
section \ref{subsec:colour}. Notice that, in particular, this will naturally split the terms in eq.~\eqref{eq:SADintrogamma} into even and odd signature contributions. 
\item We specialise the kinematic functions appearing in eq.~\eqref{eq:SADintrogamma} to the high-energy limit. First, owing to Bose symmetry, each kinematic function will acquire a definite signature symmetry, matching the symmetry of the corresponding colour operator it multiplies. Furthermore, each function will be
implicitly understood as an expansion in the high-energy logarithm $L$ defined in eq.~\eqref{eq:siglog}.
\end{itemize}
We expand the soft anomalous dimension in powers of the strong coupling, according to 
\beq
\mathbf{\Gamma}_n(\{s_{ij}\},\lambda,\alpha_s)= \sum_{\ell} \left(\frac{\alpha_s}{\pi}\right)^{\ell}
\mathbf{\Gamma}^{(\ell)}_n(\{s_{ij}\},\lambda)\,,
\eeq
where $\ell$ is the loop order. In what follows we are interested to obtain constraints on the coefficient functions appearing in $\mathbf{\Gamma}^{(4)}_4$ by using the results for the NLL, ${\cal O}(\alpha_s^4L^3)$, and the NNLL, ${\cal O}(\alpha_s^4L^2)$, in the anomalous dimension, summarised in section~\ref{subsec:sumSADreg}.
The four-loop order coefficients $\gamma_{K,R}^{(4)}$, $g_R^{(4)}$, $f^{(4)}_R$ are associated exclusively with linear and kinematically-independent contributions, ${\cal O}(L^1)$ and ${\cal O}(L^0)$, and we will not consider them in this section. Their high-energy limit is considered instead in appendix \ref{sec:ReggeBasisA}, and we will return to analyse the resulting ${\cal O}(L^1)$ terms in section~\ref{SADgenrep}.
This leaves the terms proportional to ${\cal F}$, ${\cal G}_R$ and ${\cal H}_1$, i.e. we consider 
\begin{align}
\begin{split}
\label{eq:SADNNLL} 
\mathbf{\Gamma}_4^{(4)}\left(\{s_{ij}\},\lambda\right) 
& = \mathbf{\Gamma}_{\rm 4T-4L}^{(4)}\left(\{\beta_{ijkl}\}\right) 
+\mathbf{\Gamma}_{\rm Q4T-4L}^{(4)}\left(\{\beta_{ijkl}\}\right)
+\mathbf{\Gamma}_{\rm 5T-4L}^{(4)}\left(\{\beta_{ijkl}\}\right) 
+\mathcal{O}(L)\,,
\end{split}
\end{align}
where for individual terms we drop the subscript indicating the number of partons, since we focus exclusively on the $n=4$ case below. The remaining subscripts are the defining characteristics of the colour operator as in eq.~(\ref{eq:SADintrogamma}).

\subsubsection{Four-generator four-line term ($\mathbf{4T}$$-$$\mathbf{4L}$)}
We start by considering the first term in eq.~\eqref{eq:SADNNLL}, i.e.   
\begin{align}
\begin{split}
\label{eq:DeltaFi}
\mathbf{\Gamma}_{\rm{4T-4L}}(\{\beta_{ijkl}\},\alpha_s)
&\equiv\sum_{(i,j,k,l)} f^{ade}f^{bce} \T_i^a\T_j^b\T_k^c\T_l^d 
\,{\cal F}(\beta_{ijlk},\beta_{iklj};\alpha_s).
\end{split}
\end{align}
An expression for this term that is specialised to two-parton scattering in the high-energy limit has been discussed already in refs.~\cite{Caron-Huot:2017fxr,Almelid:2017qju}, however, we provide here a short derivation for pedagogical purposes, in order to introduce useful notation for the elaboration of the other terms in eq.~\eqref{eq:SADNNLL}.

The colour structure in eq.~\eqref{eq:DeltaFi} is antisymmetric under the exchange of $i\leftrightarrow l$ or \hbox{$j\leftrightarrow k$}. Due to Bose symmetry, the function ${\cal F}(\as)$ must be antisymmetric under the exchange of the same indices. Under this exchange one has ${\cal F}(\beta_{ijlk},\beta_{iklj};\alpha_s) = - {\cal F}(\beta_{iklj},\beta_{ijlk};\alpha_s)$.  Using this property, we write eq.~\eqref{eq:DeltaFi} for the case of two-parton scattering as 
\begin{align}
\begin{split}
\label{eq:Fallf}
\mathbf{\Gamma}_{\rm 4T-4L}(\{\beta_{ijkl}\},\alpha_s)=&\,8 \T_1^a\T_2^b\T_3^c\T_4^d\bigg[f^{abe}f^{cde} {\cal F}(\beta_{1324},\beta_{1423};\alpha_s)\\
&\hspace{2.0cm}+f^{ace}f^{bde}   {\cal F}(\beta_{1234},\beta_{1432};\alpha_s)\\
&\hspace{2.0cm}+f^{ade}f^{bce}  {\cal F}(\beta_{1243},\beta_{1342};\alpha_s)\bigg].
\end{split}
\end{align}
As we have seen, in the high-energy limit,  signature symmetry plays a major role. In eq.~(\ref{eq:Fallf}) it can be implemented by considering symmetric and antisymmetric combinations under the exchange $2 \leftrightarrow 3$. This leads us to introduce the following two functions: 
\begin{align}
\begin{split}
\label{eq:Fsigeven}
{\cal F}^{(+)}(\{\beta_{ijkl}\},\alpha_s)&\equiv  
\frac{1}{2}\bigg\{{\cal F}(\beta_{1324},\beta_{1423};\alpha_s)
+ {\cal F}(\beta_{1234},\beta_{1432};\alpha_s)\bigg\},\\
{\cal F}^{(-)}(\{\beta_{ijkl}\},\alpha_s)&\equiv  
\frac{1}{2}\bigg\{{\cal F}(\beta_{1234},\beta_{1432};\alpha_s)
-{\cal F}(\beta_{1324},\beta_{1423};\alpha_s)\bigg\}
+{\cal F}(\beta_{1243},\beta_{1342};\alpha_s),
\end{split}
\end{align}
such that eq.~\eqref{eq:Fallf} becomes 
\begin{align}
\label{eq:Fallfsig}
\begin{split}
\mathbf{\Gamma}_{\rm 4T-4L}(\{\beta_{ijkl}\},\alpha_s)=
&\,8 \T_1^a\T_2^b\T_3^c\T_4^d\bigg[\left(f^{abe}f^{cde} 
+f^{ace}f^{bde}\right){\cal F}^{(+)}(\{\beta_{ijkl}\},\alpha_s) \\
&\hspace{3.0cm}
+\,f^{ade}f^{bce} {\cal F}^{(-)}(\{\beta_{ijkl}\},\alpha_s)\bigg].
\end{split}
\end{align} 
Due to Bose symmetry, the symmetry of ${\cal F}^{(\pm)}$ must be mirrored into the colour structure. This becomes evident when expressing the colour operators in eq.~\eqref{eq:Fallfsig} in our Regge-limit basis. Using the colour algebra identity of eq.~\eqref{eq:LieAlgebra}, i.e. $f^{abc}\T^c = -i [\T^a,\T^b]$, we have for instance
\beqa 
\label{eq:4Ttsubi} \nn
f^{abe} f^{cde}{\bf T}_1^{a} {\bf T}_2^{b} {\bf T}_3^{c} {\bf T}_4^{d} 
&=& - \Big[{\bf T}_1\cdot {\bf T}_2,
[{\bf T}_3\cdot {\bf T}_4, {\bf T}_1\cdot {\bf T}_3]\Big] \\ \nn 
&=& -\frac{1}{8} \Big[\Ts ,[\Ts,\Tu]\Big] \\ 
&=& \frac{1}{16}\left(\Big[\tts,[\tts,\tsu]\Big]
+2\Big[\tsu,[\tsu,\tts]\Big]\right),
\eeqa
where we used the definitions in eqs.~\eqref{TtTsTu} and~\eqref{eq:stucc}, 
and any Casimirs arising vanish in the commutators. With similar steps, the two other colour operators in eq. \eqref{eq:Fallfsig} are written as 
\begin{subequations}
\label{eq:4Ttsucdi}
\begin{align}
\label{eq:4Ttsuci}
f^{ace}f^{bde}\T_1^a\T_2^b\T_3^c\T_4^d&= \frac{1}{16}\left(-\Big[\tts,[\tts,\tsu]\Big]
+2\Big[\tsu,[\tsu,\tts]\Big]\right)\,,\\
\label{eq:4Ttsudi}
f^{ade}f^{bce}\T_1^a\T_2^b\T_3^c\T_4^d&
= -\frac{1}{8}\Big[\tts,[\tts,\tsu]\Big].
\end{align}
\end{subequations}
Inserting the expressions in eqs.~\eqref{eq:4Ttsubi} and \eqref{eq:4Ttsucdi} into 
eq.~\eqref{eq:Fallfsig} we get 
\begin{align}
\label{eq:Ffingen}
\begin{split}
\mathbf{\Gamma}_{\rm 4T-4L}(\{\beta_{ijkl}\},\as)
&=2 {\cal F}^{(+)}(\{\beta_{ijkl}\},\as) \Big[\tsu, [\tsu,\tts]\Big]\\&\quad\mbox{}
-{\cal F}^{(-)}(\{\beta_{ijkl}\},\as) \Big[\tts, [\tts,\tsu]\Big]\,.
\end{split}
\end{align}
It is easy to see that the symmetry properties of ${\cal F}^{(\pm)}$ are nicely mirrored into the colour structure: the first nested commutator is signature-even, containing two $\tsu$ operators, while the second is odd, having a single $\tsu$. 

At three loops, using the properties of the variables $z_{ijkl}$ introduced in \eqref{eq:rhoz}: 
\begin{equation}
\label{eq:zrel}
z_{ijkl}=\frac{1}{z_{ikjl}}=1-z_{ilkj}=\frac{z_{ijlk}}{z_{ijlk}-1},
\end{equation}
one can write the functions ${\cal F}^{(\pm)}$ as 
\begin{align}
\begin{split}
{\cal F}^{(+,3)}\left(\{\beta_{ijkl}\}\right)
&=\frac{1}{64}F_1(z_{1234}), \\ 
{\cal F}^{(-,3)}\left(\{\beta_{ijkl}\}\right)
&=\frac{1}{64}\Big(F_2(z_{1234})-F_3(z_{1234})\Big),
\end{split}
\end{align}
where the functions $F_1$, $F_2$ and $F_3$ have been introduced in ref.~\cite{Almelid:2017qju}, and read  
\begin{align}
\begin{split}
\label{eq:F123}
F_1(z)&\equiv F(1-1/z)-F(1/z)+F(1-z)-F(z),\\
F_2(z)&\equiv F(1/z)-F(1-1/z)+F(1/(1-z))-F(z/(z-1)),\\
F_3(z)&\equiv F(z)-F(1-z)+F(z/(z-1))-F(1/(1-z))=-F_1(z)-F_2(z).
\end{split}
\end{align} 

Here we consider the four-loop contribution to eq.~\eqref{eq:Ffingen}. Taking into account the perturbative expansion introduced in eq.~\eqref{eq:Fexp} one has  
\begin{align}
\begin{split}
\label{eq:Ffin4loop}
\mathbf{\Gamma}_{\rm{4T-4L}}^{(4)}(\{\beta_{ijkl}\})
&=2\left(\sum_R {\cal F}^{(+,4)}_R(\{\beta_{ijkl}\})\right) 
\Big[\tsu, [\tsu,\tts]\Big]\\
&\hspace*{100pt}-\left(\sum_R{\cal F}^{(-,4)}_R(\{\beta_{ijkl}\}) \right)
\Big[\tts, [\tts,\tsu]\Big] ,
\end{split}
\end{align} 
where we recall that the sum over representations starts at this order due to an additional internal loop, which gives rise to either a factor of $C_A$, or $n_f T_F$, depending on the particles propagating in the loop\footnote{Here we have only considered QCD particle types: adjoint gluons and fundamental quarks. The factor will change depending on the gauge theory considered.}.  
The first term in eq.~\eqref{eq:Ffin4loop} is signature even, and the second signature odd. It is worthwhile recalling that, upon expansion, the soft anomalous dimension in eq.~\eqref{eq:Ffin4loop} will be multiplied by the odd tree-level amplitude in eq.~\eqref{eq:treelevel}: hence, odd signature in the amplitude corresponds to even signature in the soft anomalous dimension. Taking this into account, we can already make a few observations. At NLL accuracy there are only gluonic contributions in the even amplitude, as calculated in ref.~\cite{Caron-Huot:2017zfo}. Therefore, only ${\cal F}^{(-,4)}_A|_{\rm NLL}$ will be non-zero in eq.~\eqref{eq:Ffin4loop}, while ${\cal F}^{(-,4)}_F|_{\rm NLL} = 0$. Similarly, the 
NNLL contribution to the odd amplitude, first presented in ref.~\cite{Falcioni:2020lvv} and discussed in detail in section~\ref{sec:results} above, is also given in terms of gluonic contributions only. Following the reasoning above, we expect that
${\cal F}^{(+,4)}_A|_{\rm NNLL}$ may be non-zero and ${\cal F}^{(+,4)}_F|_{\rm NNLL} = 0$. No predictions for ${\cal F}^{(-,4)}_R|_{\rm NNLL}$ can be made at this stage, however, given that the even amplitude is still unknown at this logarithmic accuracy. 

\subsubsection{Quartic Casimir four-generator four-line term ($\mathbf{Q4T}$$-$$\mathbf{4L}$)} 
\label{Gsec}
The quartic Casimir term only appears starting at four loops. Restricting to the case of $2\to 2$ scattering and writing explicitly the colour structure, eq.~\eqref{Gamman-4T4LQ} becomes 
\begin{align}
\begin{split}
\label{eq:line4a}
\mathbf{\Gamma}_{\rm Q4T-4L}(\{\beta_{1234}\},\alpha_s) =&
\sum_R \,{\cal G}_R(\beta_{1243},\beta_{1342};\alpha_s) \times \\&
\hspace{2.0cm} \sum_{\sigma \in S_4} \operatorname{Tr}_R\left(T^{\sigma(a)}T^{\sigma(b)}
T^{\sigma(c)}T^{\sigma(d)}\right)   \T_1^a\T_2^b\T_3^c\T_4^d\,,
\end{split}
\end{align}
where again there is a sum over the representations $R$ propagating in the loop. Here we extracted the function ${\cal G}_R$ out of the sum over permutations of the legs $(i,j,k,l)$ using its symmetry:
the colour structure is symmetric under the exchange of any pair of indices due to the symmetrised trace.  Having done that, we performed the sum over permutations $(i,j,k,l)$ on the colour structure.
Because ${\cal G}_R(\beta_{ijlk},\beta_{iklj};\alpha_s)$ is a completely symmetric function under permutations, ${\cal G}^{(-)}_R= 0$, and we can identify ${\cal G}^{(+)}_R={\cal G}_R$.

In order to conveniently express eq.~(\ref{eq:line4a}), we first introduce some new colour notation for terms involving a symmetrised trace over four generators attached to four numbered partonic generators.
The colour structures can be expressed using the colour-flow channels defined in 
eq.~\eqref{TtTsTu}, with
\begin{align}
\begin{split}
\label{eq:dppppp}
\bm{\mathcal{D}}^R_{pppp}\equiv&\,\frac{1}{4!}\sum_{\sigma \in S_4} \operatorname{Tr}_R\left(T^{\sigma(a)}T^{\sigma(b)}T^{\sigma(c)}T^{\sigma(d)}\right) \T_p^a\T_p^b\T_p^c\T_p^d,
\end{split}
\end{align}
where $p\in \{s,t,u\}$, for example
\begin{align}
\begin{split}
\label{eq:dssssp}
\bm{\mathcal{D}}^R_{ssss}=&\,\frac{1}{4!}\sum_{\sigma \in S_4} \operatorname{Tr}_R\left(T^{\sigma(a)}T^{\sigma(b)}T^{\sigma(c)}T^{\sigma(d)}\right) \T_s^a\T_s^b\T_s^c\T_s^d\\=\,&\bm{\mathcal{D}}^R_{1111}+4\bm{\mathcal{D}}^R_{1112}+6 \bm{\mathcal{D}}^R_{1122}+4\bm{\mathcal{D}}^R_{1222}+\bm{\mathcal{D}}^R_{2222}.
\end{split}
\end{align} 
A general formula is 
\beq
\label{eq:dppppm}
\bm{\mathcal{D}}^R_{pppp}=\,\bm{\mathcal{D}}^R_{kkkk}+4\bm{\mathcal{D}}^R_{kkkl}+6 \bm{\mathcal{D}}^R_{kkll}+4\bm{\mathcal{D}}^R_{klll}+\bm{\mathcal{D}}^R_{llll},
\eeq where 
\beq
\label{eq:dppppcasm}
\bm{\mathcal{D}}^R_{pppp}=
\begin{cases}
\bm{\mathcal{D}}^R_{ssss}& (k, l) \in \{(1,2),(3,4)\}\\
\bm{\mathcal{D}}^R_{uuuu}&(k, l) \in \{(1,3),(2,4)\}\\
\bm{\mathcal{D}}^R_{tttt}&(k, l) \in \{(1,4),(2,3)\}.
\end{cases}
\eeq
The expression in eq.~\eqref{eq:dppppm} is symmetric under $k\leftrightarrow l$. For each of the channels, corresponding to the respective Mandelstam invariants $p \in \{s,t,u\}$, the indices $(k,l)$ can be assigned to be either of the two combinations shown in eq.~\eqref{eq:dppppcasm}.
 
Using colour conservation, 
we can write the colour structure of eq.~\eqref{eq:line4a} as
\begin{align}
\label{D1234stu} \nn
\sum_{\sigma \in S_4} \operatorname{Tr}_R\left(T^{\sigma(a)}T^{\sigma(b)}
T^{\sigma(c)}T^{\sigma(d)}\right)   \T_1^a\T_2^b\T_3^c\T_4^d\mtree& \\ 
&\hspace{-6.0cm}=\bigg[2\Big(\bm{\mathcal{D}}^R_{ssss}+\bm{\mathcal{D}}^R_{uuuu}+\bm{\mathcal{D}}^R_{tttt}\Big)-4\left(\frac{d_{RR_i}}{N_{R_i}}+ \frac{d_{RR_j}}{N_{R_j}}\right)\bigg]\mtree,
\end{align}
where the quartic Casimirs correspond to the projectile $i$ (partons 1 and 4) and the target~$j$ (partons 2 and 3). The whole expression is signature-even as expected. This expression is useful as it holds for any representation. We will see in eqs.~(\ref{eq:BN4sig+}) and~(\ref{eq:BN4sig-2}) and in appendix~\ref{sec:ReggeBasisA} that the colour structures multiplying the quartic component of the cusp anomalous dimension $g_R$ can be expressed in a similar way.

\paragraph{Adjoint representation.} 
In the following we restrict our attention to the four-loop coefficient ${\cal G}^{(+,4)}_R$ in the adjoint representation, $R=A$. 
The reason we focus specifically on this representation was already explained at the end of the previous section considering ${\cal F}^{(+,4)}_R$, that is: the result for the signature-odd amplitude at NNLL accuracy, presented in section~\ref{sec:results}, only receives a contribution from purely gluonic diagrams. 
Thus, only ${\cal G}^{(+,4)}_A$ contributes to the sum over $R$ in eq.~\eqref{eq:line4a} at NNLL accuracy. It is then possible to use the identity~\cite{Becher:2019avh}
\begin{align}
\begin{split}
\label{tracerelA}
\sum_{\sigma \in S_4} 
\operatorname{Tr}\left(F^{\sigma(a)}F^{\sigma(b)}
F^{\sigma(c)}F^{\sigma(d)}\right) & = 
12\left[\mbox{Tr}\big(F^aF^bF^cF^d\big) +\mbox{Tr}\big(F^dF^cF^bF^a\big)\right] \\
&\hspace{2.0cm}
+4C_A \left( f^{abe} f^{cde}- f^{ade} f^{bce}\right),
\end{split}
\end{align}
to write 
\begin{align}
\begin{split} 
\label{eq:line4b}
\mathbf{\Gamma}_{\rm Q4T-4L,A}^{(4)}(\{\beta_{ijkl}\})&=
\big( {\bf Q}_1^{(4),A} + {\bf Q}_2^{(4),A} \big)\,
{\cal G}_A^{(+,4)} (\beta_{1234}, \beta_{1432}),
\end{split}
\end{align}
where we have defined 
\begin{subequations}
\beqa
\label{ColQ41}
{\bf Q}_1^{(4),A} &=& 
12 \, \T^a_1\T^b_2\T^c_3\T^d_4\,
\left[\mbox{Tr}\big(F^aF^bF^cF^d\big) 
+\mbox{Tr}\big(F^dF^cF^bF^a\big)\right], \\ 
\label{ColQ42}
{\bf Q}_2^{(4),A} &=& 4 C_A
\, \T^a_1\T^b_2\T^c_3\T^d_4
\left( f^{abe} f^{cde}- f^{ade} f^{bce} \right),
\eeqa
\end{subequations}
and $(F^{x})^{ab}\equiv if^{axb}$. The second term, i.e. ${\bf Q}_2^{(4),A}$, can readily be written in a Regge-limit basis by using the identities in eqs.~(\ref{eq:4Ttsubi}) and~(\ref{eq:4Ttsudi}). We get 
\beq \label{eq:Gcas}
{\bf Q}_2^{(4),A} = \frac{C_A}{4}  \left(3\Big[\tts,[\tts,\tsu]\Big]+2\Big[\tsu,[\tsu,\tts]\Big] \right).
\eeq
Concerning ${\bf Q}_1^{(4),A}$, repeated use of the commutator relation eq.~\eqref{eq:LieAlgebra} allows us to write it as 
\beq\label{eq:trace1}
{\bf Q}_1^{(4),A} = 
12\bigg\{ \Big[ \big[ \T^{b}_1, \T^{e}_1\big], \T^{f}_1\Big]  \T_2^b
\Big[ \big[ \T^{d}_3, \T^{f}_3\big], \T^{e}_3 \Big] \T_4^d
+ \, \T_1^a  \Big[ \T^{e}_2, \big[ \T^{f}_2, \T^{a}_2\big] \Big] 
\T_3^c \Big[ \T^{f}_4, \big[ \T^{e}_4, \T^{c}_4\big] \Big] \bigg\}.
\eeq
Notice that we apply the commutator relation such as to obtain an expression with manifest target-projectile symmetry (where, as usual, partons $1$ and $4$ represent the projectile while partons $2$ and $3$ the target).
At this point we recall that the colour operator in the soft anomalous dimension acts on the 
tree-level amplitude, to give the part of the four-loop amplitude from which the single-pole singularities are extracted. Therefore, it is sufficient to obtain a representation for the colour operator ${\bf Q}_1^{(4),A}$ when acting on the tree-level colour structure ${\bf T}_i\cdot {\bf T}_j$, as defined in eq.~\eqref{eq:treelevel}.
The commutators in eq.~\eqref{eq:trace1} can then be expressed as attachments to the projectile ($i$) or target ($j$), as in section~\ref{sec:col4}, 
so eq.~\eqref{eq:trace1} becomes
\beq
\label{eq:Ttrans}
{\bf Q}_1^{(4),A} 
(\T_i\cdot \T_j)=12
\left(\T^{([[b,e],f],x,d)}\right)_i\left( \T^{(b,x,[[d,f],e])}\right)_j
\,\,+ \,\, i\leftrightarrow j.
\eeq
It is now in a suitable form to apply the identities in section~\ref{subsec:colour} and appendix~\ref{sec:colourappendix}, converting the
operator to the Regge-limit basis:
\begin{align}\begin{split}
\label{eq:tracefin}
{\bf Q}_1^{(4),A} (\T_i\cdot \T_j)
& =\bigg\{2\left(\frac{d_{AA}}{N_A} -\frac{C_A^4}{24}\right) - \frac{3C_A}{4}\Big[\tts,[\tts,\tsu]\Big]  \\
&\hspace{0.5cm}-\,\frac{1}{2}\tts[(\tsu)^2,\tts]
+\frac{3}{2}[\tsu,\tts]\tts\tsu \bigg\} \,(\T_i\cdot \T_j).
\end{split}
\end{align}
Inserting eqs.~(\ref{eq:Gcas}) and~(\ref{eq:tracefin}) into eq.~\eqref{eq:line4b} we have 
\begin{align}
\begin{split}
\label{eq:Gfin}
\mathbf{\Gamma}_{\rm Q4T-4L,A}^{(4)}(\{\beta_{ijkl}\}) \mtree
&={\cal G}_A^{(+,4)} (\beta_{1234}, \beta_{1432})
\bigg\{2\left(\frac{d_{AA}}{N_A} -\frac{C_A^4}{24}\right)
-\frac{1}{2}\tts[(\tsu)^2,\tts]\\
&\quad\mbox{}+\frac{3}{2}[\tsu,\tts]\tts\tsu+\frac{C_A}{2} \Big[\tsu,[\tsu,\tts]\Big]\bigg\} 
\, \mtree.
\end{split}
\end{align}
Notice that after a cancellation of the signature-odd commutator term between ${\bf Q}_1^{(4),A}$ and ${\bf Q}_2^{(4),A}$, the resulting colour operator in eq.~(\ref{eq:Gfin}) is manifestly signature-even, as anticipated at the beginning of the section. 
Importantly, we observe that the quartic four-generator four-leg term $\mathbf{\Gamma}_{\rm Q4T-4L,A}^{(4)}$ is entirely non-planar, given that the commutators in eq.~\eqref{eq:Gfin} and the combination $d_{AA}/N_A - C_A^4/24$ are separately non-planar (see eq.~\eqref{eq:dAAminusCa4}).
$\mathbf{\Gamma}_{\rm Q4T-4L,A}^{(4)}$ now is expressed in the Regge-limit basis, and  eq.~(\ref{eq:Gfin}) will be used in section~\ref{subsec:Constraints}, along with the other terms, to derive constraints based on the explicit NNLL results of section~\ref{sec:results}. 

Finally, we can also equate eq.~\eqref{D1234stu} to eq.~\eqref{eq:Gfin} in the adjoint representation to express the previously-unknown signature-even combination of quartic $s$ and $u$ channel operators acting on the tree amplitude, in terms of nested commutators:
\begin{align}
\label{Dsusigeven}
\begin{split}
\Big(\bm{\mathcal{D}}^A_{ssss}+\bm{\mathcal{D}}^A_{uuuu}\Big)\mtree&= \bigg( 2\left(\frac{d_{AR_i}}{N_{R_i}}+ \frac{d_{AR_j}}{N_{R_j}}\right)-\frac{C_A^4}{24}
-\frac{1}{4}\tts[(\tsu)^2,\tts]\\
&\quad\hspace{-0.2cm}+\frac{3}{4}[\tsu,\tts]\tts\tsu+\frac{C_A}{4} \Big[\tsu,[\tsu,\tts]\Big] \bigg)\mtree,
\end{split}
    \end{align}
while the quartic $t$-channel operator acting on the tree amplitude simply gives
 \beq
 \label{DAt}
\bm{\mathcal{D}}^A_{tttt}\mtree= \frac{d_{AA}}{N_A}\mtree.
\eeq
These results will be useful in appendix~\ref{sec:ReggeBasisA}, where we analyse the colour structures multiplying the quartic component of the cusp anomalous dimension $g_R$.

\subsubsection{Five-generator four-line term ($\mathbf{5T}$$-$$\mathbf{4L}$)}
The third term in eq.~\eqref{eq:SADNNLL} reads 
\begin{align}
\label{eq:H1i_mainText}
\begin{split}
\mathbf{\Gamma}_{\rm 5T-4L}\left(\{\beta_{ijkl}\},\alpha_s\right)
&=\sum_{(i,j,k,l)} \, \bm{\mathcal{T}}_{ijkli}\,
{\cal H}_1(\beta_{ijlk},\beta_{iklj};\alpha_s)\\
&=\sum_{(i,j,k,l)} if^{adg}f^{bch} f^{egh} \, 
\{\T_i^a,\T_i^e\}_+\T_j^b\T_k^c\T_l^d\, 
{\cal H}_1(\beta_{ijlk},\beta_{iklj};\alpha_s).
\end{split}
\end{align}
The colour structure is antisymmetric under $j\leftrightarrow k$ so $\bm{\mathcal{T}}_{ijkli}=-\bm{\mathcal{T}}_{ikjli}$ and therefore ${\cal H}_1$ is antisymmetric under a swap of its arguments, due to Bose symmetry. We want to write an expression with manifest symmetry under $s\leftrightarrow u$, which can be achieved by exploiting the symmetries under swaps of $2\leftrightarrow 3$ or $1\leftrightarrow 4$ of $\bm{\mathcal{T}}_{ijkli}$ and ${\cal H}_1$.
Similarly to ${\cal F}(\beta_{ijlk},\beta_{iklj},\alpha_s)$, let us introduce symmetric and antisymmetric combinations under $2\leftrightarrow 3$ of ${\cal H}_1$:
\begin{subequations}
\label{Hplusminuesdef}
\begin{align}
{\cal H}_1^{(+)}(\{\beta_{ijkl}\}, \alpha_s)
&\equiv\frac{1}{2}\Big\{ {\cal H}_1(\beta_{1324},\beta_{1423};\alpha_s)
+ {\cal H}_1(\beta_{1234},\beta_{1432};\alpha_s)\Big\}, \\
{\cal H}_1^{(-)}(\{\beta_{ijkl}\}, \alpha_s)
&\equiv\frac{1}{2}\Big\{ {\cal H}_1(\beta_{1324},\beta_{1423};\alpha_s)
- {\cal H}_1(\beta_{1234},\beta_{1432};\alpha_s)\Big\}, \\[0.1cm]
\tilde{{\cal H}}_1^{(-)}(\{\beta_{ijkl}\}, \alpha_s)
&\equiv {\cal H}_1(\beta_{1243},\beta_{1342};\alpha_s).
\end{align}
\end{subequations}

As shown in appendix~\ref{sec:ReggeBasisA}, with these definitions we can write  eq.~\eqref{eq:H1i_mainText} as
\begin{align}\begin{split}
\label{eq:H1fin_main_text}
\mathbf{\Gamma}_{\rm 5T-4L}^{(4)}(\{\beta_{ijkl}\})\mtree&=\Bigg[ {\cal H}_1^{(+,4)}(\{\beta_{ijkl}\})\bigg(-\frac{C_A}{2}\Big[\tsu,[\tsu,\tts]\Big]+C_A \tsu [\tsu,\tts]\\&\hspace{-3.5cm} -\frac{1}{6}\tts[(\tsu)^2,\tts]\bigg)+\frac{1}{4}\tilde{{\cal H}}_1^{(-,4)}(\{\beta_{ijkl}\})\bigg[\tts,\Big[\tts,\big[\tts,\tsu\big]\Big]\bigg]\\&\hspace{-3.5cm}
+{\cal H}_1^{(-,4)}(\{\beta_{ijkl}\})\bigg(-\frac{1}{2}\bigg[\tsu,\Big[\tsu,\big[\tsu,\tts\big]\Big]\bigg]+\frac{1}{8}\bigg[\tts,\Big[\tts,\big[\tts,\tsu\big]\Big]\bigg]\bigg)\Bigg]\mtree,\end{split}
\end{align}
with all colour operators expressed in terms of nested commutators. Thus, as expected, all three terms in eq.~(\ref{eq:SADNNLL}), given in eqs.~(\ref{eq:Ffin4loop}), (\ref{eq:Gfin}) and (\ref{eq:H1fin_main_text}), are strictly non-planar.

\subsubsection{The Regge limit of the soft anomalous dimension}\label{subsec:Reggelimit}

We have now specialised eq.~\eqref{eq:SADNNLL} to the case of two-parton scattering, and expressed the colour operators involved in these terms in a Regge-limit basis
in eqs.~(\ref{eq:Ffin4loop}), (\ref{eq:Gfin}) and (\ref{eq:H1fin_main_text}).
In order to compare the resulting expression for the soft anomalous dimension with the high-energy limit calculation summarised in section~\ref{subsec:sumSADreg} we 
formally
consider each of the kinematic functions as an expansion in the high-energy logarithm $L$, for instance: 
\beq\label{formalLexpansion}
{\cal F}^{(-,4)}_A(L)= {\cal F}^{(-,4,3)}_A L^3 + {\cal F}^{(-,4,2)}_A L^2+{\cal F}^{(-,4,1)}_A L +{\cal F}^{(-,4,0)}_A,
\eeq
with unknown coefficients ${\cal F}^{(-,\ell,n)}_A$ for $\ell=4$ and $n = 3, 2, 1, 0$, which we expect on general grounds to be transcendental numbers of weight $2\ell-1-n=7-n$ or lower.

Separating the four-loop soft anomalous dimension $\bm{\Gamma}^{(4)}$ into components with definite signature symmetry, we have at four loops 
\beq
\bm{\Gamma}^{(4)}_{\text{Regge}}=\bm{\Gamma}^{(+,4)}_{\text{Regge}}+\bm{\Gamma}^{(-,4)}_{\text{Regge}}\,,
\eeq
where we added the subscript $\text{Regge}$, to indicate that the Regge limit has been taken. Explicitly, using the results in eqs.~(\ref{eq:Ffin4loop}), (\ref{eq:Gfin}) and (\ref{eq:H1fin_main_text}) we obtain
\beq \label{eq:gammaNNLL+}
\begin{aligned}
\mathbf{\Gamma}^{(+,4)}_{\text{Regge}}(L)\mtree
=&\Bigg\{2{\cal F}_{A}^{(+,4)} (L)\Big[\tsu, [\tsu,\tts]\Big]  
\\&\quad\mbox{}+  {\cal G}_{A}^{(+,4)}(L)\bigg(2\left(\frac{d_{AA}}{N_A} 
-\frac{C_A^4}{24}\right) -\frac{1}{2}\tts[(\tsu)^2,\tts]
\\&\hspace{2.5cm}+\frac{3}{2}[\tsu,\tts]\tts\tsu +\frac{C_A}{2} \Big[\tsu,[\tsu,\tts]\Big]\bigg) \\
&\quad\mbox{}
+{\cal H}_{1}^{(+,4)}(L)\bigg(-\frac{1}{2}C_A\Big[\tsu,[\tsu,\tts]\Big]+C_A \tsu [\tsu,\tts] \\
&\hspace{2cm}
-\frac{1}{6}\tts[(\tsu)^2,\tts]\bigg)\Bigg\}\mtree+ {\cal O}(L),
\end{aligned}
\eeq
for the signature-even part, while the odd component reads
\beq
\label{eq:gammaNNLL-}\begin{aligned}
\mathbf{\Gamma}^{(-,4)}_{\text{Regge}}(L)\mtree
&=\Bigg\{-\left(\sum_R{\cal F}_{R}^{(-,4)}(L)\right) \Big[\tts, [\tts,\tsu]\Big] \\ 
&\hskip-4pt +{\cal H}_{1}^{(-,4)}(L)
\bigg(-\frac{1}{2}\bigg[\tsu\Big[\tsu,\big[\tsu,\tts\big]\Big]\bigg]
+\frac{1}{8}\bigg[\tts,\Big[\tts,\big[\tts,\tsu\big]\Big]\bigg]\bigg) \\ 
&\hskip-4pt+\frac{1}{4}\tilde{{\cal H}}_{1}^{(-,4)}(L) 
\bigg[\tts,\Big[\tts,\big[\tts,\tsu\big]\Big]\bigg]
\Bigg\}\mtree
+ {\cal O}(L).
\end{aligned}
\eeq
Functions that do not contribute through NNLL, i.e., only generate ${\cal O}(L)$ and ${\cal O}(L^0)$ contributions, are not shown in eqs.~\eqref{eq:gammaNNLL+} and~\eqref{eq:gammaNNLL-}. We discuss these in section~\ref{SADgenrep} and 
appendix~\ref{sec:ReggeBasisA}.

\subsection{Constraints on the kinematic functions in the soft anomalous dimension} \label{subsec:Constraints}

We are now ready to compare the general parametrisation of the four-loop soft anomalous dimension to the explicit results of our calculation in the high-energy limit.
Before considering the four-loop case, where the kinematic functions are unknown, it is useful to conduct a similar exercise at three loops, where the functions are known~\cite{Almelid:2015jia} and their high-energy limit has been previously obtained~\cite{Caron-Huot:2017fxr,Almelid:2017qju}. 

Using eqs.~(\ref{eq:Ffingen}) and~(\ref{eq:falphfin}) we have
\begin{subequations}
\beq
\label{eq:ThreeLoopgammaNNLL+}\begin{aligned}
\mathbf{\Delta}^{(+,3)}_{\text{Regge}}(L)\mtree
&=
\bigg\{2\Big({\cal F}^{(+,3,2)}L^2+{\cal F}^{(+,3,1)}L+ {\cal F}^{(+,3,0)}\Big) \Big[\tsu, [\tsu,\tts]\Big]\\&
+f^{(3)}\bigg(\Big[\tsu,[\tsu,\tts]\Big] + \frac{C_A^3}{2} - 6\frac{d_{AR_i}}{N_{R_i} C_i} -6\frac{d_{AR_j}}{N_{R_j} C_j} \bigg)\bigg\}\mtree
\end{aligned}
\eeq
\beq
\label{eq:ThreeLoopgammaNNLL-}\begin{aligned}
\mathbf{\Delta}^{(-,3)}_{\text{Regge}}(L)\mtree
&=-\Big({\cal F}^{(-,3,2)}L^2+{\cal F}^{(-,3,1)}L+ {\cal F}^{(-,3,0)}\Big) \Big[\tts, [\tts,\tsu]\Big]\mtree
\end{aligned}
\eeq
\end{subequations}
Matching these expressions to $\dd$ in appendix~\ref{sec:SAD3loopa} we obtain the following expansion coefficients: 
\begin{align}
\label{Three_loop_calF_limits}
\begin{array}{lll}
{\displaystyle {\cal F}^{(+,3,2)}=0},\hspace{1cm}
&{\displaystyle {\cal F}^{(+,3,1)}=0},\hspace{1cm}
&{\displaystyle {\cal F}^{(+,3,0)}=\frac{1}{8} \left(4 \zeta_3 \zeta_2 -\zeta_5\right)}
    \\
  {\displaystyle  {\cal F}^{(-,3,2)}=0},\hspace{1cm}
&{\displaystyle {\cal F}^{(-,3,1)}=-\frac{i\pi}{4} \zeta_3},\hspace{1cm}
&{\displaystyle {\cal F}^{(-,3,0)}=-\frac{11i\pi}{4} \zeta_4}\,,
\end{array}
\end{align}
consistently with refs.~\cite{Caron-Huot:2017fxr,Almelid:2017qju,Almelid:2015jia,Henn:2016jdu}.
A similar procedure will be followed below at four loops where there are more functions contributing, all of which are yet unknown. To this end we consider the expressions in eqs.~(\ref{eq:gammaNNLL+}) and~(\ref{eq:gammaNNLL-}) in the Regge-limit basis, which we comapre with the explicit results we obtained through NNLL accuracy in eq.~\eqref{eq:softADdef4loopsymD6}.

\paragraph{Constraints at four loops NLL for signature-odd functions.}

We start by considering the signature-odd contribution to the soft anomalous dimension at NLL accuracy. We expand the functions in eq.~(\ref{eq:gammaNNLL-}) as in eq.~(\ref{formalLexpansion}), and match it to  eq.~(\ref{eq:gammaNNLL-}) order by order in the high-energy logarithm $L$. At $\ord(L^3)$, equating eq.~\eqref{eq:gammaNNLL-} to eq.~\eqref{eq:delta43min6}, we have
\begin{align}
\begin{split}
-i\pi\frac{\zeta_3}{24} \Big[\tts,[\tts,\tsu]\Big]\tts
 \,\,\overset{!}{=} \,\,& -\Big[\tts,[\tts,\tsu]\Big]\,{\cal F}^{(-,4,3)}_A \\
&-\frac{1}{2}\bigg[\tsu,\Big[\tsu,\big[\tsu,\tts\big]\Big]\bigg]{\cal H}^{(-,4,3)}_1 \\
&+\frac{1}{8}\bigg[\tts,\Big[\tts,\big[\tts,\tsu\big]\Big]\bigg]
\left(2\tilde{{\cal H}}^{(-,4,3)}_1+{\cal H}^{(-,4,3)}_1\right).
\end{split}
\end{align}

Firstly, ${\cal H}^{(-,4,3)}_1$ is the only term involving a colour operator $\propto (\tsu)^3$, which does not appear on the left-hand side, so we conclude that ${\cal H}^{(-,4,3)}_1=0$. Next, $\tilde{{\cal H}}^{(-,4,3)}_1$ multiplies a fully nested commutator, which also cannot be matched to the colour operators on the left-hand side, so it must vanish as well. In order to match the single term that remains, we recall that at four loops the soft anomalous dimension acts directly on the tree-level amplitude, so we can use $\tts \mathcal{M}_{\text{tree}}= C_A\mathcal{M}_{\text{tree}}$. This is consistent with the expectation that ${\cal F}^{(-,4,3)}_A$ should contain a factor of $C_A$, while ${\cal F}^{(-,4,3)}_F$ does not contribute at NLL accuracy (see the discussion following eq.~\eqref{eq:Ffin4loop}). We deduce
\begin{subequations}
\label{const_NLL_minus}
\begin{align}
{\cal F}^{(-,4,3)}_A&=i\pi C_A\frac{\zeta_3}{24} 
\hspace{2cm}{\cal F}^{(-,4,3)}_F=0 \\
{\cal H}^{(-,4,3)}_1&=0 \hspace{3.15cm}
\tilde{{\cal H}}^{(-,4,3)}_1=0 \,.
\end{align}
\end{subequations}

We note that the even amplitude at four loops for NNLL (and beyond) is still unknown. As a consequence, $\mathbf{\Gamma}^{(-,4,m)}=\text{Im}[\mathbf{\Gamma}^{(4,m)}]$, with $m=\{0,1,2\}$ remain unconstrained.

\paragraph{Constraints at four loops NLL for signature-even functions.}

Consider now kinematic functions multiplying signature-even colour structures, which must be real and symmetric under $s\leftrightarrow u$. 
 
We express eq.~\eqref{eq:gammaNNLL+} at $L^3$ order and equate it to the relevant $L^3$ coefficient in eq.~(\ref{eq:delta43plus6}), which vanishes identically, getting
\beq
\begin{aligned}         
0& \,\,\overset{!}{=} \,\,\Big[\tsu,[\tsu,\tts]\Big]
\bigg\{2{\cal F}^{(+,4,3)}_A+\frac{C_A}{2}\left({\cal G}^{(+,4,3)}_A
-{\cal H}^{(+,4,3)}_1\right)\bigg\} \\
&\quad\mbox{} +\left( 2\frac{d_{AA}}{N_A}-\frac{C_A^4}{12} -\frac{1}{2}\tts[(\tsu)^2,\tts]
+\frac{3}{2}[\tsu,\tts]\tts\tsu\right){\cal G}^{(+,4,3)}_A \\
&\quad\mbox{}
+\bigg(\tsu[\tsu,\tts]\tts 
- \frac{1}{6}\tts[(\tsu)^2,\tts]\bigg) {\cal H}^{(+,4,3)}_1.
\end{aligned}
\eeq
The only function that multiplies quartic Casimir terms is ${\cal G}^{(+,4,3)}_A$ so it must be zero. In the last line ${\cal H}^{(+,4,3)}_1$ multiplies linearly independent colour structures so it must vanish as well. While ${\cal F}^{(+,4,3)}_A$ appears in combination with other functions, those vanish hence so does ${\cal F}^{(+,4,3)}_A$. At $L^3$ order we thus obtain  the following constraints:
\begin{subequations}
\label{const_NLL_plus}
\begin{align}
{\cal F}^{(+,4,3)}_A&=0 \hspace{2.5cm}{\cal F}^{(+,4,3)}_F=0 \\
{\cal G}^{(+,4,3)}_A&=0 \hspace{2.6cm}{\cal G}^{(+,4,3)}_F=0 \\
{\cal H}^{(+,4,3)}_1&=0.
\end{align}
\end{subequations}
These results are of course in line with the fact that the signature-even NLL anomalous dimension is two-loop exact. 

\paragraph{Constraints at four loops NNLL for signature-even functions.} 

At $L^2$ order, upon equating the relevant terms of eq.~\eqref{eq:gammaNNLL+} to eq.~\eqref{eq:delta42} we have 
\beq\label{ConstraintsGamma42}
\begin{aligned}         
\zeta_2 \zeta_3 C_{\dd}^{(+,4,2)}
& \,\,\overset{!}{=} \,\,2C_{\dd}^{(+,4,2)} {\cal G}^{(+,4,2)}_A+
\bigg(\tsu[\tsu,\tts]\tts -\frac{1}{6} \tts[(\tsu)^2,\tts]\bigg)
{\cal H}^{(+,4,2)}_1\\
&\hspace{0.3cm}+\Big[\tsu, [\tsu,\tts]\Big]\bigg\{2{\cal F}^{(+,4,2)}_A
+\frac{C_A}{2}\left({\cal G}^{(+,4,2)}_A-{\cal H}^{(+,4,2)}_1\right)\bigg\}.   
\end{aligned}
\eeq
We immediately see the same colour term $C_{\dd}^{(+,4,2)}$, defined in eq.~(\ref{eq:delta42}), on the left-hand side and multiplying  ${\cal G}^{(+,4,2)}_A$ on the right-hand side. This fixes ${\cal G}^{(+,4,2)}_A$ and implies that the combination of the other terms must be zero. ${\cal H}^{(+,4,2)}_1$ multiplies colour operators that are linearly independent of the others, and must vanish. Finally the combination of functions in the second line of eq.~\eqref{ConstraintsGamma42} multiplying the fully-nested commutator must vanish, and since ${\cal H}^{(+,4,2)}_1=0$, it implies a simple relation between ${\cal F}^{(+,4,2)}_A$ and ${\cal G}^{(+,4,2)}_A$. The constraints at $L^2$ order for even-signature functions are then
\begin{subequations}
\label{const_NNLL_plus}
\begin{align}
{\cal F}^{(+,4,2)}_A&=-C_A\frac{\zeta_2\zeta_3}{8} \hspace{2.5cm}{\cal F}^{(+,4,2)}_F=0 \label{eq:cstF+4}\\
{\cal G}^{(+,4,2)}_A&=\frac{\zeta_2\zeta_3}{2} \hspace{3.4cm}{\cal G}^{(+,4,2)}_F=0 \label{eq:cstG+4}\\
{\cal H}^{(+,4,2)}_1&=0 \label{eq:cstH+4}.
\end{align}
\end{subequations}
The expressions for $\dd^{(+,4,m)}=\text{Re}[\dd^{(4,m)}]$, $m=\{0,1\}$ are currently not known, so our firm constraints for the even signature part of the soft anomalous dimension at four loops end at NNLL accuracy. 
The $m=1$ term, however, has a rather special status due to its connection with the cusp anomalous dimension, which we discuss in the next section before summarising the complete set of constraints.

\subsection{The soft anomalous dimension at four loops}
\label{SADgenrep}

In this section, we present expressions parametrising the four-loop soft anomalous dimension in the high-energy limit through all powers of the high-energy logarithm $L$. 
Although this goes beyond the logarithmic accuracy of any explicit calculation of the amplitude, we also discuss here the generalisation of the relation in eq.~(\ref{affirmed_eqality_of_sing_part_of_tilde_alpha_g_and_K}) between the cusp anomalous dimension and the singularities of the gluon Regge trajectory to four loops. We show that this generalisation is natural despite the presence of quartic Casimir contributions and it leads to an extra set of constraints on the soft anomalous dimension.

\paragraph{The soft anomalous dimension at four loops.}
To begin, it is useful to define an operator representation of the cusp anomalous dimension 
\beq
\label{Gamma_cusp_p}
\bm{\Gamma}_p^{\rm{cusp}}\equiv\frac{1}{2} \gamma_K(\alpha_s) \T_p^2+ \sum_R g_R(\alpha_s) \bm{\mathcal{D}}^R_{pppp}+ {\cal O}(\alpha_s^5), 
\eeq
associated with a channel $p \in \{s,t,u\}$, where we suppress corrections containing sextic and higher Casimir operators. The quartic operator $\bm{\mathcal{D}}^R_{pppp}$ is defined in eq.~\eqref{eq:dppppm}.

When the $t$-channel operators act on the tree amplitude, they reproduce it, multiplied by the respective adjoint Casimir, namely 
 \beq 
 \label{t_op_on_Mtree}
 \tts \mtree=C_A \mtree, \hspace{1.5cm} \bm{\mathcal{D}}^R_{tttt}\mtree= \frac{d_{RA}}{N_A}\mtree, 
 \eeq 
 which yields
 \beq
 \label{cuspttree}
 \bm{\Gamma}_t^{\rm{cusp}}\mtree={\Gamma}_A^{\rm{cusp}}\mtree\,,
 \eeq 
where ${\Gamma}_A^{\rm{cusp}}$ on the right-hand side is simply the cusp anomalous dimension defined by an adjoint Wilson line. In contrast, when $\bm{\mathcal{D}}^R_{ssss}$ and $\bm{\mathcal{D}}^R_{uuuu}$ act on the tree amplitude they generate mixing into other colour states, similarly to their quadratic counterparts~$\T_s^2$ and~$\T_u^2$. In particular, their adjoint signature-even combination is given in eq.~\eqref{Dsusigeven} and signature-odd  combination is given in eq.~\eqref{Dsusigodd}.

With these definitions and properties in place we are ready to present the general form of the soft anomalous dimension for $2\to2$ scattering in the high-energy limit. The signature-even part reads
\beq
\label{eq:BN4sig+}\begin{aligned}
\bm{\Gamma}^{(+,4)}_{ij\to ij}\left(L, \frac{-t}{\lambda^2}\right)\mtree&= \Bigg\{
 L\, \bm{\Gamma}_t^{\rm{cusp},(4)} + \log\frac{-t}{\lambda^2} \left(\Gamma_i^{\rm{cusp},(4)}+ \Gamma_j^{\rm{cusp},(4)}\right) + 2\gamma_i^{(4)}+ 2\gamma_j^{(4)}
\\
&\hspace{-1.8cm} + \left(\sum_R f^{(4, R)}\right) \bigg(\Big[\tsu,[\tsu,\tts]\Big] + \frac{C_A^3}{2} - 6\frac{d_{AR_i}}{N_{R_i} C_i} -6\frac{d_{AR_j}}{N_{R_j} C_j}\bigg) \\
&\hspace{-1.8cm}
+ 2\sum_R \left({\cal G}^{(4)}_R(L)-\frac{g^{(4)}_R}{6} L \right) \left(\bm{\mathcal{D}}^R_{tttt}+\bm{\mathcal{D}}^R_{ssss}+\bm{\mathcal{D}}^R_{uuuu}-2\left(\frac{d_{RR_i}}{N_{R_i}}+\frac{d_{RR_j}}{N_{R_j}}\right)\right) \\
&\hspace{-1.8cm}+2\left(\sum_R{\cal F}^{(+,4)}_R(L)\right) \Big[\tsu, [\tsu,\tts]\Big] 
+{\cal H}_1^{(+,4)}(L)\bigg(C_A \tsu [\tsu,\tts]
\\&\hspace{-1.8cm}-\frac{1}{2}C_A\Big[\tsu,[\tsu,\tts]\Big]
 -\frac{1}{6}\tts[(\tsu)^2,\tts]\bigg)\Bigg\} \mtree,
\end{aligned}
\eeq
while the signature-odd part is
\beq
\label{eq:BN4sig-2}\begin{aligned}
\bm{\Gamma}^{(-,4)}_{ij\to ij}(L)\mtree&=\Bigg\{
\frac{i\pi}{2}\bigg[\bm{\Gamma}_s^{\rm{cusp},(4)}-\bm{\Gamma}_u^{\rm{cusp},(4)}\bigg]
-\left(\sum_R{\cal F}^{(-,4)}_R (L)\right)\Big[\tts, [\tts,\tsu]\Big] \\
&\hspace{-1.8cm}
+{\cal H}^{(-,4)}_1(L)\bigg(-\frac{1}{2}\bigg[\tsu,\Big[\tsu,\big[\tsu,\tts\big]\Big]\bigg]+\frac{1}{8}\bigg[\tts,\Big[\tts,\big[\tts,\tsu\big]\Big]\bigg]\bigg)\\ 
&\hspace{-1.8cm}+\tilde{{\cal H}}^{(-,4)}_1(L)\,\frac{1}{4}\bigg[\tts,\Big[\tts,\big[\tts,\tsu\big]\Big]\bigg] \Bigg\} \mtree.
\end{aligned}
\eeq
These two expressions generalise eqs.~(\ref{eq:gammaNNLL+}) and (\ref{eq:gammaNNLL-}), respectively, to include ${\cal O}(L^1)$ and ${\cal O}(L^0)$ terms. The derivation of these contributions is presented in appendices~\ref{sec:cuspappendix} and~\ref{sec:ReggeBasisA}.  
      
In both the signature-even expression of eq.~\eqref{eq:BN4sig+} and the odd one in eq.~\eqref{eq:BN4sig-2} we have restored the full $p$-channel ($p \in \{s,t,u\}$) cusp anomalous dimension $\bm{\Gamma}_p^{\rm{cusp}}$ of eq.~(\ref{Gamma_cusp_p}), consisting of both the quadratic and quartic components. Specifically, the function $\gamma_K$, which was originally used to express the dipole term in eq.~(\ref{eq:SADintrogamma}), only appears now as part of the full cusp anomalous dimension  $\bm{\Gamma}_p^{\rm{cusp}}$, along with its quartic counterpart $g_R$. The way the full $\bm{\Gamma}_p^{\rm{cusp}}$ gets restored is explained in appendix~\ref{sec:cuspappendix}. 

Note that, in line with the general expectation, all the contributions that survive in the planar limit -- the terms in the first line of eq.~(\ref{eq:BN4sig+}) and the first term in the first line of eq.~(\ref{eq:BN4sig-2}) --  involve just one or two partons, while all those involving three or four partons are non-planar. 
For most terms the behaviour in the large-$N_c$ limit is already manifest in the above equations owing to the (nested) commutator structure, which is inherently non-planar, or the behaviour of the quartic Casimir contributions given in eq.~(\ref{eq:dAAlargeNc}). 
There are a couple of terms for which a closer examination is required: the first of these is the 
third line in eq.~\eqref{eq:BN4sig+}, where in the adjoint representation one may use eq.~\eqref{eq:Gfin} to obtain a manifestly non-planar expression (while the fundamental representation contribution is automatically subleading in the large-$N_c$ limit). 
The second is the first term in the first line of eq.~(\ref{eq:BN4sig-2}), which contains planar as well as non-planar contributions, as one may verify using  eq.~(\ref{gammalinA-}) to express the linear terms.

For pure Yang-Mills, or SYM,  where only the adjoint representation is relevant, one may substitute eq.~\eqref{eq:Gfin} and~(\ref{gammalinA-}) into the above equations to obtain more explicit expressions the soft anomalous dimension in the high-energy limit, separated by signature, including all powers of $L$. The signature-even part is
\beq
\label{eq:BN4sig+YM}\begin{aligned}
\bm{\Gamma}^{(+,4)}_{ij\rightarrow ij,\, {\rm (S)YM}}(L)\mtree&= \Bigg\{L\, \bm{\Gamma}_t^{\rm{cusp},(4)} + \log\frac{-t}{\lambda^2} \left(\Gamma_i^{\rm{cusp},(4)}+ \Gamma_j^{\rm{cusp},(4)}\right) + 2\gamma_i^{(4)}+ 2\gamma_j^{(4)} \\
&\hspace{-1.0cm} + f^{(4, A)} \bigg(\Big[\tsu,[\tsu,\tts]\Big] + \frac{C_A^3}{2} - 6\frac{d_{AR_i}}{N_{R_i} C_i} -6\frac{d_{AR_j}}{N_{R_j} C_j}\bigg) \\
&\hspace{-1.0cm} 
+ \left({\cal G}^{(4)}_{A}(L)-\frac{g^{(4)}_{A}}{6} L \right) \bigg(2\left(\frac{d_{AA}}{N_A} -\frac{C_A^4}{24}\right)-\frac{1}{2}\tts[(\tsu)^2,\tts] \\
&\hspace{-1.0cm}  +\frac{3}{2}[\tsu,\tts]\tts\tsu+\frac{C_A}{2} \Big[\tsu,[\tsu,\tts]\Big]\bigg) \\
&\hspace{-1.0cm} +2{\cal F}^{(+,4)}_{A} (L) \Big[\tsu, [\tsu,\tts]\Big] 
+{\cal H}^{(+,4)}(L)\bigg(-\frac{1}{2}C_A\Big[\tsu,[\tsu,\tts]\Big] \\
&\hspace{-1.0cm}+C_A \tsu [\tsu,\tts] -\frac{1}{6}\tts[\tsu,\tts]\tsu\bigg)\Bigg\}\mtree,
\end{aligned}
\eeq
and the signature-odd part is\beq
\label{eq:BN4sig-YM}\begin{aligned}
\bm{\Gamma}^{(-,4)}_{ij\rightarrow ij, \,{\rm (S)YM}}(L)\mtree&=\Bigg\{\frac{i\pi}{C_A} \Gamma_A^{\rm{cusp}, (4)}\tsu-{\cal F}^{(-,4)}_{A} (L)\Big[\tts, [\tts,\tsu]\Big]
\\&\hspace{-2.0cm} 
+ 2\,{i \pi} g_A^{(4)}\left(\frac{d_{AR_i}}{C_iN_{R_i}}+\frac{d_{AR_j}}{C_jN_{R_j}}-\frac{2d_{AA}}{C_AN_A}- \frac{C_A^3}{16}\right)
\tsu\\&\hspace{-2.0cm}-  \frac{i\pi g_A^{(4)}}{16}
\bigg(   3\Big[\tts,\big[\tts, [\tts,\tsu]\big]\Big]+\Big[\tts,[\tts,\tsu]\Big]\tts-3 \tts[ \tts,\tsu]\tts\bigg)\\&\hspace{-2.0cm}+{\cal H}^{(-,4)}_1(L)\bigg(-\frac{1}{2}\bigg[\tsu,\Big[\tsu,\big[\tsu,\tts\big]\Big]\bigg]+\frac{1}{8}\bigg[\tts,\Big[\tts,\big[\tts,\tsu\big]\Big]\bigg]\bigg)\\&\hspace{-2.0cm}+\frac{1}{4}\tilde{{\cal H}}^{(-,4)}_1(L)\bigg[\tts,\Big[\tts,\big[\tts,\tsu\big]\Big]\bigg]\Bigg\}\mtree.
\end{aligned}
\eeq
These expressions make manifest the fact that the planar contributions are all captured by the $\Gamma^{\rm{cusp}}$ and collinear anomalous dimension terms.
 
\paragraph{The singularities of the Regge trajectory and the cusp anomalous dimension.}
We have seen that the connection~\cite{Korchemskaya:1994qp,Korchemskaya:1996je} between the infrared singularities of the gluon Regge trajectory and the integral of the cusp anomalous dimension~\eqref{affirmed_eqality_of_sing_part_of_tilde_alpha_g_and_K}, namely 
\begin{equation}
\label{affirmed_eqality_of_sing_part_of_tilde_alpha_g_and_K2}
\tilde{\al}_g(t)  =  K +\mathcal{O}(\eps^0)\,,
\end{equation}
 where 
\begin{align}
    \label{eq:Kdef2}
     K(\al_s(\mu^2))&\equiv -\frac{1}{4}\int_0^{\mu^2}\frac{d\lambda^2}{\lambda^2}\gamma_K(\al_s(\lambda^2)) =
     \sum_{n = 0}^{\infty} 
\left( \frac{\al_s(\mu^2)}{\pi} \right)^n K^{(n)}
=
     \frac{1}{2\epsilon}
     \frac{\alpha_s(\mu^2)}{\pi}+\ldots ,
\end{align}
extends to three loops, despite the presence of a Regge cut contribution at this order, i.e.~${\cal O}(\alpha_s^3L^1)$. For clarity we recall that this is contingent on defining the trajectory $\tilde\alpha_g$ in the Regge-cut scheme, which we defined by absorbing all planar contributions generated at two and three loops into the Regge-pole term. Specifically, the trajectory $\tilde{\alpha}_g$ was related to the one in the MRS scheme by eq.~(\ref{eq:reggeTilde}), and the resulting cut-scheme subtracted trajectory is finite (see eq.~(\ref{hat_tilde_alpha_g})):
\begin{align}
\hat{\tilde{\al}}_g(t)  \equiv  \tilde{\al}_g(t)  -  K = {\cal O}(\epsilon^0)\,. \,
\end{align}

From the perspective of the soft anomalous dimension this entails a remarkably simple structure through three loops, namely its even-signature part takes the form
\begin{align}
\label{Gammaplus3loop}
\begin{split}
\mathbf{\Gamma}^{(+)}_{ij\to ij}\left(\alpha_s,L,\frac{-t}{\lambda^2}\right) = \frac12 \gamma_{K}(\alpha_s) L\tts  &\,+\, \Gamma_i\left(\alpha_s,\frac{-t}{\lambda^2}\right)
\,+\,\Gamma_j\left(\alpha_s,\frac{-t}{\lambda^2}\right) 
\\&\,\,\, +\, \dd^{(+,3,0)} \left(\frac{\alpha_s}{\pi}\right)^3 + {\cal O}(\alpha_s^4)\,,
\end{split}
\end{align}
where, crucially, we used the fact that\footnote{Note that its signature-odd counterpart, $\dd^{(-,3,1)}$ is non-vanishing, see appendix~\ref{sec:SAD3loopa} for details} $\dd^{(+,3,1)} =0$~\cite{Almelid:2015jia,Caron-Huot:2017fxr}. We thus see that the only term in the (signature-even) soft anomalous dimension which is linear in the high-energy logarithms $L$ is the one proportional to the cusp anomalous dimension. Consequently, the exponentiation of the singularities via eq.~(\ref{eq:Zfactor}) directly determines the singularities of the exponent of $(s/(-t))$, which is precisely the singular part of the gluon Regge trajectory. 

This naturally generalises to four loops, where quartic Casimir contributions become relevant, as displayed in the definition of the cusp anomalous dimension eq.~\eqref{cuspAD_Sec_anom_dim}.  In order to write an equation such as eq.~\eqref{affirmed_eqality_of_sing_part_of_tilde_alpha_g_and_K2} valid  through four loops (and beyond), let us define
\begin{equation}
\label{eq:Kcuspdef}
K_{\rm cusp}(\alpha_s(\mu^2))\equiv -\frac{1}{2}\int_0^{\mu^2}\frac{d\lambda^2}{\lambda^2}
\Gamma^{\rm{cusp}}_A(\alpha_s(\lambda^2))\,.
\end{equation}
Generalising eq.~\eqref{affirmed_eqality_of_sing_part_of_tilde_alpha_g_and_K2}, we have
 \beq
 \label{eq:AlphaVsGamma}
C_A\tilde{\alpha}_g (t)=K_{\rm cusp} +{\cal O}(\epsilon^0)
\eeq 
through four loops, now including the quartic Casimir contributions.
The Regge-pole exponential in eq.~\eqref{Schemes_Cut} can then be expressed as $\exp(K_{\rm cusp} L)$. 

\paragraph{Linear terms in the soft anomalous dimension at four loops.}
Let us now analyse the implications of this relation from the perspective of the soft anomalous dimension. 
We do that by directly comparing the two exponentiation pictures, that of the singularities via eq.~(\ref{eq:Zfactor}) on the one hand and that of the high-energy logarithms as a Regge pole, on the other. The exponents in the two pictures take the form:
\begin{equation}
\label{exp_sing_lin}
 -\frac{1}{2}\int_0^{\mu^2}\frac{d\lambda^2}{\lambda^2}
\mathbf{\Gamma}^{(+)}_{ij\to ij}\left(\alpha_s,L,\frac{-t}{\lambda^2}\right)
\quad\longleftrightarrow\quad C_A\tilde{\alpha}_g (t) L\,.
\end{equation}
 For the two to agree 
 for the terms that are simultaneously ${\cal O}(1/\epsilon)$ and ${\cal O}(L^1)$, one requires, just as in eq.~(\ref{Gammaplus3loop}) at three loops, that the linear term in $L$ within $\mathbf{\Gamma}^{(+)}_{ij\to ij}$ would be given precisely by $L \bm{\Gamma}_t^{\rm{cusp}}$, where $\bm{\Gamma}_t^{\rm{cusp}}$ is defined in eq.~(\ref{Gamma_cusp_p}). 
Upon acting on the tree amplitude, the $t$-channel operators produce Casimirs in the adjoint representation according to eq.~(\ref{t_op_on_Mtree}), and one recovers the cusp anomalous dimension in the adjoint representation as in eq.~(\ref{cuspttree}).
In this way the singularities of the gluon Regge trajectory satisfy eq.~(\ref{eq:AlphaVsGamma}). 

We thus conclude that the natural generalisation of the relation of eq.~(\ref{eq:AlphaVsGamma}) between the gluon Regge trajectory and the cusp anomalous dimension amounts to the requirement that the terms linear in $L$ within the signature-even part of the soft anomalous dimension would simply be $L \bm{\Gamma}_t^{\rm{cusp}}$. This conjecture can also be formulated as
\begin{align}
\label{Soft_cusp_relatioN_der}
      \left.  \frac{d}{dL} \mathbf{\Gamma}^{(+)}_{ij\to ij}\left(\alpha_s(\mu^2),\frac{-t}{\mu^2}\right) 
      \right\vert_{L=0} {\cal M}^{\rm tree}_{ij\to ij}
     \, =\, &\,\Gamma^{\rm{cusp}}_A(\alpha_s(-t))\, {\cal M}^{\rm tree}_{ij\to ij}\,,
\end{align}
which of course holds through three loops using eq.~(\ref{Gammaplus3loop}) and (\ref{Gamma_cusp_p}), where only the quadratic Casimir term $\T_t^2$ is present in $\bm{\Gamma}_t^{\rm{cusp}}$. In contrast, at four loops also the $\bm{\mathcal{D}}^R_{tttt}$ becomes relevant.

With this in mind, let us examine the general structure of the signature-even soft anomalous dimension at four loops in eq.~(\ref{eq:BN4sig+}). The expected $L\bm{\Gamma}_t^{\rm{cusp},(4)}$ term is indeed there. So, for our conjecture to hold, all other terms which depend on $L$ must not contain any further linear contribution. Differentiating eq.~(\ref{eq:BN4sig+}) and suppressing higher logarithms one finds:
\begin{align}
\label{eq:BN4sig+_dL3p}
\begin{split}
\frac{d\bm{\Gamma}^{(+,4)}_{ij\to ij,\text{Regge}}}{dL}\bigg|_{L=0}\!\! \mtree
&= \Bigg\{\bm{\Gamma}_t^{\rm{cusp}, (4)} \, + \,2\left(\sum_R  {\cal F}^{(+,4,1)}_R \right) \Big[\tsu, [\tsu,\tts]\Big]\\
&  \hskip-30pt+ 2\sum_R \left({\cal G}^{(4,1)}_R  -\frac{g^{(4)}_R}{6} \right) \bigg(\bm{\mathcal{D}}^R_{tttt}+\bm{\mathcal{D}}^R_{ssss}+\bm{\mathcal{D}}^R_{uuuu}-2\left(\frac{d_{RR_i}}{N_{R_i}}+\frac{d_{RR_j}}{N_{R_j}}\right)\bigg)\\
& \hskip-30pt
+{\cal H}^{(+,4,1)}\bigg(-\frac{1}{2}C_A\Big[\tsu,[\tsu,\tts]\Big]+C_A \tsu [\tsu,\tts] \\&\quad\mbox{}\hspace{5cm}-\frac{1}{6}\tts[(\tsu)^2,\tts]\bigg)\Bigg\}\mtree\,,
\end{split}
\end{align}
which satisfies the conjectured relation in eq.~(\ref{Soft_cusp_relatioN_der}) subject to the following constraints 
\begin{align}
\label{eq:ConjConstr}
{\cal F}^{(+,4,1)}_R=0,\hspace{1cm}{\cal G}^{(4,1)}_R& =\frac{g^{(4)}_R}{6},\hspace{1cm}
{\cal H}^{(+,4,1)}_1=0\, .
\end{align}
The coefficients $g_R^{(4)}$ are known in QCD~\cite{Boels:2017ftb,Boels:2017skl,Moch:2017uml,Grozin:2017css,Henn:2019swt,Huber:2019fxe,vonManteuffel:2020vjv,Agarwal:2021zft}; see  eq.~(\ref{g_R_values}).
We stress that the constraints in eq.~(\ref{eq:ConjConstr}), 
which concern the uncharted territory of N$^3$LLs, have a very different status as compared to those of 
eqs.~(\ref{const_NLL_minus}), (\ref{const_NLL_plus}) and (\ref{const_NNLL_plus}): while the latter are based on explicit calculations in the high-energy limit, the former relies on a conjectured generalisation of the relation between the gluon Regge trajectory and the cusp anomalous dimension.  
One may note the intriguing similarity between the constraint on ${\cal G}^{(4)}_R$ in eq.~(\ref{eq:ConjConstr}) and the collinear constraint~\cite{Becher:2019avh} on the same function in eq.~(\ref{eq:Gcol}). We stress that these are different kinematic limits. Whereas in the high-energy limit the function ${\cal G}^{(4)}_A$ does have a double logarithmic contribution -- see eq.~(\ref{eq:cstF+4}) -- in the collinear limit eq.~(\ref{eq:Gcol}) forbids any non-linear dependence on the relevant logarithm.

\paragraph{Summary: Regge-limit constraints on  $\mathbf{\Gamma}_n^{(4)}$. } We derived the soft anomalous dimension in the high-energy limit and used it to constrain the kinematic functions parametrising this quantity in general kinematics as proposed in ref.~\cite{Becher:2019avh}. The computed four-loop result, taking into account NLLs of both even and odd signature, along with the 
newly-computed NNLL of even signature, appears in eq.~\eqref{eq:softADdef4loopsymD6}. In turn, upon taking the general-kinematics parametrisation and specialising it to $2\to 2$ kinematics in the Regge limit, we obtained eqs.~(\ref{eq:gammaNNLL+}) and~(\ref{eq:gammaNNLL-}). Having chosen a common basis of colour operators for both the computed result and the parametrised one, the values of the expansion coefficients of the unknown kinematic functions in powers of $L$ can be directly deduced, and are summarised in eqs.~(\ref{const_NLL_minus}), (\ref{const_NLL_plus}) and (\ref{const_NNLL_plus}). 

In addition, analysing the connection between the gluon Regge trajectory and the cusp anomalous dimension originally proposed in ref.~\cite{Korchemskaya:1994qp,Korchemskaya:1996je}, we conjectured that the linear term in the soft anomalous dimension in the Regge limit is given precisely by $\mathbf{\Gamma}_t^{\rm cusp} L$ such that eq.~(\ref{Soft_cusp_relatioN_der}) holds. This directly implies an additional set of constraints on the N$^3$LL signature-even contributions to the soft anomalous dimension according to eq.~(\ref{eq:ConjConstr}).

\begin{table}[ht]
\begin{center}
\begin{tabular}{ ||l|l|l|l||l|l|l|l|| } 
\hline
\multicolumn{4}{||c||}{Signature even}  
& \multicolumn{4}{c||}{Signature odd}\\
 \hline
\qquad& $L^3$\hspace*{30pt}& $L^2$\hspace*{30pt}&$L^1$  (conj.)&\qquad &  $L^3$\hspace*{30pt}&  $L^2$ \hspace*{30pt}&$L^1$\hspace*{30pt}\\
\hline
  ${\cal F}^{(+,4)}_A$ & 0 & $-\frac{C_A}{8}  \zeta_2\zeta_3$ & 0
  &${\cal F}^{(-,4)}_A$ & $ i \pi  \frac{C_A}{24}\zeta_3$& ? &?  \\ 
   ${\cal F}^{(+,4)}_F$ & 0 & 0 &0 & ${\cal F}^{(-,4)}_F$ & 0 & ?& ?\\ 
   \hline
   ${\cal G}^{(+,4)}_A$ & 0 & $\frac{1}{2} \zeta_2\zeta_3$ & $\frac16 g_A^{(4)}$
   & &  & & \\  
${\cal G}^{(+,4)}_F$ & 0 & 0 &$\frac16 g_F^{(4)}$&  &   &  & \\ 
       \hline
   ${\cal H}^{(+,4)}_1$  & 0 &0 &0&${\cal H}^{(-,4)}_1$ & 0 &? &?\\
     &   &  & &$\tilde{{\cal H}}^{(-,4)}_1$ & 0 & ?&?\\
   \hline
\end{tabular}
 \caption{Constraints on the high-energy limit of the kinematic functions  entering the soft anomalous dimension at four loops, separated by signature. Note that ${\cal G}$ only has a signature-even component, and ${\cal H}_1^{(4)}$ is purely gluonic. All constraints at order $L^3$ and $L^2$ in this table are based on explicit computations in the high-energy limit, while those for order~$L^1$ are based on the conjectured generalisation of the relation between cusp singularities and the Regge pole to four loops.  The coefficients $g_R^{(4)}$ are known in QCD~\cite{Boels:2017ftb,Boels:2017skl,Moch:2017uml,Grozin:2017css,Henn:2019swt,Huber:2019fxe,vonManteuffel:2020vjv,Agarwal:2021zft} and are quoted in eq.~(\ref{g_R_values}).} \label{tab:reggeconstraints}
 \end{center}
\end{table}
The full set of constraints on the four-loop kinematic functions is summarised in Table~\ref{tab:reggeconstraints}.
Here, the left half of the table summarises the constraints on signature-even (real) functions, while the right half the signature-odd (imaginary) ones. While the function ${\cal G}_R$ multiplying the quartic four-line term is by construction  signature-even, the two other kinematic functions ${\cal F}_R$ and ${\cal H}_1$ have both even and odd components and their decomposition 
is given in eqs.~(\ref{eq:Fsigeven}) and (\ref{Hplusminuesdef}), respectively. We note that our current knowledge of the signature-even contributions is far greater than that of the odd. In the table we represented unknown expansion coefficients by question marks.

As a final note we emphasise that the above constraints are fully consistent with the result by Vladimirov~\cite{Vladimirov:2017ksc}, that only colour operators consisting of an even number of generators can appear in the soft anomalous dimension. 
This implies that the functions multiplying the five generator term ${\cal H}_1$ and ${\cal H}_2$ in the soft anomalous dimension in eq.~\eqref{eq:SADintrogamma} vanish identically. This is in line with the last row in Table~\ref{tab:reggeconstraints}, as well as the collinear-limit constraints on these functions in ref.~\cite{Becher:2019avh}.

\section{Conclusion}

In this work we take a step forward in the understanding of $2\to 2$ gauge-theory amplitudes in the high-energy limit, by studying the tower of NNLL in the signature-odd (real) amplitude and computing these explicitly through four loops. 
This tower of corrections is particularly interesting for the analysis of the Regge limit, because amplitudes at this logarithmic accuracy develop a rich structure, featuring both a Regge pole and a Regge cut.
Furthermore, taking the high-energy limit gives us access to properties of four-loop amplitudes, which are beyond the reach of perturbative calculations with state-of-the-art techniques. Chief among these is the long-distance singularity structure in fixed-angle scattering, for which the high-energy limit is highly constraining. 

In order to compute amplitudes in the Regge limit, we employ the method described in refs.~\cite{Caron-Huot:2013fea,Caron-Huot:2017fxr}. In essence this approach allows us to compute transition amplitudes between the projectile and the target at widely separated rapidities, each described by a state consisting of a given numbers of Reggeons.
The Balitsky-JIMWLK Hamiltonian is then used to evolve the Reggeon states to the same rapidity. At NNLL accuracy this involves, beyond the single Reggeon state, also triple-Reggeon states and mixing amongst these. The sum of all transition amplitudes defines a {\textit{reduced amplitude}}, which is related to the full amplitude by simple multiplicative factors, eq.~(\ref{Mreduced}).

We classify the transition amplitudes entering the NNLL of the reduced amplitude to all loop orders, in eq.~(\ref{eq:mhatNNLL}). These fall into two distinct categories, one being a purely {\textit{Single Reggeon State}} (SRS) transition and the other including four transitions involving {\textit{Multiple Reggeon States}} (MRS). The former features a single Reggeon in both the projectile and the target, undergoing trivial rapidity evolution as a single Reggeon across the entire rapidity interval. 
The latter include all transitions in which a triple-Reggeon state is generated at any stage during the evolution, be it 
at the projectile or the target ends, or  during the course of rapidity evolution. 
Specifically, the aforementioned four are: $3\to3$ transitions, $1\to 3$ or $3\to 1$ and $1\to1$ which are mediated by a triple-Reggeon state in the evolution. We show that at NNLL accuracy MRS transition amplitudes can be computed to any perturbative order by iterating the \emph{leading-order} Balitsky-JIMWLK Hamiltonian. Thus, the MRS are universal quantities in any gauge theory, which do not depend on the matter content. 

We computed the reduced amplitudes through four loops, given in eqs.~(\ref{eq:M20}), (\ref{eq:M31}) and (\ref{eq:M42}), providing a detailed derivation of results presented in ref.~\cite{Falcioni:2020lvv}. In particular, we developed a new method to calculate the colour factor of the amplitude, when the target and the projectile belong to general representations of the gauge group. This allowed us to derive new colour identities and obtain expressions of the reduced amplitudes in an operator form, which is suitable to investigate universal features of both the infrared and of the high-energy factorisation. We found that only the $3\to3$ transitions feature non-trivial colour structure, where different colour components mix during evolution. All the other MRS transitions are proportional to the colour octet exchange to all perturbative orders. We observed that $3\to1$ and $1\to3$ transitions at three and at four loops, in eqs.~(\ref{eq:resJ1H31I3}) and (\ref{eq:resJ1H31H33I3simb}) respectively, cancel exactly against corresponding terms (which involve quartic Casimirs associated with the representations of the projectile and the target) in the $3\to3$ exchanges of eqs.~(\ref{eq:resJ3H33I3}) 
and~(\ref{eq:reshat33hat33simb}). We conjecture that such a mechanism is in place to all perturbative orders and that it completely removes all contributions to the amplitude from $3\to 1$ and $1\to 3$ transitions. As a result, only the $1\to1$ transition generates mixing between states with one and three Reggeons, as we check explicitly at four loops. There, a further cancellation takes place: the $1\to1$ contribution in eq.~(\ref{eq:resJ1H31H13I1simb}) cancels the planar terms of $3\to3$ transitions in eq.~(\ref{eq:reshat33hat33simb}), to all orders in $\epsilon$. This renders the reduced amplitude at four loops, eq.~(\ref{eq:M42simb}), manifestly non-planar. 

The complete cancellation of the contributions emerging from mixing between single and triple Reggeon states, against corresponding terms associated with quartic Casimirs in the $3\to3$ evolution, is highly suggestive of a general pattern, extending to all orders in this tower of logarithms. As we have seen, it leads to a partial cancellation of planar contributions in the reduced amplitude at three loops, and a complete cancellation of such at four loops. Our expectation is that the reduced amplitude will be non-planar at any order beyond four loops. The only planar contributions in the reduced amplitude then occur at two and three loops, before the full set of single-triple Reggeon transitions opens~up.

The non-planar nature of the total contribution to the reduced amplitude from multiple Reggeon states at four loops (and likely beyond) points to a simple relation between these quantities and Regge cuts, which are known to arise only from non-planar diagrams~\cite{Mandelstam:1963cw,Eden:1966dnq,Collins:1977jy}. However, the separation between single-Reggeon state (SRS) and multiple-Reggeon state (MRS) contributions to the amplitude as defined in our calculation, is not in one-to-one correspondence with the separation between the Regge pole and the Regge cut contributions. 
This is already clear at two and at three loops, where MRS do contain planar contributions.
Hence, the MRS give rise to both pole and cut contributions, while the SRS contributes exclusively to the Regge-pole exchange. 

In order to elucidate the separation between Regge cut and pole, we rely again on the structure of the reduced amplitudes in the planar limit. We find that MRS contributions that are leading in $N_c$ appear only in the colour octet component and are independent of the process, both at two loops, eq.~(\ref{eq:M20planar}), and at three loops, eq.~(\ref{eq:M31planar}).
Following this analysis of colour factors, we show that both the SRS contribution and the planar terms of the MRS contribution may be described by Regge-pole factorisation, while all remaining non-planar MRS terms define a cut contribution, as done in eq.~(\ref{eq:Mdecomp_cut_scheme}). We name this separation of the amplitude the {\textit{Regge-cut scheme}}. It departs from the one adopted in ref.~\cite{Caron-Huot:2017fxr}, dubbed {\textit{MRS scheme}}, where the SRS contribution alone is factorised as a Regge pole. 
The change of scheme modifies the definition of the impact factors and of the Regge trajectory, which determine the Regge pole contribution, by the planar part of the MRS contribution. 
Notably, the two-loop impact factors and the three-loop Regge trajectory completely characterise the Regge-pole contribution to the NNLL to all orders. At four loops and beyond there is no parameter which could allow one to shuffle planar MRS contributions to the Regge pole. Therefore, starting at four loops the MRS transition amplitudes must contribute exclusively to the cut and must be entirely non-planar. This is indeed what we find in our four-loop calculation, eqs.~(\ref{eq:M42simb}) and (\ref{eq:M42}).

In section~\ref{sec:pole_cut_explicit_rep} we construct the complete amplitudes and then we distinguish pole and cut contributions to the amplitude according to the Regge-cut scheme (\ref{eq:Mdecomp_cut_scheme}).
At two loops, we provide in eq.~(\ref{M20cut}) the definition of the Regge cut coefficient $\mathcal{M}^{(-,2,0),\,\text{cut}}_{ij\to ij}$ in an operator form, which is valid to all orders in $\epsilon$ for every colour component in any process. This coincides with the MRS contribution of eq.~(\ref{eq:M20}), with its planar limit, eq.~(\ref{eq:M20planar}), subtracted. We find that, in the octet component, $\mathcal{M}^{(-,2,0),\,\text{cut}}_{ij\to ij}$ agrees with the Regge-pole factorisation breaking term $R^{(2),0,[8]}_{ij}$, defined in refs.~\cite{DelDuca:2013ara,DelDuca:2014cya} on the basis of infrared factorisation.
We determine the corresponding quark and gluon impact factors in this scheme, eq.~(\ref{eq:Ctilde}), by giving their relation with the results in the MRS scheme~\cite{Caron-Huot:2017fxr}.
Remarkably, it is possible to move into the impact factors further terms that appear in $\mathcal{M}^{(-,2,0),\,\text{cut}}_{ij\to ij}$ and are subleading in $N_c$, as done in eq.~(\ref{FLimpactFactor}). This follows from the structure of the non-planar terms in the reduced amplitude at two loops, given in eq.~(\ref{eq:TsuNc2}). By following this redefinition, we obtain a new cut, $\mathcal{M}^{(-,2,0),\,\text{FL-cut}}_{ij\to ij}$, defined in eq.~(\ref{FLM20cut}), which agrees with the two-loop Regge cut $A_{\text{eik}}\,C_{ij}^C$ computed by Fadin and Lipatov~\cite{Fadin:2016wso,Fadin:2017nka}.

At three loops, the Regge cut $\mathcal{M}^{(-,3,1),\,\text{cut}}_{ij\to ij}$ takes the form of eq.~(\ref{M31cut}). It includes a term proportional to $\mathcal{M}^{(-,2,0),\,\text{cut}}_{ij\to ij}$ plus the reduced amplitude at three loops $\hat{\mathcal{M}}^{(-,3,1)}_{ij\to ij}$, eq.~(\ref{eq:M31}), with its planar part subtracted. The latter is assigned to the Regge pole and thus it enters the Regge trajectory at three loops. Eq.~(\ref{eq:reggeTilde}) provides the relation between the three-loop trajectory in the Regge-cut scheme and in the MRS scheme of ref.~\cite{Caron-Huot:2017fxr}. In that work, the three-loop trajectory was determined in the MRS scheme for $\mathcal{N}=4$ SYM. There, it was also pointed out that the MRS scheme breaks a well-known relation~\cite{Korchemskaya:1994qp,Korchemskaya:1996je} between the infrared singularities of the gluon Regge trajectory and $K(\alpha_s)$ of eq.~(\ref{eq:Kdef}), the integral over the  lightlike cusp anomalous dimension. This relation holds for the two-loop Regge trajectory, but it is violated at three loops in the MRS scheme. In contrast, we find that the three-loop Regge trajectory in the Regge-cut scheme, $\tilde{\alpha}_g^{(3)}$, features precisely the singularities predicted by the cusp anomalous dimension, as shown in eq.~(\ref{eq:alphaTilde3}). 

We compute also the finite contribution to $\tilde{\alpha}_g^{(3)}$ in $\mathcal{N}=4$ SYM in full colour. Notably, we find that the latter agrees with the known result in the planar theory~\cite{Drummond:2007aua,Naculich:2007ub}, without any non-planar correction. In other words, the trajectory features a maximally non-Abelian colour factor, which is in line with the expected eikonal origin for this 
quantity~\cite{Korchemskaya:1994qp,Korchemskaya:1996je,Falcioni:2019nxk}.

Our three-loop analysis suggests that the Regge-cut scheme captures the analytic structure of high-energy amplitudes. As a confirmation, we find that, in this scheme,
the Regge cut agrees with the function $R^{(3),1,[8]}_{ij\to ij}$ of refs.~\cite{DelDuca:2013ara,DelDuca:2014cya}, which contains the factorisation-breaking singularities in the octet component. However, different choices are also possible. In particular, as mentioned above, using eq.~(\ref{eq:TsuNc2}) we identify a specific set of non-planar terms in the reduced two-loop amplitude that are consistent with Regge-pole factorisation. Absorbing these into the Regge-pole term at two loops (eq.~(\ref{FLM20cut})) modifies the contribution of the Regge cut at three loops of eq.~(\ref{M31cut}), only by replacing $\mathcal{M}^{(-,2,0),\,\text{cut}}_{ij\to ij}$ with the expression of the cut in the new scheme, $\mathcal{M}^{(-,2,0),\,\text{FL-cut}}_{ij\to ij}$. We verify that the three-loop cut defined in this way coincides with the cut contribution, $-A_{\text{eik}}C_{ij}^C\,(C_R+C_3)$, in refs.~\cite{Fadin:2016wso,Fadin:2017nka}. Notably, the three-loop Regge trajectory is not affected by colour subleading terms and, even with the FL definition of the cut, it maintains its relation with the lightlike cusp anomalous dimension, as well as its maximally non-Abelian colour factor. Therefore, our new analysis of the colour factors in the reduced amplitudes, allows us to find the precise relation between the computational scheme introduced in refs.~\cite{Caron-Huot:2013fea,Caron-Huot:2017fxr} and the study of factorisation breaking and of the Regge cut, performed respectively in refs.~\cite{DelDuca:2013ara,DelDuca:2013dsa,DelDuca:2014cya} and \cite{Fadin:2016wso,Fadin:2017nka}, finding complete agreement.

Our expression for the Regge cut at four loops is given in eq.~(\ref{eq:M42cut}), in terms of the reduced amplitude at four loops and the cut contributions at two and three loops. Since the former, eq.~(\ref{eq:M42}), is non-planar by direct computation, and the latter two terms are defined in the Regge-cut scheme to be non-planar by construction, we find the four-loop cut contribution to the amplitude $\mathcal{M}^{(-,4,2),\text{cut}}_{ij\to ij}$ is non-planar as a whole. Furthermore, we show in eq.~(\ref{eq:McutAllOrder}) that by the same mechanism, in this scheme the non-planar nature of the reduced amplitude ensures that the cut remains non-planar to all loop orders.

In sections~\ref{sec:Infrared} and \ref{sec:SAD_Regge} we proceed with the investigation of infrared factorisation at four loops, employing our explicit NNLL calculation as an input. 
The comparison between the exponentiation of infrared singularities and that of high-energy logarithms is useful in several ways. First, it is a highly non-trivial check of the results. Second, it provides a rich source of constraints on the yet-unknown soft anomalous dimension at four loops (see below). Third, it allows us to extract the hard function containing finite terms in the amplitude through four loops, both in QCD and in $\mathcal{N}=4$ super Yang-Mills (SYM), finding an intriguing relation between the hard function and the finite parts of the gluon Regge trajectory, eq.~(\ref{eq:hardFunctionTheoryDependence}). The planar terms in the hard function in SYM agree with the predicted large-$N_c$ limit~\cite{Bern:2005iz,Drummond:2007aua}.

We study the soft anomalous dimension through four loops. In the high-energy limit, we separate the contributions of the dipole formula \cite{Gardi:2009qi,Gardi:2009zv,Becher:2009cu,Becher:2009qa}, from a general remainder~$\bf\Delta$ that starts at three loops, expanding both in powers of the signature-even logarithm~$L$, defined in eq.~(\ref{eq:siglog}). While dipole contributions are at most linear in $L$, the remainder contains higher powers of the logarithm, for instance its imaginary part contains terms~$L^3$ at four 
loops~\cite{Caron-Huot:2013fea,Caron-Huot:2017zfo}. Notably, the real part of the remainder at three loops ${\bf\Delta}^{(+,3)}$  does not depend on $L$~\cite{Almelid:2015jia}. In particular, since it lacks linear terms in $L$, it does not contribute to the tower of NNLLs in the soft anomalous dimension to that order~\cite{Caron-Huot:2017fxr}. Here we compute the real part of the remainder to four loops, ${\bf\Delta}^{(+,4)}$, through NNLL, i.e. ${\cal O}(\alpha_s^4 L^2)$, finding the first non-vanishing contribution to the NNLL tower, eq.~(\ref{eq:delta42}).
This quantity is manifestly non-planar: it is written in terms of commutators of the channel operators and the combination of Casimir invariants $\frac{d_{AA}}{N_A}-\frac{C_A^4}{24}$, which is subleading in the large-$N_c$ limit. This is a strong check on our calculation, because in the planar limit only diagrams connecting up to two legs can contribute to the soft anomalous dimension. 

We characterise our result for the soft anomalous dimension in eq.~(\ref{eq:delta42}) further, by comparing it with the general parametrisation for the four-loop soft anomalous dimension in general kinematics~\cite{Becher:2019avh}, which consists of all connected~\cite{Gardi:2013ita} colour structures that may arise at this order, each multiplying a yet-unknown kinematic function. 
We compute the high-energy limit of this parametrisation of the soft anomalous dimension through NNLLs, both in the real part, eq.~(\ref{eq:gammaNNLL+}), and in the imaginary part, eq.~(\ref{eq:gammaNNLL-}). Focusing on the terms in the real part of the anomalous dimension, we find only three contributions. Two of them involve colour quadrupole correlations, featuring four generators, one on each of the four lines. Of these one colour structure is of the same form that appears at three loops~\cite{Almelid:2015jia,Almelid:2017qju} multiplied by the kinematic function $\mathcal{F}_A^{(+,4)}$, while the second 
involves a quartic Casimir type structure (a symmetric trace of four adjoint generators) multiplied by~$\mathcal{G}_A^{(+,4)}$. 
In the Regge-limit both are expressed in terms of nested commutators of channel colour operators, where the second also features the combination $\frac{d_{AA}}{N_A}-\frac{C_A^4}{24}$.
The final potential contribution to the soft anomalous dimension, featuring the unknown function~$\mathcal{H}_1^{(+,4)}$, generates correlations among four lines using five colour generators. Matching the general parametrisation with our result of the anomalous dimension provides non-zero constraints on the high-energy limit of the functions $\mathcal{F}_A^{(+,4)}$, in eq.~(\ref{eq:cstF+4}), and $\mathcal{G}_A^{(+,4)}$, in eq.~(\ref{eq:cstG+4}). Interestingly, the function $\mathcal{H}^{(+,4)}_1$ must vanish to this logarithmic accuracy. This is consistent with the result of ref.~\cite{Vladimirov:2017ksc}, which shows that the correlation of an odd number of colour operators is always prohibited in the soft anomalous dimension. We determine all constraints on the parametrisation in ref.~\cite{Becher:2019avh} that can be derived from the available information on the Regge limit and we summarise our findings in Table~\ref{tab:reggeconstraints}.

We expect that the interplay between high-energy and infrared factorisation will provide further insight into gauge-theory dynamics. This conclusion is already suggested by arguments about the gluon Regge trajectory. We have pointed out that this quantity -- both its singular and its finite parts -- is expected to be associated to the anomalous dimension of Wilson-line geometries~\cite{Korchemskaya:1994qp,Korchemskaya:1996je,Falcioni:2019nxk}. Here we verified, up to three loops, the correspondence between the infrared singularities of the Regge trajectory and the terms proportional to the quadratic Casimir in the lightlike cusp anomalous dimension. We conjecture that this relation generalises to four loops and beyond as in eq.~(\ref{eq:AlphaVsGamma}), where we identify the singularities of the Regge trajectory with the integral of the complete cusp anomalous dimension, eq.~(\ref{cuspAD_Sec_anom_dim}), including quartic Casimir (and higher) contributions. This has profound implications on the structure of the soft anomalous dimension, beyond the accuracy of our calculation. Specifically, at four loops, the N$^3$LL contribution to the soft anomalous dimension must be related to the four-loop Regge trajectory.
More generally, if our conjecture holds, linear terms in $L$ in the real part of the soft anomalous dimension in the Regge limit must be simply proportional to complete cusp anomalous dimension in the adjoint representation, or equivalently we expect eq.~(\ref{Soft_cusp_relatioN_der}) should hold to all orders. At four loops this provides three new constraints on the soft anomalous dimension, given in eq.~(\ref{eq:ConjConstr}). The vanishing of $\mathcal{H}^{(+,4)}_1$ is of course consistent with the finding of ref.~\cite{Vladimirov:2017ksc}. 
The results are included in Table~\ref{tab:reggeconstraints}, which provides important input to the bootstrap program to determine the soft anomalous dimension in general kinematics, which has already been successful at the three-loop level \cite{Almelid:2017qju}. Our work paves the way for bootstrapping this quantity to four loops, where direct calculations are not yet feasible.

\vspace{20pt}
\acknowledgments

We would like to thank Simon Caron-Huot for insightful comments and Claude Duhr and Andrew McLeod for collaboration on a related project on the soft anomalous dimension.  
EG, GF and NM are supported by the STFC Consolidated Grant ‘Particle Physics at the Higgs Centre’. GF is supported by the ERC Starting Grant 715049 ‘QCDforfuture’ with Prinipal Investigator Jennifer Smillie. CM's work is supported by the Italian Ministry of University and Research (MIUR), grant PRIN 20172LNEEZ. LV is supported by Fellini, Fellowship for Innovation at INFN, funded by the European Union's Horizon 2020 research programme under the Marie Sk\l{}odowska-Curie Cofund Action, grant agreement no. 754496.

\appendix

\section{Coefficients of the Regge pole amplitude}
\label{ImpactFactors_Appendix}

In this appendix we collect the coefficients describing the Regge pole part of the two-parton scattering amplitude, namely, the Regge trajectory and impact factors. As discussed in section \ref{subsec:Regge2to2}, this component of the amplitude is 
scheme-dependent, starting at NNLL accuracy. Below we begin by compoiling the coefficients in the MRS scheme of eq.~(\ref{Schemes_MRS}) and then proceed to discuss the cut scheme of eq.~(\ref{Schemes_Cut}).

Following the definition in eq.~\eqref{alphagtildeDef}, we split the Regge trajectory into a component proportional to the integral of the cusp anomalous dimension, $K(\as)$ defined in eq.~\eqref{eq:Kdef}, and a remainder, $\hat{\al}_g$: 
\begin{equation} \label{alphagtildeDefApp}
\al_g(t) = K\left(\as(-t)\right) + \hat{\al}_g(t).
\end{equation} 
Expanding the term in this equation according to eq.~\eqref{al_g}, in QCD one has 
\begin{subequations}
\begin{align}
 \begin{split}
 \label{Kexplicit} 
K^{(1)} & = \frac{{\gamma}_K^{(1)}}{4 \epsilon}, \\ 
K^{(2)} & = \frac{{\gamma}_K^{(2)}}{8 \epsilon} -
\frac{b_0 \, {\gamma}_K^{(1)}}{32 \epsilon^2}, \\ 
K^{(3)} & = \frac{{\gamma}_K^{(3)}}{12\epsilon} 
-\frac{b_0 \, {\gamma}_K^{(2)} 
+ b_1 \,{\gamma}_K^{(1)}}{48 \epsilon^2} 
+ \frac{b_0^2 \, {\gamma}_K^{(1)}}{192 \epsilon^3},
\end{split}
\end{align}
\end{subequations}
where the coefficients $\gamma_{K}^{(i)}$ are given in \eqref{gammaK}. The remainder, cusp-subtracted trajectory $\hat{\al}_g(t)$ is known up to two loops in QCD. Its coefficients read \cite{Lipatov:1976zz,Kuraev:1976ge,Fadin:1995xg,Fadin:1996tb,Fadin:1995km,Blumlein:1998ib}
\begin{subequations}
\begin{align}
\hat\alpha_g^{(1)} &= \frac{1}{2\eps}(r_{\Gamma} - 1) = 
-\frac14\zeta_2\, \eps -\frac76\zeta_3 \,\eps^2 + {\cal O}(\eps^3),
\label{alpha1hat} \\
\hat\alpha_g^{(2)}&=C_A \left(\frac{101}{108}  
- \frac{\zeta_3}{8}\right) -\frac{7\nf}{54} +{\cal O}(\eps)\,.
\label{alpha2hat}
\end{align}
\end{subequations}
The cusp-subtracted trajectory at three loops has been calculated in ${\cal N}= 4$ SYM, \cite{Caron-Huot:2017fxr}, extracting it from the two-parton scattering amplitude obtained in ref.~\cite{Henn:2016jdu}. It reads
\beq
\label{alpha3hatSYM}
\ag{3}\rvert_{\text{SYM}}  = C_A^2\left(-\frac{\zeta_2}{144 \eps^3}+\frac{5 \zeta_4}{192}\frac{1}{\eps}+\frac{107}{144}\zeta_2\zeta_3+\frac{\zeta_5}{4}+{\cal O}\left(\eps\right)\right).
\eeq

The description of the Regge-pole component of the amplitude is completed by the information provided by the quark and gluon impact factors. Following the definition in eq.~\eqref{eq:impactIRfact}, we split the impact factors into a term $Z_{i/j}(t)$, defined as the integral of the anomalous dimension $\Gamma_{i/j}$, see eq.~\eqref{Zi}, and a collinear-subtracted remainder $D_{i/j}(t)$:
\beq\label{eq:impactIRfactApp}
C_{i/j}(t) = Z_{i/j}(t) \, D_{i/j}(t).
\eeq
In terms of the coefficients in eqs.~(\ref{gammaq}) and~(\ref{gammag}), and setting $\mu^2=-t$, the perturbative expansion of $Z_{i}(t)$ (see eq.~\eqref{impact_factors_def}) reads
\begin{align}
\label{Zi_values}
\begin{split}
Z_i^{(0)} & =  1,  \\ 
Z_i^{(1)} & =  -\, C_i\, {\gamma}_K^{(1)} \frac{1}{4\eps^2} 
+ \frac{\gamma_i^{(1)}}{\eps}, \\ 
Z_i^{(2)} & =  C^2_i \left({\gamma}_K^{(1)}\right)^2 \frac{1}{32\eps^4}\, 
+C_i\, \Bigg[ \frac{1}{\eps^3}\frac{{\gamma}_K^{(1)} }{4}\left(\frac{3b_0 }{16} 
- \gamma_i^{(1)}\right) -\,\frac{1}{\eps^2}\frac{ {\gamma}_K^{(2)}}{16} 
\Bigg] 
\\  &\hspace{0.5cm}
+ \frac{1}{\eps^2} 
\frac{\gamma_i^{(1)}}{2} \left(\gamma_i^{(1)}-\frac{b_0}{4}\right)
+\frac{\gamma_i^{(2)}}{2\eps}. 
\end{split}
\end{align}
The one- and two-loop coefficients of the quark and gluon collinear-subtracted impact factors have been calculated in the MRS scheme of eq.~(\ref{Schemes_MRS}) in \cite{Caron-Huot:2017fxr}. 
For instance, at one loop one has 
\begin{subequations}
\begin{align}
 \begin{split}
D_g^{(1)} & = - N_c \left(\frac{67}{72} - \zeta_2 \right) + \frac{5}{36} n_f 
+ \eps \bigg[ N_c \left(-\frac{101}{54} + \frac{11}{48}\zeta_2 + \frac{17}{12} \zeta_3 \right) 
+ n_f  \left( \frac{7}{27} - \frac{\zeta_2}{24} \right) \bigg] \\ 
&+ \eps^2 \bigg[ N_c \left(- \frac{607}{162} +\frac{67}{144}\zeta_2 
+ \frac{77}{72}\zeta_3 +\frac{41}{32}\zeta_4\right) 
+ n_f  \left( \frac{41}{81} - \frac{5}{72}\zeta_2 -\frac{7}{36}\zeta_3 \right) \bigg]
+ {\cal O}(\eps^3) \,,  \label{D1g}
\end{split}
\\
\begin{split}
D_q^{(1)} & = N_c \left(\frac{13}{72}  + \frac{7}{8} \zeta_2 \right)
+ \frac{1}{N_c} \left( 1 - \frac{1}{8} \zeta_2 \right) - \frac{5}{36} n_f  
+ \eps \bigg[ N_c \left(\frac{10}{27} - \frac{\zeta_2}{24} + \frac{5}{6} \zeta_3 \right)  \\ 
&\hspace{-0.7cm}+ \frac{1}{N_c} \left(2 - \frac{3}{16}\zeta_2 - \frac{7}{12} \zeta_3 \right) 
+ n_f  \left( -\frac{7}{27} + \frac{\zeta_2}{24} \right) \bigg]
+ \eps^2 \bigg[ N_c \left(\frac{121}{162} - \frac{13}{144}\zeta_2 
- \frac{7}{36} \zeta_3 + \frac{35}{64}\zeta_4 \right)  \\
&\hspace{-0.7cm}+ \frac{1}{N_c} \left(4 - \frac{\zeta_2}{2} - \frac{7}{8} \zeta_3 - \frac{47}{64}\zeta_4 \right) 
+ n_f  \left( -\frac{41}{81} + \frac{5}{72} \zeta_2 + \frac{7}{36}\zeta_3 \right) \bigg]
+ {\cal O}(\eps^3) \,. \label{D1q}
\end{split}
\end{align}
\end{subequations}
At two loops:
\begin{subequations}
\label{D2s}
\begin{align}
\begin{split}
\label{D2g} 
D_g^{(2)} & = - \frac{\zeta_2}{32\eps^2} N_c^2 + 
N_c^2 \bigg(-\frac{26675}{10368} + \frac{335}{288} \zeta_2 
+ \frac{11}{18}\zeta_3 - \frac{\zeta_4}{64} \bigg) \\ 
& + N_c n_f \bigg(\frac{2063}{3456} - \frac{25}{144} \zeta_2
+ \frac{\zeta_3}{72} \bigg) + \frac{n_f}{N_c} \bigg(-\frac{55}{384} 
+ \frac{\zeta_3}{8} \bigg) -\frac{25}{2592} n_f^2 + {\cal O}(\eps) \,,  
\end{split}
\\ 
\begin{split}
\label{D2q}
D_q^{(2)} & =  - \frac{\zeta_2}{32\eps^2} N_c^2 + 
N_c^2 \bigg(\frac{22537}{41472} + \frac{87}{64} \zeta_2 
+ \frac{41}{144}\zeta_3 - \frac{15}{256} \zeta_4 \bigg) +\frac{28787}{10368} +\frac{19}{32}\zeta_2  \\ 
& - \frac{205}{288}\zeta_3
-\frac{47}{128}\zeta_4 + \frac{1}{N_c^2} \bigg(\frac{255}{512} + \frac{21}{64} \zeta_2 
- \frac{15}{32}\zeta_3 - \frac{83}{256} \zeta_4 \bigg) 
\\
& + N_c n_f \bigg(-\frac{325}{648} - \frac{\zeta_2}{4} - \frac{23}{144} \zeta_3\bigg) 
+ \frac{n_f}{N_c} \bigg(-\frac{505}{1296} - \frac{\zeta_2}{16} - \frac{19}{144}\zeta_3 \bigg) 
+\frac{25}{864} n_f^2 + {\cal O}(\eps) \,.
\end{split}
\end{align}
\end{subequations}

The whole impact factors $C_i$ can be found inserting the results from eqs.~\eqref{Zi_values}-\eqref{D2q} into eq.~\eqref{eq:impactIRfactApp}, and expanding order by order in the strong coupling. At one loop we get
\begin{subequations}
\begin{align}
    C_q^{(1)}=& -\frac{C_F}{2 \eps^2}-\frac{3 C_F}{4 \eps}+C_A \left(\frac{3 \zeta_2}{4}+\frac{85}{72}\right)+C_F \left(\frac{\zeta_2}{4}-2\right)-\frac{5 n_f}{36}
    +\eps \bigg[C_A \left(\frac{64}{27}-\frac{11 \zeta_2}{48}+\frac{\zeta_3}{4}\right)\nn \\
    &+C_F \left(\frac{3 \zeta_2}{8}+\frac{7 \zeta_3}{6}-4\right)+n_f \left(\frac{\zeta_2}{24}-\frac{7}{27}\right)\bigg]
    +\eps^2 \bigg[C_A \left(\frac{769}{162}-\frac{85 \zeta_2}{144}-\frac{77 \zeta_3}{72}-\frac{3 \zeta_4}{16}\right)\nn \\
    &+C_F \left(\zeta_2+\frac{7 \zeta_3}{4}+\frac{47 \zeta_4}{32}-8\right)+n_f \left(\frac{5 \zeta_2}{72}+\frac{7 \zeta_3}{36}-\frac{41}{81}\right)\bigg]+\mathcal{O}(\eps^3), \\
    C_g^{(1)}=&  -\frac{C_A}{2 \eps^2} -\frac{b_0}{4 \eps} +C_A \left(\zeta_2-\frac{67}{72}\right)+\frac{5 n_f}{36}+\eps \bigg[C_A \left(\frac{11 \zeta_2}{48}+\frac{17 \zeta_3}{12}-\frac{101}{54}\right)\nn \\
    &+n_f \left(\frac{7}{27}-\frac{\zeta_2}{24}\right)\bigg]+\eps^2 \bigg[C_A \left(\frac{67 \zeta_2}{144}+\frac{77 \zeta_3}{72}+\frac{41 \zeta_4}{32}-\frac{607}{162}\right) \nn \\ 
    &+n_f \left(\frac{41}{81}-\frac{5 \zeta_2}{72}-\frac{7 \zeta_3}{36}\right)\bigg]+\mathcal{O}(\eps^3).
\end{align}
\end{subequations}

In the Regge-cut scheme, the one-loop and two-loop Regge trajectories are identical to the MRS scheme, since multiple-Reggeon exchanges do not contribute to the odd amplitude at LL and NLL. Thus, with $\hat{\tilde{\al}}_g=\tilde{\al}_g-K$, and using eq.~(\ref{eq:reggeTilde}) for the three-loop case, we have 
\begin{subequations}
\label{hat_tilde_alpha_g}
\begin{align}
\hat{\tilde{\al}}_g^{(1)} =&\, {\hat\al}_g^{(1)} = \frac{1}{2\eps}(r_{\Gamma} - 1),\\
\hat{\tilde{\al}}_g^{(2)} =&\, {\hat\al}_g^{(2)} = C_A \left(\frac{101}{108}  
- \frac{\zeta_3}{8}\right) -\frac{7\nf}{54} +O(\eps),\\
\hat{\tilde{\al}}_g^{(3)}\rvert_{\text{SYM}} =&\, N_c^2\left(\frac{5}{24}\zeta_2 \zeta_3+\frac{\zeta_5}{4} + \mathcal{O}(\eps)\right).
\end{align}
\end{subequations}

Similarly, the impact factors at one loop in the Regge-cut scheme are identical to the MRS scheme: $\tilde{C}_i^{(1)}=C_i^{(1)}$ for both quarks and gluons. However, due to eq.~(\ref{eq:Ctilde}), at two loops we have, in the cut scheme
\begin{subequations}
\begin{align}
    \tilde{C}_q^{(2)}=&\frac{C_F^2}{8 \eps^4}+\frac{1}{\eps^3}\left(\frac{11 C_A C_F}{32}+\frac{3 C_F^2}{8}-\frac{C_F n_f}{16}\right)+\frac{1}{\eps^2}\bigg[C_F^2 \left(\frac{41}{32}-\frac{\zeta_2}{8}\right)-\frac{3C_A^2\zeta_2}{32}+\frac{C_F n_f}{24}\nn\\&-C_A C_F \left(\frac{5 \zeta_2}{16}+\frac{23}{48}\right)\bigg]
    +\frac{1}{\eps}\bigg[C_A C_F \left(-\frac{19 \zeta_2}{24}+\frac{11 \zeta_3}{16}-\frac{1513}{576}\right)+C_F^2 \left(\frac{221}{64}-\frac{4 \zeta_3}{3}\right)\nn\\&+C_F n_f \left(\frac{\zeta_2}{24}+\frac{89}{288}\right)\bigg]
    +C_A^2 \left(\frac{73 \zeta_2}{32}-\frac{43 \zeta_3}{48}-\frac{19 \zeta_4}{32}+\frac{13195}{3456}\right)\nn\\&
    -C_A C_F \left(\frac{1171 \zeta_2}{576}-\frac{175 \zeta_3}{48}-\frac{17 \zeta_4}{8}+\frac{40423}{3456}\right)
    +C_F^2 \left(\frac{1151}{128}+\frac{17 \zeta_2}{32}-\frac{29 \zeta_3}{8}-\frac{65 \zeta_4}{32}\right)\nn\\&+C_F n_f \left(\frac{265}{216}+\frac{17 \zeta_2}{288}+\frac{ \zeta_3}{6}\right)-C_An_f\left(\frac{385}{432}+\frac{5\zeta_2}{16}+\frac{7\zeta_3}{24}\right)+\frac{25}{864}n_f^2+\mathcal{O}(\eps),\\
    \tilde{C}_g^{(2)}=&\frac{C_A^2}{8 \eps^4}+\frac{1}{\eps^3}\left(\frac{77 C_A^2}{96}-\frac{7 C_A n_f}{48}\right)+\frac{1}{\eps^2}\left[C_A^2 \left(\frac{103}{96}-\frac{17 \zeta_2}{32}\right)-\frac{49 C_A n_f}{144}+\frac{n_f^2}{36}\right]\nn\\&
    +\frac{1}{\eps}\left[C_A^2 \left(\frac{853}{864}-\frac{11 \zeta_2}{12}-\frac{31 \zeta_3}{48}\right)+C_A n_f \left(\frac{\zeta_2}{6}-\frac{19}{72}\right)+\frac{C_F n_f}{16}+\frac{5 n_f^2}{216}\right]\nn\\&
    +C_A^2 \left(\frac{415 \zeta_2}{576}-\frac{11 \zeta_3}{9}-\frac{\zeta_4}{2}+\frac{10525}{10368}\right)+C_A n_f \left(-\frac{\zeta_2}{16}+\frac{17\zeta_3}{36}-\frac{113}{324}\right)\nn\\&+C_Fn_f\left(\frac{55}{192}-\frac{\zeta_3}{4}\right)+n_f^2\left(\frac{29}{864}-\frac{\zeta_2}{144}\right)+\mathcal{O}(\eps).
\end{align}
\end{subequations}

\section{Anomalous dimensions}
\label{AD-coefficient_Appendix}

In this appendix we collect the coefficients of the various anomalous dimensions considered in the main text. All anomalous dimensions are expanded in powers of the strong coupling according to 
\beq
\gamma_{\phi} = \sum_{\ell = 1}^{\infty}
\bigg(\frac{\alpha_s}{\pi}\bigg)^{\ell-1}
\gamma^{(\ell)}_{\phi}.
\eeq
First of all we have the cusp anomalous dimension, defined in eq.~\eqref{cuspAD}, which involves quadratic and quartic Casimir terms, recently calculated up to four loops in QCD~\cite{Boels:2017ftb,Boels:2017skl,Moch:2017uml,Grozin:2017css,Henn:2019swt,Huber:2019fxe,vonManteuffel:2020vjv,Agarwal:2021zft,Bruser:2019auj,Bruser:2020bsh}. In QCD, the quadratic Casimir component, $\gamma_K(\alpha_s)$,  has the following expansion coefficients through three loops\footnote{Three-loop contributions to lightlike cusp anomalous dimension were first determined in \cite{Berger:2002sv,Moch:2004pa}, by using the connection between this quantity and the large-$x$ limit of non-singlet splitting functions~\cite{Korchemsky:1988si}. The complete calculation of the non-singlet three-loop splitting functions has been recently confirmed in ref.~\cite{Blumlein:2021enk}. Independent calculations of the three-loop cusp anomalous dimension were also obtained by computing form factors \cite{FormFactors,Gehrmann:2010ue} and cusped Wilson loop \cite{Grozin:2014hna,Grozin:2015kna} to this loop order. More recently, such calculations have been completed at four loops
~\cite{Boels:2017ftb,Boels:2017skl,Moch:2017uml,Grozin:2017css,Henn:2019swt,Huber:2019fxe,vonManteuffel:2020vjv,Agarwal:2021zft,Bruser:2019auj,Bruser:2020bsh}.}~\cite{Korchemsky:1985xj,Korchemsky:1987wg,Moch:2004pa}:
\beqa\nn \label{gammaK}
{\gamma}_K^{(1)} & = & 2, \\ \nn
{\gamma}_K^{(2)} & = & 
\left( \frac{67}{18} - \zeta_2 \right) C_A 
- \frac{10}{9} T_R \nf,\\ \nn
{\gamma}_K^{(3)} & = & 
\frac{C_A^2}{96} \left( 490 - \frac{1072}{3} \zeta_2 
+ 88 \zeta_3 + 264 \zeta_4 \right) 
+ \frac{C_F T_R \nf}{32} \left( - \frac{220}{3} + 64 \zeta_3 \right) \\ 
&&\hspace{0.0cm} + \, \frac{C_A T_R \nf}{96} 
\left(-\frac{1672}{9} + \frac{320}{3} \zeta_2 - 224 \zeta_3 \right) 
- \frac{2 T_R^2 \nf^2}{27},
\eeqa 
where the fundamental trace is $T_R = {\rm Tr}(t^at^a)= \frac12$. The second term, $g_R(\alpha_s)$, multiplying the quartic Casimir, starts at four loops, and depends on the gauge-group representation~$R$. Its coefficients for $R=A$ (adjoint) and $R=F$ (fundamental) in QCD read
\begin{align}
\label{g_R_values}
\begin{split}
g_A^{(4)}&=\frac{\zeta_3}{6} 
    - \frac{3\zeta_3^2}{2}+ \frac{55 \zeta_5}{12}- \frac{\zeta_2}{2} 
     - \frac{31 \zeta_6}{8},\\
g_F^{(4)}&=  n_f\left( \zeta_2
    - \frac{\zeta_3}{3} - \frac{5 \zeta_5}{3}\right)\,.
\end{split}
\end{align}
The contribution in ${\cal N}=4$ SYM, for $\gamma_K(\alpha_s)$ and $g_A(\alpha_s)$, is obtained, according to principle of maximum trascendentality, by retaining only the terms with highest trascendental weight at each order. 

Next, we have the collinear anomalous dimension $\gamma_i$ corresponding to the parton~$i$~\cite{FormFactors,DelDuca:2014cya,Falcioni:2019nxk,Dixon:2017nat}, which is part of the anomalous dimension $\Gamma_i$ and $\Gamma_j$ defined in eq.~\eqref{collAD}. The collinear anomalous dimension $\gamma_i$ has been recently calculated up to four loops~\cite{vonManteuffel:2020vjv,Agarwal:2021zft}. We provide here its coefficients up to two loops, as needed in the main text, for quarks and gluons. One has~\cite{FormFactors,Gehrmann:2010ue}
\beqa \label{gammaq} \nn
\gamma_{q}^{(1)} & = & - \frac{3}{4} \, C_F,   \\ \nn
\gamma_{q}^{(2)} & = & \frac{C_F^2}{16} 
\left( - \frac{3}{2} + 12 \zeta_2 - 24 \zeta_3 \right)  \\
&& + \, \frac{C_A C_F}{16} \left( - \frac{961}{54} - 11 \zeta_2 + 26 \zeta_3 \right)
+\frac{C_F T_R \nf}{16} \left( \frac{130}{27} + 4 \zeta_2 \right),  
\eeqa
for quarks, and
\beqa \label{gammag} \nn
\gamma_{g}^{(1)} & = & -  \frac{b_0}{4}, \\
\gamma_{g}^{(2)} & = & \frac{C_A^2}{16} 
\left( - \frac{692}{27}+ \frac{11}{3} \zeta_2 + 2 \zeta_3 \right) 
+  \frac{C_A T_R \nf}{16} \left( \frac{256}{27} - \frac{4}{3} \zeta_2 \right) 
+ \frac{C_F T_R \nf}{4},
\eeqa
for gluons.

\section{Computing colour factors for arbitrary representations}
\label{sec:colourappendix}
In this appendix we review computational techniques to evaluate the colour factors of the transition amplitudes. Universality of the Regge limit implies that the colour structure of every amplitude is given by the same colour operators, regardless whether the scattering process involves quarks or gluons in the initial and final state. In order to determine such operators, we need to develop techniques to evaluate colour tensors for general representations of the external particles. Indeed, while it is straightforward to compute directly colour Feynman rules by specialising the representations of the scattering particles, such explicit results would completely obscure universality of the Regge limit. Instead, we would like to express our results in terms of Casimir operators of the colour channel operators defined in eq.~(\ref{eq:Tsudef})
\begin{equation}
    \label{def2:TtTsu}
    \mathbf{T}^2_t,\quad \mathbf{T}^2_{s-u}=\frac{\mathbf{T}^2_s-\mathbf{T}^2_u}{2},
\end{equation}
which manifest the signature properties under $s\leftrightarrow u$ crossing. These operators emerge naturally in diagrams that feature connections among the outermost Reggeon indices, for example
\begin{align}
\begin{split}
\label{eq:T1T2}
     \includegraphics{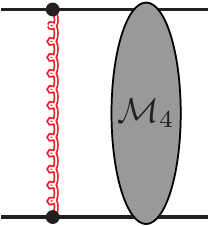}&\raisebox{30pt}{$= \left(\mathbf{T}_1^a\cdot\mathbf{T}_2^a\right)\,\mathcal{M}_4,$}\\
     &= \frac{1}{2}\left(\mathbf{T}^2_{s-u}-\frac{\mathbf{T}^2_t}{2}\right)\,\mathcal{M}_4.
\end{split}
\end{align}
where, in the second line, we applied colour conservation according to eq.~(\ref{eq:colcons}). The result above is independent on the four-point matrix element $\mathcal{M}_4$, thus providing a graphical derivation of the relation in eq.~(\ref{eq:relTsu-Tt/2}). We apply a similar procedure whenever a Reggeon is emitted from an initial-state parton and absorbed by a final-state one, according to  eq.~(\ref{eq:relTsu+Tt/2}). Therefore, colour structures up to two loops are written for general representations by
\begin{itemize}
    \item[($i$)] Using the Lie algebra, eq.~(\ref{eq:LieAlgebra}), to write three-point vertices $(F^a)_{bc}$ in terms of Reggeons connecting the target and the projectile.
    \item[($ii$)] Applying repeatedly eqs.~(\ref{eq:relTsu-Tt/2}) and (\ref{eq:relTsu+Tt/2}) to obtain the colour-channel operators of eq.~(\ref{def2:TtTsu}), acting on the tree-level amplitude.
\end{itemize}
The second step may not applicable for diagrams where all Reggeons have one or more internal attachment, namely they are all either emitted or absorbed between two other Reggon vertices. We refer to these irreducible configurations as entangled colour structures.

\begin{figure}[htb]
    \centering
    \subfloat[\label{fig:3ld1}]{\includegraphics{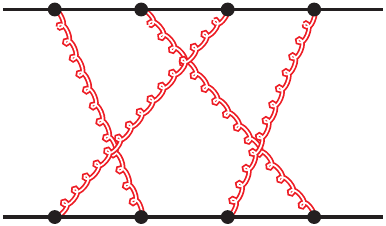}}
    \hspace{50pt}
    \subfloat[\label{fig:3ld2}]{\includegraphics{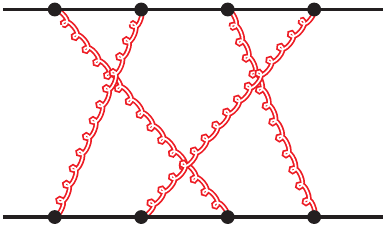}}
    \caption{Diagrammatic representation of the irreducible configurations: (a) $d_A$ in eq.~(\ref{def:dA})  and (b) $d_B$ in eq.~(\ref{def:dB}).}
\end{figure}

\begin{figure}[htb]
    \centering
    \subfloat[\label{fig:doublecross}]{\includegraphics{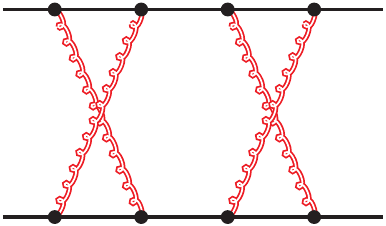}}
    \hspace{50pt}
    \subfloat[\label{fig:scottishflag}]{\includegraphics{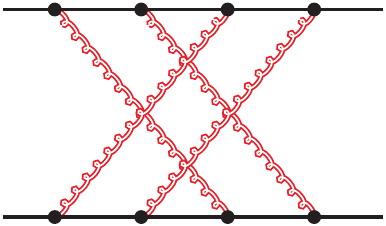}}
    \caption{Both the {\textit{double cross}} diagram (a) and the {\textit{saltire}} diagram (b) are immediately written in terms of colour dipole operators by commuting the pair of Reggeon emission vertices at the end of either the top or of the bottom line.}
\end{figure}

Indeed, entangled colour structures may occur starting at three loops. At three loops, we find two such colour tensors\footnote{At three loops there are two additional irreducible configurations, depicted in figure~\ref{fig:doublecross} and~\ref{fig:scottishflag}. These, however, can be recast into the form of eq.~(\ref{eq:T1T2}) using commutation relations.} corresponding to the graphs in figure~\ref{fig:3ld1} and~\ref{fig:3ld2}.

While these diagrams drop out the three-loop amplitude, such entangled configurations do not cancel in general and we will encounter them in the four-loop calculation. Hence, we need to extend the techniques summarised above.

\subsection*{Permutation diagrams}

We begin by introducing a compact notation for colour factors involving $k$ Reggeon attachments on both target and projectile. These configuration are naturally associated to {\textit{permutations}} of $k$ indices. Choosing the top line as target state $i$, the diagram in figure~\ref{fig:3ld1} is written as
\begin{align}
    \begin{split}\label{def:dA}
        d_A&=\Big(\T^{a_1}\T^{a_2}\T^{a_3}\T^{a_4}\Big)_i\Big(\T^{a_3}\T^{a_1}\T^{a_4}\T^{a_2}\Big)_j\equiv
        \left(
        \begin{array}{cccc}
        a_1&a_2&a_3&a_4\\
        a_3&a_1&a_4&a_2
        \end{array}
        \right).
    \end{split}
\end{align}
Similarly, the diagram in figure~\ref{fig:3ld2} is 
\begin{align}
    \begin{split}\label{def:dB}
        d_B&=\Big(\T^{a_1}\T^{a_2}\T^{a_3}\T^{a_4}\Big)_i\Big(\T^{a_2}\T^{a_4}\T^{a_1}\T^{a_3}\Big)_j\equiv
        \left(
        \begin{array}{cccc}
        a_1&a_2&a_3&a_4\\
        a_2&a_4&a_1&a_3
        \end{array}
        \right).
    \end{split}
\end{align}
We do not have an expression of $d_A$ and $d_B$ separately which is valid for general representations. However, we are interested only in the combination $d_A+d_B$, which manifests the symmetry under the interchanged of the projectile and the target. For this combination we have the following identity:
\begin{align}
\label{eq:threelooprel}
    d_A+d_B&=\left(
    \begin{array}{cccc}
    a_1&a_2&a_3&a_4\\
    a_4&a_1&a_3&a_2
    \end{array}
    \right) + 
    \left(
    \begin{array}{cccc}
    a_1&a_2&a_3&a_4\\
    a_1&a_4&a_2&a_3
    \end{array}
    \right)\,,
\end{align}
which we will prove below.
The two terms on right-hand side of eq.~(\ref{eq:threelooprel}), depicted in figures \ref{fig:d3r1} and \ref{fig:d3r2}, feature outmost Reggeon interactions represented by the indices $a_1$ and $a_4$,  respectively. Therefore, these terms are easily written in terms of colour channel operators by applying eqs.~(\ref{eq:relTsu-Tt/2}) and (\ref{eq:relTsu+Tt/2}), as described in step $(ii)$ above. The resulting two-loop graphs are again reducible, and one obtains:
\begin{figure}
    \centering
    \subfloat[\label{fig:d3r1}]{\includegraphics{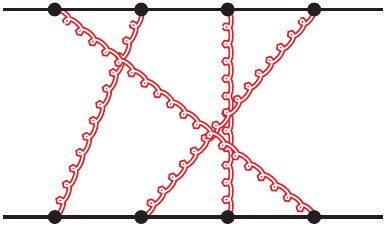}}
    \hspace{50pt}
    \subfloat[\label{fig:d3r2}]{\includegraphics{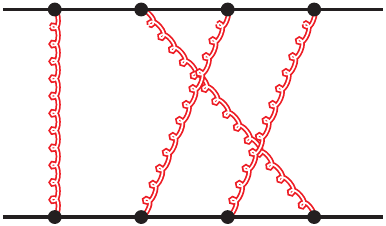}}
    \caption{Diagrammatic representation of the terms on the right hand side of eq.~(\ref{eq:threelooprel}).}
\end{figure}
\begin{align}
    \begin{split}
    \label{eq:threelooprel2}
    d_A+d_B&=\frac{1}{4}\left\{\left(\tsu\right)^3-\frac{1}{4}\left(\tts\right)^2\tsu+\frac{C_A}{4}\tts\tsu-\frac{C_A^2}{4}\tsu\right\}\T_i^a\,\T_j^a,
    \end{split}
\end{align}
Eq.~(\ref{eq:threelooprel2}) is a general expression of $d_A+d_B$ for arbitrary representations of external particles. This is the only relation needed to compute three-loop colour structures. The identity~(\ref{eq:threelooprel}) was crucial to obtain the result. This identity is conveniently derived by starting from an auxiliary three-loop configuration
\begin{align}
    \begin{split}
    \label{eq:boomerang3}
        \tilde{d}^{(3)}&=\Big(\T^{a_1}\T^{a_2}\T^{a_3}\Big)_i\,\Big(\T^x\T^{a_1}\T^{a_3}\T^x\T^{a_2}\Big)_j.
    \end{split}
\end{align}
The colour factor above is not associated to a permutation of indices, because it features a {\textit{boomerang}}, namely the contraction of a pair of indices on the same line (in this case the projectile). Using the Lie algebra relations, there are two ways of moving the $x$ on line $i$: either by commuting $x$ with $a_1$ or $x$ with $a_3$. We find respectively
\begin{subequations}
\label{eq:dtilde12}
\begin{align}
    \label{eq:dtilde1}
    \tilde{d}^{(3)}_1=&\,\Big(C_2(j)-\frac{C_A}{2}\Big)\left(\T^{a_1}\T^{a_2}\T^{a_3}\right)_i\left(\T^{a_1}\T^{a_3}\T^{a_2}\right)_j\\&\hskip4.5cm+if^{a_3 x k}\left(\T^{a_1}\T^{a_2}\T^{a_3}\right)_i\left(\T^x\T^{a_1}\T^{k}\T^{a_2}\right)_j,\nn
    \\
    \label{eq:dtilde2}
    \tilde{d}^{(3)}_2=&\,\Big(C_2(j)-\frac{C_A}{2}\Big)\left(\T^{a_1}\T^{a_2}\T^{a_3}\right)_i\left(\T^{a_1}\T^{a_3}\T^{a_2}\right)_j\\&\hskip4.5cm+if^{x a_1 k}\left(\T^{a_1}\T^{a_2}\T^{a_3}\right)_i\left(\T^{k}\T^{a_3}\T^x\T^{a_2}\right)_j.\nn
\end{align}
\end{subequations}
The two expressions are of course identical, so their difference must vanish
\begin{equation}
\label{eq:rel3loopboom}
    0=i\left(\T^{a_1}\T^{a_2}\T^{a_3}\right)_i\Big[f^{a_3 x k}\left(\T^x\T^{a_1}\T^{k}\T^{a_2}\right)_j-f^{a_1 x k}\left(\T^{x}\T^{a_3}\T^k\T^{a_2}\right)_j\Big].
\end{equation}
Finally, writing the structure constants in terms of commutators on line $i$, we obtain eq.~(\ref{eq:threelooprel}), concluding the proof.

\subsection*{Four-loop colour factors}

All the colour structures appearing at four loops are written in terms of contractions of five pairs of generators, by applying repeatedly the Lie algebra. We identify eight independent colour factors that cannot be reduced in terms of $\tsu$ and $\tts$ by following steps $(i)$ and $(ii)$ above. We choose to collect them into the following terms:
\begin{subequations}
\begin{align}
    \begin{split}
    \label{def:d1}
        d_1&=\left(
        \begin{array}{ccccc}
             a_1&a_2&a_3&a_4&a_5  \\
             a_2&a_5&a_3&a_1&a_4 
        \end{array}
        \right) + 
        \left(
        \begin{array}{ccccc}
             a_1&a_2&a_3&a_4&a_5  \\
             a_4&a_1&a_3&a_5&a_2 
        \end{array}
        \right),
        \end{split}\\
        \begin{split}
        \label{def:d2}
        d_2&=\left(
        \begin{array}{ccccc}
             a_1&a_2&a_3&a_4&a_5  \\
             a_2&a_4&a_1&a_5&a_3 
        \end{array}
        \right) + 
        \left(
        \begin{array}{ccccc}
             a_1&a_2&a_3&a_4&a_5  \\
             a_3&a_1&a_5&a_2&a_4 
        \end{array}
        \right),
    \end{split}    \\
    \begin{split}
    \label{def:d3}
        d_3&=\left(
        \begin{array}{ccccc}
             a_1&a_2&a_3&a_4&a_5  \\
             a_2&a_4&a_5&a_1&a_3 
        \end{array}
        \right) + 
        \left(
        \begin{array}{ccccc}
             a_1&a_2&a_3&a_4&a_5  \\
             a_4&a_1&a_5&a_2&a_3
        \end{array}
        \right),
    \end{split}\\
    \begin{split}
    \label{def:d4}
        d_4&=\left(
        \begin{array}{ccccc}
             a_1&a_2&a_3&a_4&a_5  \\
             a_2&a_5&a_1&a_4&a_3 
        \end{array}
        \right) + 
        \left(
        \begin{array}{ccccc}
             a_1&a_2&a_3&a_4&a_5  \\
             a_3&a_1&a_5&a_4&a_2 
        \end{array}
        \right),
    \end{split}\\
    \begin{split}
    \label{def:d5}
        d_5&=\left(
        \begin{array}{ccccc}
             a_1&a_2&a_3&a_4&a_5  \\
             a_3&a_2&a_5&a_1&a_4 
        \end{array}
        \right) +
        \left(
        \begin{array}{ccccc}
             a_1&a_2&a_3&a_4&a_5  \\
             a_4&a_2&a_1&a_5&a_3 
        \end{array}
        \right),
    \end{split}\\
    \begin{split}
    \label{def:d6}
        d_6&=\left(
        \begin{array}{ccccc}
             a_1&a_2&a_3&a_4&a_5  \\
             a_3&a_4&a_1&a_5&a_2 
        \end{array}
        \right) + 
        \left(
        \begin{array}{ccccc}
             a_1&a_2&a_3&a_4&a_5  \\
             a_3&a_5&a_1&a_2&a_4 
        \end{array}
        \right),
    \end{split}\\
    \begin{split}
    \label{def:d7}
        d_7&=\left(
        \begin{array}{ccccc}
             a_1&a_2&a_3&a_4&a_5  \\
             a_3&a_5&a_1&a_4&a_2 
        \end{array}
        \right),
    \end{split}\\
    \begin{split}
    \label{def:d8}
        d_8&=\left(
        \begin{array}{ccccc}
             a_1&a_2&a_3&a_4&a_5  \\
             a_4&a_2&a_5&a_1&a_3 
        \end{array}
        \right),
    \end{split}
\end{align}
\end{subequations}
where each term $d_1\dots d_8$ is manifestly symmetric under target-projectile exchange $i~\leftrightarrow~j$. In addition, $d_1$ is symmetric under signature symmetry, because the two terms in eq.~(\ref{def:d1}) are related to each other by reversing the order of indices on one of the lines. In order to express these colour factors in terms of channel operators, we consider again the configurations generated by operating with the Lie algebra on the corresponding boomerang diagrams. In particular, we consider the following four-loop diagrams
\begin{align}
\begin{split}
\label{eq:boomerang4}
    \tilde{d}^{(4)}_L&=\Big(\T^{a_1}\T^{a_2}\T^{a_3}\T^{a_4}\Big)_i\,\Big(\T^x\T^{a_{\sigma(1)}}\T^{a_{\sigma(2)}}\T^x\T^{a_{\sigma(3)}}\T^{a_{\sigma(4)}}\Big)_j,\\
    \tilde{d}^{(4)}_R&=\Big(\T^{a_1}\T^{a_2}\T^{a_3}\T^{a_4}\Big)_i\,\Big(\T^{a_{\sigma(1)}}\T^{a_{\sigma(2)}}\T^x\T^{a_{\sigma(3)}}\T^{a_{\sigma(4)}}\T^x\Big)_j,\\
    \tilde{d}^{(4)}_C&=\Big(\T^{a_1}\T^{a_2}\T^{a_3}\T^{a_4}\Big)_i\,\Big(\T^{a_{\sigma(1)}}\T^x\T^{a_{\sigma(2)}}\T^{a_{\sigma(3)}}\T^x\T^{a_{\sigma(4)}}\Big)_j,
\end{split}
\end{align}
which generalise the three-loop boomerang of eq.~(\ref{eq:boomerang3}) by including one more pair of indices, and the target-projectile symmetric configurations obtained from eq.~(\ref{eq:boomerang4}). Starting from these boomerang configurations, we operate as in eqs.~(\ref{eq:dtilde12}) and (\ref{eq:rel3loopboom}) and we get six independent linear relations for $d_1\dots d_8$. We derive one more constraint from the colour factor
\begin{align}
\tilde{d}^{(4)}_{P}&=\text{Tr}\left[F^xF^aF^bF^c\right]\text{Tr}\left[F^yF^aF^bF^c\right]\,\T^x_i\,\T^y_j = \left(\frac{d_{AA}}{N_A}+\frac{C_A^4}{12}\right)\,\T^x_i\,\T^x_j,
\end{align}
which can be written as a combination of $d_1\dots d_8$ by using the Lie algebra to replace traces of generators in the adjoint representation with commutators of $\T_i$ or $\T_j$. The seven identities obtained in this way determine the signature-odd contributions to $d_1\dots d_8$. In turn, this is sufficient in order to express the real part of the amplitude, thus allowing us to perform the calculations of section \ref{subsec:colour}. However, we need one more equation in order to determine also contributions of even signature. In particular, such terms were needed to compute the Regge limit of the soft anomalous dimension, discussed in section~\ref{sec:SAD_Regge}. The last constraint was determined by writing a general ansatz for $d_3$ in terms of products of Casimir operators $\tsu$ and $\tts$ acting on the tree-level amplitude $\T^x_i\,\T^x_j$. The unknown coefficients were fitted by comparing the ansatz with explicit results obtained by specialising the generators $\T_i$ and $\T_j$ in eq.~(\ref{def:d3}) either in the adjoint or in the fundamental representation. We report here the final expressions of the colour factors in eqs.~(\ref{def:d1})-(\ref{def:d8}), which all apply for any representation of the external particles. The results are:
\begin{subequations}
\begin{align}
     d_1&=\Bigg\{\frac{1}{12}\left(\frac{d_{AA}}{N_A}+\frac{5}{96}C_A^4\right)-\frac{3}{32}C_A^2\,\left(\tsu\right)^2+\frac{C_A}{32}\left[5\,\tsu\tts\tsu-\frac{7}{3}\tts\left(\tsu\right)^2\right]\nonumber
     \\
     &+\frac{1}{8}
     \left[\left(\tsu\right)^4+\frac{3}{4}\Big[\tts,\tsu\Big]\tts\tsu-\frac{5}{12}\left(\tts\right)^2\left(\tsu\right)^2\right]\Bigg\}\,\T^x_i\,\T^x_j, \label{eq:resd1}
     \\
\begin{split}
\label{eq:resd2}
    d_2&=\Bigg\{\frac{1}{12}\left(\frac{d_{AA}}{N_A}+\frac{5}{96}C_A^4\right)-\frac{C_A^2}{8}\left(\tsu\right)^2+\frac{C_A}{16}\left[3\tsu\tts\tsu-\frac{7}{6}\tts\left(\tsu\right)^2\right]\\
    &+\frac{1}{8}\left[\left(\tsu\right)^4-\frac{3}{4}\tsu\left(\tts\right)^2\tsu+\frac{1}{2}\tts\tsu\tts\tsu-\frac{1}{6}\left(\tts\right)^2\left(\tsu\right)^2\right]\\
    &-\frac{C_A^3}{64}\tsu-\frac{C_A^2}{64}\tts\tsu+\frac{C_A}{16}\left[\left(\tsu\right)^3+\frac{3}{4}\left(\tts\right)^2\tsu\right]\\
    &+\frac{1}{16}\Bigg[\Big[\tsu,\tts\Big]\left(\tsu\right)^2-\left(\tsu\right)^2\tts\tsu-\frac{1}{4}\left(\tts\right)^3\tsu\Bigg]\Bigg\}\,\T^x_i\,\T^x_j,
\end{split}\\
\begin{split}
\label{eq:resd3}
    d_3&=d_6=\Bigg\{\frac{1}{8}\left[\left(\tsu\right)^4-\frac{1}{4}\Big[\tsu,\tts\Big]\tts\tsu-\frac{1}{4}\left(\tts\right)^2\left(\tsu\right)^2\right]\\
    &+\frac{C_A^3}{64}\tsu-\frac{C_A^2}{16}\tts\tsu-\frac{C_A}{16}\left[\left(\tsu\right)^3-\frac{3}{2}\left(\tts\right)^2\tsu\right]\\
    &-\frac{1}{16}\left(\tsu\Big[\tsu,\tts\Big]\tsu+\frac{1}{2}\left(\tts\right)^3\tsu\right)\Bigg\}\,\T^x_i\,\T^x_j,
\end{split}\\
\begin{split}
\label{eq:resd4}
    d_4&=d_5=\Bigg\{\frac{1}{8}\left[\left(\tsu\right)^4-\frac{1}{4}\Big[\tsu,\tts\Big]\tts\tsu-\frac{1}{4}\left(\tts\right)^2\left(\tsu\right)^2\right]\\
    &-\frac{C_A^3}{64}\tsu+\frac{C_A^2}{16}\tts\tsu+\frac{C_A}{16}\left[\left(\tsu\right)^3-\frac{3}{2}\left(\tts\right)^2\tsu\right]\\
    &+\frac{1}{16}\left(\tsu\Big[\tsu,\tts\Big]\tsu+\frac{1}{2}\left(\tts\right)^3\tsu\right)\Bigg\}\,\T^x_i\,\T^x_j,
\end{split}\\
\begin{split}
\label{eq:resd8}
    d_7&=d_8=\Bigg\{\frac{1}{24}\left(\frac{d_{AA}}{N_A}+\frac{5}{96}C_A^4\right)-\frac{C_A^2}{16}\left(\tsu\right)^2+\frac{C_A}{32}\left[3\tsu\tts\tsu-\frac{7}{6}\tts\left(\tsu\right)^2\right]\\
    &+\frac{1}{16}\left[\left(\tsu\right)^4-\frac{3}{4}\tsu\left(\tts\right)^2\tsu+\frac{1}{2}\tts\tsu\tts\tsu-\frac{1}{6}\left(\tts\right)^2\left(\tsu\right)^2\right]\\
    &+\frac{C_A^3}{128}\tsu+\frac{C_A^2}{128}\tts\tsu-\frac{C_A}{32}\left[\left(\tsu\right)^3+\frac{3}{4}\left(\tts\right)^2\tsu\right]\\
    &-\frac{1}{32}\Bigg[\Big[\tsu,\tts\Big]\left(\tsu\right)^2-\left(\tsu\right)^2\tts\tsu-\frac{1}{4}\left(\tts\right)^3\tsu\Bigg]\Bigg\}\,\T^x_i\,\T^x_j.
\end{split}
\end{align}
\end{subequations}
\section{The reduced amplitude in an explicit colour basis \label{subsec:ExplicitRep}}
It is interesting to evaluate the NNLL reduced amplitude for different external partons. We will compute the reduced NNLL odd amplitude at two loops, three loops and four loops for $qq$, $gg$ and $qg$ scattering. We utilise the orthornormal $t$-channel basis of ref.~\cite{DelDuca:2014cya}, and use the same notation as in appendix B of ref.~\cite{Caron-Huot:2017fxr} with the relevant colour tensors given in eq.~(\ref{colourTensors}). 

Projecting the two-loop NNLL amplitude in eq.~(\ref{eq:M20}) in the octet channel for $qq$, $gg$ and $qg$ scattering we find
\begin{subequations}\begin{align}
    \mathcal{\hat{M}}^{(-,2,0),[8]}_{qq\to qq} =&\,\left[2D_q^{(2)}+D_q^{(1)}D_q^{(1)}-(i\pi)^2r_\Gamma^2S^{(2)}(\eps)\left( \frac{N_c^2}{6}-1+\frac{3}{N_c^2}\right)\right]{\cal M}^{{\rm tree},[8]}_{qq\to qq},\label{eq:M20qqOctet}\\
    \mathcal{\hat{M}}^{(-,2,0),[8_a]}_{gg\to gg} =&\,\left[2D_g^{(2)}+D_g^{(1)}D_g^{(1)}-(i\pi)^2r_\Gamma^2S^{(2)}(\eps)\left( \frac{N_c^2}{6}+6\right)\right]{\cal M}^{{\rm tree},[8_a]}_{gg\to gg},\label{eq:M20ggOctet}\\
    \mathcal{\hat{M}}^{(-,2,0),[8_a]}_{qg\to qg} =&\,\left[D_q^{(2)}+D_g^{(2)}+D_q^{(1)}D_g^{(1)}-(i\pi)^2r_\Gamma^2S^{(2)}(\eps)\left( \frac{N_c^2}{6}+1\right)\right]{\cal M}^{{\rm tree},[8_a]}_{qg\to qg}\label{eq:M20qgOctet},
\end{align}\end{subequations}
where we have normalised by the tree-amplitude octet projection defined in eq.~(\ref{eq:treelevel_using_8a}) and $S^{(2)}(\eps)$ is given in eq.~(\ref{eq:R2Res}).

The three-loop reduced amplitude of eq.~(\ref{eq:M31}) in the $t$-channel octet representation is
\begin{subequations}\begin{align}
    \mathcal{\hat{M}}^{(-,3,1),[8]}_{qq\to qq} =&\,(i\pi)^2r_{\Gamma}^3\left(N_c^2+18-\frac{18}{N_c^2}\right)\frac{N_c}{864}\left(-\frac{1}{\eps^3}+70\hat\zeta_3+\mathcal{O}(\eps^2)\right){\cal M}^{{\rm tree},[8]}_{qq\to qq},\label{eq:M31qqOctet}\\
    \mathcal{\hat{M}}^{(-,3,1),[8_a]}_{gg\to gg} =&\,(i\pi)^2r_{\Gamma}^3\left(N_c^2+36\right)\frac{N_c}{864}\left(-\frac{1}{\eps^3}+70\hat\zeta_3+\mathcal{O}(\eps^2)\right){\cal M}^{{\rm tree},[8_a]}_{gg\to gg},\label{eq:M31ggOctet}\\
    \mathcal{\hat{M}}^{(-,3,1),[8_a]}_{qg\to qg} =&\,(i\pi)^2r_{\Gamma}^3\left(N_c^2+36\right)\frac{N_c}{864}\left(-\frac{1}{\eps^3}+70\hat\zeta_3+\mathcal{O}(\eps^2)\right){\cal M}^{{\rm tree},[8_a]}_{qg\to qg},\label{eq:M31qgOctet}
\end{align}
\end{subequations}
where $\hat\zeta_3$ is defined in eq.~(\ref{def:hatz}).

At four loops we have from eq.~(\ref{eq:M42})
\begin{subequations}\begin{align}
    \mathcal{\hat{M}}^{(-,4,2),[8]}_{qq\to qq} =&\,(i\pi)^2r_{\Gamma}^4\left[\frac{1}{\eps^4}\left(\frac{N_c^2}{384}-\frac{1}{384}\right)-\frac{\hat{\zeta}_3}{\eps}\left(\frac{13  N_c^2}{24}+\frac{49 }{96}\right)+\mathcal{O}(\eps)\right]{\cal M}^{{\rm tree},[8]}_{qq\to qq}\label{eq:M42qqOctet}\\
    \mathcal{\hat{M}}^{(-,4,2),[8_a]}_{gg\to gg} =&\,(i\pi)^2r_{\Gamma}^4\left[\frac{1}{\eps^4}\left(\frac{N_c^2}{64}+\frac{1}{16}\right)-\frac{\hat{\zeta}_3}{\eps}\left(\frac{101}{32}   N_c^2+\frac{101 }{8}\right)+\mathcal{O}(\eps)\right]{\cal M}^{{\rm tree},[8_a]}_{gg\to gg},\label{eq:M42ggOctet}\\
    \mathcal{\hat{M}}^{(-,4,2),[8_a]}_{qg\to qg} =&\,(i\pi)^2r_{\Gamma}^4N_c^2\left(\frac{1}{192 \eps^4}-\frac{101 \hat{\zeta}_3}{96 \eps}+\mathcal{O}(\eps)\right){\cal M}^{{\rm tree},[8_a]}_{qg\to qg}.\label{eq:M42qgOctet}
\end{align}
\end{subequations}
As noted below eq.~(\ref{eq:M42}), $ \mExpM{4}{2}$ is non-planar, which is confirmed by the subleading-$N_c$ nature of eqs.~(\ref{eq:M42qqOctet})-(\ref{eq:M42qgOctet}).

There are other non-zero components to the odd NNLL reduced amplitudes. The singlet of $qq$ scattering
\begin{subequations}
\begin{align}
    \mathcal{\hat{M}}^{(-,2,0),[1]}_{qq\to qq}=&\,-(i\pi)^2r_\Gamma^2\left(N_c^2-4\right) \sqrt{N_c^2-1}\frac{1}{2 N_c^2}S^{(2)}(\eps){\cal M}^{{\rm tree},[8]}_{qq\to qq},\label{eq:M20qqSinglet}\\
    \mathcal{\hat{M}}^{(-,3,1),[1]}_{qq\to qq}=&\,(i\pi)^2r_{\Gamma}^3\left(N_c^2-4\right) \sqrt{N_c^2-1}\frac{1}{48N_c}  \left(\frac{1}{\eps^3} + 2  \hat\zeta_3+\mathcal{O}(\eps^2)\right){\cal M}^{{\rm tree},[8]}_{qq\to qq},\label{eq:M31qqSinglet}\\
    \mathcal{\hat{M}}^{(-,4,2),[1]}_{qq\to qq}=&\,(i\pi)^2r_{\Gamma}^4\left(N_c^2-4\right) \sqrt{N_c^2-1}\frac{1}{256}  \left(\frac{1}{\eps^4}-\frac{2\hat{\zeta}_3}{\eps}+\mathcal{O}(\eps)\right){\cal M}^{{\rm tree},[8]}_{qq\to qq},\label{eq:M42qqSinglet}
\end{align}
\end{subequations}
and the decuplet of $gg$ scattering
\begin{subequations}
\begin{align}
    \mathcal{\hat{M}}^{(-,2,0),[10+\bar{10}]}_{gg\to gg} =&\,-(i\pi)^2r_{\Gamma}^2\frac{\sqrt{N_c^2-4}}{\sqrt{2}}S^{(2)}(\eps){\cal M}^{{\rm tree},[8_a]}_{gg\to gg},\label{eq:M20ggDecuplet}\\
    \mathcal{\hat{M}}^{(-,3,1),[10+\bar{10}]}_{gg\to gg} =&\,-(i\pi)^2r_{\Gamma}^3\frac{\sqrt{N_c^2-4}}{\sqrt{2}}\frac{N_c}{12} \left(\frac{1}{\eps^3} + 38  \hat\zeta_3+\mathcal{O}(\eps^2)\right){\cal M}^{{\rm tree},[8_a]}_{gg\to gg},\label{eq:M31ggDecuplet}\\
    \mathcal{\hat{M}}^{(-,4,2),[10+\bar{10}]}_{gg\to gg} =&\,(i\pi)^2r_{\Gamma}^4\frac{\sqrt{N_c^2-4}}{\sqrt{2}}\bigg[\frac{5}{384\eps^4}\left( N_c^2+\frac{12}{5}\right)\nonumber\\&\hskip3.5cm-\frac{91 \hat{\zeta}_3}{192\eps} \left( N_c^2+\frac{1212}{91}\right)+\mathcal{O}(\eps)\bigg]{\cal M}^{{\rm tree},[8_a]}_{gg\to gg}.\label{eq:M42ggDecuplet}
\end{align}
\end{subequations}
The other components have definite even signature and so do not contribute to the odd amplitudes.
\section{The soft anomalous dimension in the high-energy limit}
\label{sec:SAD3loopa}
 This appendix provides the known values of the corrections $\dd$ to the soft anomalous dimension dipole formula in the high energy limit. $\dd$ start at three loops.
  The soft anomalous dimension in the high-energy limit is introduced in eq. \eqref{eq:gammaFactorIntro}, and at 
  three-loop order it is
  \beq
\mathbf{\Gamma}^{(3)}_{ij\to ij}\left(L,\frac{-t}{\lambda^2}\right) = \frac12 \gamma_{K}^{(3)}\left[L\tts + i\pi\tsu\right] +\Gamma_i^{(3)}\left(\frac{-t}{\lambda^2}\right)+\Gamma_j^{(3)}\left(\frac{-t}{\lambda^2}\right) +\sum_{m=0}^{2}\dd^{(3,m)} L^m
\eeq
with $\dd^{(3,m)}$ being corrections to the dipole formula at three loops.
The even signature part of the soft anomalous dimension at NLL is 
two-loop exact~\cite{Lipatov:1976zz,Kuraev:1976ge,Caron-Huot:2017zfo}, that is, the terms $\dd^{(+,\ell,\ell-1)}$ vanish at three loops and above.
 The corrections at three loops were calculated explicitly in ref.~\cite{Almelid:2015jia}, after which projections onto definite signatures and a decomposition in a basis of colour operators suitable for the Regge limit were found in ref.~\cite{Caron-Huot:2017fxr}. They read
 \begin{subequations}\label{Delta3_results}
     \begin{align}
     \label{eq:delta32+}
     \dd^{(-,3,2)}&=\text{Im}\left[\dd^{(3,2)}\right]=0\\
     \label{eq:delta32-}
     \dd^{(+,3,2)}&=\text{Re}\left[\dd^{(3,2)}\right]=0\\
     \label{eq:delta31-} \dd^{(-,3,1)}&=i \pi \Big[\tts,[\tts,\tsu]\Big] \frac{1}{4} \zeta_3\\
     \label{eq:delta31+} \dd^{(+,3,1)}&=\text{Re}\left[\dd^{(3,1)}\right]=0\\
              \label{eq:delta30min} \dd^{(-,3,0)}&=i \pi \Big[\tts,[\tts,\tsu]\Big] \frac{11}{4} \zeta_4
     \\
      \label{eq:delta30+}
     \dd^{(+,3,0)}&=\frac{1}{4} \Big[ \tsu, [\tsu,\tts]\Big]\Big(4\zeta_2 \zeta_3-\zeta_5\Big)- \frac{f^{(3)}}{2}\bigg\{ - \frac{5}{8} C_A^2 \tts+ f^{abe}f^{cde} \nn\\\quad\mbox{}& \times \Big[ \{\T_t^a, \T_t^d\}\left(\{\T_{s-u}^b, \T_{s-u}^c\}+\{\T_{s+u}^b, \T_{s+u}^c\}\right)+ \{\T_{s-u}^a,\T_{s-u}^d\}\{\T^b_{s+u}, \T_{s+u}^c\}\Big] \bigg\}.
     \end{align}
     \end{subequations}
where $f^{(3)} = \frac{1}{4} (\zeta_2 \zeta_3 + \zeta_5)$.
     The generators in eq.~(\ref{eq:delta30+}) are defined as 
     \begin{align}
     \T_{s-u}^a\equiv \frac{1}{\sqrt{2}}\left(\T_s^a-\T_u^a\right)\hspace{1cm}\T_{s+u}^a\equiv \frac{1}{\sqrt{2}}\left(\T_s^a+\T_u^a\right).
         \end{align}
These results are completely general: they require no direct action on the tree amplitude. An alternative expression for the colour factor multiplying 
$f^{(3)}$ in eq.~(\ref{eq:delta30+}) is derived in appendix~\ref{sec:ReggeBasisA}, and given by  eq.~\eqref{eq:falphfin}. The latter is only valid  upon acting directly upon the octet in colour space, $\T^a_i\T^a_j$.

In section \ref{subsec:sumSADreg} we summarise the state-of-the-art knowledge of the soft anomalous dimension at four-loop order in the high-energy limit. It may be written as 
\beq
\label{Gamma4_appendix_E}
\mathbf{\Gamma}^{(4)}_{ij\to ij}\left(L,\frac{-t}{\lambda^2}\right) = \frac12\gamma_{K}^{(4)}\left[L\tts + i\pi\tsu\right] +\Gamma_i^{(4)}\left(\frac{-t}{\lambda^2}\right)+\Gamma_j^{(4)}\left(\frac{-t}{\lambda^2}\right) +\sum_{m=0}^{3}\dd^{(4,m)} L^m.
\eeq
At NLL, there is a tower of signature-odd contributions to the soft anomalous dimension (i.e. signature-even amplitude) emanating from two-Reggeon-exchange ladder diagrams, which start at four loops~\cite{Caron-Huot:2013fea} and they have been determined to all orders~\cite{Caron-Huot:2017zfo,Caron-Huot:2020grv}. 
In the present paper we computed the signature-even component of the four-loop soft anomalous dimension through NNLL, complementing information from lower logarithmic accuracy in refs.~\cite{Caron-Huot:2017zfo,Caron-Huot:2013fea,Falcioni:2020lvv}. The results are:
 \begin{subequations}
 \label{Delta4_results}
     \begin{align}
     \label{eq:delta43plusa}
     \dd^{(+,4,3)}&=\text{Re}\left[\dd^{(4,3)}\right]=0\\
     \dd^{(-,4,3)}&=-i\pi\frac{\zeta_3}{24} \Big[\tts,[\tts,\tsu]\Big]\tts \label{dd-43}\\
     \dd^{(+,4,2)}&=\zeta_2 \zeta_3 \bigg(\frac{d_{AA}}{N_A} -\frac{C_A^4}{24}-\frac{1}{4}\tts[(\tsu)^2,\tts]+\frac{3}{4}[\tsu,\tts]\tts\tsu\bigg)\,.
     \end{align}
     \end{subequations}
Note that eqs.~(\ref{eq:delta31-}),~(\ref{eq:delta30min}) and~(\ref{dd-43}) contain a single power of $i\pi \tsu$, consistent with the odd signature of the soft anomalous dimension. The odd contribution $\dd^{(-,4,2)}$ is not yet known, and neither are the subleading logarithms $\dd^{(4,1)}$ and $\dd^{(4,0)}$.
 
 \section{The cusp and the soft anomalous dimensions}
\label{sec:cuspappendix}

In this appendix we study the relation between the soft anomalous dimension for $n$-parton scattering in general kinematics, ${\bf \Gamma}_n$, and the lightlike cusp anomalous dimension $\Gamma^{\rm{cusp}}_i$. This relation is governed by the collinear anomaly constraint of eq.~\eqref{eq:BNdipinh} and concerns only terms that are linear in the logarithm $l_{ij}=\log \frac{-s_{ij}}{\lambda^2}$. All other terms in~${\bf \Gamma}_n$ depend on the kinematics only via conformally-invariant cross ratios (CICRs) and will not be discussed here.

We have seen two particular solutions to the inhomogeneous equations~\eqref{eq:BNdipinh}. The first, given by eq.~\eqref{eq:mans_main_text}, has a trivial colour structure, while the second, given by the terms that depend on the scale in eq.~\eqref{eq:SADintrogamma}, namely  $\bm{\Gamma}_n^{\rm{dip}}$ and $\bm{\Gamma}_{n, \rm{Q4T-2,3L}}$, has a non-trivial colour structure, and it can be shown to satisfy eq.~\eqref{eq:BNdipinh} using colour conservation. 
We will verify that the difference between these two particular solutions is a function of conformally-invariant cross ratios (CICRs). Specialising to four-parton scattering and using colour conservation, the latter will be expressed in terms of colour operators that involve generators of just two of the four partons, a representation that is convenient for deriving the Regge limit of eqs.~(\ref{eq:BN4sig+}) and~(\ref{eq:BN4sig-2}), which is the final goal of this appendix. 

\paragraph{Relating particular solutions of the collinear anomaly constraints.}

The soft anomalous dimension must satisfy the collinear anomaly constraint  \cite{Becher:2009cu,Becher:2009qa,Gardi:2009qi,Gardi:2009zv} of eq.~\eqref{eq:BNdipinh}, namely the inhomogeneous differential equations
\beq
\label{eq:BNdipinha}
\sum_{j\neq i}\frac{d{\bf \Gamma}_n(\{s_{kl}\},\lambda, \alpha_s)}{d(l_{ij})}=\Gamma^{\rm{cusp}}_i(\alpha_s)\,,
\eeq
for all $i$,  where the derivatives on the left-hand side are taken with respect to the logarithms $l_{ij}= \log \frac{-s_{ij}}{\lambda^2}$ in eq.~\eqref{eq:lij}, while the right-hand side is the lightlike cusp anomalous dimension~\cite{Korchemsky:1991zp,Korchemsky:1987wg} for particle $i$. The latter has a simple dependence on the representation of the Wilson line. In particular, it is proportional to the quadratic Casimir through three loops, a property known as \emph{Casimir scaling}, and it acquires additional dependence on the representation through a quartic Casimir starting at four loops~\cite{Boels:2017ftb,Boels:2017skl,Moch:2017uml,Grozin:2017css}, as stated in eq.~\eqref{cuspAD}, namely 

\begin{equation}
\label{eq:gammacuspa}
\Gamma_i^{\rm{cusp}}(\alpha_s)=\frac{1}{2}\gamma_K(\alpha_s)C_{i}+\sum_Rg_{R}(\alpha_s)\frac{d_{R R_i}}{N_{R_i}} + \mathcal{O}(\alpha_s^5),
\end{equation} 
where $C_{i}$ is the quadratic Casimir and $\frac{d_{R R_i}}{N_{R_i}}$ the  quartic Casimir from eq.~\eqref{eq:drri}. Explicit expressions for $\frac{d_{AA}}{N_{A}},\frac{d_{AF}}{N_{F}}$ are given in eq.~\eqref{eq:dAA}. In appendix \ref{AD-coefficient_Appendix}, $\gamma_K$ is given up to three loops in eq.~\eqref{gammaK} and $g_F$ and $g_A$ are provided at four-loop order in eq. \eqref{g_R_values}.

In section~\ref{subsec:introSAD} we encountered two different particular solutions to the inhomogeneous equations (\ref{eq:BNdipinha}): the first~\cite{Mark}, in eq.~(\ref{eq:mans_main_text}), true for $n\geq 3$, is expressed directly in terms of $\Gamma_i^{\rm{cusp}}(\alpha_s)$
\begin{align}
\begin{split}
\label{eq:mans}
\Gamma_n'\left(\left\{\frac{s_{ij}}{\lambda^2}\right\}, \alpha_s\right)&\equiv\frac{1}{2(n-1)}\sum_{i=1}^{n} \sum_{j \neq i}^{n} l_{ij}\left(\Gamma^{\rm{cusp}}_i(\alpha_s) +\Gamma^{\rm{cusp}}_j(\alpha_s)-\frac{1}{(n-2)} \sum_{k \neq i, j}^{n} \Gamma_k^{\text{cusp}}(\alpha_s)\right),
\end{split}
\end{align}
and thus naturally incorporates all higher Casimirs. The second was presented in ref.~\cite{Becher:2019avh}, starting with the sum-over-dipoles formula and including quartic operators coupled to two or three partons, so as to satisfy (\ref{eq:BNdipinha}) subject to colour conservation. It reads: 
\begin{align}
\begin{split}
\label{eq:BNexp}
\bm{\Gamma}^{\rm{lin.}}_n(\{s_{ij}\},\lambda, \alpha_s)
   \equiv &\left(\bm{\Gamma}_n^{\rm{dip}}(\{s_{ij}\},\lambda, \alpha_s)- \sum_i^n\,\gamma_i(\alpha_s)\right)
   \,+\, \bm{\Gamma}_{n, \rm{Q4T-2,3L}}(\{s_{ij}\},\lambda, \alpha_s)\\=&-\frac{1}{4} \gamma_{K}(\alpha_s)\sum_{(i,j)}\,\T_i\cdot\T_j\,l_{ij}
   \\&
   -\frac{1}{2}\sum_R g_R(\alpha_s) \Bigg[ \sum_{(i,j)}\,\big( \bm{\bm{{\cal D}}}_{iijj}^R + 2 \bm{\bm{{\cal D}}}_{iiij}^R \big)  l_{ij}
   + \sum_{(i,j,k)} \bm{\bm{{\cal D}}}_{ijkk}^R\,l_{ij} \Bigg]\, ,
\end{split}
\end{align}
where 
\beq
\label{eq:Ddef}
  \bm{\bm{\mathcal{D}}}_{i j k l}^{R}=\frac{1}{4!}\sum_{\sigma \in S_4} \operatorname{Tr}_R\left[T^{\sigma(a)} T^{\sigma(b)}T^{\sigma(c)}T^{\sigma(d)}\right] \T_{i}^{a} \T_{j}^{b} \T_{k}^{c} \T_{l}^{d}. 
\eeq 
When all the partonic indices $\{i,j,k,l\}$ are the same, it is a quartic Casimir written in eq.~\eqref{eq:drri}. 

At this point we have two particular solutions to eq.~(\ref{eq:BNdipinha}), as can be checked by substitution and using colour conservation. Crucially, both these solutions are strictly linear in the logarithm $l_{ij}$ and we can directly compare them and define their difference as
\beq
\mathbf{\delta\Gamma}_n^{\rm{lin.}} \left(\{\beta_{ijkl}\}, \alpha_s\right)\equiv 
\mathbf{\Gamma}^{\rm{lin.}}_n(\{s_{ij}\},\lambda, \alpha_s)-\Gamma_n'\left(\left\{\frac{s_{ij}}{\lambda^2}\right\}, \alpha_s\right),
\eeq
which is linear in the logarithms, and at the same time only depends on combinations that are rescaling invariant, satisfying the homogeneous version of eq.~(\ref{eq:BNdipinha}). At $n=3$ we have $\mathbf{\delta\Gamma}_3^{\rm{lin.}}=0$, while for four or more partons $\mathbf{\delta\Gamma}_n^{\rm{lin.}}$ is nonzero with its form given below. Note that in this section we consider only linear terms and ignore homogeneous solutions that could have more complicated dependence on the velocities. 

Computing the difference between eq.~\eqref{eq:BNexp} and eq.~\eqref{eq:mans} and using colour conservation we obtain:
\beq
\label{eq:deltacicr}
\mathbf{\delta\Gamma}^{\rm{lin.}}_n \left(\{\beta_{ijkl}\}, \alpha_s\right)=-\sum_{(i,j,k,l)}\!\frac{\beta_{ijkl}}{n-2}\Bigg[\frac{\gamma_{K}(\alpha_s)}{
4(n-1)}\T_i\cdot \T_j+\sum_R\frac{g_{R}(\alpha_s)}{2} \left(\frac{1}{(n-1)} \bm{\mathcal{D}}^R_{iiij}+ \bm{\mathcal{D}}^R_{ijkk}\right)\Bigg],
\eeq
for $n\geq 4$. 
The first term in the square brackets is related to the quadratic Casimir. The second, which is related to the quartic Casimir, will be denoted by $\mathbf{\delta\Gamma}^{\rm{lin.}}_{n, {\rm Q}}$.

\paragraph{Four-parton scattering.}
     Specialising to four-parton scattering, $n=4$, in 
     eq.~(\ref{eq:deltacicr}), we find that further simplification is possible owing to colour conservation. The following three ways of expressing
     the quartic part of eq.~(\ref{eq:deltacicr}) are all equivalent, each using a combination of just two of the three types of colour operators $\{\bm{\mathcal{D}}^R_{i ijj}, \bm{\mathcal{D}}^R_{i iij}, \bm{\mathcal{D}}^R_{i jkk}\}$:
\beq
\label{eq:DeltaCICRn4}
\mathbf{\delta\Gamma}^{\rm{lin.}}_{4, {\rm Q}} \left(\{\beta_{ijkl}\}, \alpha_s\right)=
\begin{cases}
-\sum\limits_{R}g_R(\alpha_s)\sum\limits_{(i,j,k,l) \in S_4} \beta_{ijkl} \left(\frac{1}{12}\bm{\mathcal{D}}^R_{iiij}+\frac{1}{4}\bm{\mathcal{D}}^R_{i jkk}\right)\\[0.8cm]
-\sum\limits_Rg_R(\alpha_s)\sum\limits_{(i,j,k,l) \in S_4} \beta_{ijkl} \left(-\frac{1}{12}\bm{\mathcal{D}}^R_{iijj}+\frac{1}{3}\bm{\mathcal{D}}^R_{i jkk}\right)
\\[0.8cm]
-\sum\limits_Rg_R(\alpha_s)\sum\limits_{(i,j,k,l) \in S_4} \beta_{ijkl} \left(\frac{1}{3}\bm{\mathcal{D}}^R_{iiij}+\frac{1}{4}\bm{\mathcal{D}}^R_{i ijj}\right).
\end{cases}
\eeq
Of these three we favour using the third, where each term depends on colour generators of just two of the four legs. 
With this, we can then express eq.~\eqref{eq:BNexp} for four partons  as
\begin{align}
\label{gammalin4}
\mathbf{\Gamma}_4^{\rm{lin.}}\left(\left\{\frac{s_{ij}}{\lambda^2}\right\}, \alpha_s\right)&=
\Gamma_4'\left(\left\{\frac{s_{ij}}{\lambda^2}\right\}, \alpha_s\right) + \mathbf{\delta\Gamma}_4^{\rm{lin}} \left(\{\beta_{ijkl}\}, \alpha_s\right)\nn\\&=\frac{1}{6}\sum_{i=1}^{4} \sum_{j \neq i}^{4} l_{ij}\left(\Gamma^{\rm{cusp}}_i(\alpha_s) +\Gamma^{\rm{cusp}}_j(\alpha_s)-\frac{1}{2} \sum_{k \neq i, j}^{4} \Gamma_k^{\text{cusp}}(\alpha_s)\right)\nn\\&\quad\mbox{}
-\sum_{(i,j,k,l) \in S_4}\beta_{ijkl}\Bigg[\frac{\gamma_{K}(\alpha_s)}{
24}\T_i\cdot \T_j+\sum_Rg_R(\alpha_s)\left(\frac{1}{3}\bm{\mathcal{D}}^R_{iiij}+\frac{1}{4}\bm{\mathcal{D}}^R_{i ijj}\right)\Bigg].
\end{align}
The first line is proportional to the unit matrix in colour space while the second line involves non-trivial colour operators. In appendix~\ref{sec:ReggeBasisA}, each part of this expression will be translated to a Regge-limit colour basis and the Regge limit is then taken.

\paragraph{Linear contributions to the soft anomalous dimension in the Regge limit.} 
Let us now summarise the expressions for the cusp-related contributions in the soft anomalous dimension in the Regge limit using eq.~(\ref{gammalin4}) with results 
from appendix~\ref{sec:ReggeBasisA}.

Using the result from eq.~\eqref{gammalin4ifin}, the first line of eq.~\eqref{gammalin4} in the Regge limit is 
        \begin{align}
     \label{gamma4cusp}
     \Gamma_4'\left(L, \frac{-t}{\lambda^2},  \alpha_s\right)=\left( \frac{2}{3}L+ \log\frac{-t}{\lambda^2}\right)\left(\Gamma^{\rm{cusp}}_i(\alpha_s) +\Gamma^{\rm{cusp}}_j(\alpha_s)\right).
     \end{align}
     The subscript $i$ corresponds to the projectile (partons 1 and 4), while $j$ corresponds to the target (partons 2 and 3). 
     The whole expression is even under the exchange of $s \leftrightarrow u$.

Next we consider the second line of eq.~\eqref{gammalin4} in the Regge limit. Using results from eqs.~\eqref{deltaquadfin} and~\eqref{gammalinquartR}, and then separating the expression into signature-even and odd parts, we obtain:
\begin{align}
\begin{split}
\label{eq:mansfinib}
\delta\mathbf{\Gamma}^{\rm{lin.}}_4(L, \alpha_s)&=L\Bigg\{\frac{\gamma_K(\alpha_s)}{6}\left( 3 \tts - 2 C_i- 2 C_j\right)+ \sum_R\frac{g_{R}(\alpha_s)}{3} 
\Big(2\bm{\mathcal{D}}_{tttt}^R-\bm{\mathcal{D}}_{ssss}^R- \bm{\mathcal{D}}_{uuuu}^R\Big)\Bigg\}\\&\quad\mbox{}+\frac{i \pi}{2} \Bigg\{ \gamma_K(\alpha_s) \tsu + \sum_R
g_{R}(\alpha_s) \Big(\bm{\mathcal{D}}_{ssss}^R- \bm{\mathcal{D}}_{uuuu}^R\Big)\Bigg\}\,,
\end{split}
\end{align}
where $\tsu=\frac{1}{2}(\Ts-\Tu)$ and where we use the notation in eq.~(\ref{eq:dppppp}) for the quartic operators associated with the three channels. 
The first line in eq.~(\ref{eq:mansfinib}) is real and signature-even, while the second line is imaginary and signature-odd.

To proceed it is useful to define a cusp anomalous dimension operator associated with each channel in the following way:
\beq
\label{Gamma_cusp_channel_op}
\bm{\Gamma}_p^{\rm{cusp}}(\alpha_s)\equiv\frac{1}{2} \gamma_K(\alpha_s) \T_p^2+ \sum_R g_R(\alpha_s) \bm{\mathcal{D}}^R_{pppp}+ {\cal O}(\alpha_s), 
\eeq
 where $p \in \{s,t,u\}$.  We can then write eq.~\eqref{eq:mansfinib} as
 \begin{align}
\begin{split}
\label{eq:mansfinic}
\delta\mathbf{\Gamma}^{\rm{lin.}}_4(L, \alpha_s)&=L\bigg(\bm{\Gamma}_t^{\rm{cusp}}(\alpha_s)-\frac{\gamma_K(\alpha_s)}{3}\left( C_i+ C_j\right) -\sum_R\frac{g_{R}(\alpha_s)}{3} 
\Big(\bm{\mathcal{D}}_{ssss}^R+ \bm{\mathcal{D}}_{uuuu}^R+\bm{\mathcal{D}}_{tttt}^R\Big)\bigg)\\&\quad\mbox{}+\frac{i \pi}{2} \Big(\bm{\Gamma}_s^{\rm{cusp}}(\as)-\bm{\Gamma}_u^{\rm{cusp}}(\as)\Big).
\end{split}
\end{align}
Combining eqs.~(\ref{gamma4cusp}) and~(\ref{eq:mansfinic}) we observe that the 
quadratic Casimir terms multiplying the high-energy logarithm $L$ cancel. 
Separating the contributions by signature $\mathbf{\Gamma}_4^{\rm{lin.}} = \mathbf{\Gamma}_4^{\rm{lin.}\, (+)}+ \mathbf{\Gamma}_4^{\rm{lin.}\, (-)}$, we find
 \begin{subequations}
 \label{eq:mansfiniv}
\begin{align}
\label{eq:mansfiniv:real}
\mathbf{\Gamma}_4^{\rm{lin.}\, (+)}\left(L, \frac{-t}{\lambda^2},  \alpha_s\right)&=L \bm{\Gamma}_t^{\rm{cusp}}(\as)
\,+\, \log \frac{-t}{\lambda^2}\left(\Gamma^{\rm{cusp}}_i(\as)+\Gamma^{\rm{cusp}}_j(\as)\right) \\&+L\sum_R\frac{g_R(\alpha_s)}{3}  \left(-\bm{\mathcal{D}}^R_{tttt}-\bm{\mathcal{D}}^R_{ssss}-\bm{\mathcal{D}}^R_{uuuu}+2\left(\frac{d_{RR_i}}{N_{R_i}}+\frac{d_{RR_j}}{N_{R_j}}\right)\right)\nn\\
\mathbf{\Gamma}_4^{\rm{lin.}\, (-)}\left(\alpha_s\right)&= \frac{i \pi}{2} \Big(\bm{\Gamma}_s^{\rm{cusp}}(\as)-\bm{\Gamma}_u^{\rm{cusp}}(\as)\Big).\,
\end{align}
\end{subequations}
We thus see that quadratic Casimir coefficient $\gamma_K(\alpha_s)$ does not anymore appear in isolation, but only as part of the complete channel-related cusp anomalous dimension operator of eq.~(\ref {Gamma_cusp_channel_op}). Note that the quartic coefficient $g_R(\alpha_s)$ does feature on its own.
The two expressions in eq.~(\ref{eq:mansfiniv}) constitute part of the final expressions for the Regge-limit $\mathbf{\Gamma}_4$ in eqs.~(\ref{eq:BN4sig+}) and~(\ref{eq:BN4sig-2}), where additional contributions involving the quartic Casimir arise, which naturally combine with the second line in eq.~(\ref{eq:mansfiniv:real}).

Upon considering $\mathbf{\Gamma}_4^{\rm{lin.}}$ to be acting on the tree amplitude, and specialising to the adjoint representation $R=A$, we may use results from eqs.~(\ref{gammalin4ifin}), (\ref{deltaquadfin}), (\ref{gLtermAfin}) and (\ref{eq:gpiAfin}), to obtain explicit expressions in terms of quartic Casimirs and commutators: 
\begin{subequations}
 \begin{align}
 \label{gammalinA+}
\mathbf{\Gamma}_{4,A}^{\rm{lin.}(+)}\left(L, \frac{-t}{\lambda^2} ,\alpha_s\right)\mtree&=\Bigg\{
L {\bf \Gamma}_t^{\rm{cusp}}\,+\, 
\log \frac{-t}{\lambda^2}\left(\Gamma^{\rm{cusp}}_i+\Gamma^{\rm{cusp}}_j\right)\\&\qquad -L \frac{g_A(\alpha_s)}{6} \bigg(2\left(\frac{d_{AA}}{N_A} -\frac{C_A^4}{24}\right) -\frac{1}{2}\tts[(\tsu)^2,\tts]\nn\\&\qquad +\frac{3}{2}[\tsu,\tts]\tts\tsu+\frac{C_A}{2} \Big[\tsu,[\tsu,\tts]\Big]\bigg)\Bigg\}\mtree \nn
     \end{align}
    
  \begin{align}
     \label{gammalinA-}
     \mathbf{\Gamma}_{4,A}^{\rm{lin.}(-)}\left(\alpha_s\right)\mtree&=
    \\&\hspace{-2.5cm} 
     \frac{i \pi}{2} \Bigg\{
     \bigg[\frac{2}{C_A} \Gamma_A^{\rm{cusp}}(\as)  \,+\, 4g_A(\as)\left(\frac{d_{AR_i}}{C_iN_{R_i}}+\frac{d_{AR_j}}{C_jN_{R_j}}-\frac{d_{AA}}{2C_AN_A}- \frac{C_A^3}{16}\right)\bigg]\tsu\nn\\&\hspace{-2.5cm}-  \frac{g_A(\as)}{8}
     \bigg(   3\Big[\tts,\big[\tts, [\tts,\tsu]\big]\Big]\nn+\Big[\tts,[\tts,\tsu]\Big]\tts-3 \tts[ \tts,\tsu]\tts\bigg)\Bigg\}\mtree.\nn 
  \end{align}
  \end{subequations}

In these expressions we used the complete cusp anomalous dimension, rather than the coefficient of the quadratic component, $\gamma_K(\as)$. 
Recalling eq.~(\ref{eq:dAA}), we see that the remaining terms proportional to the quartic component $g_A(\as)$ are manifestly subleading in the large-$N_c$ limit. Thus, in the planar limit, the leading contributions come exclusively from the $\Gamma^{\rm{cusp}}$ terms and the collinear anomalous dimensions.

\section{Converting colour structures to the Regge-limit basis}
\label{sec:ReggeBasisA}
In this section, we study contributions to the soft anomalous dimension up to four loops in eq.~\eqref{eq:SADintrogamma} at $\mathcal{O}(L^0)$ and $\mathcal{O}(L)$, which have not been discussed in section~\ref{sec:factor2to2}. We will translate the respective colour structures to the Regge-limit basis and then take the Regge limit.  
We will also elaborate on how the $\bm{\Gamma}_{\rm{5T-4L}}$ colour structures in a Regge-limit basis were derived. 

First let us revisit eq.~\eqref{eq:SADintrogamma}, following the analysis we conducted in appendix~\ref{sec:cuspappendix} of the terms that are linear in the logarithms $l_{ij}=\log\frac{-s_{ij}}{\mu^2}$. 
These linear terms can be expressed in terms of the particular solution to eq.~\eqref{eq:BNdipinh} presented in eq.~\eqref{eq:mans_main_text}, plus linear homogeneous terms provided in eq.~\eqref{eq:deltacicr}. We can thus write eq.~\eqref{eq:SADintrogamma} as 
         \begin{align}
\begin{split}
\label{eq:SADintrogammaa} 
\mathbf{\Gamma}_4\left(\{s_{ij}\},\lambda, \alpha_s(\lambda^2)\right) 
& = \Gamma'_4\left(\{s_{ij}\},\lambda, \alpha_s\right) + \delta\mathbf{\Gamma}^{\rm{lin}}_4\left(\{\beta_{ijkl}\}, \alpha_s\right) 
+\mathbf{\Gamma}_{4,\rm 4T-3L}\left(\alpha_s\right) \\[0.1cm]&
+\sum_{m=1}^4\,\gamma_m(\alpha_s)+\mathbf{\Gamma}_{4,\rm 4T-4L}\left(\{\beta_{ijkl}\}, \alpha_s\right) 
+\mathbf{\Gamma}_{4,\rm Q4T-4L}\left(\{\beta_{ijkl}\}, \alpha_s\right)\\[0.1cm]
&
+\mathbf{\Gamma}_{4,\rm 5T-4L}\left(\{\beta_{ijkl}\}, \alpha_s\right).
\end{split}
\end{align}
 The first term $\Gamma'_4\left(\{s_{ij}\},\lambda, \alpha_s\right)$ is the only one with scale dependence and it is by itself a particular solution to the inhomogeneous equations~\eqref{eq:BNdipinh}. It is a unit matrix in colour space (hence it is not written in 
 boldface notation). It contributes at ${\cal O }(L)$ and ${\cal }(L^0)$ with a simple expression in the Regge limit derived below in eq.~\eqref{gammalin4ifin}. 
The second term $\delta\mathbf{\Gamma}^{\rm{lin}}_4\left(\{\beta_{ijkl}\}, \alpha_s\right)$ consists of non-trivial colour operators; it depends linearly on the logarithm of CICRs, and hence contributes at ${\cal O }(L)$ and ${\cal O}(L^0)$.
It comprises two contributions: one is written in terms of colour dipoles, while the other involves quartic Casimir operators. The former was evaluated in the Regge limit in ref.~\cite{DelDuca:2011ae} and it is re-derived here in eq.~\eqref{deltaquadfin}, while the latter is given in the Regge limit in eq.~(\ref{gammalinquartR}). The third term on the first line of eq.~(\ref{eq:SADintrogammaa}), $\mathbf{\Gamma}_{4,\rm 4T-3L}\left(\alpha_s\right)$, is independent of the kinematics, and hence only contributes at ${\cal O}(L^0)$. An expression for its colour structures in terms of operators $\mathbf{T}^a_{s-u}$, $\mathbf{T}^a_{s+u}$ and $\mathbf{T}^a_{t}$ was given in ref.~\cite{Caron-Huot:2017fxr}. Here we derive an alternative expression in the Regge-limit basis, eq.~(\ref{eq:falphfin}), involving quartic Casimirs related to the representations of the scattered particles.
The collinear anomalous dimension $\gamma_m$ remains unchanged in the Regge limit and it only contributes at ${\cal O} (L^0)$. The remaining two terms on the second line of eq.~(\ref{eq:SADintrogammaa}) were discussed in detail in section~\ref{sec:factor2to2} and will not be considered further here. Finally, the term on the last line of eq.~(\ref{eq:SADintrogammaa}) was briefly discussed in section~\ref{sec:factor2to2} with further details given in the final part of this appendix. 

\subsection*{\underline{\texorpdfstring{$\Gamma_4'$}{G4} in the Regge limit}}
\label{subsec:g4} This linear expression is a particular inhomogeneous solution to eq.~\eqref{eq:BNdipinh} and  is given in eq.~\eqref{eq:mans}. The cusp anomalous dimension is discussed in the previous appendix and its form is in eq.~\eqref{cuspAD_Sec_anom_dim}. For four partons, the particular inhomogenous solution is 
\begin{align}
\label{gammalin4i}
\Gamma_4'\left(\left\{\frac{s_{ij}}{\lambda^2}\right\}, \alpha_s\right)&=\frac{1}{6}\sum_{i=1}^{4} \sum_{j \neq i}^{4} l_{ij}\left(\Gamma^{\rm{cusp}}_i(\alpha_s) +\Gamma^{\rm{cusp}}_j(\alpha_s)-\frac{1}{2} \sum_{k \neq i, j}^{4} \Gamma_k^{\text{cusp}}(\alpha_s)\right).
\end{align}
Performing the sum and using the fact that for $2 \to 2$ scattering the logarithms are expressed in terms of Mandelstam variables
\begin{align}
\label{eq:logrel}
l_{34}&=l_{12}=\log\frac{se^{-i\pi}}{\lambda^2};\qquad
l_{24}=l_{13}=\log\frac{-u}{\lambda^2};\qquad
l_{23}=l_{14}=\log\frac{-t}{\lambda^2},
    \end{align}
    then eq.~\eqref{gammalin4i} becomes 
     \begin{align}
     \label{gammalin4ii}
     \Gamma_4'\left(\left\{\underline{s}\right\}, \lambda, \alpha_s\right)&= \frac{1}{6}\left(\log\frac{se^{-i\pi}}{\lambda^2}+\log\frac{-u}{\lambda^2}+ \log\frac{-t}{\lambda^2}\right)\bigg(\Gamma^{\rm{cusp}}_1(\alpha_s)+ \Gamma^{\rm{cusp}}_2(\alpha_s)\nn\\&\quad\mbox{} + \Gamma^{\rm{cusp}}_3(\alpha_s)+\Gamma^{\rm{cusp}}_4(\alpha_s)\bigg),
     \end{align}
      where $\left\{\underline{s}\right\}=\{s,t,u\}$.
      The subscript of $i$ corresponds to partons 1 and 4 of the projectile, while $j$ corresponds to partons 2 and 3 of the target. 
    Using target-projectile notation, we therefore make the following identification
    \beq
  \Gamma^{\rm{cusp}}_{R_1} =\Gamma^{\rm{cusp}}_{R_4}= \Gamma^{\rm{cusp}}_i\hspace{1cm} \text{and}\,\,\hspace{1cm} \Gamma^{\rm{cusp}}_{R_2} =\Gamma^{\rm{cusp}}_{R_3} = \Gamma^{\rm{cusp}}_j,
  \eeq
so eq.~\eqref{gammalin4ii} becomes
     \begin{align}
      \label{gammalin4iii}
     \Gamma_4'\left(\left\{\underline{s}\right\}, \lambda, \alpha_s\right)= \frac{1}{3}\left(\log\frac{se^{-i\pi}}{\lambda^2}+\log\frac{-u}{\lambda^2}+ \log\frac{-t}{\lambda^2}\right)\left(\Gamma^{\rm{cusp}}_i(\alpha_s) +\Gamma^{\rm{cusp}}_j(\alpha_s)\right).
     \end{align}
     Eq.~(\ref{gammalin4iii}) is even under the exchange of $s \leftrightarrow u$.
      Upon taking the Regge limit of 
      \[s \simeq-u \gg -t,\]  the logarithms are expressed in terms of the signature-even logarithm $L$ of eq. \eqref{eq:siglog} up to power-suppressed terms, getting 
\begin{align}
\begin{split}
\label{eq:lijRegi}
\log\frac{-u}{\lambda^2} \simeq \log\frac{s}{\lambda^2}\simeq L+\log\frac{-t}{\lambda^2}+\frac{i \pi}{2}.
 \end{split}
\end{align} 
In the Regge limit, eq.~\eqref{gammalin4iii} is therefore expressed as 
 \begin{align}
     \label{gammalin4ifin}
     \Gamma_4'\left(L, \frac{-t}{\lambda^2}, \alpha_s\right)=\left( \frac{2}{3}L+ \log\frac{-t}{\lambda^2}\right)\left(\Gamma^{\rm{cusp}}_i(\alpha_s) +\Gamma^{\rm{cusp}}_j(\alpha_s)\right).
     \end{align}
     This expression is signature-even and real. It is a unit matrix in colour space and contains the complete cusp anomalous dimensions for partons $i$ and $j$.
 
     \subsection*{\underline{\texorpdfstring{$\delta{\bm{\Gamma}}^{\rm {lin.}}$}{dG} in the Regge limit}}\label{subsec:deltaG}
     We now translate the colour structures multiplying CICRs in $ \delta\bm{\Gamma}^{\rm{lin.}}$ from eq.~\eqref{gammalin4} 
      \begin{align}
       \label{deltalinapp}
     \delta\bm{\Gamma}_4^{\rm{lin.}}\left(\{\beta_{ijkl}\}, \alpha_s\right)&
     =\mathbf{\delta\Gamma}_{4,{\rm 2T-2L}}^{\rm{lin.}}\left(\{\beta_{ijkl}\}, \alpha_s\right)+\mathbf{\delta\Gamma}_{4,{\rm Q}}^{\rm{lin.}}\left(\{\beta_{ijkl}\}, \alpha_s\right)\nn\\&=-\sum_{(i,j,k,l) \in S_4}\beta_{ijkl}\Bigg[\frac{\gamma_{K}(\alpha_s)}{
24}\T_i\cdot \T_j+\sum_Rg_R(\alpha_s)\left(\frac{1}{3}\bm{\mathcal{D}}^R_{iiij}+\frac{1}{4}\bm{\mathcal{D}}^R_{i ijj}\right)\Bigg]\,,
\end{align}
into the Regge-limit basis containing $\tts$ and $\tsu$ operators. 
The logarithm of the CICRs, $\beta_{ijkl}$, is defined in eq.~\eqref{eq:beta}.
The first term in eq.~(\ref{deltalinapp}) is related to the dipole term and the second to the quartic Casimir terms. We will analyse each of these terms in turn.

\paragraph{The dipole component in the Regge limit.}
The first term in eq.~(\ref{deltalinapp}) is
\beq
\label{dipa}
\mathbf{\delta\Gamma}_{4,{\rm 2T-2L}}^{\rm{lin.}} \left(\{\beta_{ijkl}\}, \alpha_s\right)\equiv-\frac{\gamma_{K}(\alpha_s)}{
24}\sum_{(i,j,k,l) \in S_4}\beta_{ijkl}\,\T_i\cdot \T_j.
\eeq
We expand the sum and we write $\beta_{ijkl}$ in terms of logarithms of the Mandelstam invariants, using eqs.~(\ref{eq:lij}), (\ref{eq:beta}) and (\ref{eq:logrel}).  The Regge limit is taken using eq.~\eqref{eq:lijRegi}. Then we express the dipole operators, $\mathbf{T}_i\cdot\mathbf{T}_j$, in terms of the channel operators defined in eq.~(\ref{TtTsTu}), obtaining
 \begin{align}
\begin{split}
\label{deltaquadfin}
\mathbf{\delta\Gamma}_{4,{\rm 2T-2L}}^{\rm{lin.}} \left(L,  \alpha_s\right)&=\frac{\gamma_K(\alpha_s)}{6} \bigg[ L\Big(3\tts- 2 C_i - 2 C_j\Big)+3i \pi\tsu\bigg],
\end{split}
\end{align}
 where $j$ corresponds to partons 2 and 3 and $i$ corresponds to partons 1 and 4 in the target-projectile notation.
 The first bracket is signature even and real. The term with $\tsu$ is signature odd and imaginary.
 
\paragraph{The quartic component  \texorpdfstring{$\delta{\bm{\Gamma}}^{\rm {lin.}}_{4,Q}$}{dG} in the Regge limit.}

We now turn to the second term on the last line of eq.~\eqref{deltalinapp} given by 
\begin{align}
\begin{split}
 \label{eq:mansgi}
 \delta\bm{\Gamma}_{4,\rm{Q}}(\{\beta_{ijkl}\}, \alpha_s)&=-\sum_R g_{R}(\alpha_s)
\sum_{(i,j,k,l)\in S_4} \beta_{ijk l}\, \bigg(\frac{1}{3}\bm{\mathcal{D}}^R_{iiij}+\frac{1}{4}\bm{\mathcal{D}}^R_{i ijj}\bigg),
\end{split}
\end{align}
where the general expression $\bm{\mathcal{D}}^R_{ijkl}$ is given in eq.~\eqref{Dijkl}.
The function $g_R(\alpha_s)$ is the quartic component of the cusp anomalous dimension of eq.~\eqref{cuspAD}; its value at four-loop order is given in eq.~\eqref{g_R_values}. 

After expanding the sum in eq.~(\ref{eq:mansgi}),  then using eq.~(\ref{eq:dppppm}), the colour structures are expressed in terms of the operators $\bm{\mathcal{D}}^R_{ssss}$, $\bm{\mathcal{D}}^R_{tttt}$ and  $\bm{\mathcal{D}}^R_{uuuu}$ introduced in eq.~(\ref{eq:dppppp}). The logarithms are expressed in terms of $\{s,t,u\}$ with eq.~(\ref{eq:logrel}), and then separating into signature-even and odd parts, we have
   \begin{align}
  \begin{split}
 \label{eq:gRDsusigapp}
\delta \bm{\Gamma}_{\rm{Q}}(\{\underline{s}\},\lambda, \alpha_s)
 &=\sum_R\frac{g_R(\alpha_s)}{6}\Bigg[
 \Big(- i \pi +\log\frac{s}{\lambda^2}-\log\frac{-u}{\lambda^2}\Big)\Big(3\bm{\mathcal{D}}^R_{uuuu}-3\bm{\mathcal{D}}^R_{ssss}\Big)\\&\quad\mbox{}\hspace{-2cm}+
\Big(- i\pi+ \log\frac{s}{\lambda^2}+\log\frac{-u}{\lambda^2}-2\log\frac{-t}{\lambda^2}\Big)\bigg( -2\bm{\mathcal{D}}^R_{ssss}-2\bm{\mathcal{D}}^R_{uuuu}+4\bm{\mathcal{D}}^R_{tttt}\bigg)\Bigg].
  \end{split}
  \end{align}
    The first line is signature odd under $s\leftrightarrow u$ in both the colour and kinematics, and the last line is even in both. Taking the Regge limit and using eq.~\eqref{eq:lijRegi}, the expression becomes 
  \begin{align}
  \begin{split}
 \label{gammalinquartR}
 \delta\bm{\Gamma}_{\rm{Q}}\left(L,\alpha_s\right)
 &=\sum_R g_{R}(\alpha_s)\bigg\{\frac{L}{3}\bigg[-\bm{\mathcal{D}}^R_{ssss}-\bm{\mathcal{D}}^R_{uuuu}+2\bm{\mathcal{D}}^R_{tttt}\bigg]+\frac{i \pi}{2} \left(\bm{\mathcal{D}}^R_{ssss}-\bm{\mathcal{D}}^R_{uuuu}\right)\bigg\}.
  \end{split}
  \end{align}
The first bracket, which contributes at $\rm{N^3LL}$ at four loops, is real and signature even, while the second is odd and imaginary.  We can split them into even and odd terms with 
 \beq
 \delta\bm{\Gamma}_{\rm{Q}}\left(L,\alpha_s\right)=\delta\bm{\Gamma}_{\rm{Q}}^{(+)}\left(L,\alpha_s\right)+\delta\bm{\Gamma}_{\rm{Q}}^{(-)}\left(\alpha_s\right) 
 \eeq 
 where
 \beq
 \label{Gamma_Q_even}
 \delta\bm{\Gamma}_{\rm{Q}}^{(+)}\left(L,\alpha_s\right)\equiv\sum_R g_{R}(\alpha_s)\frac{L}{3}\bigg[-\bm{\mathcal{D}}^R_{ssss}-\bm{\mathcal{D}}^R_{uuuu}+2\bm{\mathcal{D}}^R_{tttt}\bigg]
 \eeq and 
 \beq
 \label{Gamma_Q_odd}
 \delta\bm{\Gamma}_{\rm{Q}}^{(-)}\left(\alpha_s\right) \equiv \frac{i \pi}{2}\sum_R g_{R}(\alpha_s) \left(\bm{\mathcal{D}}^R_{ssss}-\bm{\mathcal{D}}^R_{uuuu}\right),
 \eeq 
 which we will now consider in turn in the adjoint representation.
 \paragraph{The even quartic component in the adjoint representation in the Regge limit.}
  Let us consider eq.~(\ref{Gamma_Q_even}) in the adjoint representation:
\beq
\label{gLA}
 \delta\bm{\Gamma}_{\rm{Q,A}}^{(+)}\left(\alpha_s\right) \equiv g_{A}(\alpha_s) \frac{L}{3}\bigg[-\bm{\mathcal{D}}^A_{ssss}-\bm{\mathcal{D}}^A_{uuuu}+2\bm{\mathcal{D}}^A_{tttt}\bigg],
 \eeq 
  which we wish to express in terms of the Regge-limit basis comprised of commutators of $\tts$ and $\tsu$ and quartic Casimirs. The results for the quartic channel operators acting on the tree amplitude in eqs.~(\ref{Dsusigeven}) and~(\ref{DAt}) from section~\ref{Gsec} may be substituted into eq.~\eqref{gLA} to give
 \begin{align}
\label{gLtermAfin}
\delta\bm{\Gamma}_{\rm{Q,A}}^{(+)}\left(L,\alpha_s\right)\mtree&=\frac{L}{3} g_{A}(\alpha_s)\Bigg\{3\frac{d_{AA}}{N_A}- 2\left(\frac{d_{AR_i}}{N_{R_i}}+ \frac{d_{AR_j}}{N_{R_j}}\right)-\bigg(\frac{d_{AA}}{N_A}-\frac{C_A^4}{24}\bigg)
\\
&\quad\hspace{-1.7cm}+\frac{1}{4}\tts[(\tsu)^2,\tts]-\frac{3}{4}[\tsu,\tts]\tts\tsu-\frac{C_A}{4} \Big[\tsu,[\tsu,\tts]\Big]\Bigg\}\mtree\,,\nn
\end{align}
where we note that the last term in the first line is subleading in the large-$N_c$ limit using eq.~\eqref{eq:dAA}, while the second line is readily non-planar, being written in terms of commutators of $\tts$ and $\tsu$. $\delta\bm{\Gamma}_{\rm{Q,A}}^{(+)}$ is combined the other terms that are linear in $L$ in eq.~\eqref{gammalinA-}. The adjoint quartic contributions in eq.~\eqref{gLtermAfin} are separated there, because the first term in the curly brackets, $3\frac{d_{AA}}{N_A}$, forms part of ${\mathbf \Gamma}_t^{\rm{cusp}}$ once combined with the contribution from $\gamma_K$, meanwhile the second forms part of the non-planar terms.

\paragraph{The odd quartic component in the adjoint representation in the Regge limit.} 

Next we consider eq.~(\ref{Gamma_Q_odd}) in the adjoint representation:
\beq
\delta\bm{\Gamma}^{(-)}_{\rm{Q,A}}\left(\alpha_s\right)
 =\frac{i \pi}{2} g_{A}(\alpha_s) \Big(\bm{\mathcal{D}}^A_{ssss}-\bm{\mathcal{D}}^A_{uuuu}\Big).
\eeq
In order to express it in terms of nested commutators, we first use eq.~(\ref{eq:dppppm}) to write
\begin{align}\begin{split}
\label{eq:oddDsu}
\Big(\bm{\mathcal{D}}^A_{ssss}-\bm{\mathcal{D}}^A_{uuuu}\Big)\mtree&=\Big\{\bm{\mathcal{D}}^A_{1111}+4\bm{\mathcal{D}}^A_{1112}+6 \bm{\mathcal{D}}^A_{1122}+4\bm{\mathcal{D}}^A_{1222}+\bm{\mathcal{D}}^A_{2222}\\&-\left(\bm{\mathcal{D}}^A_{1111}+4\bm{\mathcal{D}}^A_{1113}+6 \bm{\mathcal{D}}^A_{1133}+4\bm{\mathcal{D}}^A_{1333}+\bm{\mathcal{D}}^A_{3333}\right)\Big\}\mtree.
\end{split}
\end{align}
There are three types of terms to consider.
First, identifying 
\beq
\bm{\mathcal{D}}^A_{1111}=\frac{d_{AR_i}}{N_{R_i}}\,,
\hspace{1cm}
\bm{\mathcal{D}}^A_{2222}=\bm{\mathcal{D}}^A_{3333}=\frac{d_{AR_j}}{N_{R_j}},
\eeq
we observe that these projectile and target Casimir terms cancel out in eq.~\eqref{eq:oddDsu}. 
Next we consider operators with one attachment to one line and three attachments to the opposite line, as shown in  figure~\ref{fig:h13-3}, for example
\begin{align}
\label{eq:DR1112}
\bm{\bm{\mathcal{D}}}^R_{1112}&=d^{abcd}_R\left(\T^{a,b,c} \right)_i\T^{d}_j\,,
\end{align}
with the fully symmetrised trace defined as
\beq
d^{abcd}_R\equiv\frac{1}{4!}\sum_{\sigma \in S_4} \operatorname{Tr}_R\left(T^{\sigma(a)}T^{\sigma(b)}T^{\sigma(c)}T^{\sigma(d)}\right).
\eeq
We notice the colour factor $\mathbf{C}^{(3)}_{13}$, defined in eq.~(\ref{eq:C13_3}), corresponds to a particular case of eq.~(\ref{eq:DR1112}), in which the representation $R$ is the adjoint. Therefore, we compute eq.~(\ref{eq:DR1112}) by following the steps in eqs.~(\ref{eq:Simon_trick1}) and~(\ref{eq:resC13_3}), getting
\begin{align}
d^{abcd}_R \left(\T^{a,b,c} \right)_i\T^{d}_j&= A \T_i\cdot\T_j=\frac{d_{RR_i}}{N_{R_i}C_{R_i}}  \T_i\cdot\T_j
\end{align}
where the constant $A$ is computed by taking the trace of the first equation. This yields
\beq
\bm{\mathcal{D}}^R_{1112}=\frac{1}{2} \left(\tsu - \frac{\tts}{2}\right)\frac{d_{RR_i}}{N_{R_i} C_i}.
\eeq
We conclude that the terms with three attachments to one line and a single attachment to the other line in eq.~\eqref{eq:oddDsu} take the form
\begin{align}
 \label{Dklll}
  \bm{\mathcal{D}}^R_{klll}=
    \begin{cases}
     {\displaystyle \frac{1}{2} \left(\tsu - \frac{\tts}{2}\right)\frac{d_{RR_i}}{N_{R_i} C_i}} & \text{if $(k,l)\in\{(2,1),(3,4)\}$}\vspace{0.2cm}\\
     {\displaystyle \frac{1}{2} \left(\tsu - \frac{\tts}{2}\right)\frac{d_{RR_j}}{N_{R_j} C_j}} & \text{if $(k,l)\in\{(1,2),(4,3)\}$}\vspace{0.2cm}\\     
     {\displaystyle-\frac{1}{2} \left(\tsu + \frac{\tts}{2}\right)\frac{d_{RR_i}}{N_{R_i} C_i}} & \text{if $(k,l)\in\{(3,1),(2,4)\}$}\vspace{0.2cm}\\
     {\displaystyle-\frac{1}{2} \left(\tsu + \frac{\tts}{2}\right)\frac{d_{RR_j}}{N_{R_j} C_j}} & \text{if $(k,l)\in\{(1,3),(4,2)\}$}\vspace{0.2cm}\\
     \end{cases}      
\end{align}
The third type of terms in eq.~\eqref{eq:oddDsu} involves two attachments to both the target and projectile, for example
 \begin{align}
     \begin{split}
     \bm{\mathcal{D}}^A_{1122}\,\left(\T^x_i\,\T^x_j\right) &=  d^{abcd}_A \left(\T^{a,b,x}\right)_i\left(\T^{c,d,x}\right)_j\\
      &=  \bigg\{\frac{1}{2} \left(\text{Tr} \left[F^aF^bF^cF^d+ F^dF^cF^bF^a\right]\right)-\frac{C_A}{6}\left(f^{ade}f^{bce}-f^{abe}f^{cde}\right)\bigg\}\\&\times\left(\T^{a,b,x}\right)_i\left(\T^{c,d,x}\right)_j\\
      &\equiv \left(\bm{\mathcal{D}}^{A,\text{Tr}}_{1122}+\bm{\mathcal{D}}^{A, \text{ff}}_{1122}\right)\,\left(\T^x_i\,\T^x_j\right).
        \end{split}
          \end{align}
         The colour structure relation from eq.~\eqref{tracerelA} gives the second line above. The terms are then separated  with superscripts $\rm{Tr}$ for the trace terms and $\rm{ff}$ for terms with two structure constants.
          The first term is 
          \begin{align}
\begin{split}
\bm{\mathcal{D}}^{(A, \rm{Tr})}_{1122}\,\left(\T^x_i\,\T^x_j\right)&=\frac{1}{2} \left(\text{Tr} \left[F^a F^b F^c F^d+ F^dF^cF^bF^a\right]\right)\left(\T^{a,b,x}\right)_i\left(\T^{c,d,x}\right)_j\\
&=\left(\mathbf{T}^{[a,[b,c]],d,x}\right)_i\left(\mathbf{T}^{[b,[a,d]],c,x}\right)_j.
\end{split}
\end{align}
Upon using the identities of section~\ref{subsec:fourloopHard} and appendix \ref{sec:colourappendix}, it yields
\begin{align}
\begin{split}
\label{DTr1122}
    \bm{\mathcal{D}}^{(A, \rm{Tr})}_{1122}\,\left(\T^x_i\,\T^x_j\right)&=\Big\{-\frac{1}{64}C_A^4 - \frac{1}{32} \left( \Big[\tts,\big[\tts,[\tts,\tsu]\big]\Big] - \tts[\tts,\tsu]\tts\right)\\&\quad\mbox{} + \frac{1}{4} \tsu[\tsu,\tts],\tts + \frac{1}{16} \tsu\tts\tts\tsu\Big\} \,\left(\T^x_i\,\T^x_j\right) .
\end{split}
\end{align}
Focusing on the term involving two structure constants now, we find 
    \begin{align}
        \bm{\mathcal{D}}^{A, \text{ff}}_{1122}\,\left(\T^x_i\,\T^x_j\right)&=-\frac{C_A}{6}\left(f^{ade}f^{bce}-f^{abe}f^{cde}\right)\left(\T^{a,b,x}\right)_i\left(\T^{c,d,x}\right)_j\nn\\
     &  =\frac{C_A}{6}\left(\T^{[d,e],b,x}\right)_i\left(\T^{[e,b],d,x}\right)_j-\frac{C_A^3}{48}\left(\tsu-\frac{\tts}{2}\right)\,\left(\T^x_i\,\T^x_j\right)\label{eq:dquad},
       \end{align}
       where the second term is reduced to a dipole using 
       \beq
       f^{abg}\T^a_i\T^b_i= \frac{i C_A}{2} \T_i^g
      \eeq 
      and then eq.~\eqref{eq:relTsu-Tt/2} is employed. The commutator relation eq.~\eqref{eq:LieAlgebra} is used to express the first term as commutators, then upon using eqs.~(\ref{eq:relTsu-Tt/2}) and~(\ref{eq:relTsu+Tt/2}), it becomes
       \begin{align}
       \begin{split}
       \left(\T^{[d,e],b,x}\right)_i\left(\T^{[e,b],d,x}\right)_j&=\frac{1}{24}\bigg\{(\tsu)^2 \tts + \frac{1}{2} \tsu \tts \tsu - \frac{1}{4} \tsu( \tts)^2  \\&\quad\mbox{}+ \frac{1}{2} \tts \tsu \tts - \frac{1}{4} (\tts)^2 \tsu + \frac{1}{8}(\tts)^3 \bigg\} \,\left(\T^x_i\,\T^x_j\right).
       \end{split}
       \end{align}
         Substituting into eq.~\eqref{eq:dquad}, it then reads
         \begin{align}
         \begin{split}
         \label{Dff1122}
                   \bm{\mathcal{D}}^{A, \text{ff}}_{1122}\,\left(\T^x_i\,\T^x_j\right)&= \bigg\{\frac{C_A^4}{64}-\frac{C_A^3}{32}\tsu+\frac{C_A}{48}\tsu \tts \tsu  -\frac{C_A^2}{24
                   } (\tsu)^2+ \frac{C_A^2}{48}\tts \tsu\\&\quad \mbox{} -\frac{C_A}{96}(\tts)^2 \tsu \bigg\}\,\left(\T^x_i\,\T^x_j\right).              
                   \end{split}
              \end{align}
The contributions to $\bm{\mathcal{D}}^{A}_{1133}$ can be found by simply applying crossing $\tsu \to -\tsu$ to the $\bm{\mathcal{D}}^{A}_{1122}$ terms.
Substituting the expressions from eqs.~(\ref{Dklll}), (\ref{DTr1122}) and~(\ref{Dff1122}) into eq.~\eqref{eq:oddDsu}, we have 
\begin{align}
\begin{split}
\label{Dsusigodd}
\Big(\bm{\mathcal{D}}^A_{ssss}-\bm{\mathcal{D}}^A_{uuuu}\Big)\mtree&=\bigg\{4\tsu \left(\frac{d_{AR_i}}{N_{R_i} C_i} +\frac{d_{AR_j}}{N_{R_j} C_j}\right)+ \frac{1}{8} \Big(-2 C_A^3 \tsu \\&\hspace{-2.5cm}- \Big[\tts,[\tts,\tsu]\tts\Big]\Big)-\frac{3}{8} \left( \Big[\tts,\big[\tts,[\tts,\tsu]\big]\Big] - \tts[\tts,\tsu]\tts\right)\bigg\}\mtree.
\end{split}
\end{align}
The imaginary contribution multiplying the $g_A$ is  
\begin{align}
\begin{split}
\label{eq:gpiAfin}
\delta\mathbf{\Gamma}_{\rm{Q,A}}^{(-)}\left(\alpha_s\right)\,\mtree=&\frac{i \pi}{2} g_{A}(\alpha_s)\bigg\{4 \tsu\left(\frac{d_{AR_i}}{N_{R_i} C_i} +\frac{d_{AR_j}}{N_{R_j} C_j}\right)+ \frac{1}{8} \bigg(-2C_A^3\tsu \\&\hspace{-2.5cm} -3\Big[\tts,\big[\tts, [\tts,\tsu]\big]\Big]-\Big[\tts,[\tts,\tsu]\Big]\tts+3 \tts[ \tts,\tsu]\tts \bigg)\bigg\}\,\mtree,
\end{split}
\end{align}
with $j$ being the target and $i$ the projectile. 

\subsection*{\underline{The four-generator three-line term (\texorpdfstring{$\mathbf{4T}$$-$$\mathbf{3L}$}{4T-3L}) in the 
Regge-limit basis}}\label{subsec:appendixFTerm}
The third term in eq.~\eqref{eq:SADintrogammaa} is given by eq.~(\ref{Gamman-4T3L}) and reads
 \beq
 \label{eq:gammafi}
\mathbf{\Gamma}_{n,\rm 4T-3L}(\{s_{ij}\},\lambda, \alpha_s)= f(\alpha_s) \sum_{(i,j,k)}\,\bm{{\cal T}}_{iijk} = -f(\alpha_s)\sum_{(i,j,k)} f^{abe}f^{cde}\{\T_i^a,\T_i^d\}_+\T_j^b\T_k^c,
 \eeq
 with individual colour structures being symmetric under $j\leftrightarrow k$ and the sum being symmetric under the interchange of any two of the three partons.
 For these to contribute at four-loop order they must be multiplied by either $C_A$ or $n_f T_F$ so there is some dependence on the matter content of the theory.
 
Considering $2\to 2$ scattering, we expand the sum over four distinct partonic legs and eq.~\eqref{eq:gammafi} becomes 
 \beqa 
 \label{eq:gammaf4inum}\nn
\hskip-1cm\mathbf{\Gamma}_{\rm 4T-3L}(\alpha_s
)=-  f(\alpha_s) f^{abe} f^{cde}
\Big\{&&\!\! \Big(\T_1^a \T_1^d + \T_1^d \T_1^a \Big) 
    \Big(\T_2^b \T_3^c + \T_2^b \T_4^c + \T_3^b \T_4^c \Big)   \\ \nn
&&+\, \Big(\T_2^a \T_2^d + \T_2^d \T_2^a \Big) 
    \Big(\T_1^b \T_3^c + \T_1^b \T_4^c + \T_3^b \T_4^c \Big)   \\ 
&&+\, \Big(\T_3^a \T_3^d + \T_3^d \T_3^a \Big) 
    \Big(\T_1^b \T_2^c + \T_1^b \T_4^c + \T_2^b \T_4^c \Big)   \\ \nn
&&+\, \Big(\T_4^a \T_4^d + \T_4^d \T_4^a \Big) 
    \Big(\T_1^b \T_2^c + \T_1^b \T_3^c + \T_2^b \T_3^c \Big) \Big\}\,,
\eeqa
where we drop the number of legs ($4$) in the subscript. 
By repeatedly using the commutation relation in eq.~\eqref{eq:LieAlgebra}, it is possible to write eq.~\eqref{eq:gammafi} in terms of the product
of three generators associated to the projectile and three generators associated to the target.

As an example of following these rules, the first line of eq.~\eqref{eq:gammaf4inum} is
\beqa\label{Color1} 
f^{abe} f^{cde}
\Big(\T_1^a \T_1^d + \T_1^d \T_1^a \Big) 
\Big(\T_2^b  \T_3^c + \T_2^b \T_4^c + \T_3^b \T_4^c \Big) &&  \\ \nn
&&\hspace{-8.0cm} =\, - 
\Big(\T_1^a [\T_1^e, \T_1^c] [\T_2^e, \T_2^a]  \T_3^c 
+ \T_1^a \T_1^d \Big[ [\T_2^c, \T_2^d ]  , \T_2^a\Big]  \T_4^c 
\\ \nn
&&\hspace{-7.0cm} + \T_1^a \T_1^d \Big[ [\T_3^c, \T_3^d], \T_3^a \Big]  \T_4^c  
+ [\T_1^e, \T_1^c] \T_1^a [\T_2^e, \T_2^a]  \T_3^c 
\\ \nn
&&\hspace{-7.0cm} +\T_1^d \T_1^a \Big[ [\T_2^c, \T_2^d], \T_2^a\Big]  \T_4^c 
+ \T_1^d \T_1^a \Big[ [\T_3^c, \T_3^d], \T_3^a \Big]  \T_4^c \Big).
\eeqa
Similar expressions can be found for the remaining three lines of eq.~\eqref{eq:gammaf4inum}.
Using these rules we write eq.~\eqref{eq:gammafi} as
\beqa
\label{eq:gammaf4sum}\nn
\mathbf{\Gamma}_{\rm 4T-3L}(\alpha_s)&=& -f(\alpha_s)\bigg( \sum_{(i,j,k)\in i_f}\T_i^a [\T_i^e, \T_i^c] [\T_j^e, \T_j^a]  \T_k^c +[\T_i^e, \T_i^c] \T_i^a [\T_j^e, \T_j^a]\T_k^c \\&+&\sum_{(i,j,k)\in j_f} \T_i^a\T_i^d \Big[[\T_j^c, \T_j^d], \T_j^a\Big]  \T_k^c + \Big[[\T_j\cdot \T_k,\T_i\cdot \T_j], \T_i\cdot\T_j\Big]\bigg) \hspace{1cm}
\eeqa
where \[i_f=\{(1,2,3),(2,1,4),(3,1,4),(4,2,3)\},\] \[j_f=\{(2,1,3),(3,1,2),(1,2,4),(4,1,2),(1,3,4),(4,1,3),(2,3,4),(3,2,4)\}.\] Then when $\mathbf{\Gamma}_{4,\rm 4T-3L}(\alpha_s)$ acts on the tree-level amplitude we can rewrite the expression as chains of attachments on the target and projectile. We then use the identities of eqs.~(\ref{eq:relTsuPMTt/2}) and~(\ref{eq:Simon_trick}), to express eq.~\eqref{eq:gammafi} as
\begin{align}
\label{eq:falphfin}
\bm{\Gamma}_{\rm 4T-3L}(\alpha_s)\mtree=
f(\alpha_s)\bigg(\Big[\tsu,[\tsu,\tts]\Big] + \frac{C_A^3}{2} - 6\frac{d_{AR_i}}{N_{R_i} C_i} -6\frac{d_{AR_j}}{N_{R_j} C_j} \bigg)\mtree,
\end{align}
where, as usual, $j$ is the target and $i$ the projectile. We see that $\bm{\Gamma}_{\rm 4T-3L}(\alpha_s)$ is given by even colour structures and that it is non-planar owing to the property of commutators and eq.~(\ref{eq:dAAlargeNc}).
  
\subsection*{\underline{The five-generator four-line term (\texorpdfstring{$\mathbf{5T}$$-$$\mathbf{4L}$}{5T-4L}) in the Regge-limit basis}}\label{subsec:appendixHTerm}
The last term in eq.~\eqref{eq:SADintrogammaa} is given by eq.~(\ref{eq:H1i_mainText}) which we quote again here 
\begin{align}
\label{eq:H1i}
\begin{split}
\mathbf{\Gamma}_{\rm 5T-4L}\left(\{\beta_{ijkl}\},\alpha_s\right)
&=\sum_{(i,j,k,l)} \, \bm{\mathcal{T}}_{ijkli}\,
{\cal H}_1(\beta_{ijlk},\beta_{iklj};\alpha_s)\\
&=\sum_{(i,j,k,l)} if^{adg}f^{bch} f^{egh} \, 
\{\T_i^a,\T_i^e\}_+\T_j^b\T_k^c\T_l^d\, 
{\cal H}_1(\beta_{ijlk},\beta_{iklj};\alpha_s).
\end{split}
\end{align}
We specialise to $2\to 2$ scattering and our aim is to eq.~(\ref{eq:H1i}) in terms of expressions with definite symmetry under $s\leftrightarrow u$.
After introducing the functions ${\cal H}_1^{(+)}$, ${\cal H}_1^{(-)}$ and $\tilde{{\cal H}}_1^{(-)}$ in eq.~(\ref{Hplusminuesdef}), we expand the sum in eq.~\eqref{eq:H1i} and then use the properties of $\beta_{ijkl}$ in eq.~\eqref{eq:betasym} and the antisymmetry of $\bm{\mathcal{T}}_{ijkli}$ and ${\cal H}_1$, to obtain
\begin{align}
\begin{split}
\label{eq:H1ib}
\mathbf{\Gamma}_{\rm 5T-4L}^{(4)}(\{\beta_{ijkl}\})
=&\, 2 {\cal H}_1^{(+,4)}(\{\beta_{ijkl}\})\Big\{\bm{\mathcal{T}}_{13421}
+\bm{\mathcal{T}}_{24312}+\bm{\mathcal{T}}_{31243}+\bm{\mathcal{T}}_{42134}\\
&\hspace{3cm}+\big(\bm{\mathcal{T}}_{12431}+\bm{\mathcal{T}}_{21342}
+\bm{\mathcal{T}}_{34213}+\bm{\mathcal{T}}_{43124}\big)\Big\}\\
&+ 2{\cal H}_1^{(-,4)}(\{\beta_{ijkl}\})\Big\{\bm{\mathcal{T}}_{13421}
+\bm{\mathcal{T}}_{24312}+\bm{\mathcal{T}}_{31243}+\bm{\mathcal{T}}_{42134}\\
&\hspace{3cm}-\left(\bm{\mathcal{T}}_{12431}+\bm{\mathcal{T}}_{21342}
+\bm{\mathcal{T}}_{34213}+\bm{\mathcal{T}}_{43124}\right)\Big\}\\
&+ 2\tilde{{\cal H}}_1^{(-,4)}(\{\beta_{ijkl}\})\Big\{\bm{\mathcal{T}}_{12341}
+\bm{\mathcal{T}}_{21432}+\bm{\mathcal{T}}_{34123}+\bm{\mathcal{T}}_{43214}\Big\}.
\end{split}
\end{align}
The colour coefficient of ${\cal H}_1^{(+)}$ is even under the exchange of $s\leftrightarrow u$, while the coefficients of  both ${\cal H}_1^{(-)},\tilde{{\cal H}}_1^{(-)}$ are odd. We decompose the $\bm{\mathcal{T}}_{ijkli}$ into two terms as
\beq
\label{eq:fcol2i}
\bm{\mathcal{T}}_{ijkli}=if^{adg}f^{bch}f^{egh}\T_i^a\T_i^e\T_j^b  \T_k^c\T_l^d-\frac{C_A}{4}f^{ade}f^{bce}\T_i^a\T_j^b  \T_k^c\T_l^d.
\eeq
By repeatedly expressing the structure constants in eq.~(\ref{eq:fcol2i}) as commutators according to eq.~\eqref{eq:LieAlgebra}, it is possible to write an expression with three structure constants and five generators, in terms of four generators on the projectile line $i$ (corresponding to partons $1$ and $4$), and four generators on the target line $j$ (corresponding to partons $2$ and $3$) acting on the tree-level amplitude. For the four indices  $(i,j,k,l)$, there are two independent cases to consider. First,  we fix $i$ and $k$ on one line  and then require $j$ and $l$ to attach to the opposite line with the following expression:
\begin{align}
\label{eq:fffTTTTTac2i}
\begin{split}i f^{adg}f^{bch}f^{egh}\T_i^a\T_i^e\T_j^b  \T_k^c\T_l^d&= [\T_i^g, \T_i^{d}]\T_i^e\Big[\T_j^c, [\T_j^e, \T_j^{g}] \Big] \T_k^c\T_l^d.
\end{split}
\end{align} 
The second case when rewriting the first term in eq.~\eqref{eq:fcol2i} is to fix $i$ and $l$ on the same line and $j$ and $k$ on the opposite line. This time we apply the Jacobi identity \beq
\label{eq:jacobirel}
f^{ace}f^{bde}=f^{abe}f^{cde}+ f^{ade}f^{bce},
\eeq
in such a way to avoid operators with dotted colour generators attached with both ends on the same line which do not allow for a straightforward translation to the Regge-limit basis. This ensures that there are four generators attached to the upper and lower lines with the form
\begin{align}
\label{eq:fffTTTTTac2ii}
if^{adg}f^{bch}f^{egh}\T_i^a\T_i^e\T_j^b  \T_k^c\T_l^d=&\, \T_i^a[\T_i^c, \T_i^{h} \,]\Big[\T_j^{h}, [\T_j^a, \T_j^{d}] \Big] \T_k^c\T_l^d\nn\\&+\T_i^a[\T_i^{h}, \T_i^{b}]\T_j^b\Big[\T_k^{h}, [\T_k^a, \T_k^{d}] \Big] \T_l^d.
\end{align}
Using the above relations we can write eq.~\eqref{eq:H1ib} as a sum over the lists
\[l_1=\{(1,3,4,2),(2,4,3,1),(3,1,2,4),(4,2,1,3)\}\]
\[l_2=\{(1,2,3,4),(2,1,4,3),(3,4,1,2),(4,3,2,1)\}\]
to give
\begin{align}
 &\mathbf{\Gamma}_{\rm 5T-4L}^{(4)}(\{\beta_{ijkl}\})\mtree=2\Bigg\{{\cal H}_1^{(-,4)}(\{\beta_{ijkl}\} )
\bigg[C_A f^{ade}f^{bce}\T_1^a\T_2^b\T_3^c\T_4^d
\\
&\hspace*{10pt}
+\!\!\sum_{(i,j,k,l)\in l_1}\bigg([\T_i^d, \T_i^{g}]\T_i^e\Big[\T_j^c, [\T_j^g, \T_j^{e}] \Big] \T_k^c\T_l^d-
[\T_i^d, \T_i^{g}]\T_i^e\Big[\T_l^c, [\T_l^g, \T_l^{e}] \Big] \T_k^c\T_j^d\bigg)\bigg]\nn\\
&\hspace*{10pt}
+{\cal H}_1^{(+,4)}(\{\beta_{ijkl}\} )\bigg[-C_A \left(f^{ace}f^{bde}+f^{abe}f^{cde}\right)\T_1^a\T_2^b\T_3^c\T_4^d \nn\\
&\hspace*{10pt}
+\!\! \sum_{(i,j,k,l)\in l_1}\bigg([\T_i^d, \T_i^{g}]\T_i^e\Big[\T_j^c, [\T_j^g, \T_j^{e}] \Big] \T_k^c\T_l^d+
[\T_i^d, \T_i^{g}]\T_i^e\Big[\T_l^c, [\T_l^g, \T_l^{e}] \Big] \T_k^c\T_j^d\bigg)\bigg]\nn \\
&\hspace*{10pt}
+\tilde{{\cal H}}_1^{(-,4)}(\{\beta_{ijkl}\} )
\bigg[-C_A f^{ade}f^{bce}\T_1^a\T_2^b\T_3^c\T_4^d\nn\\
&\hspace*{10pt}
+\!\!\sum_{(i,j,k,l) \in l_2}\bigg(\T_i^a[\T_i^c, \T_i^{g} \,]\Big[\T_j^{g}, [\T_j^a, \T_j^{d}] \Big] \T_k^c\T_l^d+\T_i^a[\T_i^{g}, \T_i^{b}]\T_j^b\Big[\T_k^{g}, [\T_k^a, \T_k^{d}] \Big] \T_l^d\bigg)\bigg]\Bigg\}\mtree.\nn
\end{align}
The terms without commutators can be written in the Regge-limit basis using eqs.~(\ref{eq:4Ttsubi})-(\ref{eq:4Ttsudi}). The commutators can be expanded and divided into two types of terms. Firstly, those that are just products of four dipoles can immediately be translated into the Regge limit basis with eq.~\eqref{TtTsTu}.  The second type is those that appear in the entangled combinations of $d_1,\dots d_8$ in appendix  \ref{sec:colourappendix}. They are then written in a similar way to eq.~\eqref{eq:Ttrans} to use the identities in section~\ref{subsec:twoloopHard} and appendix \ref{sec:colourappendix}. Putting everything together we can write  eq.~\eqref{eq:H1ib} as
\begin{align}\begin{split}
\label{eq:H1stu}
\mathbf{\Gamma}_{\rm 5T-4L}^{(4)}(\{\beta_{ijkl}\})\mtree&=\Bigg[ {\cal H}_1^{(+,4)}(\{\beta_{ijkl}\})\bigg(\frac{3C_A}{4} \tsu [\tsu,\tts]-\frac{C_A}{2}\Big[\tsu,[\tsu,\tts]\Big]\\&\hspace{-3.5cm} -\frac{1}{4}\tts[\tsu,\tts]\tsu\bigg)+\frac{1}{4}\tilde{{\cal H}}_1^{(-,4)}(\{\beta_{ijkl}\})\Big[\tts,\big[\tts,[\tts,\tsu]\big]\Big]\\&\hspace{-3.5cm}
+{\cal H}_1^{(-,4)}(\{\beta_{ijkl}\})\bigg(-\frac{1}{2}\Big[\tsu,\big[\tsu,[\tsu,\tts]\big]\Big]+\frac{1}{8}\Big[\tts,\big[\tts,[\tts,\tsu]\big]\Big]\bigg)\Bigg]\mtree.\end{split}
\end{align}
Finally, using eq.~(\ref{eq:zerorel}) we can eliminate one of the even-signature commutator terms in eq.~\eqref{eq:H1stu} by substituting
\beq
\tts\big[\tsu,\tts\big]\tsu\mtree=\bigg(\frac{2}{3}\tts\left[\left(\tsu\right)^2,\tts\right]-\tsu\big[\tsu,\tts\big]\tts\bigg)\mtree,
\eeq
where the two commutators on the right-hand side have already been used in eq.~(\ref{eq:M42simb}) to express the four-loop reduced amplitude and are part of our basis.
This gives
\begin{align}\begin{split}
\label{eq:H1fina}
\mathbf{\Gamma}_{\rm 5T-4L}^{(4)}(\{\beta_{ijkl}\})\mtree&=\Bigg[ {\cal H}_1^{(+,4)}(\{\beta_{ijkl}\})\bigg(-\frac{C_A}{2}\Big[\tsu,[\tsu,\tts]\Big]+C_A \tsu [\tsu,\tts]\\&\hspace{-3.5cm} -\frac{1}{6}\tts[(\tsu)^2,\tts]\bigg)+\frac{1}{4}\tilde{{\cal H}}_1^{(-,4)}(\{\beta_{ijkl}\})\Big[\tts,\big[\tts,[\tts,\tsu]\big]\Big]\\&\hspace{-3.5cm}
+{\cal H}_1^{(-,4)}(\{\beta_{ijkl}\})\bigg(-\frac{1}{2}\Big[\tsu,\big[\tsu,[\tsu,\tts]\big]\Big]+\frac{1}{8}\Big[\tts,\big[\tts,[\tts,\tsu]\big]\Big]\bigg)\Bigg]\mtree,\end{split}
\end{align}
which is eq.~(\ref{eq:H1fin_main_text}).

\bibliographystyle{JHEP}
\bibliography{ReggeRefs.bib}

\end{document}